\documentclass[12pt]{article}
\makeatletter
\def\ps@pprintTitle{%
	\let\@oddhead\@empty
	\let\@evenhead\@empty
	\def\@oddfoot{\centerline{\thepage}}%
	\let\@evenfoot\@oddfoot}
\makeatother

\usepackage[utf8]{inputenc}
\usepackage{graphicx}
\usepackage{xcolor}
\usepackage{booktabs}
\usepackage{amsmath,amsthm,amssymb,amsfonts,amscd,amsbsy,amsxtra}
\usepackage{wasysym}
\usepackage{ae} 
\usepackage{pgf}
\usepackage{subcaption}
\usepackage[normalem]{ulem}
\usepackage[pagewise]{lineno}
\usepackage{siunitx}
\usepackage{bbm}
\usepackage{tabularx}
\usepackage{longtable}
\usepackage[longtable]{multirow}
\usepackage{bm}
\usepackage{titling}
\usepackage[norule]{footmisc}
\usepackage{chemformula}
\usepackage{enumerate}

\usepackage[doi=false, isbn=false, url=false, citestyle=numeric-comp, backend=biber, style=ieee]{biblatex}
\RequirePackage[colorlinks,citecolor=blue,urlcolor=blue]{hyperref}
\usepackage[left=1.5cm,right=1.5cm,top=1cm,bottom=1cm,includeheadfoot]{geometry}

\newcommand{\R}{\mathbb{R}}

\newcommand{\myquote}[1]{``#1''}
\DeclareMathOperator*{\argmin}{argmin}

\newcommand\blfootnote[1]{%
  \begingroup
  \renewcommand\thefootnote{}\footnote{#1}%
  \addtocounter{footnote}{-1}%
  \endgroup
}

\begin{document}
\setlength{\tabcolsep}{0.05em} 
\def\and{
  \end{tabular}%
  \hskip 0em
  \begin{tabular}[t]{c}}
\makeatother

\setlength\parindent{0pt}
\setlength{\droptitle}{-2cm}

\providecommand{\keywords}[1]{{\textbf{\textit{Keywords --}} #1}}

\title{\Large{Quantitative comparison of different approaches for reconstructing the carbon-binder domain from tomographic image data of cathodes in lithium-ion batteries and its influence on electrochemical properties}}

\author{\large Benedikt~Prifling\textsuperscript{1,$\ast$}, \and Matthias~Neumann$^{1}$, \and Simon~Hein$^{2,3}$, \and Timo~Danner$^{2,3}$, \and Emanuel~Heider$^{4}$, \and Alice~Hoffmann$^{4}$, \and Philipp~Rieder$^{1}$, \and Andr\'{e}~Hilger$^{5}$, \and Markus~Osenberg$^{6}$, \and Ingo~Manke$^{5}$, \and Margret~Wohlfahrt-Mehrens$^{4}$, \and Arnulf~Latz$^{2,3,7}$, \and Volker~Schmidt$^{1}$}

\date{}
\maketitle

\begin{center}
    \vspace{-2em}
	\it \normalsize{
	$^{1}$ Institute of Stochastics, Ulm University, 89081 Ulm, Germany\\
	$^{2}$ German Aerospace Center (DLR), Institute of Engineering Thermodynamics,705696 Stuttgart, Germany\\
	$^{3}$ Helmholtz Institute for Electrochemical Energy Storage (HIU), 89081 Ulm, Germany\\
	$^{4}$ ZSW-Zentrum f\"{u}r Sonnenenergie- und Wasserstoff-Forschung Baden-W\"{u}rttemberg, 89081 Ulm, Germany\\
	$^{5}$ Institute of Applied Materials, Helmholtz-Zentrum Berlin f\"{u}r Materialien und Energie, 14109 Berlin, Germany\\
	$^{6}$ Department of Materials Science and Technology, TU Berlin, 10623 Berlin, Germany\\
	$^{7}$ Institute of Electrochemistry, Ulm University, 89081 Ulm, Germany\\}
\end{center}

\begin{abstract}
\noindent It is well known that the spatial distribution of the carbon-binder domain (CBD) offers a large potential to further optimize lithium-ion batteries. However, it is challenging to reconstruct the CBD from tomographic image data obtained by synchrotron tomography. In the present paper,  we consider several approaches to segment 3D image data of two different cathodes into three phases, namely active material, CBD and pores. More precisely, we focus on global thresholding, a local closing approach based on EDX data, a k-means clustering method, and a procedure based on a neural network that has been trained by correlative microscopy, i.e., based on data gained by synchrotron tomography and FIB-SEM data representing the same electrode. We quantify the impact of the considered segmentation approaches on morphological characteristics as well as on the resulting performance by spatially-resolved transport simulations. Furthermore, we use experimentally determined electrochemical properties to identify an appropriate range for the effective transport parameter of the CBD. The developed methodology is applied to two differently manufactured cathodes, namely an ultra-thick unstructured cathode and a two-layer cathode with varying CBD content in both layers. This comparison elucidates the impact of a specific structuring concept on the 3D microstructure of cathodes.
\end{abstract}


\blfootnote{$^\ast$\ Corresponding author, \textit{Email addresses:} benedikt.prifling@uni-ulm.de}

\setlength{\tabcolsep}{0.4em} 

\section{Introduction}\label{sec:introduction}

Because of their outstanding energy density, low self-discharge rate and high power density, lithium-ion batteries are the most widely used technology for storing electrical energy  \cite{scrosati.2013, korthauer.2018, goodenough.2013, newman.2004}. However, further optimization of the performance is necessary due to the continuously growing requirements for electric vehicles and a general need for reducing carbon dioxide emissions to mitigate global warming \cite{tran.2012,cano.2018}. Since it is well known that the 3D microstructure of battery electrodes strongly influences the resulting electrochemical performance \cite{cho.2015, huang.2009, lenze.2018, li.1997, shin.2004, wiedemann.2013}, tailoring the morphology of the 3D microstructure by specifically developed structuring concepts seems to be a promising approach. Obviously, the manufacturing process consisting among others of mixing \cite{kremer.2020, bockholt.2013}, drying \cite{huttner.2020, kumberg.2021} and calendering \cite{zheng.2012, haselrieder.2013, kuchler.2018b} has a significant impact on the electrode morphology \cite{bockholt.2016}. Although the carbon-binder domain (CBD) is regarded as passive constituent of the electrode morphology its spatial distribution is particularly crucial for the resulting electrochemical properties of cathodes \cite{kremer.2020, kroll.2021, hutzenlaub.2013, dominko.2003} and anodes \cite{morasch.2018, landesfeind.2018}. Thus, the segmentation of tomographic image data into three phases, namely active material, CBD and pores, is necessary to adequately describe the 3D microstructure of battery electrodes. On the one hand, a high resolution of 3D image data up to the nanometer scale, which can be achieved by FIB-SEM tomography, enables for the application of segmentation techniques, which distinguish between these three phases. Disadvantageously, FIB-SEM tomography provides only a small field of view such that the resulting 3D image of the electrode is often not sufficiently representative. On the other hand, $X$-ray based imaging techniques such as synchrotron tomography allow for a non-destructive measurement of a comparatively large cutout of the electrode. The technique has been applied successfully for the analysis of a wide range of electrode materials including transition metal oxides ~\cite{cadiou.2020, wagner.2020, xu.2021},  lithium-iron phosphates~\cite{cooper.2014} as well as organic active materials~\cite{neumann.2022}. However, the contrast between CBD and pores is comparatively low in many cases such that a a frequently used approach is to segment only the active material and its complement, see \cite{cooper.2014, ebner.2013a, ebner.2013b, shearing.2010, yan.2012}. Several studies then use modeling approaches for inserting the CBD in a subsequent step, see \cite{kuchler.2018b, zielke.2014, trembacki.2017, hein.2020}.\\

In the present paper, we consider four conceptually different data-driven approaches to reconstruct the microstructure of two differently manufactured cathodes using tomographic image data. While in~\cite{hein.2020}, the CBD is virtually included based on different geometric models for a given segmentation of active material, the novelty of the present paper consists of the quantitative comparison between data-driven three-phase reconstructions. These segmentation approaches include global thresholding, $k$-means clustering, machine learning trained by correlative microscopy, and a reconstruction based on EDX data. This comparison elucidates the impact of different segmentation approaches on morphological and electrochemical properties of the resulting electrode microstructures. Moreover, we determine the effective transport parameter of the CBD for each segmentation approach by validating the output of spatially-resolved half-cell simulations with experimentally determined electrochemical data. This approach allows us to specify a range in which the effective transport parameter is located. Thereby, the presented approach takes the important aspect of uncertainty during the reconstruction process~\cite{krygier.2021} into account when analyzing the microstructure of battery electrodes based on 3D image data. \\

This paper is organized as follows. In Section~\ref{sec:experimental}, we describe the manufacturing process of two different cathodes as well as the tomographic imaging procedure. Next, we present four different approaches of segmenting active material, CBD and pores from 3D image data in Section~\ref{sec:segmentation}. The computation of electrochemical properties by spatially-resolved numerical simulations is described in Section~\ref{sec:electrochemistry}. In Section~\ref{sec:results}, the influence of the different trinarization approaches on the 3D microstructure is quantitatively investigated by means of statistical image analysis. In addition, we present results regarding simulated electrochemical properties, where a particular focus is put on the effective transport parameter of the carbon-binder domain, which is fitted via experimentally determined lithiation curves. Finally, the paper is concluded with a summary of the main results and an outlook to possible future research.

\section{Experimental}\label{sec:experimental}

In this section, manufacturing, material composition as well as the tomographic imaging of the cathode materials considered in the present paper are described.

\subsection{Materials and cathode manufacturing}\label{subsec:manufacturing}

We investigate two different cathode samples, the 3D microstructure of which is quantitatively characterized based on different segmentation approaches. Moreover, an additional electrode is considered, which is solely used for the trinarization approach based on correlative microscopy in Section~\ref{sec:segmentation}. In the following, we describe four different suspensions, denoted by A, B, C and D, which were used to manufacture these samples. Note that one of the electrodes is a two-layer electrode, where the two layers are prepared with different suspensions. All suspensions share the underlying materials, but differ with regard to their composition.\\

Commercially available  $\mathrm{Li}\mathrm{Ni}_{0.6}\mathrm{Co}_{0.2} \mathrm{Mn}_{0.2}\mathrm{O}_{2}$ (BASF), shortly denoted by NMC, was mixed and dispersed with carbon black (SuperP, Imerys) and graphite (SFG6L, Imerys) as conducting additive and polyvinylidene fluoride (PVdF, Solvay Solexis) as a binder, where the union of carbon black, graphite and binder forms the carbon-binder domain (CBD). N-methyl-2-pyrrolidone (NMP, Sigma Aldrich) was used as a solvent. Note that all materials were utilized as delivered without further treatment. Because two suspensions were needed simultaneously for the  manufacturing of the two-layer electrode, different mixers applying the same working principle were used for the preparation of the cathode suspensions. To be precise, a $\SI{10}{\dm^{3}}$ planetary mixer (Netzsch, Germany) and a $\SI{1.6}{\dm^{3}}$ planetary mixer (Grieser, Germany) were used. Both mixers were equipped with two agitators, a cross-bar stirrer (CS) and a butterfly stirrer (BS) running at low and high speed, respectively. In the case of the $\SI{10}{\dm^{3}}$ mixer, an axially double butterfly stirrer was used while the $\SI{1.6}{\dm^{3}}$ mixer contained a single butterfly stirrer. Transport of the components into the mixing zone was ensured by a wall scraper rotating at slow speed. For each suspension, the solid material composition as well as the type of mixer used for the preparation are given in Table~\ref{tab:manufacturing}. \\

\begin{table}[ht]
    \centering
    \begin{tabular}{lcccc}
    \toprule
        Suspension & A & B & C & D\\ \midrule
        Content in solid mass $[w\%]$ & & & \\
        NMC & 93.5 & 91.5 & 95.5 & 93.0\\
        carbon black & 2.0 & 2.0 & 2.0 & 2.0\\
        graphite & 1.0 & 1.0 & 1.0 & 1.0\\
        PVdF & 3.5 & 5.5 & 1.5 & 4.0 \\ \midrule
        Mixer & Grieser & Netzsch & Grieser & Netzsch \\
        \bottomrule
    \end{tabular}
    \caption{Material compositions and mixers for the four different suspensions, which are used to manufacture the cathode samples considered in the present paper.}
    \label{tab:manufacturing}
\end{table}

The suspensions were prepared starting from a binder solution containing 7 to 10 $w\%$ of PVdF which was dissolved in NMP at room temperature. First, carbon black and then graphite was added to the binder solution and dispersed, respectively. After that, NMC was added stepwise and dispersed after each addition. Finally, the viscosity of each suspension was adjusted for application by thinning with NMP. From the suspensions, ultra-thick electrodes were produced using a pilot line coating machine (LACOM, Germany). A single-layer electrode (abbreviated by SL) was prepared from the suspension A using a single slot die. A two-layer electrode (abbreviated by TL) was prepared by simultaneous slot die coating with a double slot die applying suspension B at the bottom and suspension C at the top. The suspensions were cast onto an aluminum foil (Korff, Switzerland). A drying oven with a total length of $\SI{8}{\meter}$, separated into four drying stages, independently adjustable in temperature, was used for evaporation of the solvent. The belt speed was $\SI{0.8}{\meter\per\minute}$ and the temperatures of the ovens were $\SI{50}{\celsius}$, $\SI{70}{\celsius}$, $\SI{95}{\celsius}$ and $\SI{110}{\celsius}$ for both electrodes. The mass loading resulted in $\SI{51}{\milli\gram\per\cm^{2}}$ and $\SI{53}{\milli\gram\per\cm^{2}}$ for the single-layer and the two-layer electrode, respectively. After drying, the electrodes were calendered using a pilot line calender (KKA, Germany) with a line pressure restricted to a maximum value of $\SI{208.3}{\pascal \per\meter}$ and rolls heated to $\SI{100}{\celsius}$. The final density of the electrode composites was $\SI{3.1}{\gram\per\cubic\centi\metre}$ for both electrodes. Note that the single-layer and the two-layer cathode share the following volume fractions: 59.54\% active material, 11.54\% CBD and 28.92\% pore space.\\

In addition, a third cathode sample is considered, image data of which is solely used in Section~\ref{sec:segmentation} for establishing a trinarization approach based on correlative microscopy. This sample is manufactured with suspension D analogously to the single-layer and the two-layer cathode, except for a slower belt speed of $\SI{0.6}{\meter\per\minute}$ and a slightly lower mass loading of $\SI{49.1}{\milli\gram\per\centi\metre^{2}}$.

\subsection{Tomographic imaging}\label{subsec:imaging}

First, we describe the imaging procedure of the single-layer as well as the two-layer cathode. The tomography measurements of these cathode samples have been conducted at the P05 beamline (Petra III, DESY, Germany) \cite{khokhriakov, wilde.2016}. More precisely, a monochromatic nearly parallel X-ray beam is guided on the rotating sample without the use of X-ray focusing optics. Behind the sample, the transmitting beam is detected with a setup consisting of a $\mathsf{CdWO}_{4}$ scintillator for X-ray to light transformation, an optical microscope and a CMOS camera. 
The samples have been measured with an energy of $\SI{28}{\kilo\eV}$ to assure an optimal image contrast, where a double crystal monochromator is used for selection. Both samples have been measured as close as possible to the scintillator screen to reduce phase contrast. During the tomography each sample was constantly rotated while 2401 images have been captured using a KIT CMOS camera (5120 $\times$ 3840 pixel) with an exposure time of $\SI{130}{\milli\second}$. Combined with the 10 times optics this resulted in a voxel size of $\SI{0.642}{\micro\meter}$. For the reconstruction the normalized data was denoised using a total variation minimization filter \cite{rudin.1992} and then reconstructed using the gridrec routine  based on the filtered back projection \cite{dowd.1999}. Note that all subsequent results regarding the single-layer as well as the two-layer sample are based on three non-overlapping equally size cutouts, where the entire thickness is used in through-plane direction.  \\

With regard to the third cathode sample, which is used for establishing the neural network approach based on correlative microscopy, imaging by synchrotron tomography as well as by FIB-SEM tomography has been carried out. First, synchrotron tomography has been conducted at the P05 beamline (Petra III, DESY, Germany) using the $\mu$-CT setup. For the tomography, a beam energy of $\SI{25}{\kilo\eV}$ was found to yield optimal transmission contrast. The energy was filtered using a double multilayer monochromator. The sample that was fixated on the translation/rotation stage was positioned $\SI{15}{\milli\metre}$ away from the $\mathsf{CdWO}_{4}$ scintillator. Behind the scintillator the portion of the signal that has been transformed into visible light was magnified (10 times magnification) by the microscope optics and redirected into the camera system. A KIT CMOS camera equipped with a CMOSIS CMV 20000 sensor (5120 $\times$ 3840 pixel) was then used to capture the signal with an exposure time of $\SI{130}{\milli \second}$. The whole tomography consisted of 3000 projections, for ring artefact reduction a center of rotation variation protocol was used. The whole setup yielded a $\SI{0.642}{\micro\meter}$ raw pixel size. The synchrotron tomography was reconstructed using the P05 in-house reconstruction tools based on the filtered back projection algorithm. After reconstruction an additional non-local means denoising step was performed \cite{buades.2005, coupe.2008}.\\

The FIB-SEM tomography has been conducted at Helmholtz-Zentrum Berlin (HZB) using the ZEISS Crossbeam 340. For this purpose, the sample that previously was measured at P05 has been fixated on an aluminum sample holder. For better orientation on the sample, a first low resolution large scale surface scan was performed. The scan was then aligned with the 3D synchrotron tomography reconstruction using the SIFT algorithm \cite{lowe.2004}. Afterwards, using the synchrotron tomography, a suitable ROI has been selected for FIB-SEM tomography. For the FIB-SEM tomography, a Gallium ion milling source with $\SI{30}{\kilo\eV}$ and $\SI{300}{\pico\ampere}$ ion current was used. The Gemini electron gun was operated at $\SI{2}{\kilo\eV}$. For imaging the SE2 chamber detector with an image capture rate of 30 seconds per image was used. The pixel size was set to $\SI{10}{\nano\metre}$. Finally, the 3D image data obtained by FIB-SEM tomography was manually aligned with the synchrotron tomography data set using Fiji/ImageJ \cite{schindelin.2012}. \\

In addition, 2D EDX data has been gathered for the local closing approach described in Section~\ref{subsec:edx}. For this purpose, cross sections of electrodes were prepared perpendicular to the electrode surface by a broad $\mathsf{Ar}^{+}$ ion beam milling device (Hitachi IM4000Plus) at an accelerating voltage of $\SI{5}{\kilo\volt}$ for 2-3 hours depending on the electrode thickness. A subsequent analysis of the electrode microstructure was conducted with scanning electron microscopy (SEM) using a LEO1530VP (Zeiss) equipped with a thermal field emission gun. To determine the locally resolved elemental distribution of fluorine, energy dispersive EDX spectroscopy (X-Max50, Aztec Advanced Software, Oxford Instruments) was used. Characteristic X-rays of fluorine were used as a measure for the spatial distribution of PVdF within the electrode.

\section{Phase-based segmentation}\label{sec:segmentation}

This section covers four different approaches to reconstruct the 3D image data obtained by synchrotron tomography. Each of these trinarization methods is designed in such a way that the experimentally determined volume fractions of all three phases can be matched. However, it is not possible to resolve the inner structure of the CBD based on synchrotron image data since the resolution is too low. Thus, we assume that each CBD voxel contains an inner porosity, sometimes called nano-porosity, of 50\%, which is close to inner porosities of 47\% and 58\% reported in \cite{zielke.2015} and \cite{vierrath.2015}, respectively. Finally, a voxel-based analysis is carried out to obtain a first impression about potential differences between the four segmentation approaches.   

\subsection{Global thresholding}\label{subsec:thresholding}

To begin with, we consider the trinarization of 3D image data by two global thresholds \cite{burger.2016,gonzalez.2008,russ.2007}, which are chosen in such a way that the experimentally determined volume fractions of all three phases are matched. For this purpose, we choose a sufficiently large sampling window, which does not contain void space outside the electrodes to avoid edge effects. The size of this cutout is given by $1500 \times 900 \times 250$ voxels (two-layer cathode) and $1000 \times 800 \times 220$ voxels (single-layer cathode), respectively. In the following, we refer to this approach as Thresholding. A visualization of the grayvalue histogram together with two thresholds as vertical lines is shown in Figure~\ref{fig:visualization}.

\subsection{Clustering approach}\label{subsec:kmeans}

A further method for the segmentation of 3D image data representing three-phase materials is based on a hard clustering approach, such as $k$-means clustering with $k=3$ \cite{jain.2010, james.2013, kroese.2019}. In particular, this kind of unsupervised learning has been successfully applied to cathodes in lithium-ion batteries \cite{prifling.2019b}. In the present paper, we slightly modify the algorithm considered in \cite{prifling.2019b} in order to ensure that the experimentally determined volume fraction of each phase is matched. In general, each voxel will be classified based on the grayvalues in its $3\times 3\times 3$ neighborhood. However, arranging these 27 values in a fixed order is not meaningful since, e.g., rotating or flipping the $3\times 3\times 3$ neighborhood would significantly change the feature vector. To overcome this problem, we sort the grayvalues in ascending order. To additionally increase the information content of the feature vector, we further group the voxels in the local neighborhood by their distance to the currently considered voxel. Thus, the first entry of the feature vector contains the grayvalue of the current voxel, the next six entries correspond to the sorted grayvalues of the $6$-neighborhood, the subsequent 12 entries belong to the voxels with distance $\sqrt{2}$ and the remaining 8 entries correspond to the voxels with distance $\sqrt{3}$. The $i$-th cluster $C_{i}$ with $i\in \{1,2,3\}$ (corresponding to the three phases active material, CBD, and pores) is now given by
\begin{equation*}C_{i} = \{ v_{j} \ : \ i = \argmin_{\ell=1,2,3} w_{\ell} \cdot \sum\limits_{m=1}^{27} x_{m} \cdot (f_{j}^{(m)} - \mu_{\ell}^{(m)})^{2} \},
\end{equation*}
where $v_{j}$ denotes the $j$-th voxel, $f_{j}=(f_{j}^{(1)},...,f_{j}^{(27)})\in \R^{27}$ the corresponding feature vector, and $\mu_{\ell} = (\mu_{\ell}^{(1)},...,\mu_{\ell}^{(27)}) \in \R^{27}$ the cluster centroids in the feature space. The phase weights $w_{1},w_{2},w_{3}>0$ as well as the feature weights $x_{1},...,x_{27}>0$ can now be chosen in such a way that we match the experimentally determined volume fractions of each phase. For this purpose, we choose $w_{1}=1$ and $x_{1}=1$ as reference. Moreover, we further reduce the number of parameters which have to be optimized by assuming equal weights for voxels with the same distance to the currently considered voxel, i.e., we assume that $x_{2}=...=x_{7}$, $x_{8}= ... = x_{19}$ and $x_{20}=...=x_{27}$. This leads to five parameters, which are computed by minimizing the cost function $\sum\limits_{\ell=1}^{3}(\varepsilon_{\ell,\text{exp}} - \hat{\varepsilon_{\ell}})^{2}$, where $\varepsilon_{\ell,\text{exp}}$ denotes the experimentally determined volume fraction of phase $\ell$ and $\hat{\varepsilon_{\ell}}$ equals the volume fraction of phase $\ell$ estimated on the segmented 3D image data obtained by running the $k$-means algorithm. This optimization is carried out with Powell's BOBYQA algorithm \cite{powell.2009}. Since the segmentation result depends on the initial cluster centroids \cite{celebi.2013}, we initialize the active material cluster by the feature vector associated with the brightest voxel and the pore cluster by the one associated with the darkest voxel. The CBD cluster is initialized with the feature vector that is most similar to the average of the feature vectors of the initial active material and pore centroid. In the following, we refer to this approach as $k$-means. In Figure~\ref{fig:visualization}, a two-dimensional sketch of this segmentation approach is shown, where three different colors are used to highlight the three clusters, whose centroid is marked with a large blue dot.

\subsection{Neural network}\label{subsec:nn}

In order to train a neural network that classifies each voxel according to the grayvalues in the synchrotron images, we make use of correlative microscopy. More precisely, a small cutout of the electrode has been imaged by FIB-SEM tomography after measuring the whole electrode sample by synchrotron tomography as described in Section~\ref{subsec:imaging}. This approach relies on the fact that a three-phase reconstruction of 3D FIB-SEM data is possible due to the better contrast compared to image data obtained by synchrotron tomography. More precisely, a global threshold determined by Otsu's method is used to  segment the active material \cite{otsu.1979}, whereas a U-Net is trained to distinguish between pores and CBD \cite{ronneberger.2015}. Finally, a slicewise flood-filling algorithm has been applied to the active material phase in order to remove inclusions of CBD or pores \cite{gonzalez.2008,russ.2007}. Due to the different voxel sizes of both kinds of image data, each synchrotron voxel corresponds to $128 \times 128 \times 128$ voxels in the FIB-SEM data. Thus, we can compute the material composition -- i.e. a three-dimensional vector containing the volume fractions of active material, CBD and pore space -- for each synchrotron voxel, for which FIB-SEM data is available. This information serves as ground truth for training a feed-forward neural network, which uses the gray values of an input voxel and its $5\times 5 \times 5$ neighborhood. The neural network is a multilayer perceptron consisting of five hidden layers with 75 units each and a softmax output layer with three units representing the predicted material composition of the input voxel \cite{rojas.2013, goodfellow.2016}.\\

Since the physical size of the FIB-SEM cutout is comparatively small (only 2541 voxels as training data), we make use of a data augmentation for the training data, where we flip and/or rotate the $5\times 5 \times 5$ neighborhood. Since these kind of transformations do not change the material composition, we increase the size of the training data by a factor of 48, which corresponds to the number of elements of the symmetry group of a hexahedron \cite{coxeter.1973}. The data points are randomly shuffled and split into 60\% training data, 20\% validation data and 20\% test data. The validation data is used for early stopping in case of ten subsequent epochs with a non-decreasing error on the validation set. The network consists of 5 hidden layers with 75 nodes each \cite{rojas.2013, goodfellow.2016}. The mean squared error, which is used as loss function, has been optimized using Nesterov's accelerated stochastic gradient descent~\cite{sutskever.2013} with a learning rate of 0.01 and a momentum coefficient of 0.99. After training the network is applied to the synchrotron image data of the single-layer as well as the two-layer sample, respectively. For each sample, this results in a 3D image, where for each voxel the material composition is predicted. This kind of information can be either interpreted as fuzzy membership or as probability of belonging to a certain phase \cite{dubois.2000, zimmermann.2001}. The top left plot in Figure~\ref{fig:visualization} shows the prediction accuracy on the test set of the trained neural network for each of the three phases, which indicates that the material composition can be reliably predicted. \\

In order to transform the output of the neural network into a segmentation with three classes, we consider two procedures. The first approach relies on the experimentally determined material composition as well as on a predefined ordering of the three phases, denoted by $\mathsf{P}_{1},\mathsf{P}_{2}$ and $\mathsf{P}_{3}$. More precisely, we assign the voxels with the highest predicted probability of belonging to phase $\mathsf{P}_{1}$ to $\mathsf{P}_{1}$ until the target volume fraction of $\mathsf{P}_{1}$ is matched. This procedure is then repeated for $\mathsf{P}_{2}$, except that we no longer consider voxels already classified as $\mathsf{P}_{1}$. In the following, this approach will be abbreviated as NN-$\mathsf{P}_{1}$-$\mathsf{P}_{2}$-$\mathsf{P}_{3}$ with $\mathsf{P}_{1},\mathsf{P}_{2},\mathsf{P}_{3}\in\{\text{AM,CBD,P}\}$. For example, first segmenting the active material, then assigning the CBD leads to the trinarization NN-$\mathsf{AM}$-$\mathsf{CBD}$-$\mathsf{P}$.  The second possibility for transforming the material composition by the neural network to a trinarization is based on conditional probabilities, where the first phase $\mathsf{P}_{1}$ is obtained analogously to the first approach. However, we then compute the conditional probabilities of voxels belonging to $\mathsf{P}_{2}$ and $\mathsf{P}_{3}$ conditioned on the event that these voxels are not classified as $\mathsf{P}_{1}$. Since these two conditional probabilities add up to one, there is -- given that the phase $\mathsf{P}_{1}$ is fixed -- exactly one possibility to obtain a trinarization, which matches the experimentally determined material composition. This trinarization method will be denoted by NN-$\mathsf{P}_{1}$-Cond in the following. For example, first classifying the active material and then assigning the CBD and pore space based on the conditional probability, that a certain voxel is not classified as active material, leads to the trinarization NN-$\mathsf{AM}$-Cond. In total, there exist six different orderings of the three phases required for the first approach, as well as three different trinarizations based on the conditional probability approach, leading to nine different neural network segmentations. 

\subsection{Local closing based on EDX data}\label{subsec:edx}

Similar to \cite{kremer.2020}, 2D image data obtained by energy-dispersive X-ray spectroscopy (EDX) is used to estimate the corresponding CBD gradient along the transport direction, which is then fitted by a linear function, see Figure~\ref{fig:CBD_gradient}. 

\begin{figure}[!htbp]
	\begin{subfigure}[c]{0.45\textwidth}
		\centering
		\includegraphics[width=0.8\textwidth]{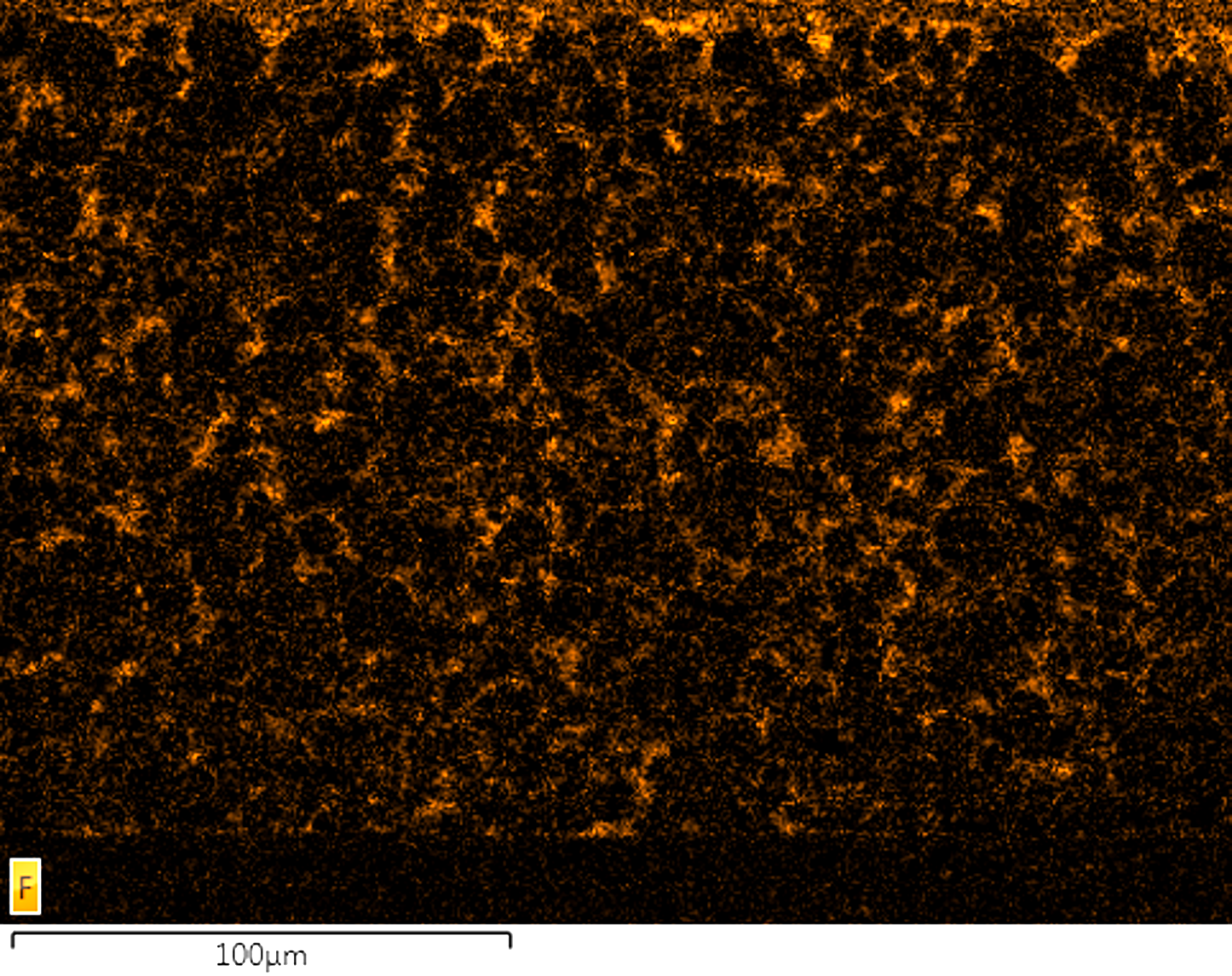}	
	\end{subfigure}
	\begin{subfigure}[c]{0.45\textwidth}
		\centering
		\includegraphics[width=0.8\textwidth]{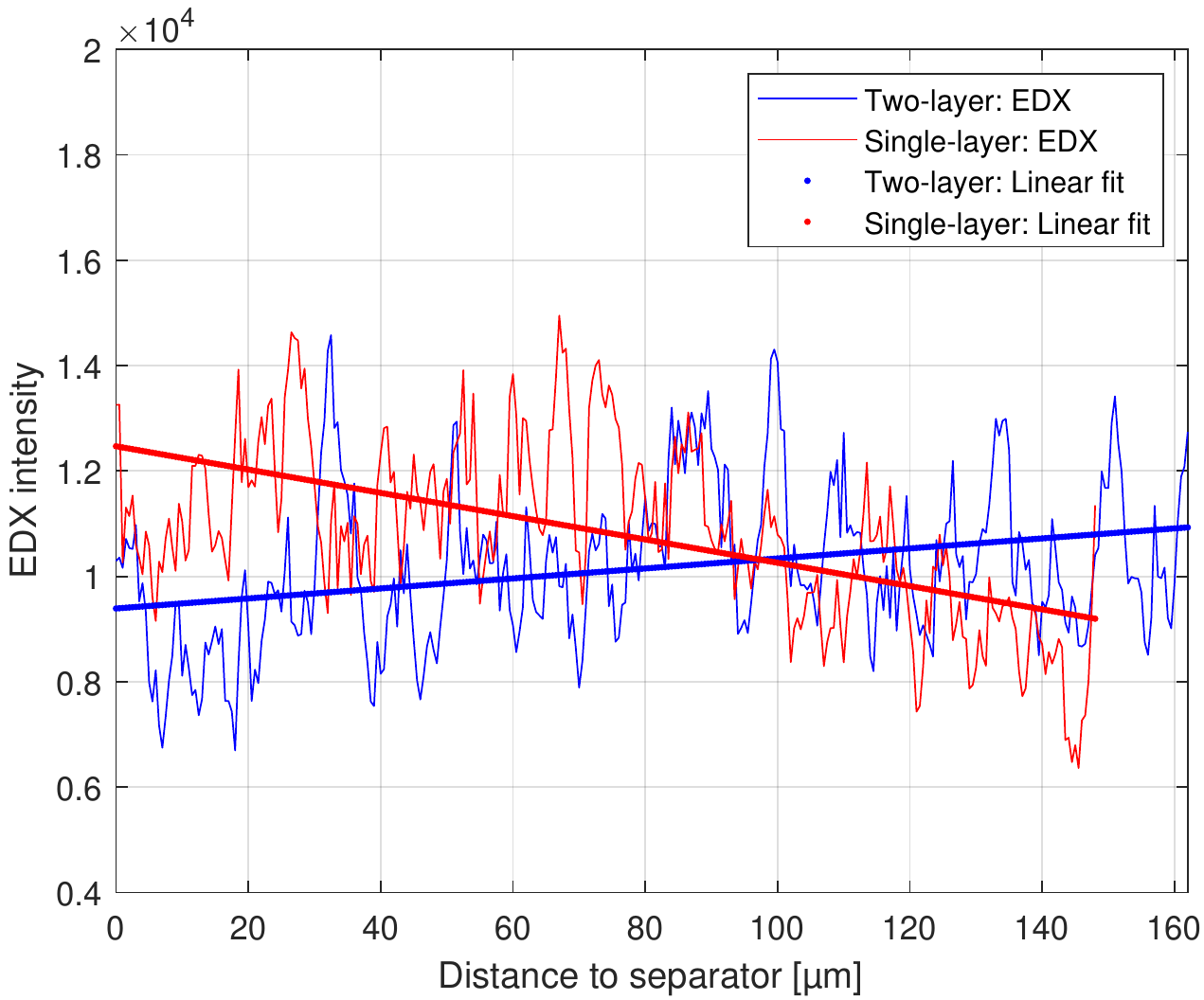}
	\end{subfigure}
	\caption{Left: EDX image (flourine mapping) of the single-layer cathode. Right: CBD gradient computed from EDX data (dots) and corresponding linear fit (solid line) for SL and TL.}
	\label{fig:CBD_gradient}
\end{figure}

The first step to obtain a 3D segmentation that reflects the linear CBD gradient is to use the active material obtained by the $k$-means segmentation. Afterwards, the CBD is inserted by a morphological closing of the active material phase, where the structuring element is given by a ball with some location-dependent radius $r>0$ \cite{serra.1982, soille.2003}. Note that it has been shown in \cite{hein.2020} that using a morphological closing is an appropriate model for inserting the CBD. As described in \cite{kremer.2020}, the closing radius $r$ depends on the distance to the separator such that the slice-dependent amount of CBD is proportional to the estimated CBD gradient, where the known CBD volume fraction is matched by multiplying the EDX intensity values by a constant that is computed with the bisection method \cite{kincaid.2009}. In the following, we refer to this approach as EDX-Closing. 

\begin{figure}[!htbp]
    \centering
    \includegraphics[width=\textwidth]{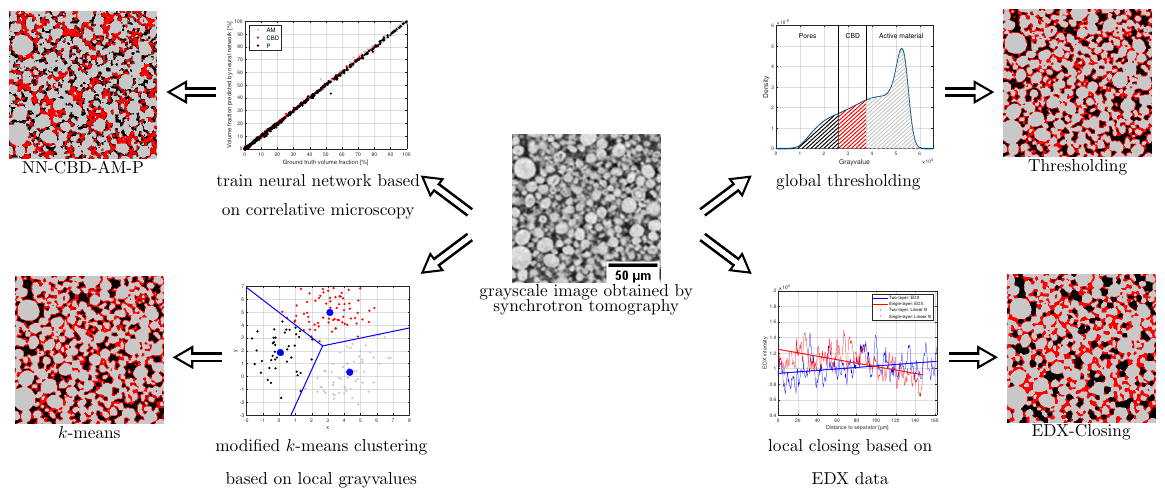}
	\caption{Comparison of different trinarization approaches for a 2D slice.}
	\label{fig:visualization}
\end{figure}

\subsection{Voxel-based comparison of trinarization approaches}\label{subsec:voxel_comparison}

Before investigating the influence of the different trinarization approaches on morphological and electrochemical properties in Section~\ref{sec:results}, we perform a quantitative voxel-based analysis to obtain a first impression regarding the potential differences between the segmentation approaches described above. At first, selected results obtained by the four different trinarization approaches are visualized in Figure~\ref{fig:visualization}. Before we quantify the influence of the trinarization approach on geometric descriptors of the resulting 3D microstructures in Section~\ref{subsec:results_morphology}, we first quantify the difference between the presented three-phase reconstructions by the fraction of equally assigned voxels as well as the Jaccard index \cite{hassaballah.2020}, see Figure~\ref{fig:similarity}. Both measures take values between zero and one, where lower values correspond to more pronounced differences between two trinarizations. In the present setting, the Jaccard index compares the spatial distribution of a predefined phase between two different trinarizations by computing the ratio of the intersection volume and the volume of the union. Note that the fraction of equally assigned voxels as well as the Jaccard index corresponding to a certain phase are symmetric characteristics such that the entries below the main diagonal in Figure~\ref{fig:similarity} contain the information regarding the single-layer cathode, whereas the entries above the main diagonal correspond to the two-layer cathode. On the one hand, the top left plot shows that there exist non-negligible differences between the neural network approaches, i.e., the method for converting the output of the neural network to a trinarization has an influence on the resulting three-phase reconstruction. On the other hand, there are even more pronounced differences between the neural network trinarizations and the remaining three approaches, namely $k$-means, EDX-Closing and Thresholding. In addition, the remaining three plots in Figure~\ref{fig:similarity} indicate that the least differences between the trinarization approaches are observed with regard to the segmentation of active material, which is most likely caused by the high contrast between active material and the remaining two phases. Furthermore, there are negligible differences between the single-layer and the two-layer cathode, except for the trinarization obtained by global thresholding.

\begin{figure}[!htbp]
	\centering
	\includegraphics[width=0.9\textwidth]{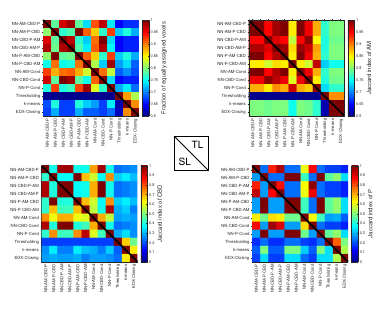}
	\caption{Fraction of equally assigned voxels (top left) as well as Jaccard index for active material (top right), CBD (bottom left) and pores (bottom right). Note that the entries above the main diagonal correspond to the two-layer cathode, whereas the entries below the main diagonal refer to the single-layer cathode. Due to the high accordance with regard to the spatial distribution of active material, the corresponding color bar only ranges from 0.6 to 1.}
	\label{fig:similarity}
\end{figure}

\section{Simulation of electrochemical properties}\label{sec:electrochemistry}

The electrochemical simulations are conducted using the research branch of the framework BEST, which is developed in collaboration between the DLR Institute of Engineering Thermodynamics and the Fraunhofer Institute for Industrial Mathematics (ITWM)\footnote{https://www.itwm.fraunhofer.de/best}. Focus of this work is on the influence of the CBD on electrochemical reactions and transport. Therefore, we will describe our CBD model and assumptions in more detail in subsequent paragraphs. A derivation of the governing equations and a description of our numerical framework can be found in previous publications \cite{hein.2020,latz.2011,latz.2015}. To provide a systematic overview of the electrochemical simulation approach we summarize the model equations, boundary conditions, initial conditions and parameters in the supporting information. More specifically, the governing equations in the different phases are listed in Table~\ref{tab:governing}. Interface and boundary conditions are given in Table~\ref{tab:interfaceandboundary}.
Interface models between active materials and electrolyte are listed in Table~\ref{tab:reaction}.\\

As described in the previous section, the 3D image data of both cathodes is segmented into three distinct phases, namely cathode active material, CBD and porosity. However, the inner structure of the CBD can not be resolved by means of synchrotron tomography, which has been also discussed at the beginning of Section~\ref{sec:segmentation}. Therefore, the CBD in our simulations on the electrode scale actually contains two materials, namely, the solid carbon and binder matrix as well as liquid electrolyte. Similarly, the porous separator contains both the glass-fiber material and liquid electrolyte. In our simulations we do not resolve the actual microstructure of these materials. We rather use a homogenization approach~\cite{Schmitt2020} to simulate the effective transport through these mixed domains. This approach is computationally much more efficient and enables simulations on the cell scale, however, requires additional input parameters for our model.\\
The relevant transport coefficients which need to be corrected due to the internal microstructure of the materials are the diffusion coefficient of lithium in the electrolyte ($D_e$) and the ionic and the electronic conductivity ($\kappa_e$ and $\sigma_s$)) of the electrolyte and solid phase, respectively. 
We determine the effective transport parameters based on the concept of effective tortuosity using the general expression given by Equation~\ref{eq:effective}.
\begin{align}
    X_{p}^{d,eff} &= \gamma_{p}^{d} \cdot X_{p}^{bulk} \text{ with }X \in \{D, \sigma, \kappa\} \label{eq:effective}\\
\end{align}
The effective transport parameter $X_{p}^{d,eff}$ is defined for a phase $p$, which can be electrolyte (e) or solid (s), in a domain $d$, which is the CBD or the separator.
The effective parameter $\gamma_{p}^{d} $ is defined using the respective volume fraction $\varepsilon_{p}^{d} $ and the effective tortuosity $\tau_{p}^{d} $ by
\begin{align}
    \gamma_{p}^{d}  &=\frac{\varepsilon_{p}^{d} }{\tau_{p}^{d}}\text{ with phase }p\in\{e,s\}\text{ and domain }d\in\{CBD,Sep\}\,.
\end{align}
We assume that the inner porosity of the CBD is equal to $50\%$. Hence, the effective tortuosity of the electrolyte part of the CBD $\tau_{e}^{CBD}$ can be computed based on $\gamma_{e}^{CBD}$ using the relationship 
\begin{align}
    \tau_{e}^{CBD} &=\frac{ \varepsilon_{e}^{CBD}}{\gamma_{e}^{CBD} } = \frac{1}{2\cdot \gamma_{e}^{CBD}}.
\end{align}
The effective tortuosity of the solid part in the CBD and the electrolyte part of the separator are computed likewise. The electrochemical parameters used in the simulations within this paper are listed in the Supporting Information, see Table~\ref{tab:parameters}. \\

In the previous paragraph we provide a qualitative description for the influence of the porous phases on the transport phenomena. 
Additionally, these porous materials also have an impact on the reactive surface effective at the interface to the active material. 
At the interfaces, where the active material is in contact with an porous electrolyte domain, we multiply the intercalation current with the porosity of the electrolyte phase.
In the case of the interface between CBD domain and active material domain the reaction current is given by the equation~(\ref{eq:reaction:cbdam}).
\begin{align}
 i_{react}^{CBD-AM} &= i_{react} \varepsilon_{e}^{CBD}\label{eq:reaction:cbdam}
\end{align}
The list of all interface conditions can be found in Table~\ref{tab:interfaceandboundary}.

To evaluate the impact of different methods for CBD reconstruction we performed two different types of virtual experiments:
\begin{enumerate}[i)]
\item Constant current lithiation in half-cell configuration with six different currents (1, 3, 6, 8, 10 and $\SI{12}{\milli\ampere\per\cm^{2}}$),
\item Impedance spectroscopy in symmetrical cell configuration under blocking condition.
\end{enumerate}

The simulation domains for the lithiation as well as the symmetrical impedance simulations are shown in Figure~\ref{fig:SimDom}.

\begin{figure}[!htbp]
		\centering
		\includegraphics[width=0.98\textwidth]{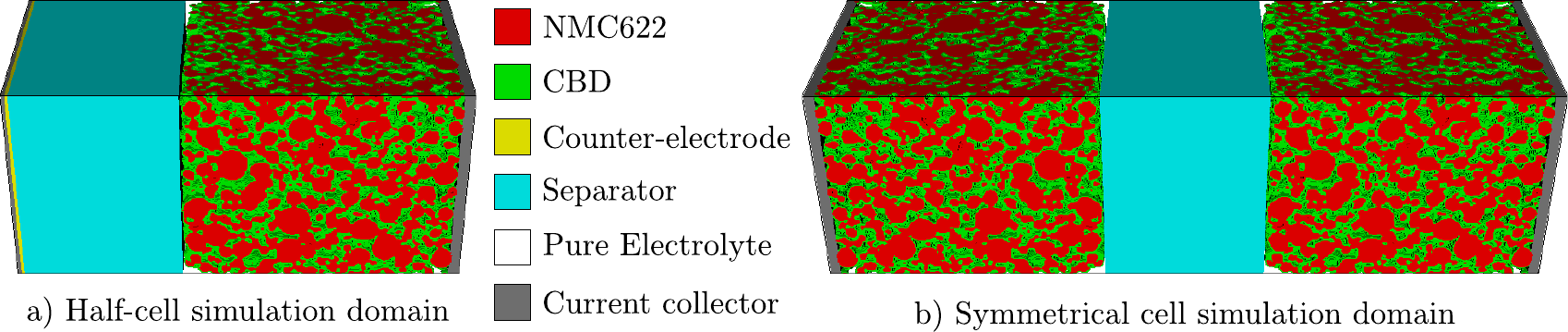}
	\caption{Simulation domains used for the two different types of electrochemical simulations.}
	\label{fig:SimDom}
\end{figure}

Three different cutouts of the electrode tomography are used as simulation domain for each trinarization approach and electrode type. 
The trinarized 3D microstructures are cropped to a lateral size of 200 voxels for the electrochemical simulations due to computational constraints.
This modifications keeps the thickness of the electrode and areal capacity unchanged.\\
Impedance spectra are calculated using the step excitation method. Details of the approach are also provided in \cite{hein.2020}. All electrochemical simulations are conducted using the HPC resources of JUSTUS2.

\section{Results and discussion}\label{sec:results}

This section covers the quantitative analysis of the different trinarization approaches with regard to their morphological properties by means of statistical microstructure analysis as well as the resulting electrochemical behaviour based on spatially and temporally-resolved numerical simulations.  

\subsection{Influence of selected trinarization approach on morphological descriptors}\label{subsec:results_morphology}

In this section, we discuss the influence of the different trinarization approaches described in Section~\ref{sec:segmentation} on the morphology of the resulting three-phase microstructures. For the sake of clarity, we only discuss the three trinarizations corresponding to the conditional probability approach, whereas the results for the remaining six neural network trinarizations can be found in the supporting information. Considering the 2D slices in Figure~\ref{fig:visualization}, one can already observe visual differences with regard to the morphological properties of the three phases. The CBD-phase determined by morphological closing based on EDX data is accumulated around the active material, which in turn leads to the formation of relatively large pores. Thereby, this approach differs clearly from the other approaches. On the other hand, the neural network approach results in a finely structured pore space. Moreover, by visual inspection it is hard to detect differences between the segmentation based on global thresholds and the one obtained by $k$-means clustering. Recall from Section~\ref{sec:segmentation} that all segmentation approaches are calibrated such that the volume fractions of active material and CBD-phase coincide with the experimentally determined values. In order to quantitatively evaluate the different trinarization approaches, we consider several microstructure characteristics for each of the three phases, which are considered as random closed sets \cite{chiu.2013}, denoted by $\Xi_{\text{AM}}, \Xi_{\text{CBD}}$, and $\Xi_{\text{P}}$.\\

We begin with the surface area per unit volume. This quantity is estimated from voxelized 3D image data as described in \cite{schladitz.2007}. Besides the surface area per unit volume of each phase, denoted by $S_{\text{AM}}, S_{\text{CBD}}$ and $S_{\text{P}}$, see Table~\ref{tab:my_label}, the surface area per unit volume of the interface between active material and the pore space is of interest from an electrochemical point of view since the intercalation takes place at this surface. Due to the inner porosity of the CBD, this characteristic, denoted by $S_{\text{Int}}$, is given by $S_{\text{Int}} = S_{\text{AM,P}} + 0.5 \cdot S_{\text{AM,CBD}}$. For this purpose, the surface area per unit volume of the interface between two phases is computed as described in \cite{lautenschlaeger.2022}. Interestingly, the surface area per unit volume of all three phases does not depend on the underlying trinarization approach. Thus, there are only minor differences between the values of $S_{\text{Int}}$.\\

\begin{table}[!htbp]
    \centering
    \begin{tabular}{lccccccc}
    \toprule
        & Sample & AM-Cond & CBD-Cond & P-Cond & Thresholding & $k$-means & EDX-Closing\\ \midrule
        $S_{\text{AM}}\ [\mu m^{-1}]$ & TL & 0.925 & 0.925 & 0.925 & 0.925 & 0.918 & 0.918 \\
        $S_{\text{AM}}\ [\mu m^{-1}]$ & SL & 0.925 & 0.925 & 0.924 & 0.918 & 0.924 & 0.924 \\ \midrule
        $S_{\text{CBD}}\ [\mu m^{-1}]$ & TL & 0.360 & 0.361 & 0.36 & 0.359 & 0.369 & 0.374 \\
        $S_{\text{CBD}}\ [\mu m^{-1}]$ & SL & 0.362 & 0.362 & 0.362 & 0.360 & 0.361 & 0.370 \\ \midrule
        $S_{\text{P}}\ [\mu m^{-1}]$ & TL & 0.272 & 0.271 & 0.273 & 0.273 & 0.27 & 0.265 \\
        $S_{\text{P}}\ [\mu m^{-1}]$ & SL & 0.271 & 0.271 & 0.272 & 0.279 & 0.272 & 0.263 \\ \midrule
        $S_{\text{Int}} [\mu m^{-1}]$ & TL & 0.672 & 0.672 & 0.672 & 0.672 & 0.664 & 0.662 \\
        $S_{\text{Int}} [\mu m^{-1}]$ & SL & 0.671 & 0.671 & 0.671 & 0.668 & 0.671 & 0.667 \\ \midrule
        $r_{\text{min,AM}}\ [\mu m]$ & TL & 1.81 & 1.78 & 1.86 & 1.90 & 2.12 & 2.12 \\
        $r_{\text{min,AM}}\ [\mu m]$ & SL  & 1.83 & 1.80 & 1.84 & 1.93 & 2.17 & 2.17 \\ \midrule
        $r_{\text{min,CBD}}\ [\mu m]$ & TL  & 0.66 & 0.75 & 0.23 & 0.21 & 0.22 & 0.78 \\
        $r_{\text{min,CBD}}\ [\mu m]$ & SL  & 0.60 & 0.73 & 0.24 & 0.21 & 0.21 & 0.78 \\ \midrule
        $r_{\text{min,P}}\ [\mu m]$ & TL & 0.21 & 0.21 & 0.70 & 0.76 & 0.80 & 0.00 \\
        $r_{\text{min,P}}\ [\mu m]$ & SL & 0.22 & 0.21 & 0.69 & 0.78 & 0.78 & 0.00 \\ \midrule
        $r_{\text{max,AM}}\ [\mu m]$ & TL & 3.11 & 2.88 & 3.13 & 3.63 & 3.74 & 3.74 \\
        $r_{\text{max,AM}}\ [\mu m]$ & SL & 3.19 & 3.12 & 3.17 & 3.66 & 3.80 & 3.80 \\ \midrule
        $r_{\text{max,CBD}}\ [\mu m]$ & TL & 1.03 & 1.23 & 0.67 & 0.48 & 0.52 & 1.26 \\
        $r_{\text{max,CBD}}\ [\mu m]$ & SL & 0.96 & 1.14 & 0.70 & 0.46 & 0.47 & 1.27 \\ \midrule
        $r_{\text{max,P}}\ [\mu m]$ & TL & 0.93 & 0.64 & 1.16 & 1.53 & 1.57 & 2.61 \\
        $r_{\text{max,P}}\ [\mu m]$ & SL & 0.97 & 0.67 & 1.15 & 1.47 & 1.55 & 2.58 \\ 
        \bottomrule
    \end{tabular}
    \caption{Scalar microstructure characteristics for different trinarization approaches.}
    \label{tab:my_label}
\end{table}

Additionally, the microstructure descriptors $r_{\text{max}}$ and $r_{\text{min}}$ are given in Table~\ref{tab:my_label}, where the descriptor $r_{\text{max}}$ denotes the $50 \%$-quantile of the so-called continuous pore size distribution. Similarly, the descriptor $r_{\text{min}}$ denotes the $50 \%$-quantile of a phase size distribution obtained by a geometric simulation of mercury intrusion and can be considered as the radius of the typical bottleneck. By means of $r_{\text{max}}$ and $r_{\text{min}}$, the constrictivity $\beta =r_{\text{min}}^2/r_{\text{max}}^2 \in [0,1]$ can be defined, which is a measure for the strength of bottleneck effects and a meaningful characteristic for effective transport properties \cite{holzer.2013b, neumann.2020, prifling.2021}. With respect to these microstructure descriptors, formally defined in \cite{neumann.2019TauBeta}, clear differences between the considered trinarization approaches can be observed, whereas there are no significant differences between the single-layer and the two-layer cathode. In particular, Figure~\ref{fig:cpsd} shows that EDX-Closing leads to significantly larger pores, which in turn leads to the largest value of $r_{\text{max}}$. Furthermore, $k$-means and Thresholding lead to nearly identical continuous phase size distributions for all three phases, whereas the neural network trinarizations differ from each other with regard to the continuous phase size distribution of the CBD as well as the pores. It is also interesting to note that with regard to the CBD as well as the pore space, the neural network segmentation based on conditioning on the respective phase leads to larger clusters of this phase. With regard to the simulated mercury intrusion porosimetry, see Figure~\ref{fig:mip}, we observe that the curves corresponding to the CBD as well as the pores are prone to discretization errors. Considering the active material, there are only slight differences, which, in turn, leads to similar values for $r_{\text{min}}$. Furthermore, the approach based on EDX data is the only case, where clear differences between the single-layer and the two-layer cathode can be observed. These differences are quantified by means of the simulated mercury intrusion porosimetry of the pore space. In Figures~\ref{fig:cpsd} and ~\ref{fig:mip}, the curves corresponding to the segmentation approaches based on correlative microscopy are shifted to the left compared to the remaining three-phase reconstructions when considering the pore space.

\begin{figure}[!htbp]
	\begin{subfigure}[c]{0.32\textwidth}
		\centering
		\includegraphics[width=0.9\textwidth]{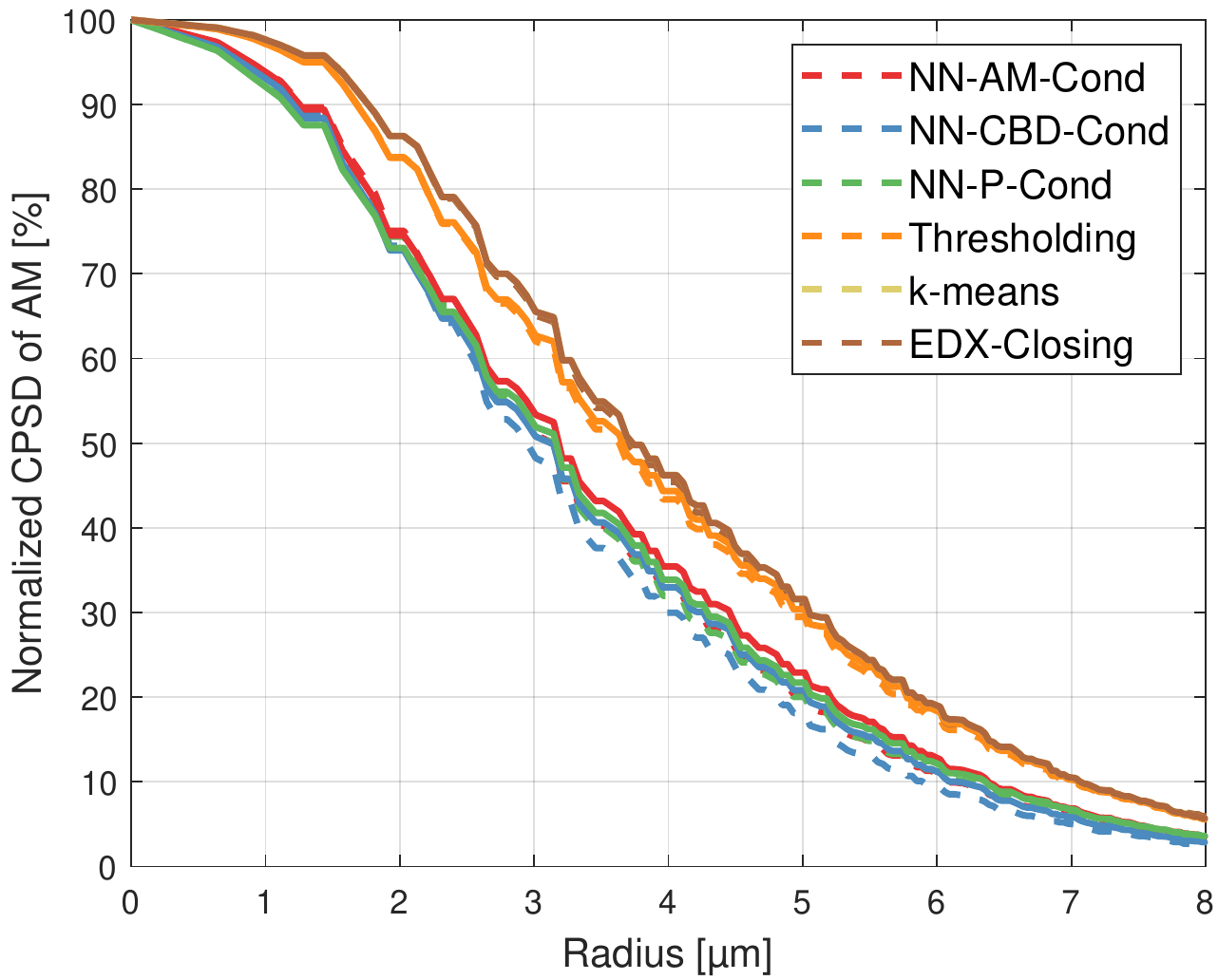}
	\end{subfigure}
	\hfill
	\begin{subfigure}[c]{0.32\textwidth}
		\centering
		\includegraphics[width=0.9\textwidth]{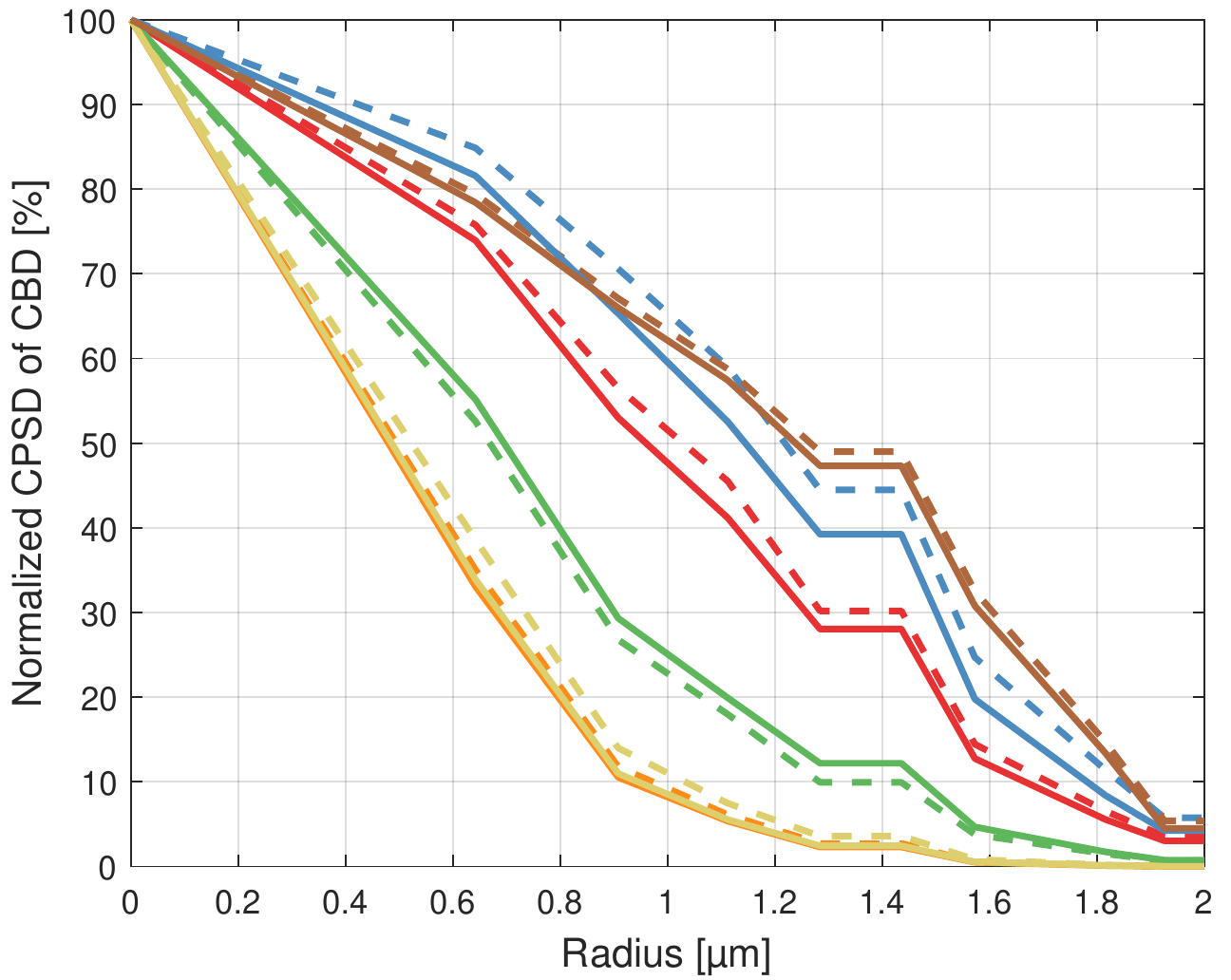}	
	\end{subfigure}
	\begin{subfigure}[c]{0.32\textwidth}
		\centering
		\includegraphics[width=0.9\textwidth]{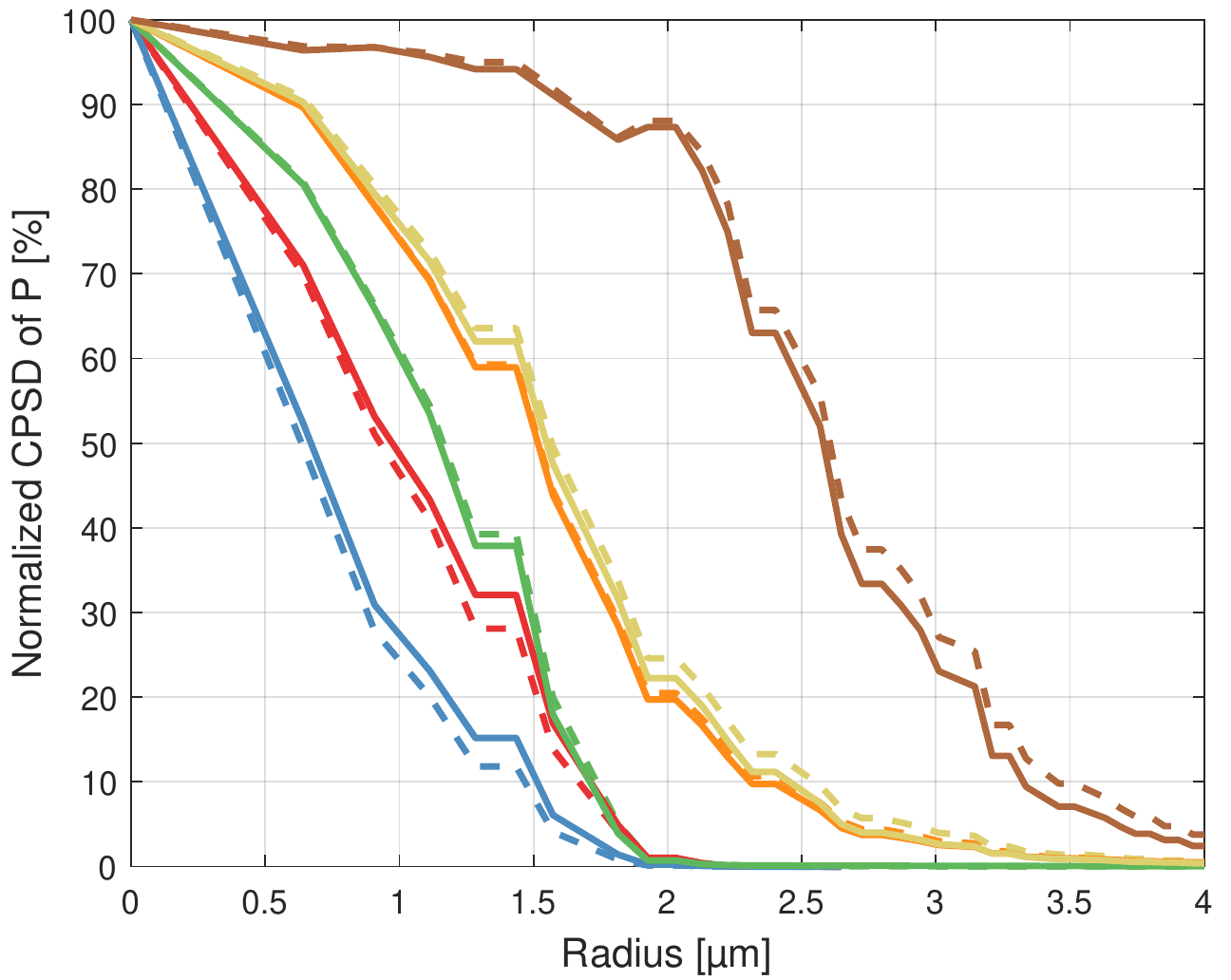}
	\end{subfigure}
	\caption{Continuous phase size distribution of active material (left), CBD (center), and pore space (right) for the two-layer cathode (dashed curves) and the single-layer cathode (solid curves).}
	\label{fig:cpsd}
\end{figure}

Moreover, the distribution of geodesic tortuosity is considered. This is a purely geometric quantity, in contrast to the effective tortuosity considered in Section~\ref{sec:electrochemistry}, providing the distribution of the length of shortest paths through a predefined phase in the electrode divided by the thickness of the electrode, see \cite{neumann.2019TauBeta} for a formal definition. Note that different concepts of tortuosity exist in the literature \cite{clennell.1997, thorat.2009, ghanbarian.2013, tjaden.2018}, where in the case of geodesic tortuosity Dijkstra's algorithm is used to estimate this quantity from voxelized image data \cite{jungnickel.2007}. As shown in Figure~\ref{fig:geodesic_tortuosity}, the distribution of geodesic tortuosity of the active material neither depends on the selected trinarization approach nor on the considered cathode sample. In contrast, the length of shortest paths through the CBD as well as the pore space is larger for the trinarizations obtained by the neural networks compared to the remaining three segmentation approaches. These differences between the four trinarization approaches considered in this paper are stronger than the differences between the single-layer and the two-layer cathode.\\

\begin{figure}[!htbp]
	\begin{subfigure}[c]{0.32\textwidth}
		\centering
		\includegraphics[width=0.9\textwidth]{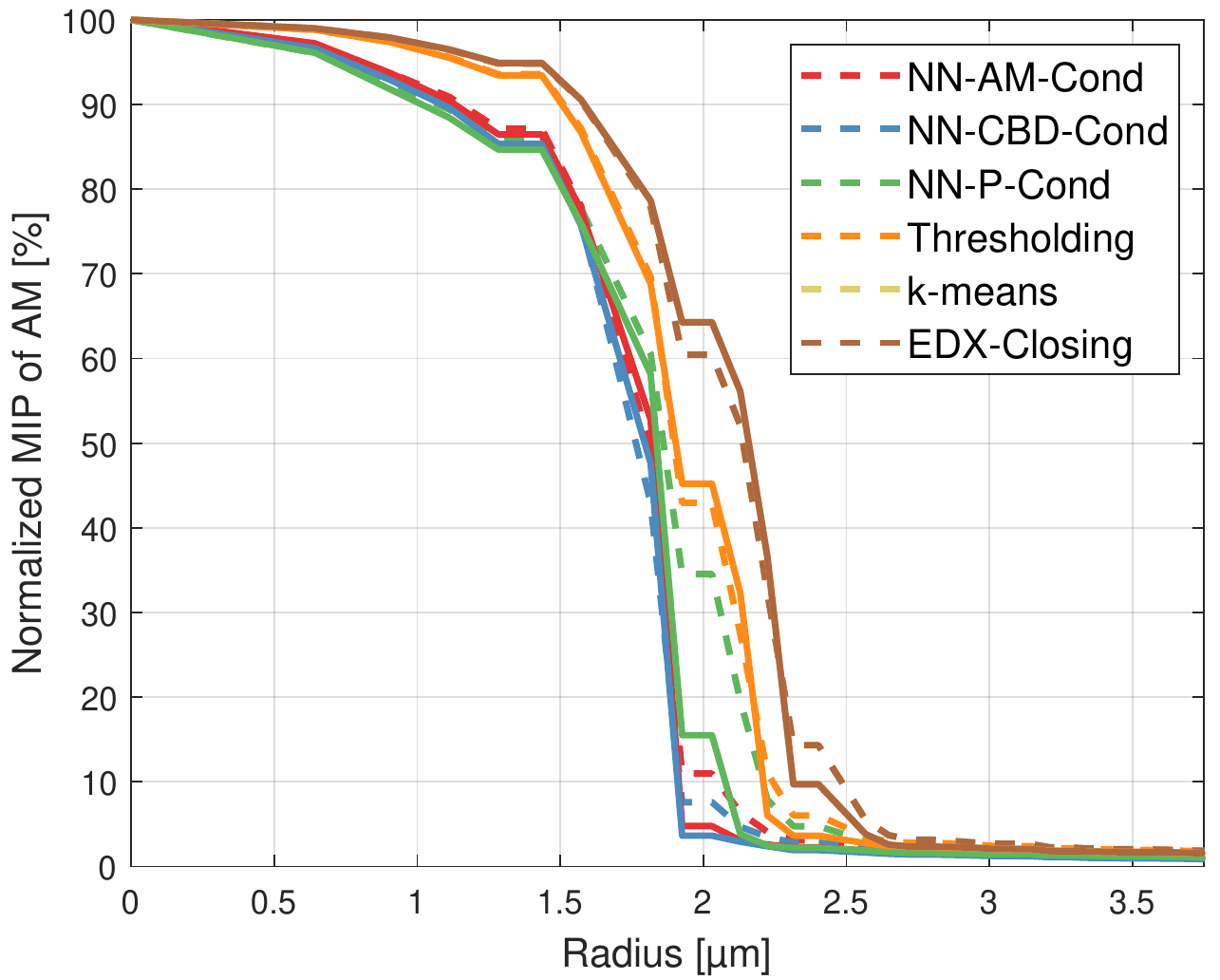}
	\end{subfigure}
	\hfill
	\begin{subfigure}[c]{0.32\textwidth}
		\centering
		\includegraphics[width=0.9\textwidth]{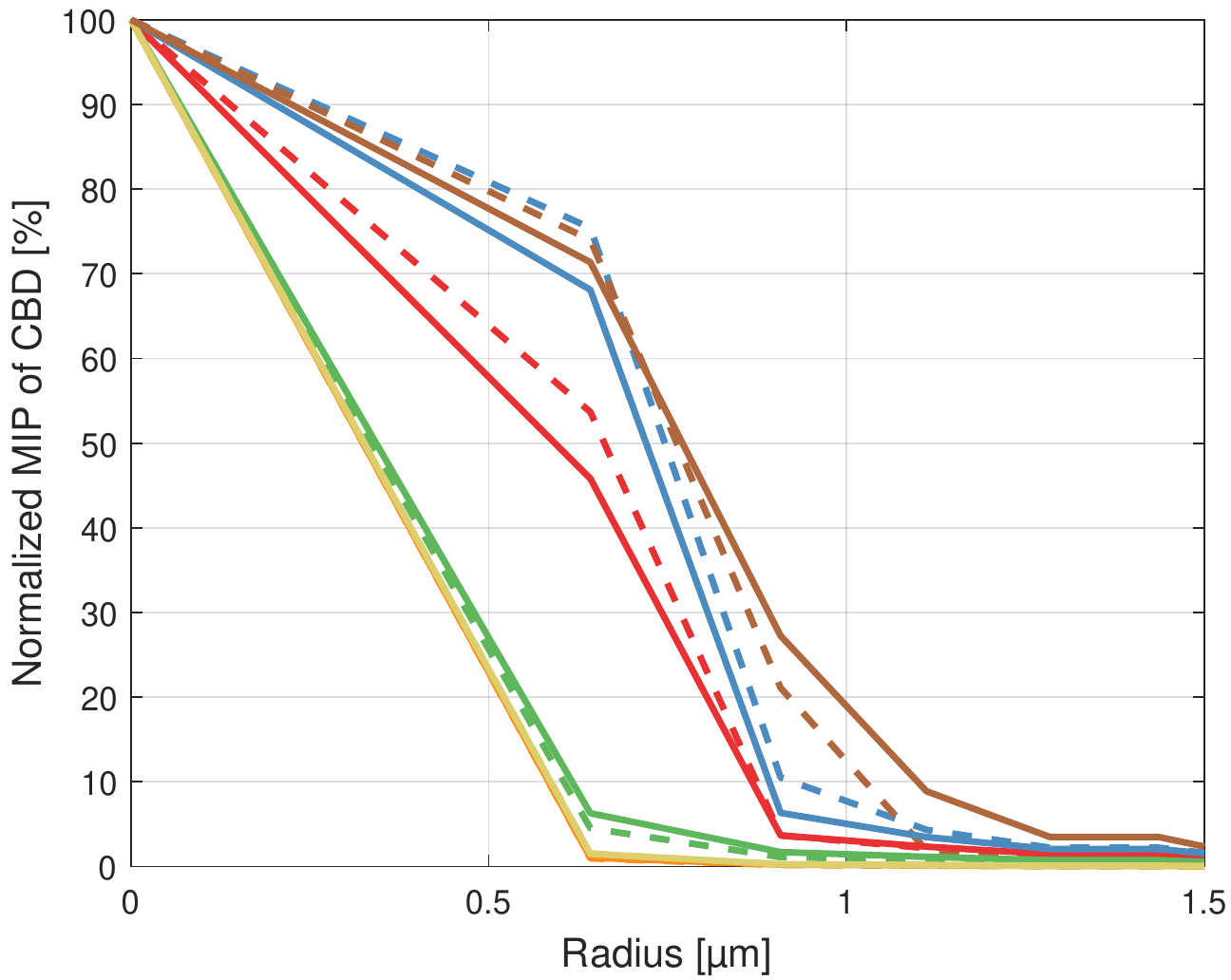}	
	\end{subfigure}
	\begin{subfigure}[c]{0.32\textwidth}
		\centering
		\includegraphics[width=0.9\textwidth]{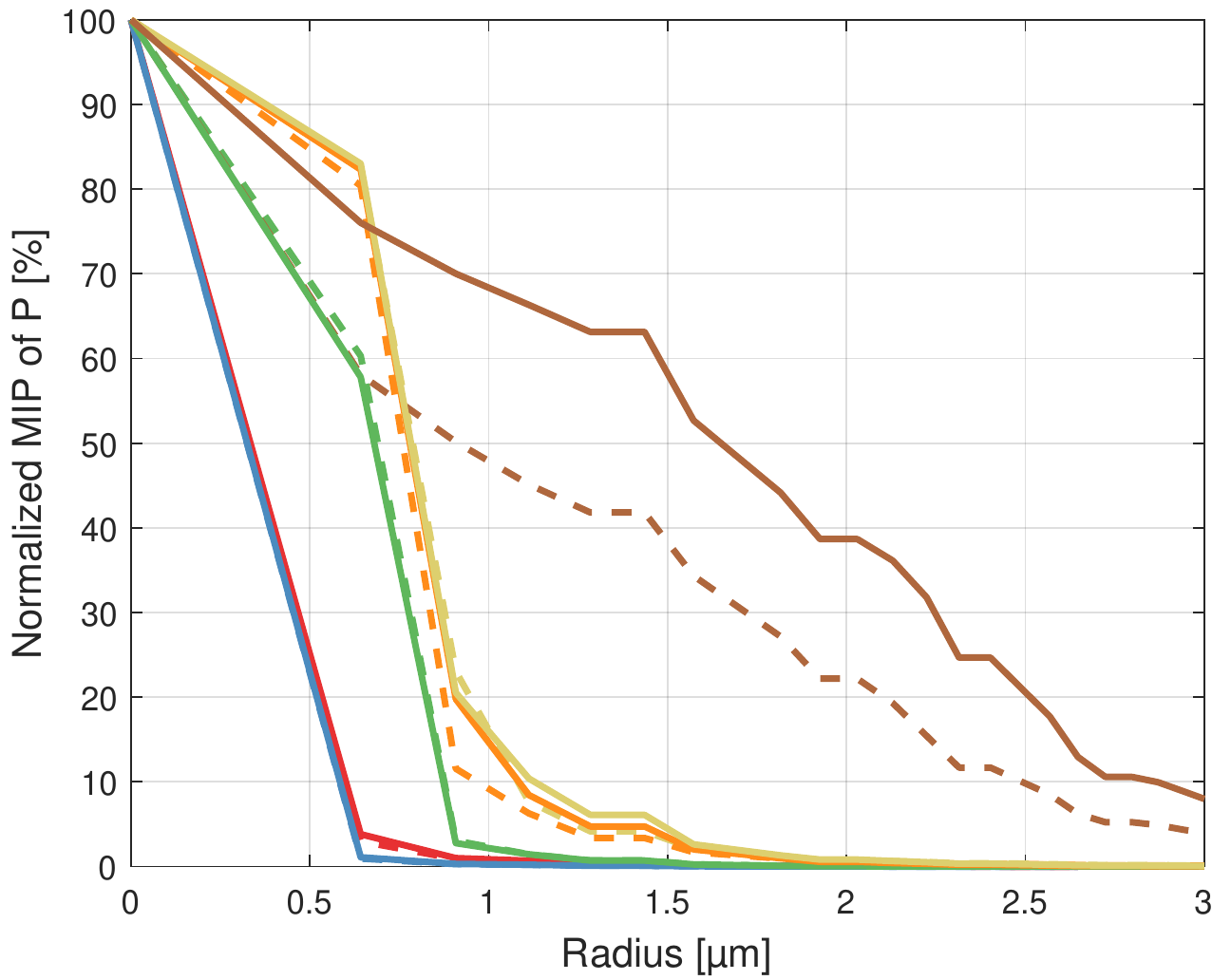}
	\end{subfigure}
	\caption{Simulated mercury intrusion porosimetry of active material (left), CBD (center) and pore space (right) for the two-layer cathode (dashed curves) and the single-layer cathode (solid curves).}
	\label{fig:mip}
\end{figure}

\begin{figure}[!htbp]
	\begin{subfigure}[c]{0.32\textwidth}
		\centering
		\includegraphics[width=0.9\textwidth]{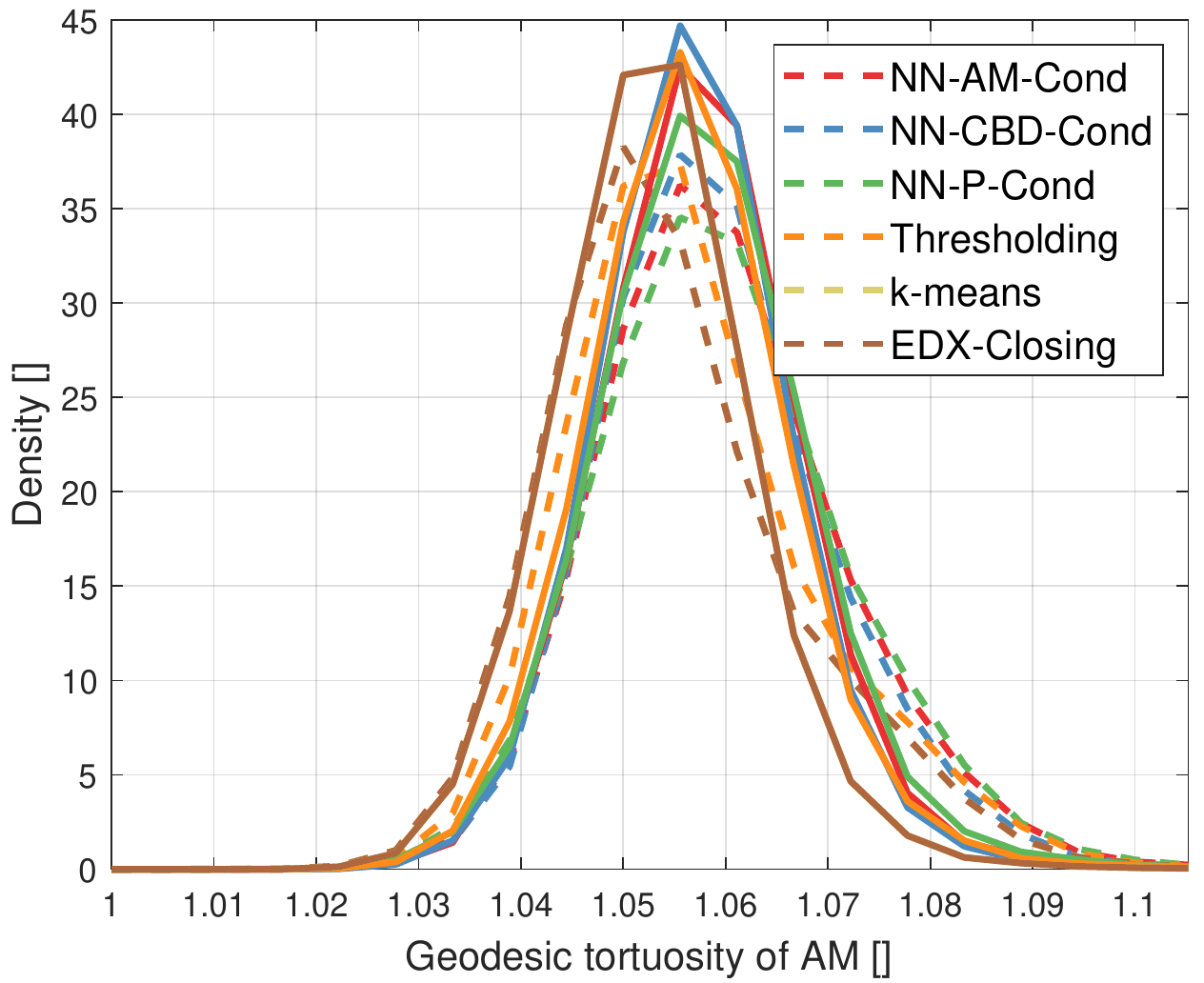}
	\end{subfigure}
	\hfill
	\begin{subfigure}[c]{0.32\textwidth}
		\centering
		\includegraphics[width=0.9\textwidth]{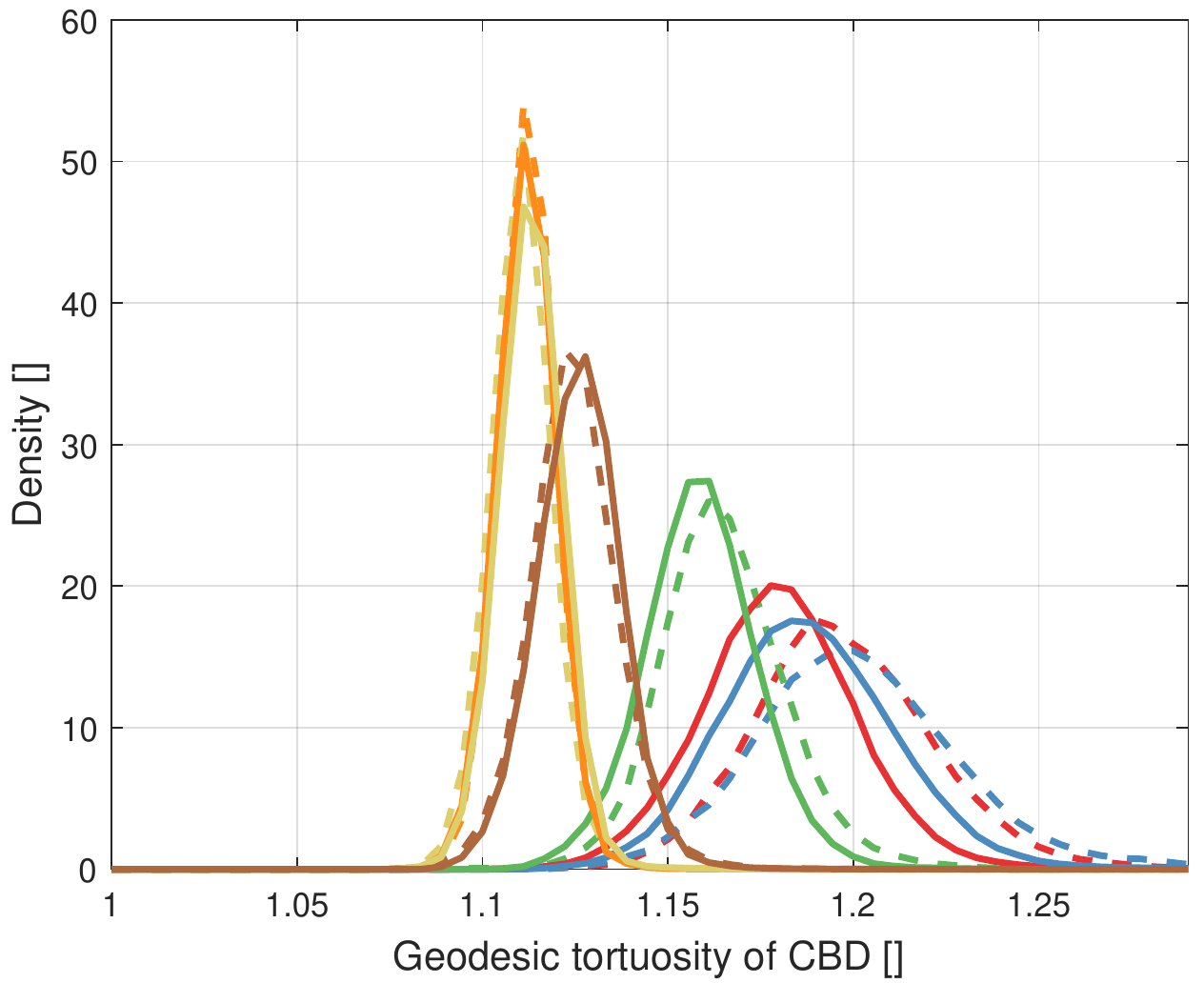}	
	\end{subfigure}
	\begin{subfigure}[c]{0.32\textwidth}
		\centering
		\includegraphics[width=0.9\textwidth]{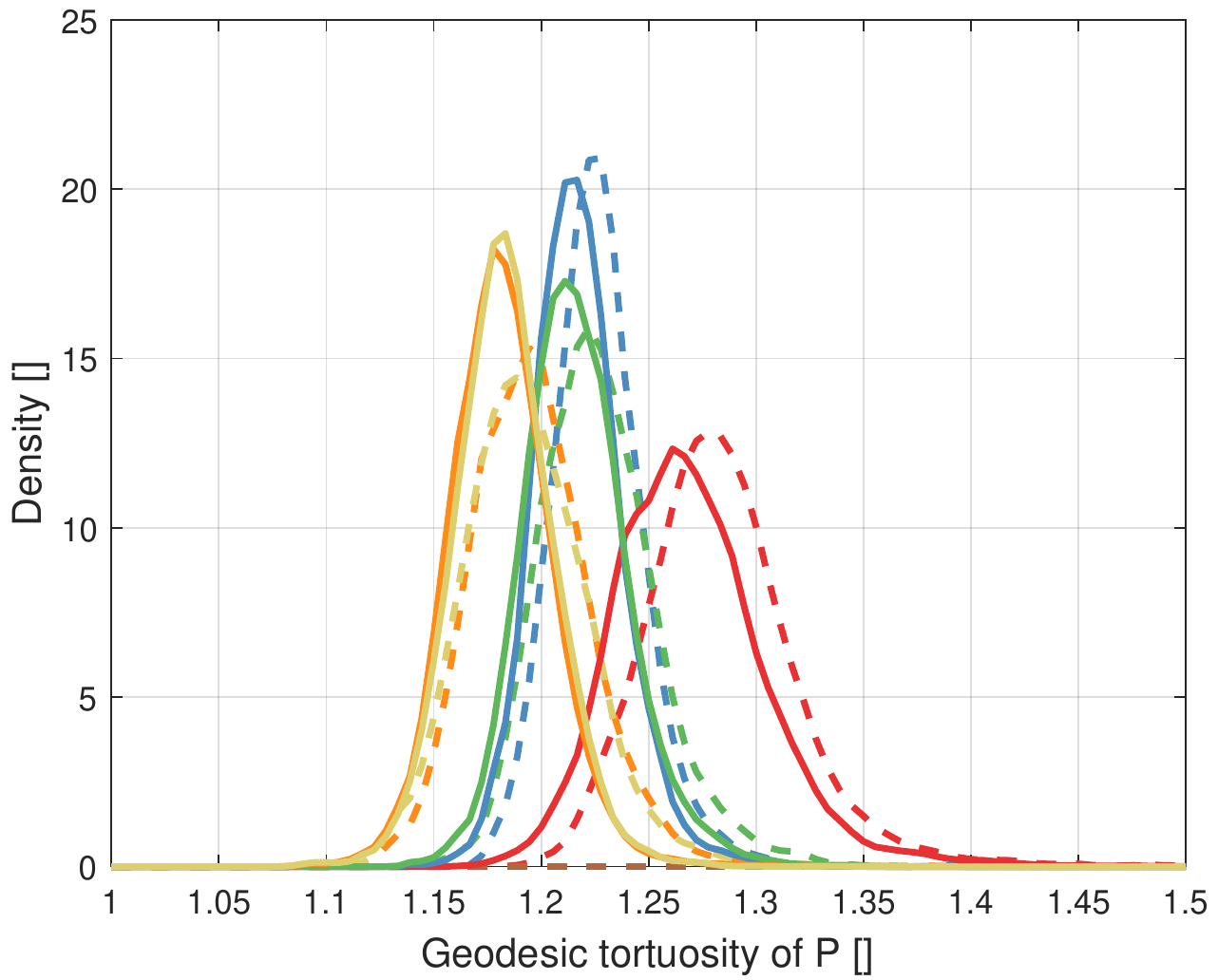}
	\end{subfigure}
	\caption{Geodesic tortuosity of active material (left), CBD (center) and pore space (right) for the two-layer cathode (dashed curves) and the single-layer cathode (solid curves).}
	\label{fig:geodesic_tortuosity}
\end{figure}

In addition, the centered two-point coverage probability function is considered, see Figure~\ref{fig:tpp}. For stationary and isotropic random closed sets $\Xi_{i},\Xi_{j}$ in the three-dimensional Euclidean space $\mathbb{R}^{3}$ with $i,j\in \{\text{AM}, \text{CBD}, \text{P}\}$, this characteristic is defined via $C_{ij}(r)=\mathbb{P}(0 \in \Xi_{i}, x \in \Xi_{j}) - \varepsilon_{i} \varepsilon_{j}$  for any $x\in\R^3$ and $r=|x|\ge0$ , where $\varepsilon_i,\varepsilon_j$ denotes the volume faction of $\Xi_i, \Xi_j$, respectively. This function is also called covariance function in the literature \cite{matheron.1975, serra.1982}. Due to the normalization by subtracting the product of the volume fractions, a value of zero implies that the events $0 \in \Xi_{i}$ and $x\in \Xi_{j}$ are stochastically independent. Positive values of $C_{i,j}(r)$ can be interpreted as a positive correlation between those two events, whereas negative values correspond to a negative correlation. Typically, choosing equal phases (i.e. $i=j$, see top row of Figure~\ref{fig:tpp}) leads to a monotonously decreasing function taking non-negative values, which approaches zero for large radii $r$. On the other hand, considering two different phases (i.e. $i\neq j$, see bottom row of Figure~\ref{fig:tpp}) leads in most cases to a monotonously increasing function approaching zero from below. Figure~\ref{fig:tpp} shows that there are no differences between both samples regardless of the phases under consideration. Furthermore, the curves in the top row of Figure~\ref{fig:tpp} show the same qualitative behavior as the continuous phase size distribution in Figure~\ref{fig:cpsd}. The most noticeable effect is the unique behaviour of the closing approach based on EDX data with regard to the bottom right plot in Figure~\ref{fig:tpp}. More precisely, the remaining segmentation approaches show a peak at around $\SI{2}{\micro\meter}$, which corresponds to an increased likelihood of observing CBD and pores $\SI{2}{\micro\meter}$ away from each other. The curves corresponding to EDX-Closing show a steadily increasing two-point coverage probability function instead. \\

\begin{figure}[!htbp]
    \centering
	\begin{subfigure}[c]{0.32\textwidth}
		\centering
		\includegraphics[width=0.9\textwidth]{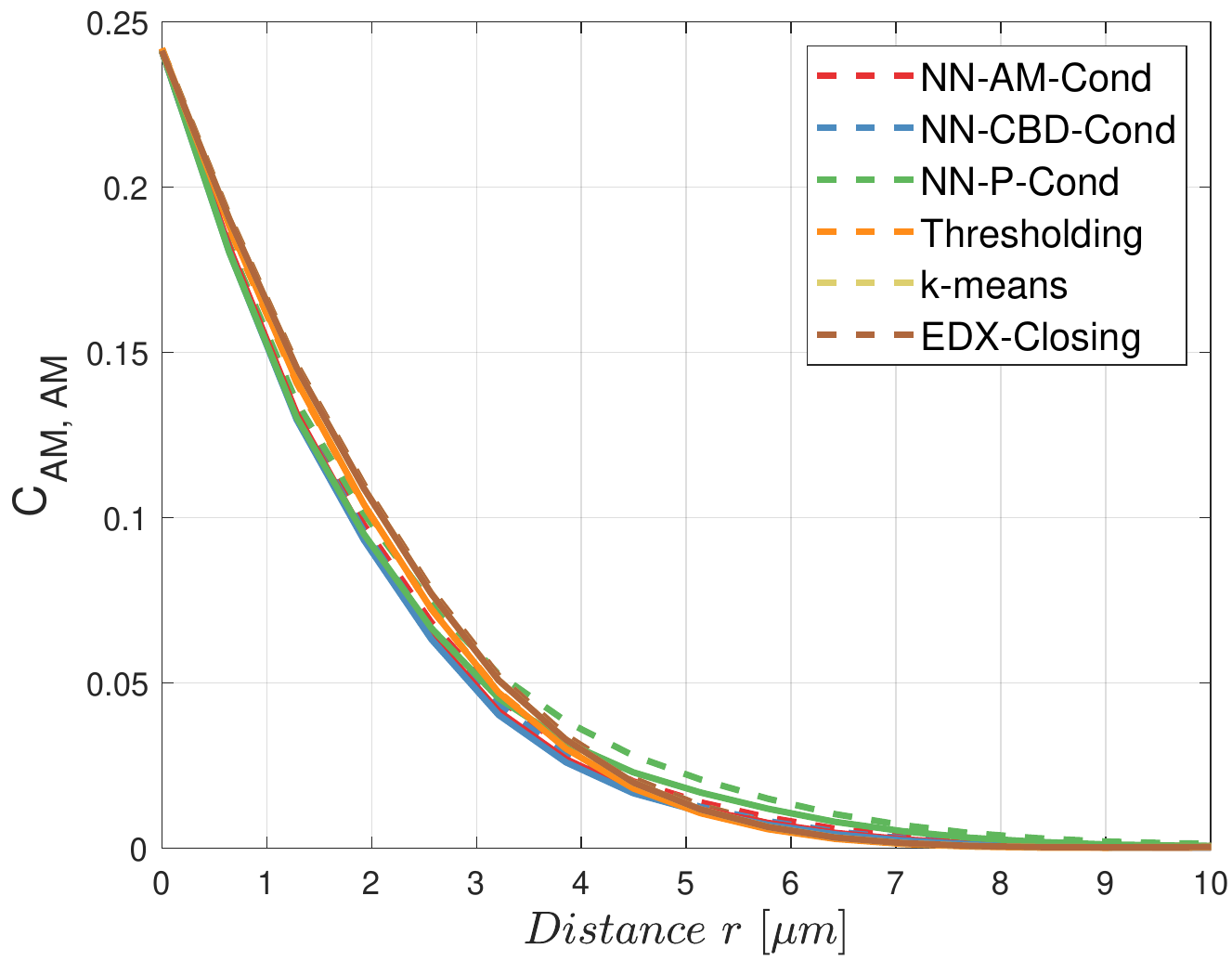}
	\end{subfigure}
	\hfill
	\begin{subfigure}[c]{0.32\textwidth}
		\centering
		\includegraphics[width=0.9\textwidth]{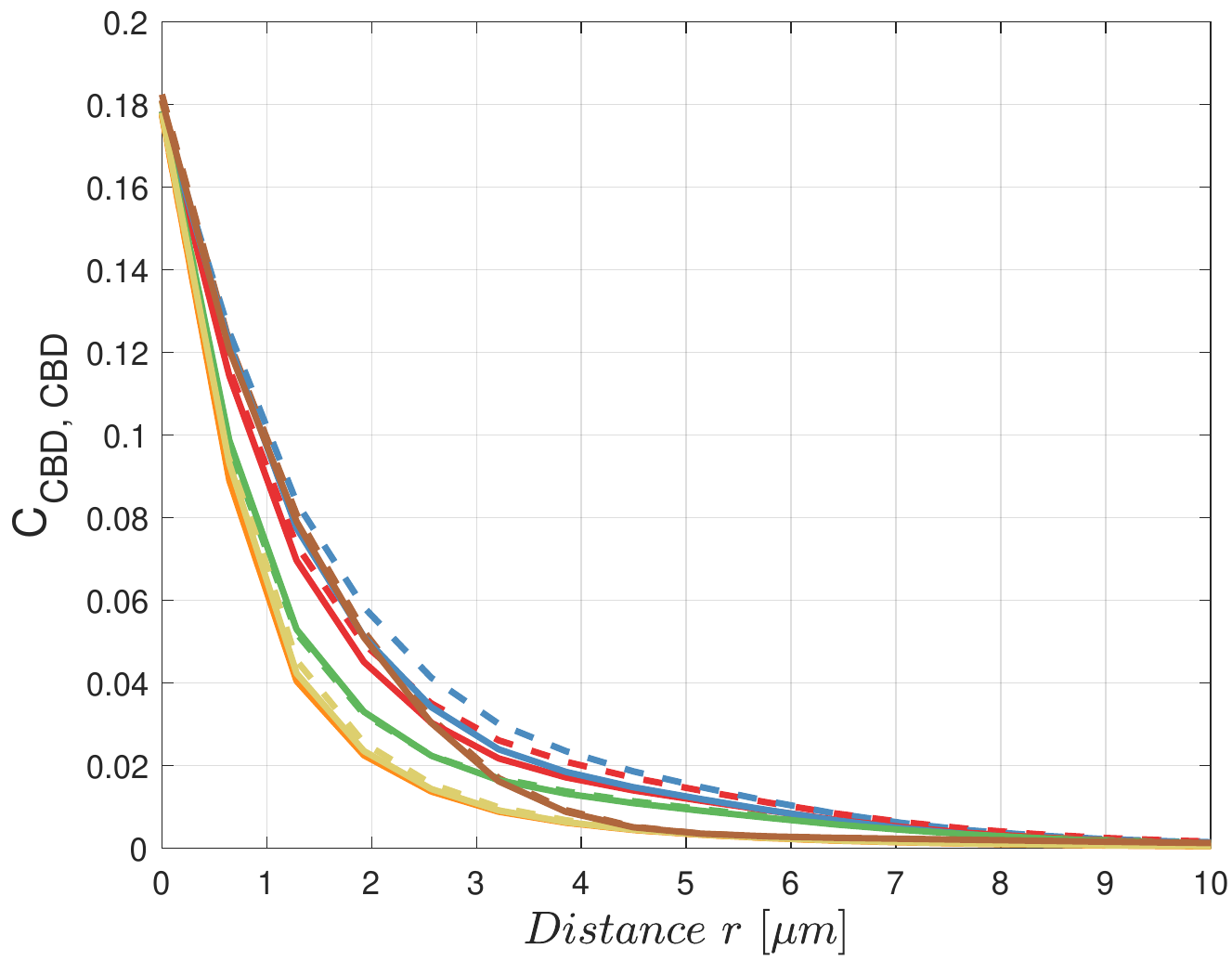}	
	\end{subfigure}
	\hfill
	\begin{subfigure}[c]{0.32\textwidth}
		\centering
		\includegraphics[width=0.9\textwidth]{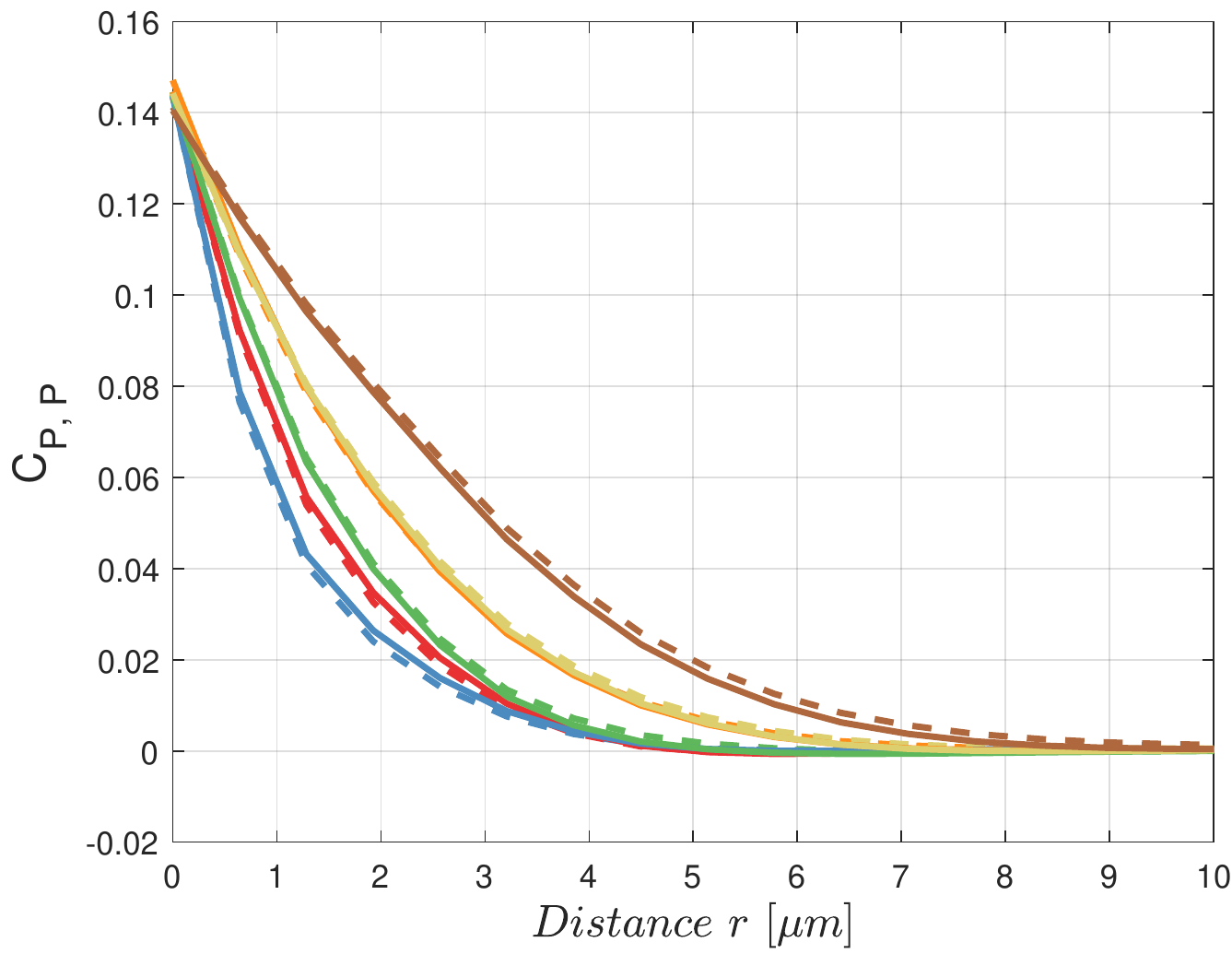}
	\end{subfigure}

	\begin{subfigure}[c]{0.32\textwidth}
		\centering
		\includegraphics[width=0.9\textwidth]{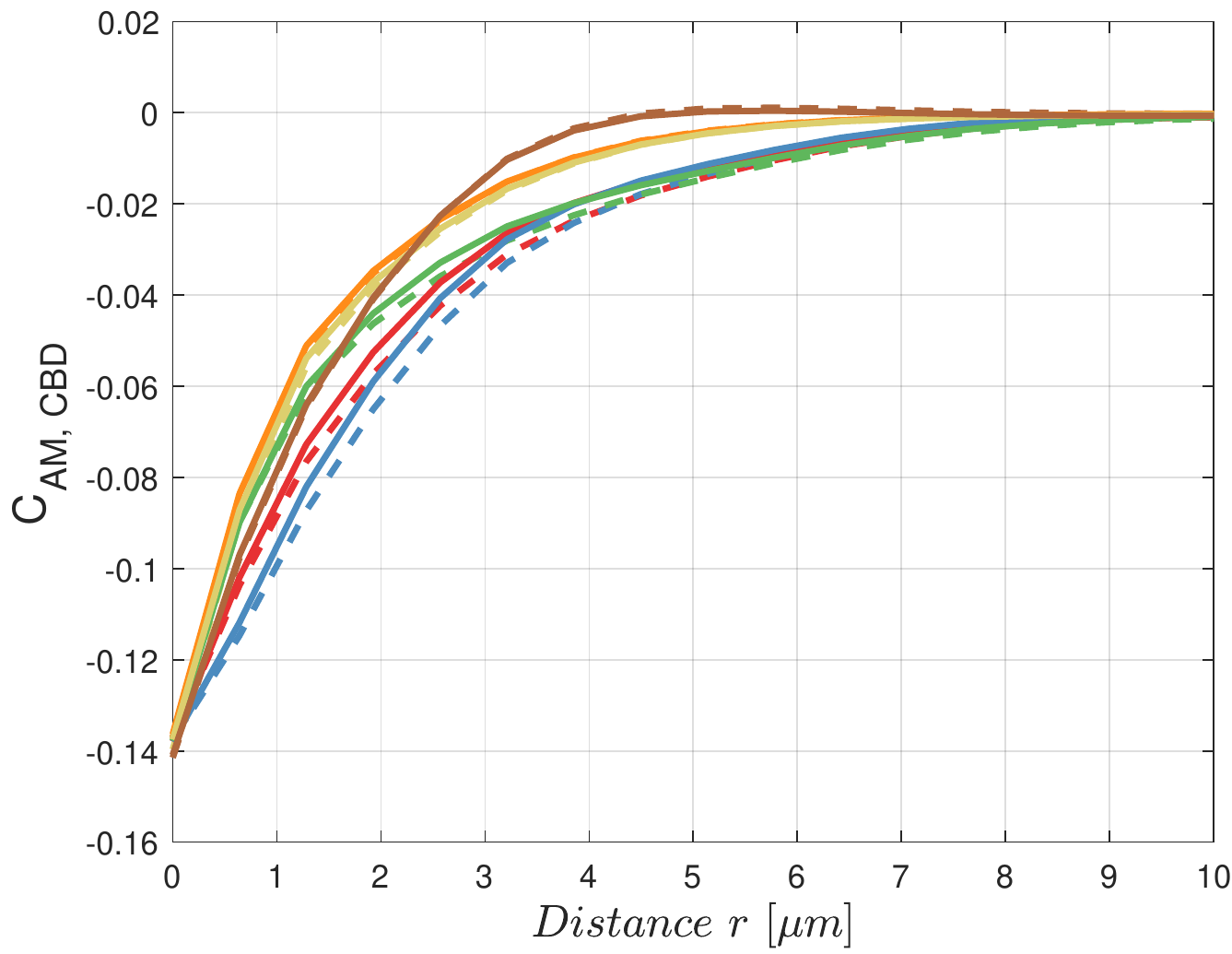}	
	\end{subfigure}
	\hfill
	\begin{subfigure}[c]{0.32\textwidth}
		\centering
		\includegraphics[width=0.9\textwidth]{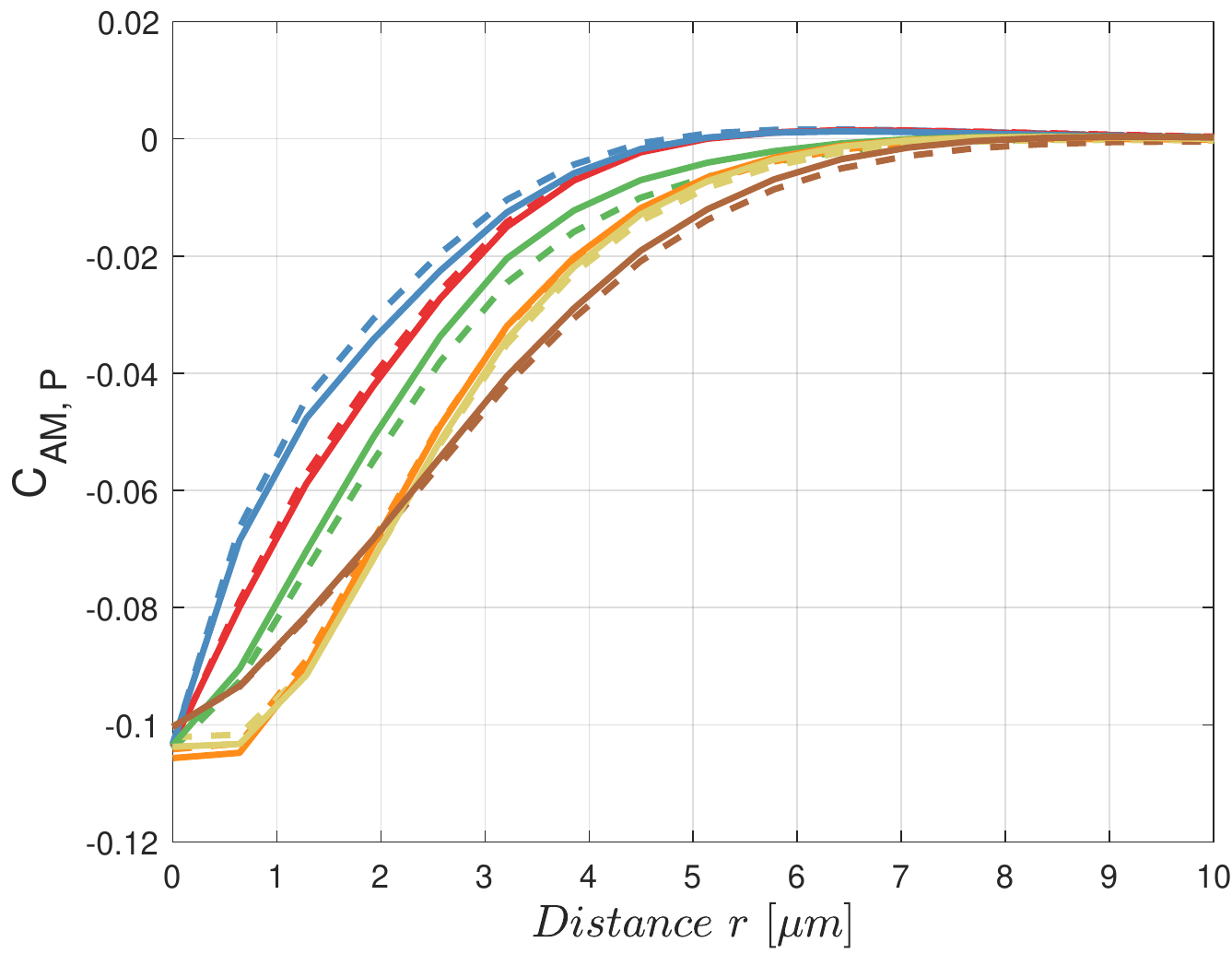}
	\end{subfigure}
	\hfill
	\begin{subfigure}[c]{0.32\textwidth}
		\centering
		\includegraphics[width=0.9\textwidth]{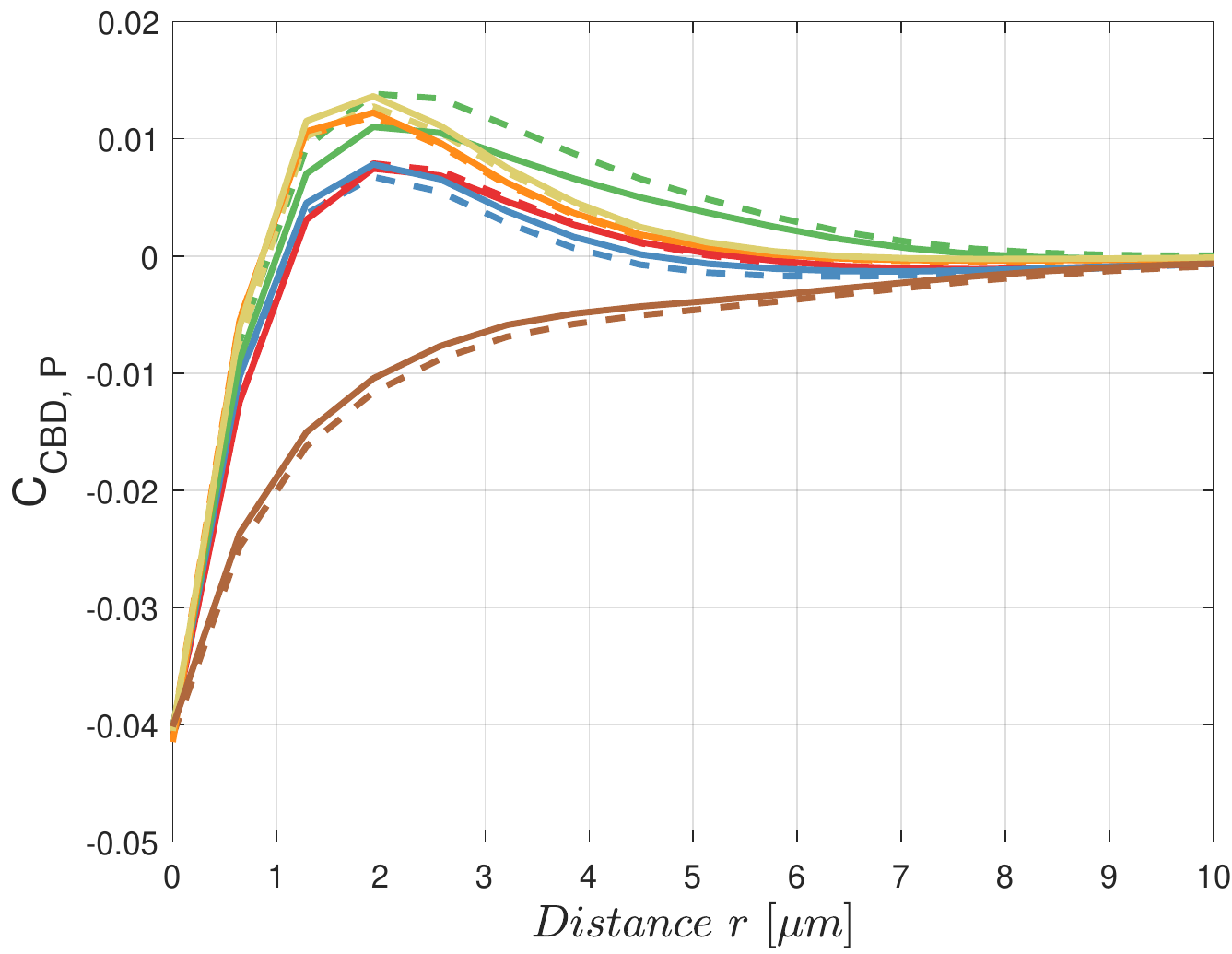}
	\end{subfigure}
	
	\caption{Top row: Centered two-point coverage probability function of active material, CBD and pore space (from left to right). Bottom row: Centered two-point coverage probability functions $C_{\text{AM,CBD}}$, $C_{\text{AM,P}}$ and $C_{\text{CBD,P}}$ (from left to right). Note that dashed curves are used for the two-layer cathode, whereas the solid curves correspond to the single-layer cathode.}
	\label{fig:tpp}
\end{figure}

Finally, we consider the volume fraction of each phase in dependence of the distance to the separator, see Figure~\ref{fig:volfrac_dist}. With respect to the spatial distribution of active material, there is a clear difference between the single-layer and the two-layer cathode regardless of the trinarization approach. More precisely, the two-layer sample shows a pronounced drop of the volume fraction of active material at $\SI{80}{\micro\meter}$, i.e. at the interface between both layers. With regard to the CBD, there are clear differences between the results obtained for each of the trinarization approaches, where all three-phase reconstructions except EDX-Closing indicate a larger amount of CBD at the interface. This peak is most pronounced for $k$-means and Thresholding. Obviously, EDX-Closing reflects the linear gradient estimated from EDX data. Note that this linear gradient is estimated from a single 2D EDX image and is thus subject to a larger uncertainty compared to the information extracted from 3D image data. Therefore, segmentation approaches not reflecting the linear gradient observed in EDX data are not automatically considered as unrealistic. Interestingly, this does not lead to a linear behaviour of the distance-dependent porosity. Except for EDX-Closing, there are comparatively small differences between both samples, see the plots on the right-hand side of Figure~\ref{fig:volfrac_dist}. 

\begin{figure}[!htbp]
    \centering
	\begin{subfigure}[c]{0.32\textwidth}
		\centering
		\includegraphics[width=0.9\textwidth]{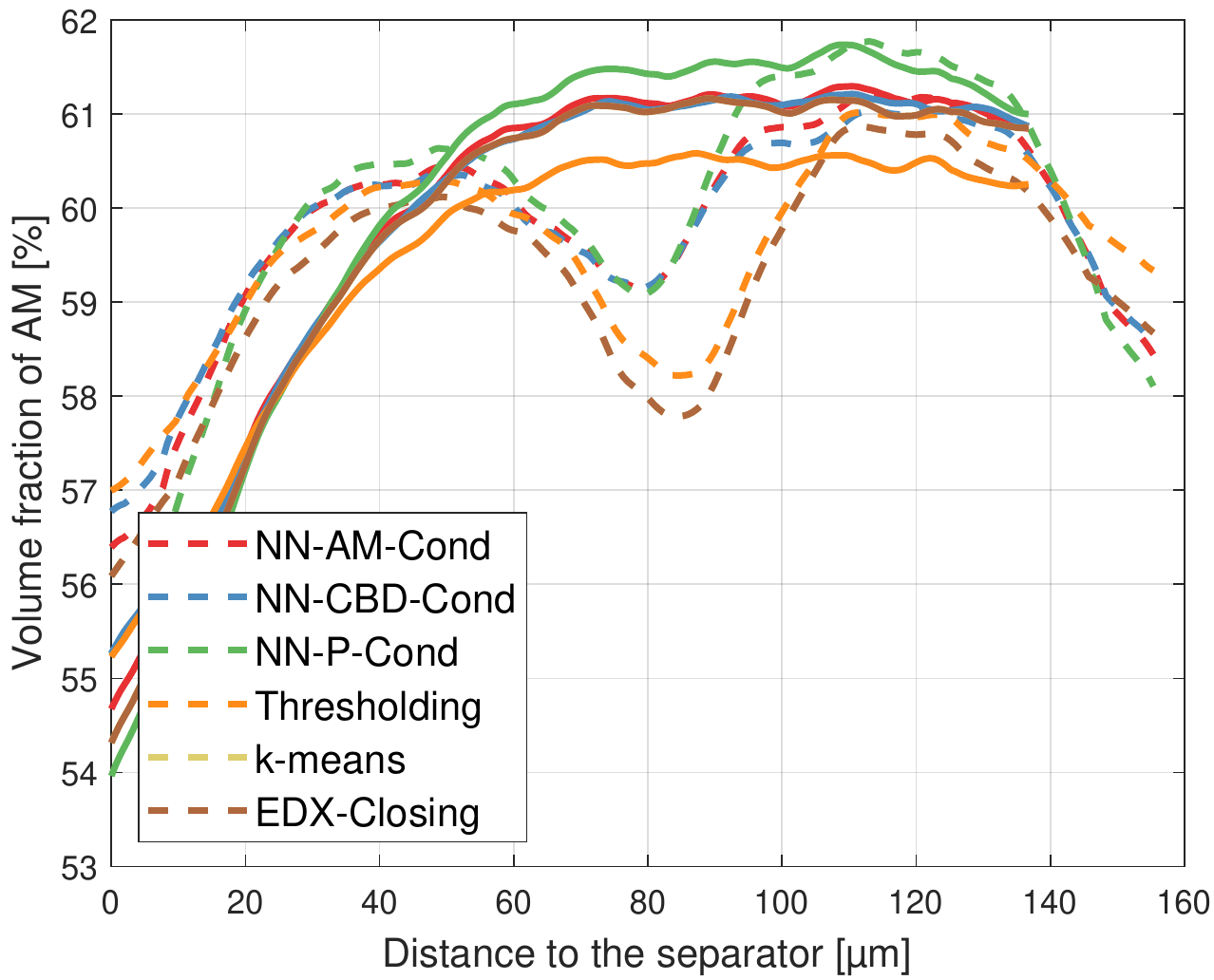}
	\end{subfigure}
	\hfill
	\begin{subfigure}[c]{0.32\textwidth}
		\centering
		\includegraphics[width=0.9\textwidth]{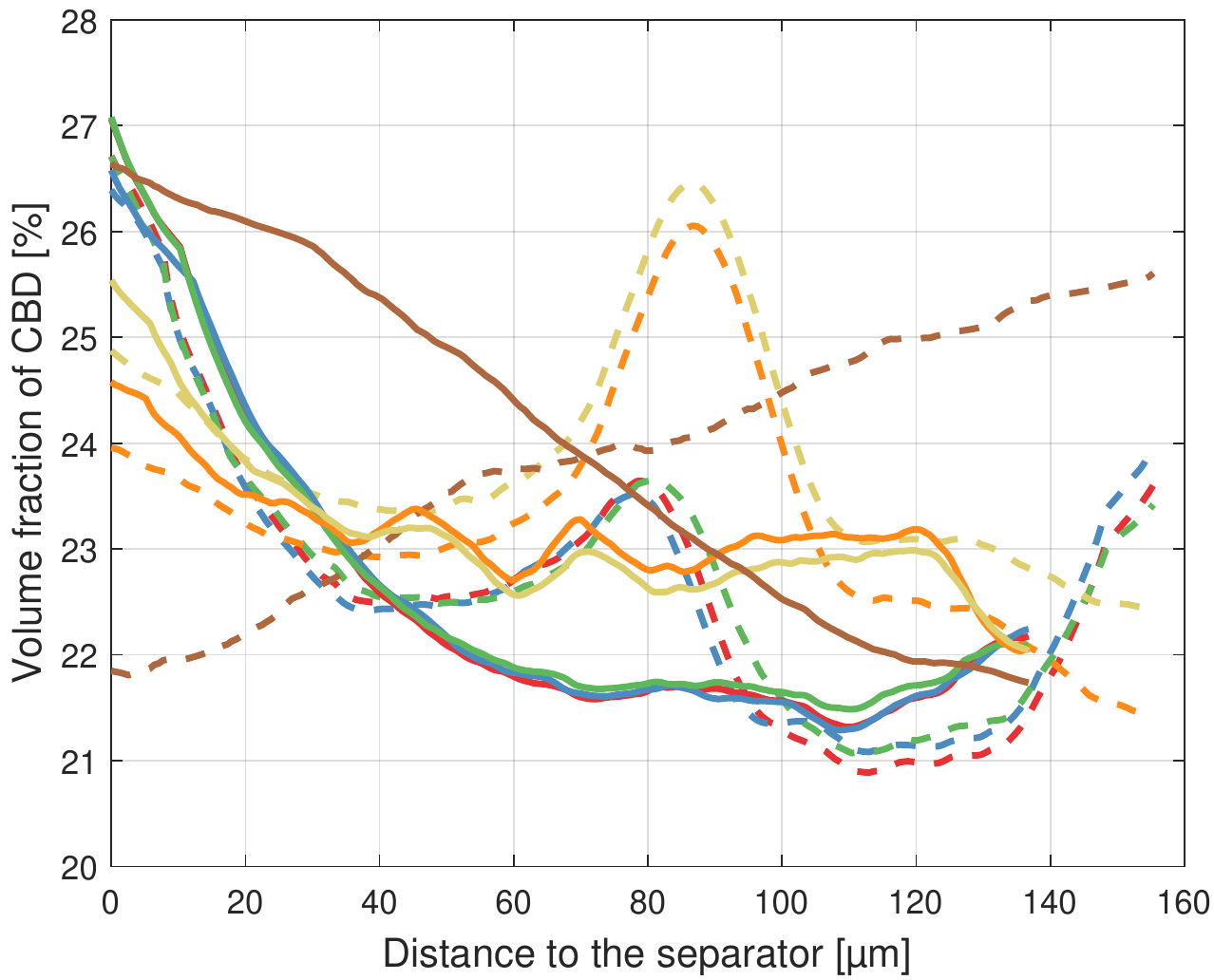}	
	\end{subfigure}
	\hfill
	\begin{subfigure}[c]{0.32\textwidth}
		\centering
		\includegraphics[width=0.9\textwidth]{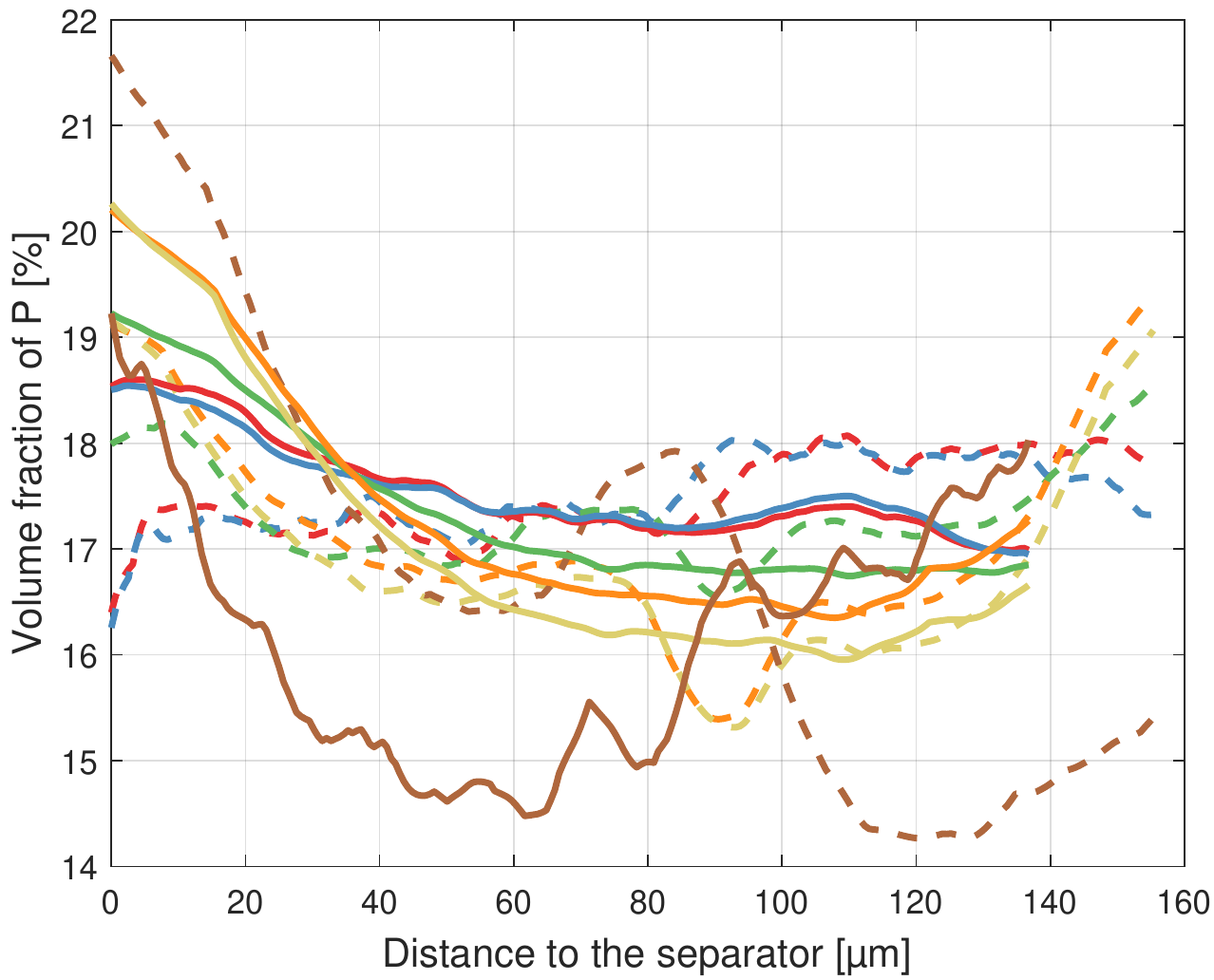}
	\end{subfigure}
	\caption{Volume fraction of active material (left), CBD (center) and pores (right) in dependence of the distance to the separator for the two-layer cathode (dashed curves) and the single-layer cathode (solid curves).}
	\label{fig:volfrac_dist}
\end{figure}

\subsection{Influence of selected trinarization on electrochemical properties}\label{subsec:results_ec}

The influence of the selected trinarization approach on the electrochemical simulations is investigated using half-cell lithiation simulations and symmetrical impedance simulations. The only free parameter to achieve a good agreement between experiments and simulations is the effective transport parameter $\tau_{\mathsf{CBD}}$ within the electrolyte part of the CBD.
Relevant transport mechanisms in a thick NMC electrode are the electronic conductivity through the solid phase and the lithium transport through the electrolyte. Both quantities strongly depend on the distribution and morphology of the CBD. Especially, the electrolyte transport depends on the local effective tortuosity in the CBD. The electronic conductivity depends both on the conductive network of the CBD and the conductivity of the active material, were the latter is additionally dependent on the state of charge. However, at larger CBD contents losses due to electronic transport are minor compared to transport losses in the electrolyte. Therefore, we use the lithiation simulation at a current of $\SI{6}{\milli\ampere\per\cm^{2}}$ to identify the local effective tortuosity of the CBD that leads to the best agreement between experiment and simulations. Note, at low CBD contents this assumption can be invalid and contribution of the two processes cannot be deconvoluted unambiguously.
The best matching effective transport parameters are identified for both electrode types (single-layer cathode and two-layer cathode) with three cutouts each and all trinarization approaches except for NN-CBD-P-AM and NN-P-CBD-AM.
In previous studies we have shown that the EDX-closing trinarization is able to provide a reasonable agreement between electrochemical measurements and simulations \cite{kremer.2020,Kremer.2020b}. Figure~\ref{fig:EC:EDX_Gamma} visualizes the impact of the effective $\gamma$-parameter on the lithiation simulation for the EDX-closing trinarizations of the two-layer electrode in comparison to the experimental results. Lithiation curves with a current of $\SI{6}{\milli\ampere\per\cm^{2}}$ serving as target for our parameter optimization are highlighted by the green symbols. \\
\begin{figure}[!htbp]
	\begin{subfigure}[c]{0.48\textwidth}
		\centering
		\includegraphics[width=0.9\textwidth]{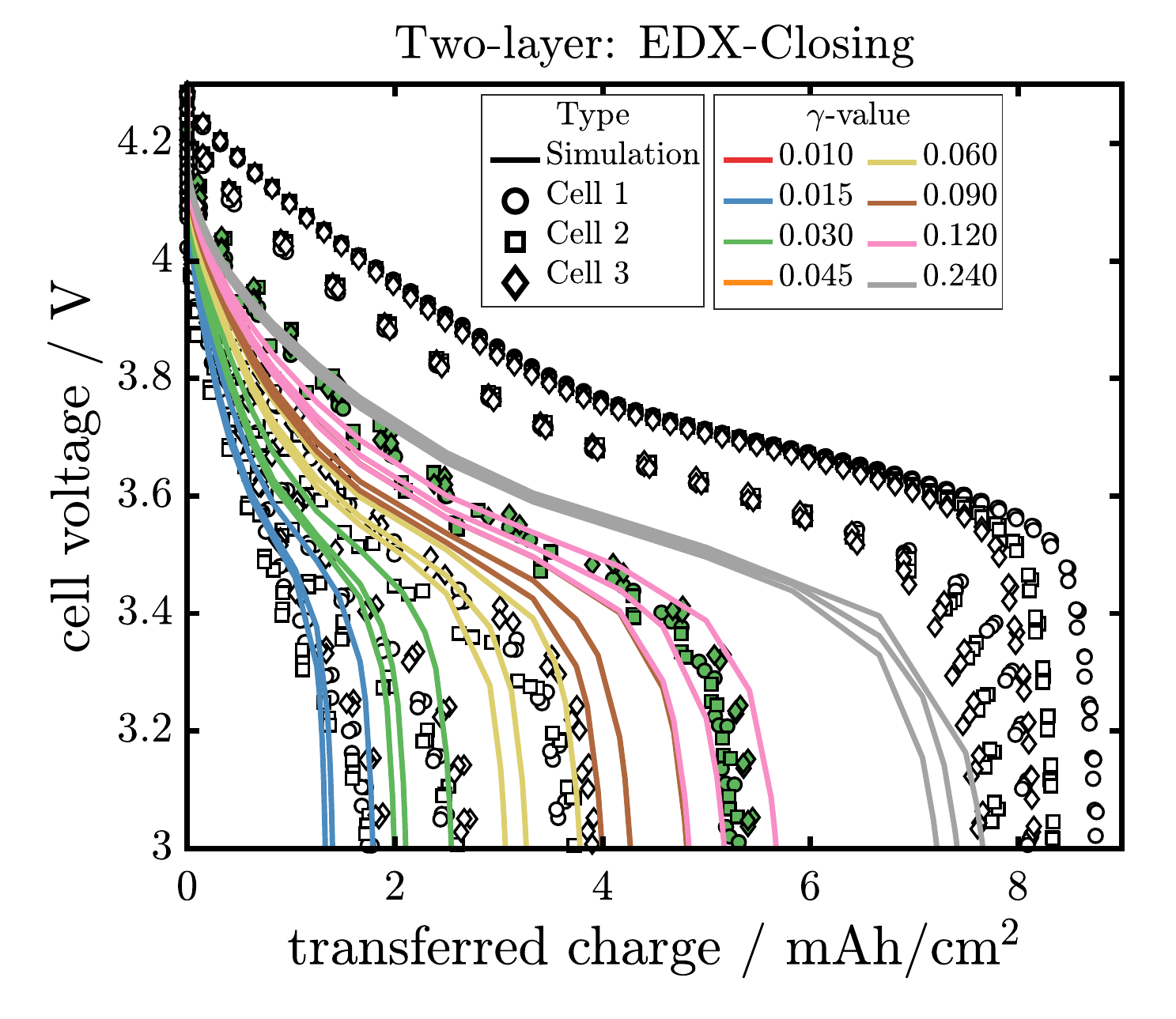}	
		\caption{Impact of the effective parameter $\gamma$ on the simulated cell voltage for the EDX-Closing trinarization at $\SI{6}{\milli\ampere\per\cm^{2}}$. The target experiments are highlighted with filled green symbols.}
	\label{fig:EC:EDX_Gamma}
	\end{subfigure}
	\hfill
	\begin{subfigure}[c]{0.48\textwidth}
		\centering
		\includegraphics[width=0.9\textwidth]{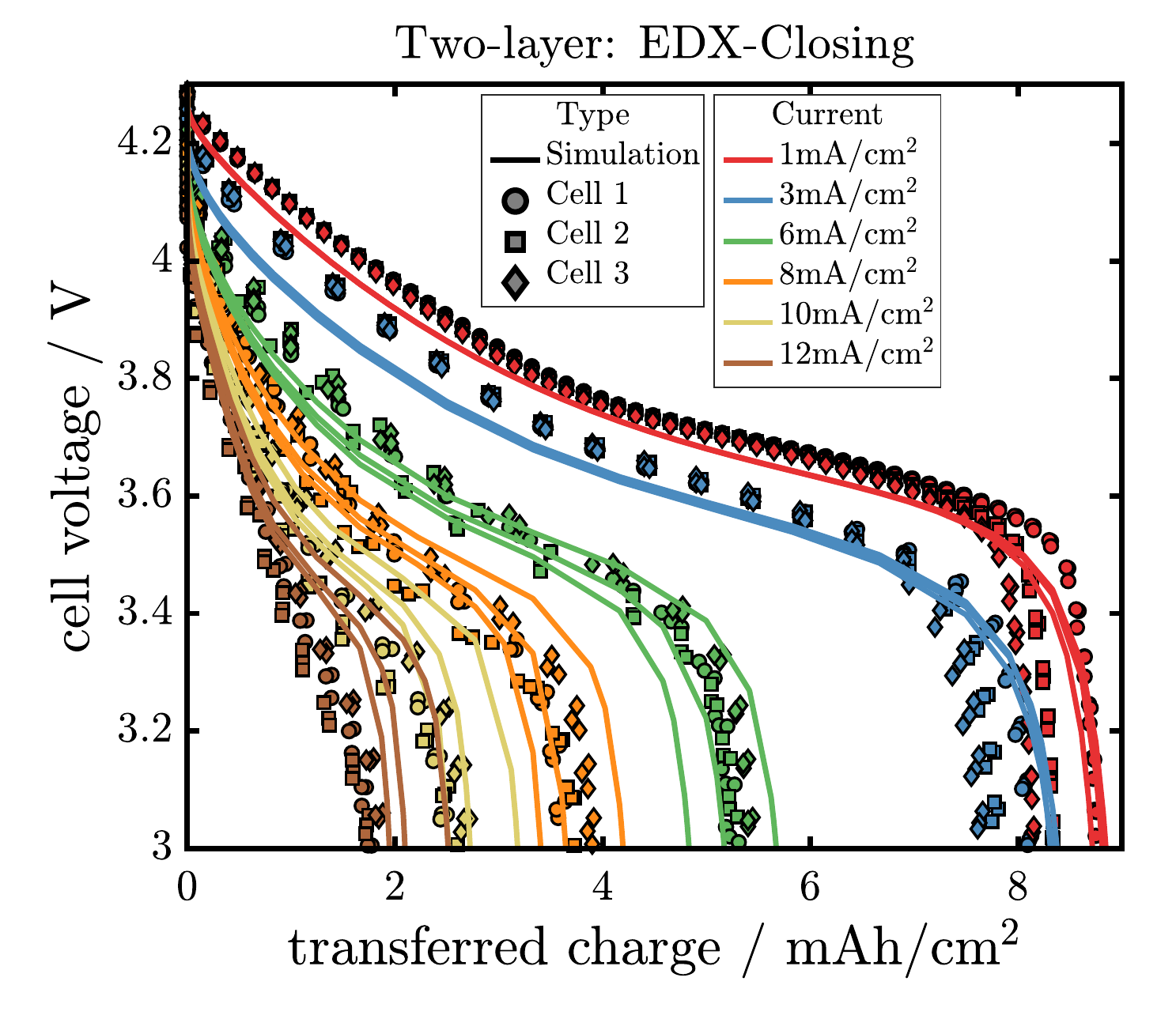}	
		\caption{The simulated cell voltage of the EDX-Closing trinarization for the best matching effective tortuosity ($\gamma=0.24$) for all currents.}
	\label{fig:EC:EDX}
	\end{subfigure}
	\\
	\hfill
	\begin{subfigure}[c]{0.48\textwidth}
		\centering
		\includegraphics[width=0.9\textwidth]{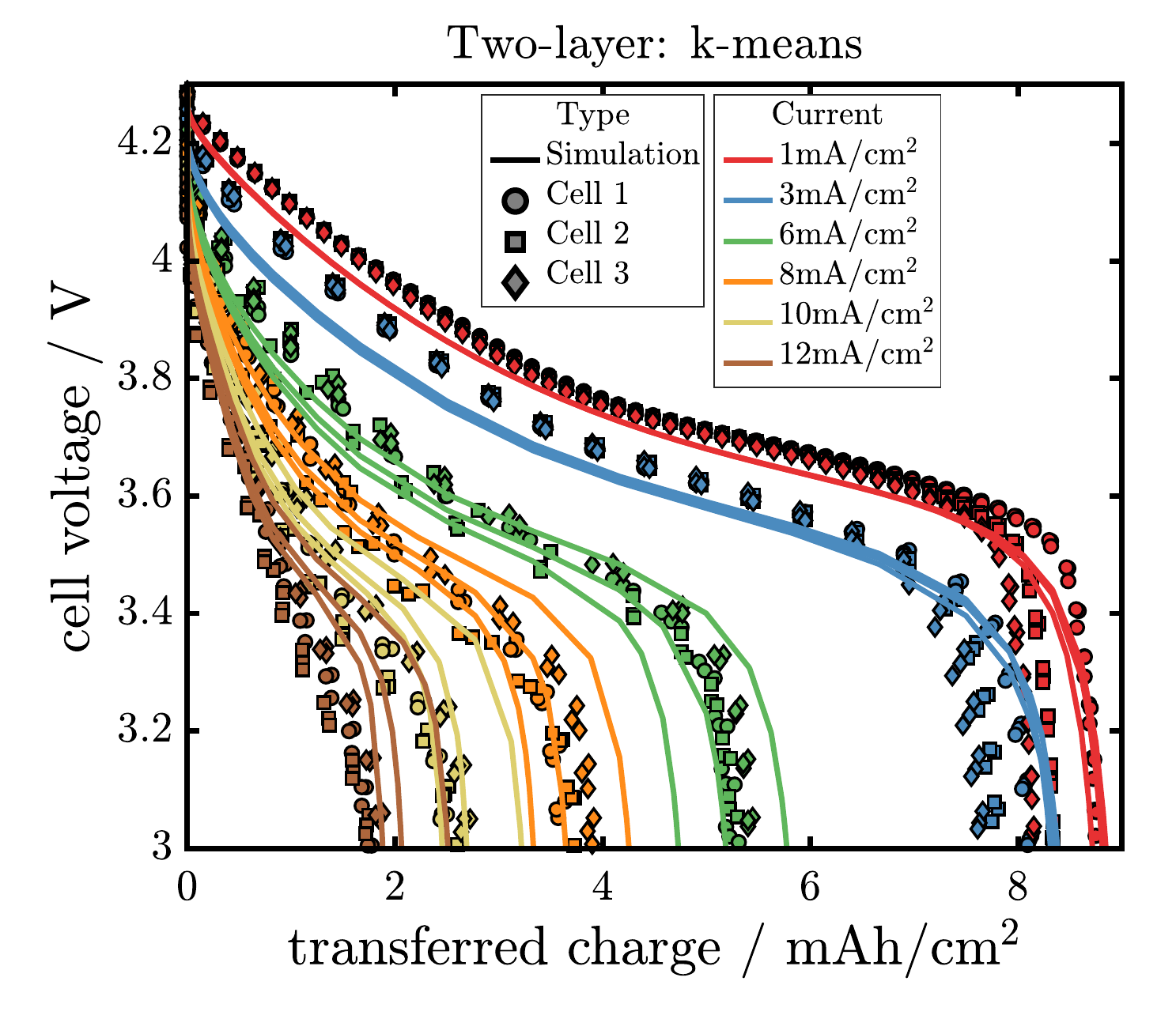}	
		\caption{The simulated cell voltage for the best matching effective tortuosity ($\gamma=0.06$) for all currents for the $k$-means trinarization.}
	\label{fig:EC:KMEANS_Gamma}
	\end{subfigure}
	\hfill
	\caption{Impact of the effective transport through the CBD for the two-layer electrode.}
	\label{fig:EC:selectionexample}
\end{figure}

The impact of the spatial distribution of the CBD on the cell voltage and the achievable lihtiation capacity is apparent.
A smaller value for the effective transport parameter $\gamma$ will reduce the achievable lihtiation capacity of the simulated electrode. 
In turn increasing the value of $\gamma$ reduces the transport resistance in the electrolyte allowing to access larger electrode capacity. 
A value of $\gamma=0.12$ provides the best match between simulations and experiments for the two-layer electrode created using the EDX-closing method presented in Figure~\ref{fig:EC:selectionexample}. 
This parameter value corresponds to a local effective tortuosity of the electrolyte phase of the CBD of $4.2$.\\

The simulation results for all six currents for the selected effective transport parameter ($\gamma_{EDX}^{two-layer}=0.12$) within the CBD are shown in Figure~\ref{fig:EC:EDX}.
The numerical results show some spread for higher currents due to local fluctuations in the three electrode cutouts. 
Nevertheless, the simulated cell voltages are in excellent agreement with the experimental data for all currents. 
However, as shown in Figure~\ref{fig:EC:KMEANS_Gamma} applying the same procedure to the k-means trinarization will result in a similar match between experiments and simulations. 
In this case, the resulting effective tortuosity of the CBD is somewhat larger ($\gamma=0.06$, $\tau$=8.3). 
Similar results can be reported for all cases studied in this work. The figures used for both electrodes and all trinarizations to select the best matching effective tortuosity are shown in the Supplementary Information, see Figure~\ref{SI:fig:EC:variTau}.
The corresponding values for the ten different trinarizations and the two different electrode types are also listed in Table~\ref{tab:tau}.
\begin{table}[!htbp]
    \centering
    \begin{tabular}{lcccc}
    \toprule
    Method & \multicolumn{2}{c}{Single-layer} & \multicolumn{2}{c}{Two-layer}\\
      & $\frac{1}{\tau_{SL}}$& $\tau_{SL}$ &$\frac{1}{\tau_{TL}}$ & $\tau_{TL}$ \\ \midrule
    $k$-means       & 0.02 & 50   & 0.12 & 8.3\\
    EDX-Closing          & 0.18 & 5.6  & 0.24 & 4.2\\
    Thresholding & 0.02 & 50   & 0.12 & 8.3\\
    NN-AM-CBD-P       & 0.09 & 11.1 & 0.18 & 5.6\\
    NN-AM-Cond       & 0.09 & 11.1 & 0.18 & 5.6\\
    NN-AM-P-CBD       & 0.03 & 33.3 & 0.18 & 5.6\\
    NN-CBD-AM-P       & 0.09 & 11.1 & 0.18 & 5.6\\
    NN-CBD-Cond        & 0.09 & 11.1 & 0.18 & 5.6\\
    NN-P-AM-CBD       & 0.03 & 33.3 & 0.18 & 5.6\\
    NN-P-Cond        & 0.03 & 33.3 & 0.18 & 5.6\\
    \bottomrule
    \end{tabular}
    \caption{List of effective tortuosity values providing the best match to the corresponding electrochemical data.}
    \label{tab:tau}
\end{table}
The impact of the trinarization on the electrode performance differs between the methods investigated in this work. Yet, the two-layer electrode and the single-layer electrode exhibit the same trends. A smaller effective tortuosity indicates that the CBD is distributed in the electrode such that even a small local transport resistance will reduce the overall transport through the electrode. The EDX-Closing trinarization leads to the smallest effective tortuosities for the single-layer and two-layer electrodes due to the distribution of the CBD at the bottlenecks of the active material microstructure. The $k$-means and thresholding approach, on the other hand, result in the largest effective tortuosities, which implies that the spatial distribution of CBD created by these trinarization methods is not fully covering the bottlenecks for the electrolyte transport. The results obtained for the trinarization based on neural networks are qualitatively in between these two extremes.\\
Additional analytical techniques are required to probe the influence of the CBD distribution. 
As shown above the distribution and corresponding effective  tortuosity values have a significant influence on lithium ion transport in the electrolyte. 
Impedance spectroscopy on symmetrical cells in blocking conditions has become a standard tool for the characterization of the pore transport resistance.\cite{Usseglio-Viretta.2018,Sandherr.2022,Luo.2022} 
Therefore, we additionally performed impedance simulations on symmetrical cells to investigate the impact of the different trinarization methods. 
The corresponding impedance spectra for the single-layer and two-layer cathode are shown in Figure~\ref{fig:EC:imp:sl} and ~\ref{fig:EC:imp:tl}, respectively. However, the different trinarizations result in very similar impedance spectra which will not allow to discern distribution related effects in corresponding impedance measurements. Therefore, also the electrode impedance does not provide a hint on the most favorable trinarization method.\\
In summary, we demonstrate that it is possible to identify one effective tortuosity per electrode type and trinarization method such that the simulations are in fair agreement with the experimental data for all currents. However, there are large variations in the effective tortuosity of the CBD between the different trinarization methods. None of the individual techniques is able to provide a consistent representation for all electrode samples investigated in this work. Hence, we could not determine the trinarization method providing the best representation of the electrode microstructure. High-resolution image data of the CBD might yield additional information on the effective CBD conductivity which then eventually will allow to choose the most suitable trinarization technique.

\section{Conclusion and outlook}\label{sec:conclusion}

In the present paper, 3D image data of a single-layer and a two-layer cathodes obtained by synchrotron tomography has been segmented into active material, the carbon-binder domain and the pore space by four different approaches, where the approach based on correlative microscopy allows for nine different trinarizations by altering the way of converting the material composition predicted by the neural network to a three-phase reconstruction. 
The different segmentation approaches, which are designed to match the experimentally determined volume fractions, are quantitatively compared by means of statistical image analysis as well as spatially and temporally resolved simulations of electrochemical properties. 
It turns out that there are non-negligible differences between the proposed trinarization approaches. Among others, the geodesic tortuosity as well as the continuous phase size distribution of both - the CBD and the pores - depend on the chosen segmentation approach. 
Furthermore, it has been shown that there are clear differences between the trinarizations obtained by correlative microscopy. 
Thus, the rule for converting the material composition predicted by the neural network to a three-phase reconstruction is of importance, even though the differences compared to the remaining three approaches are more pronounced. 
However, a high level of agreement between the experimental measurements and the lithiation simulations can be achieved for all trinarization methods by adjusting the effective transport parameter of the carbon-binder domain. 
Note that using a fixed current for fitting this parameter allows us to match the experimental curves for five different currents, which indicates that each trinarization approach is reasonable. 
By doing so, the effective tortuosity within the CBD is restricted to the interval $[4.2,50]$.
This large range indicates that further research is required to determine the best trinarization approach. For example, the high-resolution 3D FIB-SEM data could be used to quantitatively investigate ionic transport within the nanopores. Nevertheless, this approach based on spatially resolved numerical simulations allows to predict the optimal spatial distribution of the CBD in lithium-ion battery electrodes, leading to an improved electrochemical performance

\section*{Acknowledgements}

The presented work was financially supported by the German Ministry \myquote{Bundesministerium f\"{u}r Bildung und Forschung} within the projects HighEnergy and HiStructures under the reference numbers 03XP0073C and 03XP0243C/D/E as well as within the framework of the program \myquote{Vom Material zur Innovation}. This study contributes to the research performed at CELEST (Center for Electrochemical Energy Storage Ulm Karlsruhe). The work by MN was partially funded by the German Research Foundation (DFG) under Project ID 390874152 (POLiS Cluster of Excellence, EXC 2154).
The authors acknowledge support by the state of Baden-W\"{u}rttemberg through bwHPC and the German Research Foundation (DFG) through grant no INST 40/575-1 FUGG (JUSTUS 2 cluster). We thank Christian
Dreer for working out the production process for the single-layer and two-layer electrodes and their manufacturing and Claudia Pfeifer for the preparation and EDX analysis of the electrode cross sections. All
responsibility for the content of this publication is assumed by the authors.

\section*{Data availability}
The raw/processed data required to reproduce these findings cannot be shared at this time as the data also forms part of an ongoing study.

\DeclareFieldFormat{pages}{#1}
\printbibliography

@article{Schmitt2020,
author = {Schmitt, Tobias and Latz, Arnulf and Horstmann, Birger},
doi = {10.1016/j.electacta.2019.135491},
issn = {00134686},
journal = {Electrochimica Acta},
pages = {135491},
title = {{Derivation of a local volume-averaged model and a stable numerical algorithm for multi-dimensional simulations of conversion batteries}},
volume = {333},
year = {2020}
}

@article{bockholt.2013,
author = {Bockholt, Henrike and Haselrieder, Wolfgang and Kwade, Arno},
year = {2013},
pages = {25--35},
title = {Intensive Dry and Wet Mixing Influencing the Structural and Electrochemical Properties of Secondary Lithium-Ion Battery Cathodes},
volume = {50},
journal = {ECS Transactions},
}

@article{bockholt.2016,
title={The interaction of consecutive process steps in the manufacturing of lithium-ion battery electrodes with regard to structural and electrochemical properties},
author={Bockholt, H. and Indrikova, M. and Netz, A. and Golks, F. and Kwade, A.},
year={2016},
journal={Journal of Power Sources},
volume={325},
pages={140--151},
}

@inproceedings{buades.2005,
author = {Buades, Antoni and Coll, Bartomeu and Morel, J.-M},
year = {2005},
pages = {60--65},
title = {A non-local algorithm for image denoising},
volume = {2},
booktitle = {Proceedings of the IEEE Computer Society Conference on Computer Vision and Pattern Recognition},
organization={IEEE Computer Society},
address={San Diego},
}

@book{burger.2016,
	title={{D}igital {I}mage {P}rocessing: {A}n {A}lgorithmic {I}ntroduction {U}sing {J}ava},
	author={Burger, W. and Burge, M.J.},
	publisher={Springer},
	year={2016},
	address={London},
	edition = {2\textsuperscript{nd} ed.},
}

@article{cano.2018,
author= {Cano, Zachary P. and Banham, Dustin and Ye, Siyu and Hintennach, Andreas and Lu, Jun and Fowler, Michael and Chen, Zhongwei},
year={2018},
title = {Batteries and fuel cells for emerging electric vehicle markets},
journal = {Nature Energy},
pages ={279--289},
volume={3},
number={4},
}

@article{celebi.2013,
title = {A comparative study of efficient initialization methods for the k-means clustering algorithm},
journal = {Expert Systems with Applications},
volume = {40},
number = {1},
pages = {200--210},
year = {2013},
author = {M. Emre Celebi and Hassan A. Kingravi and Patricio A. Vela},
}

@book{chiu.2013,
 author = {Chiu, S. N. and Stoyan, D. and Kendall, W. S. and Mecke, J.},
 title = {Stochastic Geometry and its Applications},
 year={2013},
 edition={3\textsuperscript{rd} ed.},
 publisher = {J. Wiley \& Sons},
 address = {Chichester},
}

@article{cho.2015,
	author = {Cho, S. and Chen, C.-F. and Mukherjee, P. P.}, 
	title = {Influence of Microstructure on Impedance Response in Intercalation Electrodes},
	volume = {162}, 
	number = {7}, 
	pages = {A1202--A1214}, 
	year = {2015}, 
	journal = {Journal of The Electrochemical Society},
}

@article{clennell.1997,
	title={Tortuosity: a guide through the maze},
	author={Clennell, M. B.},
	journal={Geological Society, London, Special Publications},
	volume={122},
	number={1},
	pages={299--344},
	year={1997},
	publisher={Geological Society of London},
}

@article{cooper.2014,
title = {Image based modelling of microstructural heterogeneity in {LiFePO}\textsubscript{4} electrodes for {Li-ion} batteries},
journal = {Journal of Power Sources},
volume = {247},
pages = {1033--1039},
year = {2014},
author = {S.J. Cooper and D.S. Eastwood and J. Gelb and G. Damblanc and D.J.L. Brett and R.S. Bradley and P.J. Withers and P.D. Lee and A.J. Marquis and N.P. Brandon and P.R. Shearing},
}

@article{coupe.2008,
author={P. Coup\'{e} and P. Yger and S. Prima and P. Hellier and C. Kervrann and C. Barillot},
journal={{IEEE} Transactions on Medical Imaging},
title={An Optimized Blockwise Nonlocal Means Denoising Filter for {3-D} Magnetic Resonance Images},
year={2008},
volume={27},
number={4},
pages={425--441},
}

@book{coxeter.1973,
  title={{R}egular {P}olytopes},
  author={Coxeter, H.S.M.},
  year={1973},
  publisher={Dover Publications},
  address = {New York},
}

@article{dominko.2003,
title = {The role of carbon black distribution in cathodes for {Li} ion batteries},
journal = {Journal of Power Sources},
volume = {119--121},
pages = {770--773},
year = {2003},
author = {Robert Dominko and Miran Gaberscek and Jernej Drofenik and Marjan Bele and Stane Pejovnik and Janko Jamnik},
}

@inproceedings{dowd.1999,
author = {Betsy A. Dowd and Graham H. Campbell and Robert B. Marr and Vivek V. Nagarkar and Sameer V. Tipnis and Lisa Axe and D. Peter Siddons},
title = {Developments in synchrotron {X-ray} computed microtomography at the {N}ational {S}ynchrotron {L}ight {S}ource},
volume = {3772},
booktitle = {Developments in X-Ray Tomography II},
editor = {Ulrich Bonse},
organization = {International Society for Optics and Photonics},
publisher = {SPIE},
pages = {224--236},
year = {1999},
}

@book{dubois.2000,
  title={{F}undamentals of {F}uzzy {S}ets},
  author={Dubois, D. and Prade, H.M.},
  year={2000},
  publisher={Kluwer Academic Publishers},
  address={Boston},
}

@article {ebner.2013a,
	author = {Ebner, Martin and Marone, Federica and Stampanoni, Marco and Wood, Vanessa},
	title = {Visualization and Quantification of Electrochemical and Mechanical Degradation in {Li} Ion Batteries},
	volume = {342},
	number = {6159},
	pages = {716--720},
	year = {2013},
	publisher = {American Association for the Advancement of Science},
	journal = {Science}
}

@article{ebner.2013b,
author = {Ebner, Martin and Geldmacher, Felix and Marone, Federica and Stampanoni, Marco and Wood, Vanessa},
title = {{X-}Ray Tomography of Porous, Transition Metal Oxide Based Lithium Ion Battery Electrodes},
journal = {Advanced Energy Materials},
volume = {3},
number = {7},
pages = {845--850},
year = {2013}
}

@article{ghanbarian.2013,
	title={Tortuosity in porous media: a critical review},
	author={Ghanbarian, B. and Hunt, A.~G. and Ewing, R. P. and Sahimi, M.},
	journal={Soil Science Society of America Journal},
	volume={77},
	number={5},
	pages={1461--1477},
	year={2013},
}

@book{gonzalez.2008,
  title={{D}igital {I}mage {P}rocessing},
  author={Gonzalez, Rafael C. and Woods, Richard E.},
  year={2018},
  publisher={Pearson},
  address={New York},
  edition={4\textsuperscript{rd} ed.},
}

@article{goodenough.2013,
author = {Goodenough, John B. and Park, Kyu-Sung},
title = {The {Li}-Ion Rechargeable Battery: {A} Perspective},
journal = {Journal of the American Chemical Society},
volume = {135},
number = {4},
pages = {1167--1176},
year = {2013},
}

@book{goodfellow.2016,
author={I. Goodfellow and Y. Bengio and A. Courville},
year={2016},
title = {{D}eep {L}earning},
publisher={MIT Press},
address={Cambridge},
}

@article{haselrieder.2013,
author={Wolfgang Haselrieder and Stoyan Ivanov and Daniel Klaus Christen and Henrike Bockholt and Arno Kwade},
title={Impact of the Calendering Process on the Interfacial Structure and the Related Electrochemical Performance of Secondary Lithium-Ion Batteries},
journal={ESC Transactions},
year={2013},
volume={50},
number={26},
pages={59--70},
}

@book{hassaballah.2020,
  title={{D}eep {L}earning in {C}omputer {V}ision: {P}rinciples and {A}pplications},
  author={Hassaballah, M. and Awad, A.I.},
  year={2020},
  publisher={CRC Press},
  address={Boca Raton},
}

@article{hein.2020,
	doi = {10.1149/1945-7111/ab6b1d},
	year = {2020},
	publisher = {The Electrochemical Society},
	volume = {167},
	number = {1},
	pages = {013546},
	author = {Simon Hein and Timo Danner and Daniel Westhoff and Benedikt Prifling and Rares Scurtu and Lea Kremer and Alice Hoffmann and Andr{\'{e}} Hilger and Markus Osenberg and Ingo Manke and Margret Wohlfahrt-Mehrens and Volker Schmidt and Arnulf Latz},
	title = {Influence of Conductive Additives and Binder on the Impedance of Lithium-Ion Battery Electrodes: Effect of Morphology},
	journal = {Journal of The Electrochemical Society},
}

@article{holzer.2013b,
 author = {Holzer, L. and Wiedenmann, D. and M\"{u}nch, B. and Keller, L. and Prestat, M. and Gasser, P. and Robertson, I. and Grob\'{e}ty, B.},
 year = {2013},
 title = {The influence of constrictivity on the effective transport properties of porous layers in electrolysis and fuel cells},
 pages = {2934--2952},
 volume = {48},
 journal = {Journal of Materials Science},
}

@article{huang.2009,
title = {Morphology effect on the electrochemical performance of {NiO} films as anodes for lithium ion batteries},
journal = {Journal of Power Sources},
volume = {188},
number = {2},
pages = {588--591},
year = {2009},
author = {X.H. Huang and J.P. Tu and X.H. Xia and X.L. Wang and J.Y. Xiang and L. Zhang and Y. Zhou},
}

@article{huttner.2020,
author = {Huttner, Fabienne and Haselrieder, Wolfgang and Kwade, Arno},
title = {The Influence of Different Post-Drying Procedures on Remaining Water Content and Physical and Electrochemical Properties of Lithium-Ion Batteries},
journal = {Energy Technology},
volume = {8},
number = {2},
pages = {1900245},
year = {2020}
}

@article{hutzenlaub.2013,
title = {{FIB/SEM}-based calculation of tortuosity in a porous {LiCoO}\textsubscript{2} cathode for a {Li-ion} battery},
journal = {Electrochemistry Communications},
volume = {27},
pages = {77--80},
year = {2013},
author = {T. Hutzenlaub and A. Asthana and J. Becker and D.R. Wheeler and R. Zengerle and S. Thiele},
}

@article{jain.2010,
title = {Data clustering: 50 years beyond {K}-means},
journal = {Pattern Recognition Letters},
volume = {31},
number = {8},
pages = {651--666},
year = {2010},
author = {Anil K. Jain},
}

@book{james.2013,
year={2013},
title={{A}n {I}ntroduction to {S}tatistical {L}earning},
author={Gareth James and Daniela Witten and Trevor Hastie and Robert Tibshirani},
publisher={Springer},
address={New York},
}

@book{jungnickel.2007,
  title={{G}raphs, {N}etworks and {A}lgorithms},
  author={Jungnickel, D.},
  year={2007},
  publisher={Springer},
  address={Berlin},
  edition={3\textsuperscript{rd} ed.},
}

@inproceedings{khokhriakov,
author = {Igor Khokhriakov and Lars Lottermoser and Rainer Gehrke and Thorsten Kracht and Eugen Wintersberger and Andreas Kopmann and Matthias Vogelgesang and Felix Beckmann},
title = {Integrated control system environment for high-throughput tomography},
volume = {9212},
booktitle = {Developments in {X-}Ray Tomography {IX}},
editor = {Stuart R. Stock},
organization = {International Society for Optics and Photonics},
publisher = {SPIE},
pages = {307--317},
year = {2014},
doi = {10.1117/12.2060975},
URL = {https://doi.org/10.1117/12.2060975}
}

@book{kincaid.2009,
  title={{N}umerical {A}nalysis: {M}athematics of {S}cientific {C}omputing},
  author={Kincaid, D. and Cheney, E.W.},
  year={2009},
  publisher={American Mathematical Society},
  address={Providence},
}

@book{korthauer.2018,
  title={{L}ithium-{I}on {B}atteries: {B}asics and {A}pplications},
  author={Korthauer, R.},
  year={2018},
  publisher={Springer},
  address={Berlin},
}

@article{kremer.2020,
author = {Kremer, Lea Sophie and Hoffmann, Alice and Danner, Timo and Hein, Simon and Prifling, Benedikt and Westhoff, Daniel and Dreer, Christian and Latz, Arnulf and Schmidt, Volker and Wohlfahrt-Mehrens, Margret},
title = {Manufacturing Process for Improved Ultra-Thick Cathodes in High-Energy Lithium-Ion Batteries},
journal = {Energy Technology},
pages = {1900167},
volume={8},
number = {2},
year={2020},
}

@article{Kremer.2020b,
author = {Kremer, Lea Sophie and Danner, Timo and Hein, Simon and Hoffmann, Alice and Prifling, Benedikt and Schmidt, Volker and Latz, Arnulf and Wohlfahrt‐Mehrens, Margret},
doi = {10.1002/batt.202000098},
issn = {2566-6223},
journal = {Batteries {\&} Supercaps},
month = {nov},
number = {11},
pages = {1172--1182},
title = {Influence of the Electrolyte Salt Concentration on the Rate Capability of Ultra‐Thick {NCM 622} Electrodes},
url = {https://onlinelibrary.wiley.com/doi/10.1002/batt.202000098},
volume = {3},
year = {2020}
}

@book{kroese.2019,
  title={{D}ata {S}cience and {M}achine {L}earning: {M}athematical and {S}tatistical {M}ethods},
  author={Kroese, D.P. and Botev, Z. and Taimre, T. and Vaisman, R.},
  year={2019},
  publisher={CRC Press},
  address={Boca Raton},
}

@article{kuchler.2018b,
	author = {Klaus Kuchler and Benedikt Prifling and Denny Schmidt and Henning Mark\"{o}tter and Ingo Manke and Timo Bernthaler and Volker Knoblauch and Volker Schmidt},
	title = {Analysis of the {3D} microstructure of experimental cathode films for lithium-ion batteries under increasing compaction},
	year={2018},
	journal={Journal of Microscopy},
	volume={272},
	number={2},
	pages={96--110},
}

@article{kumberg.2021,
author = {Kumberg, Jana and Baunach, Michael and Eser, Jochen C. and Altvater, Andreas and Scharfer, Philip and Schabel, Wilhelm},
title = {Investigation of Drying Curves of Lithium-Ion Battery Electrodes with a New Gravimetrical Double-Side Batch Dryer Concept Including Setup Characterization and Model Simulations},
journal = {Energy Technology},
volume = {9},
number = {2},
pages = {2000889},
year = {2021}
}

@article{landesfeind.2018,
author = {Landesfeind, Johannes and Eldiven, Askin and Gasteiger, Hubert A.}, 
title = {Influence of the Binder on Lithium Ion Battery Electrode Tortuosity and Performance},
volume = {165}, 
number = {5}, 
pages = {A1122--A1128}, 
year = {2018}, 
journal = {Journal of The Electrochemical Society} 
}

@article{landesfeind.2019,
author = {Landesfeind, Johannes and Gasteiger, Hubert A.},
doi = {10.1149/2.0571912jes},
issn = {0013-4651},
journal = {Journal of The Electrochemical Society},
number = {14},
pages = {A3079--A3097},
title = {Temperature and Concentration Dependence of the Ionic Transport Properties of Lithium-Ion Battery Electrolytes},
url = {http://jes.ecsdl.org/lookup/doi/10.1149/2.0571912jes},
volume = {166},
year = {2019}
}

@article{latz.2011,
title = {Thermodynamic consistent transport theory of Li-ion batteries},
journal = {Journal of Power Sources},
volume = {196},
number = {6},
pages = {3296--3302},
year = {2011},
author = {A. Latz and J. Zausch},
}

@article{latz.2015,
  title={Multiscale modeling of lithium ion batteries: thermal aspects},
  author={Latz, Arnulf and Zausch, Jochen},
  journal={Beilstein Journal of Nanotechnology},
  volume={6},
  number={1},
  pages={987--1007},
  year={2015},
}

@article{lautenschlaeger.2022,
author = {Lautenschl\"{a}ger, Martin P. and Prifling, Benedikt and Kellers, Benjamin and Weinmiller, Julius and Danner, Timo and Schmidt, Volker and Latz, Arnulf},
title = {Understanding Electrolyte Filling of Lithium-Ion Battery Electrodes on the Pore Scale Using the Lattice {B}oltzmann Method},
journal = {Batteries \& Supercaps},
pages = {e202200090},
doi = {https://doi.org/10.1002/batt.202200090},
year={2022},
}

@article{lenze.2018,
title={Impacts of Variations in Manufacturing Parameters on Performance of Lithium-Ion Batteries},
author={Georg Lenze and Henrike Bockholt and Christiane Schilcher and Linus Frob\"{o}se and Dietmar Jansen and Ulrike Krewer and Arno Kwade},
journal={Journal of The Electrochemical Society},
volume={165},
number={2},
pages={A314--A322},
year={2018},
}

@article{li.1997,
author={Li, W. and Currie, J. C.},
year={1997},
volume={144},
number={8},
journal = {Journal of The Electrochemical Society} ,
pages={2773--2779},
title={Morphology Effects on the Electrochemical performance of {LiNi}\textsubscript{1-x}{Co}\textsubscript{x}{O}\textsubscript{2}},
}

@article{lowe.2004,
author={Lowe, David G.},
year = {2004},
title = {Distinctive Image Features from Scale-Invariant Keypoints},
journal = {International Journal of Computer Vision},
pages = {91--110},
volume = {60},
number = {2},
}

@book{matheron.1975,
author={G. Matheron},
year ={1975},
title={{R}andom {S}ets and {I}ntegral {G}eometry},
publisher={J. Wiley \& Sons},
address={New York},
}

@article{morasch.2018,
author = {Morasch, Robert and Landesfeind, Johannes and Suthar, Bharatkumar and Gasteiger, Hubert A.}, 
title = {Detection of Binder Gradients Using Impedance Spectroscopy and Their Influence on the Tortuosity of {Li-}Ion Battery Graphite Electrodes},
volume = {165}, 
number = {14}, 
pages = {A3459--A3467}, 
year = {2018}, 
journal = {Journal of The Electrochemical Society} 
}

@article{neumann.2020,
author={Neumann, M. and Stenzel, O. and Willot, F. and Holzer, L. and Schmidt, V.},
title={Quantifying the influence of microstructure on effective conductivity and permeability: Virtual materials testing},
journal={International Journal of Solids and Structures},
year={2020},
volume={184},
pages={211--220},
}

@article{neumann.2019TauBeta,
author = {Neumann, M. and Hirsch, C. and Stan\v{e}k, J. and Bene\v{s}, V. and Schmidt, V.},
title = {Estimation of geodesic tortuosity and constrictivity in stationary random closed sets},
journal = {Scandinavian Journal of Statistics},
volume = {46},
number = {3},
pages = {848--884},
year={2019},
doi = {10.1111/sjos.12375},
}

@book{newman.2004,
  title={{E}lectrochemical {S}ystems},
  author={Newman, J. and Thomas-Alyea, K.E.},
  year={2004},
  publisher={J. Wiley \& Sons},
  address={Hoboken},
  edition={3\textsuperscript{rd} ed.},
}

@article{otsu.1979,
  author={Otsu, Nobuyuki},
  journal={IEEE Transactions on Systems, Man, and Cybernetics}, 
  title={A Threshold Selection Method from Gray-Level Histograms}, 
  year={1979},
  volume={9},
  number={1},
  pages={62--66},
}

@article{powell.2009,
  title={The {BOBYQA} algorithm for bound constrained optimization without derivatives},
  author={Powell, Michael J. D.},
  journal={Cambridge NA Report NA2009/06},
  note={University of Cambridge, Cambridge},
  pages={26--46},
  year={2009},
}

@article{prifling.2019b,
title = {Analysis of structural and functional aging of electrodes in lithium-ion batteries during rapid charge and discharge rates using synchrotron tomography},
journal = {Journal of Power Sources},
volume = {443},
pages = {227259},
year = {2019},
author = {Benedikt Prifling and Alexander Ridder and André Hilger and Markus Osenberg and Ingo Manke and Kai Peter Birke and Volker Schmidt},
}

@article{prifling.2021,
author={Prifling, Benedikt and R\"{o}ding, Magnus and Townsend, Philip and Neumann, Matthias and Schmidt, Volker},   
title={Large-Scale Statistical Learning for Mass Transport Prediction in Porous Materials Using 90,000 Artificially Generated Microstructures},    
journal={Frontiers in Materials},
volume={8},  
year={2021}, 
DOI={10.3389/fmats.2021.786502},
pages={786502},
}

@book{rojas.2013,
  title={{N}eural {N}etworks: {A} {S}ystematic {I}ntroduction},
  author={Rojas, R.},
  year={2013},
  publisher={Springer},
  address={Berlin},
}

@InProceedings{ronneberger.2015,
author={Ronneberger, Olaf and Fischer, Philipp and Brox, Thomas},
editor={Navab, Nassir and Hornegger, Joachim and Wells, William M. and Frangi, Alejandro F.},
title={{U-Net:} Convolutional Networks for Biomedical Image Segmentation},
booktitle={Medical Image Computing and Computer-Assisted Intervention -- MICCAI 2015},
year={2015},
publisher={Springer International Publishing},
address={Cham},
pages={234--241},
}

@article{rudin.1992,
title = {Nonlinear total variation based noise removal algorithms},
journal = {Physica D: Nonlinear Phenomena},
volume = {60},
number = {1},
pages = {259--268},
year = {1992},
author = {Leonid I. Rudin and Stanley Osher and Emad Fatemi},
}

@book{russ.2007,
  title={The Image Processing Handbook},
  author={Russ, John C.},
  year={2007},
  publisher={CRC Press},
  address ={Boca Raton},
  edition={5\textsuperscript{th} ed.},
}

@article{schindelin.2012,
author = {Schindelin, Johannes and Arganda-Carreras, Ignacio and Frise, Erwin and Kaynig, Verena and Longair, Mark and Pietzsch, Tobias and Preibisch, Stephan and Rueden, Curtis and Saalfeld, Stephan and Schmid, Benjamin and Tinevez, Jean-Yves and White, Daniel James and Hartenstein, Volker and Eliceiri, Kevin and Tomancak, Pavel and Cardona, Albert},
year = {2012},
title = {Fiji: an open-source platform for biological-image analysis},
journal = {Nature Methods},
pages = {676--682},
volume = {9},
}

@inproceedings{schladitz.2007,
author={Katja Schladitz and Joachim Ohser and Werner Nagel},
year={2007},
title={Measuring intrinsic volumes in digital {3D} images},
booktitle={13th International Conference Discrete Geometry for Computer Imagery},
editor={Attila Kuba and L\'{a}szl\'{o} Ny\'{u}l and K\'{a}lm\'{a}n Pal\'{a}gyi},
pages={247--258},
publisher={Springer},
address={Berlin},
}

@book {scrosati.2013,
title = {{L}ithium {B}atteries: {A}dvanced {T}echnologies and {A}pplications},
author = {Scrosati, Bruno and Abraham, K. M. and {van Schalkwijk}, Walter and Hassoun, Jusef},
address = {Hoboken},
publisher = {J. Wiley \& Sons},
year = {2013},
pages = {374},
}

@book{serra.1982,
author={J. Serra},
title={{I}mage {A}nalysis and {M}athematical {M}orphology},
year={1982},
publisher={Academic Press},
address={London},
}

@article{shearing.2010,
title = {Characterization of the 3-dimensional microstructure of a graphite negative electrode from a {Li-ion} battery},
journal = {Electrochemistry Communications},
volume = {12},
number = {3},
pages = {374--377},
year = {2010},
author = {P.R. Shearing and L.E. Howard and P.S. J{\o}rgensen and N.P. Brandon and S.J. Harris},
}

@article{shin.2004,
title = {Influence of microstructure on the electrochemical performance of {LiMn}\textsubscript{2-y-z}{Li}\textsubscript{y}{Ni}\textsubscript{z}{O}\textsubscript{4} spinel cathodes in rechargeable lithium batteries},
journal = {Journal of Power Sources},
volume = {126},
number = {1},
pages = {169--174},
year = {2004},
author = {Youngjoon Shin and A Manthiram},
}

@book{soille.2003,
	title={{M}orphological {I}mage {A}nalysis: {P}rinciples and {A}pplications},
	author={Soille, P.},
	year={2003},
	edition={2\textsuperscript{nd} ed.},
	publisher={Springer},
	address = {New York},
}

@inproceedings{sutskever.2013,
  title = 	 {On the importance of initialization and momentum in deep learning},
  author = 	 {Ilya Sutskever and James Martens and George Dahl and Geoffrey Hinton},
  booktitle = 	 {Proceedings of the 30th International Conference on Machine Learning},
  pages = 	 {1139--1147},
  year = 	 {2013},
  editor = 	 {Sanjoy Dasgupta and David McAllester},
  volume = 	 {28},
  series = 	 {Proceedings of Machine Learning Research},
  address = 	 {Atlanta},
  publisher =    {PMLR},
}

@article{thorat.2009,
title = {Quantifying tortuosity in porous {L}i-ion battery materials},
journal = {Journal of Power Sources},
volume = {188},
pages = {592 -- 600},
year = {2009},
author = {Indrajeet V. Thorat and David E. Stephenson and Nathan A. Zacharias and Karim Zaghib and John N. Harb and Dean R. Wheeler},
}

@article{tjaden.2018,
author = {B. Tjaden and D. J. L. Brett and P. R. Shearing},
title = {Tortuosity in electrochemical devices: a review of calculation approaches},
journal = {International Materials Reviews},
volume = {63},
number = {2},
pages = {47--67},
year  = {2018},
publisher = {Taylor & Francis},
}

@article{tran.2012,
author={Tran, Martino and Banister, David and Bishop, Justin D. K. and McCulloch, Malcolm D.},
year={2012},
title={Realizing the electric-vehicle revolution},
journal = {Nature Climate Change},
pages = {328--333},
volume={2},
number={5},
}

@article{trembacki.2017,
title = {Mesoscale Effective Property Simulations Incorporating Conductive Binder},
author = {Trembacki, Bradley L. and Noble, David R. and Brunini, Victor E. and Ferraro, Mark E. and Roberts, Scott A.},
journal = {Journal of the Electrochemical Society},
number = {11},
volume = {164},
year = {2017},
pages={E3613--E3626}
}

@article{vierrath.2015,
title = {Morphology of nanoporous carbon-binder domains in {Li-ion} batteries -- A {FIB-SEM} study},
journal = {Electrochemistry Communications},
volume = {60},
pages = {176--179},
year = {2015},
author = {Severin Vierrath and Lukas Zielke and Riko Moroni and Andrew Mondon and Dean R. Wheeler and Roland Zengerle and Simon Thiele},
}

@article{wiedemann.2013,
	title = {Effects of three-dimensional cathode microstructure on the performance of lithium-ion battery cathodes},
	journal = {Electrochimica Acta},
	volume = {88},
	pages = {580--588},
	year = {2013},
	author = {A. H. Wiedemann and G. M. Goldin and S. A. Barnett and H. Zhu and R. J. Kee},
}

@article{wilde.2016,
author = {Wilde, Fabian and Ogurreck, Malte and Greving, Imke and Hammel, J\"{o}rg U. and Beckmann, Felix and Hipp, Alexander and Lottermoser, Lars and Khokhriakov, Igor and Lytaev, Pavel and Dose, Thomas and Burmester, Hilmar  and M\"{u}ller, Martin  and Schreyer, Andreas},
title = {{Micro-CT} at the imaging beamline {P05} at {PETRA III}},
journal = {AIP Conference Proceedings},
volume = {1741},
number = {1},
pages = {030035},
year = {2016},
doi = {10.1063/1.4952858},
}

@article{yan.2012,
author = {Yan, Bo and Lim, Cheolwoong and Yin, Leilei and Zhu, Likun}, 
title = {Three Dimensional Simulation of Galvanostatic Discharge of {LiCoO}\textsubscript{2} Cathode Based on {X}-ray Nano{-CT} Images},
volume = {159}, 
number = {10}, 
pages = {A1604--A1614}, 
year = {2012}, 
journal = {Journal of The Electrochemical Society} 
}

@article{zheng.2012,
journal = {Journal of Power Sources},
title={Calendering effects on the physical and electrochemical properties of {Li}[{Ni}\textsubscript{1/3}{Mn}\textsubscript{1/3}{Co}\textsubscript{1/3}]{O}\textsubscript{2} cathode},
volume = {208},
pages = {52--57},
year = {2012},
author = {Honghe Zheng and Li Tan and Gao Liu and Xiangyun Song and Vincent S. Battaglia},
}

@article{zielke.2014,
author = {Zielke, Lukas and Hutzenlaub, Tobias and Wheeler, Dean R. and Manke, Ingo and Arlt, Tobias and Paust, Nils and Zengerle, Roland and Thiele, Simon},
title = {A Combination of {X-}Ray Tomography and Carbon Binder Modeling: Reconstructing the Three Phases of {LiCoO}\textsubscript{2} {Li-}Ion Battery Cathodes},
journal = {Advanced Energy Materials},
volume = {4},
number = {8},
pages = {1301617},
year = {2014}
}

@article{zielke.2015,
author = {Zielke, Lukas and Hutzenlaub, Tobias and Wheeler, Dean R. and Chao, Chien-Wei and Manke, Ingo and Hilger, Andr\'{e} and Paust, Nils and Zengerle, Roland and Thiele, Simon},
title = {Three-Phase Multiscale Modeling of a {LiCoO}\textsubscript{2} Cathode: Combining the Advantages of {FIB-SEM} Imaging and {X}-Ray Tomography},
journal = {Advanced Energy Materials},
volume = {5},
number = {5},
pages = {1401612},
year = {2015},
}

@book{zimmermann.2001,
  title={{F}uzzy {S}et {T}heory and {I}ts {A}pplications},
  author={Zimmermann, H.J.},
  year={2001},
  publisher={Kluwer Academic Publishers},
  address={Boston},
  edition={4\textsuperscript{th} ed.},
}

@article{neumann.2022,
title = {{3D} microstructure characterization of polymer battery electrodes by statistical image analysis based on synchrotron {X-ray} tomography},
journal = {Journal of Power Sources},
volume = {542},
pages = {231783},
year = {2022},
author = {Neumann, M. and Ademmer, M. and Osenberg, M. and Hilger, A. and Wilde, F. and Muench, S. and Hager, M.~D. and Schubert, U. S. and Manke, I. and Schmidt, V.},
}

@article{cadiou.2020,
	title={Multiscale characterization of composite electrode microstructures for high density lithium-ion batteries guided by the specificities of their electronic and ionic transport mechanisms},
	author={Cadiou, F. and Douillard, T. and Besnard, N. and Lestriez, B. and Maire, E.},
	journal={Journal of The Electrochemical Society},
	volume={167},
	number={10},
	pages={100521},
	year={2020},
}

@article{xu.2021,
	title={Guiding the Design of Heterogeneous Electrode Microstructures for {Li}-Ion Batteries: Microscopic Imaging, Predictive Modeling, and Machine Learning},
	author={Xu, H. and Zhu, J. and Finegan, D. P. and Zhao, H. and Lu, X. and Li, W. and Hoffman, N. and Bertei, A. and Shearing, P. and Bazant, M. Z.},
	journal={Advanced Energy Materials},
	volume={11},
	number={19},
	pages={2003908},
	year={2021},
}

@article{wagner.2020,
	title = {Hierarchical structuring of {NCM111}-cathode materials in lithium-ion batteries: An in-depth study of the influence of primary and secondary particle size effects on electrochemical performance},
	author = {Wagner, A. and Bohn, N. and Ge{\ss}wein, H. and Neumann, M. and Osenberg, M. and Hilger, A. and Manke, I. and Schmidt, V. and Binder, J. R.},
	journal={ACS Applied Energy Materials},
	year={2020},
	pages={12565--12574},
	volume={3},
}

@article{kroll.2021,
  title={Three-Phase Reconstruction Reveals How the Microscopic Structure of the Carbon-Binder Domain Affects Ion Transport in Lithium-Ion Batteries},
  author={Kroll, M. and Karstens, S. L. and Cronau, M. and H\"{o}ltzel, A. and Schlabach, S. and Nobel, N. and Redenbach, C. and Roling, B. and Tallarek, U.},
  journal={Batteries \& Supercaps},
  volume={4},
  number={8},
  pages={1363--1373},
  year={2021},
}

@article{krygier.2021,
  title={Quantifying the unknown impact of segmentation uncertainty on image-based simulations},
  author={Krygier, M. C. and LaBonte, T. and Martinez, C. and Norris, C. and Sharma, K. and Collins, L. N. and Mukherjee, P. P. and Roberts, S. A.},
  journal={Nature Communications},
  volume={12},
  pages={5414},
  year={2021}
}

@article{Usseglio-Viretta.2018,
author = {Usseglio-Viretta, Francois L. E. and Colclasure, Andrew and Mistry, Aashutosh N. and Claver, Koffi Pierre Yao and Pouraghajan, Fezzeh and Finegan, Donal P. and Heenan, Thomas M. M. and Abraham, Daniel and Mukherjee, Partha P. and Wheeler, Dean and Shearing, Paul R. and Cooper, Samuel J. and Smith, Kandler},
doi = {10.1149/2.0731814jes},
issn = {0013-4651},
journal = {Journal of The Electrochemical Society},
number = {14},
pages = {A3403--A3426},
title = {Resolving the Discrepancy in Tortuosity Factor Estimation for Li-Ion Battery Electrodes through Micro-Macro Modeling and Experiment},
url = {http://jes.ecsdl.org/lookup/doi/10.1149/2.0731814jes},
volume = {165},
year = {2018}
}

@article{Sandherr.2022,
author = {Sandherr, Jens and Nester, Sara and Kleefoot, Max-Jonathan and Bolsinger, Marius and Weisenberger, Christian and Haghipour, Amir and Harrison, David K. and Ruck, Simon and Riegel, Harald and Knoblauch, Volker},
doi = {10.2139/ssrn.4135362},
journal = {SSRN Electronic Journal},
title = {Improving the Ionic Transport Properties of Graphite Anodes for Lithium Ion Batteries by Surface Modification Using Nanosecond Laser},
year = {2022}
}

@article{Luo.2022,
author = {Luo, Yuting and Bai, Yang and Mistry, Aashutosh and Zhang, Yuwei and Zhao, Dexin and Sarkar, Susmita and Handy, Joseph V. and Rezaei, Shahed and Chuang, Andrew Chihpin and Carrillo, Luis and Wiaderek, Kamila and Pharr, Matt and Xie, Kelvin and Mukherjee, Partha P. and Xu, Bai Xiang and Banerjee, Sarbajit},
doi = {10.1038/s41563-021-01151-8},
issn = {14764660},
journal = {Nature Materials},
number = {2},
pages = {217--227},
pmid = {34824396},
publisher = {Springer US},
title = {{Effect of crystallite geometries on electrochemical performance of porous intercalation electrodes by multiscale operando investigation}},
volume = {21},
year = {2022}
}

\newpage

\section*{Supporting Information}

\setcounter{table}{0}
\renewcommand{\thetable}{S\arabic{table}}%
\setcounter{figure}{0}
\renewcommand{\thefigure}{S\arabic{figure}}%
\setcounter{section}{1}
\renewcommand{\thesection}{SI \arabic{section}}

\subsection{Further trinarizations}

\begin{table}[!htbp]
    \centering
    \begin{tabular}{lccccccc}
    \toprule
        & Sample & AM-C-P & AM-P-C & C-P-AM & C-AM-P & P-AM-C & P-C-AM \\ \midrule
        $S_{\text{AM}}\ [\mu m^{-1}]$ & TL & 0.925 & 0.925 & 0.925 & 0.925 & 0.925 & 0.925 \\
        $S_{\text{AM}}\ [\mu m^{-1}]$ & SL & 0.925 & 0.925 & 0.925 & 0.925 & 0.925 & 0.924 \\ \midrule
        $S_{\text{CBD}}\ [\mu m^{-1}]$ & TL & 0.361 & 0.359 & 0.361 & 0.361 & 0.359 & 0.360 \\
        $S_{\text{CBD}}\ [\mu m^{-1}]$ & SL & 0.362 & 0.361 & 0.362 & 0.362 & 0.361 & 0.362 \\ \midrule
        $S_{\text{P}}\ [\mu m^{-1}]$ & TL & 0.271 & 0.273 & 0.271 & 0.271 & 0.273 & 0.273 \\
        $S_{\text{P}}\ [\mu m^{-1}]$ & SL & 0.271 & 0.272 & 0.271 & 0.271 & 0.272 & 0.272 \\ \midrule
        $S_{\text{Int}}$ & TL & 0.671 & 0.672 & 0.672 & 0.671 & 0.672 & 0.672 \\
        $S_{\text{Int}}$ & SL & 0.671 & 0.671 & 0.671 & 0.671 & 0.671 & 0.671 \\ \midrule
        $r_{\text{min,AM}}\ [\mu m]$ & TL & 1.81 & 1.81 & 1.77 & 1.81 & 1.81 & 1.86 \\
        $r_{\text{min,AM}}\ [\mu m]$ & SL  & 1.83 & 1.83 & 1.79 & 1.83 & 1.83 & 1.84 \\ \midrule
        $r_{\text{min,CBD}}\ [\mu m]$ & TL  & 0.75 & 0.21 & 0.75 & 0.75 & 0.21 & 0.23 \\
        $r_{\text{min,CBD}}\ [\mu m]$ & SL  & 0.73 & 0.22 & 0.73 & 0.73 & 0.22 & 0.25 \\ \midrule
        $r_{\text{min,P}}\ [\mu m]$ & TL & 0.21 & 0.70 & 0.21 & 0.21 & 0.70 & 0.70 \\
        $r_{\text{min,P}}\ [\mu m]$ & SL & 0.21 & 0.69 & 0.21 & 0.21 & 0.69 & 0.69 \\ \midrule
        $r_{\text{max,AM}}\ [\mu m]$ & TL & 3.11 & 3.11 & 2.92 & 3.10 & 3.11 & 3.00 \\
        $r_{\text{max,AM}}\ [\mu m]$ & SL & 3.19 & 3.19 & 3.01 & 3.19 & 3.19 & 3.06 \\ \midrule
        $r_{\text{max,CBD}}\ [\mu m]$ & TL & 1.23 & 0.48 & 1.23 & 1.23 & 0.48 & 0.70 \\
        $r_{\text{max,CBD}}\ [\mu m]$ & SL & 1.14 & 0.50 & 1.14 & 1.14 & 0.50 & 0.74 \\ \midrule
        $r_{\text{max,P}}\ [\mu m]$ & TL & 0.56 & 1.16 & 0.65 & 0.57 & 1.16 & 1.16 \\
        $r_{\text{max,P}}\ [\mu m]$ & SL & 0.60 & 1.15 & 0.69 & 0.61 & 1.15 & 1.15 \\
        \bottomrule
    \end{tabular}
    \caption{Scalar microstructure characteristics for different trinarization approaches.}
    \label{tab:scalar_si}
\end{table}

\begin{figure}[!htbp]
	\begin{subfigure}[c]{0.32\textwidth}
		\centering
		\includegraphics[width=0.9\textwidth]{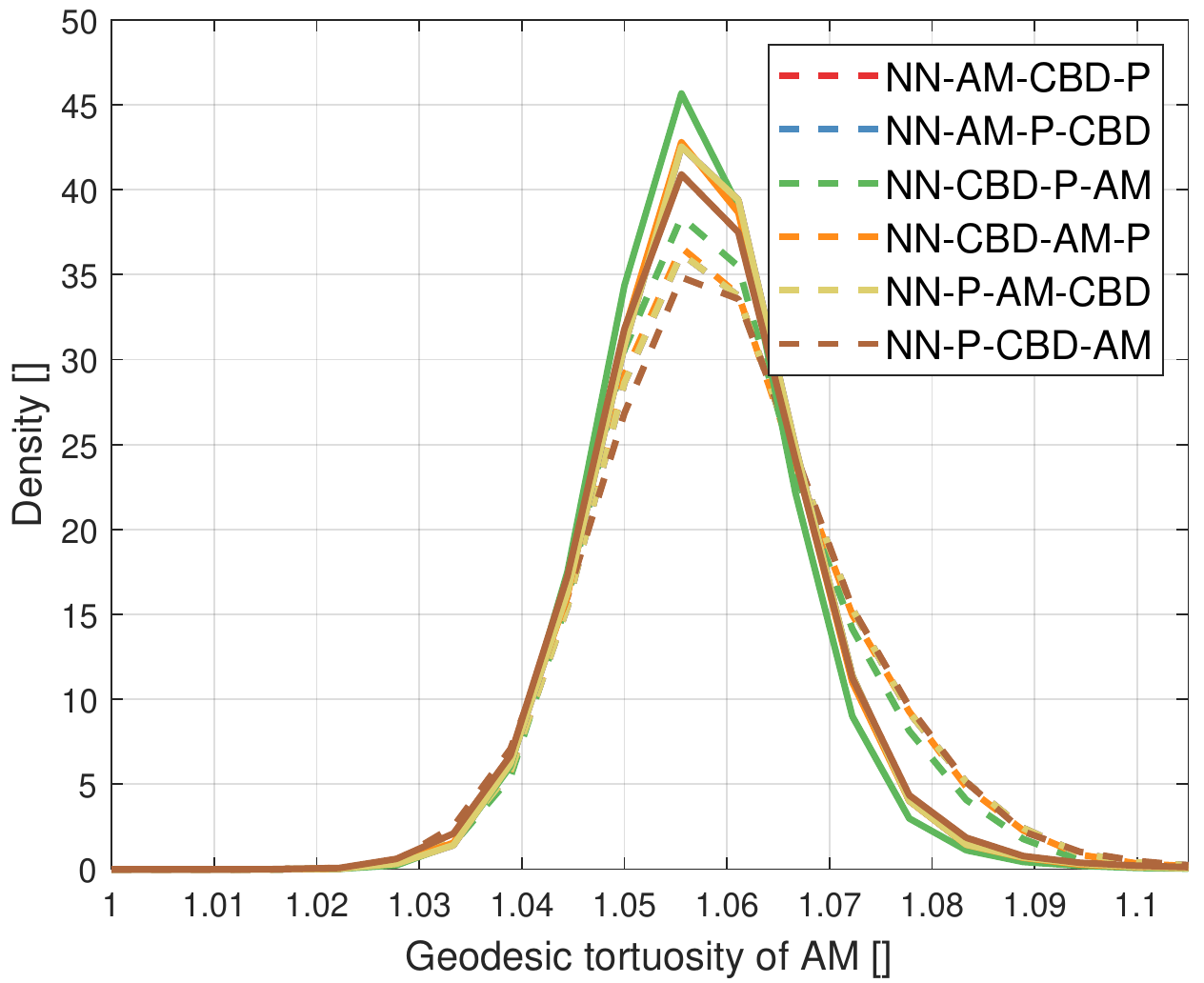}
	\end{subfigure}
	\hfill
	\begin{subfigure}[c]{0.32\textwidth}
		\centering
		\includegraphics[width=0.9\textwidth]{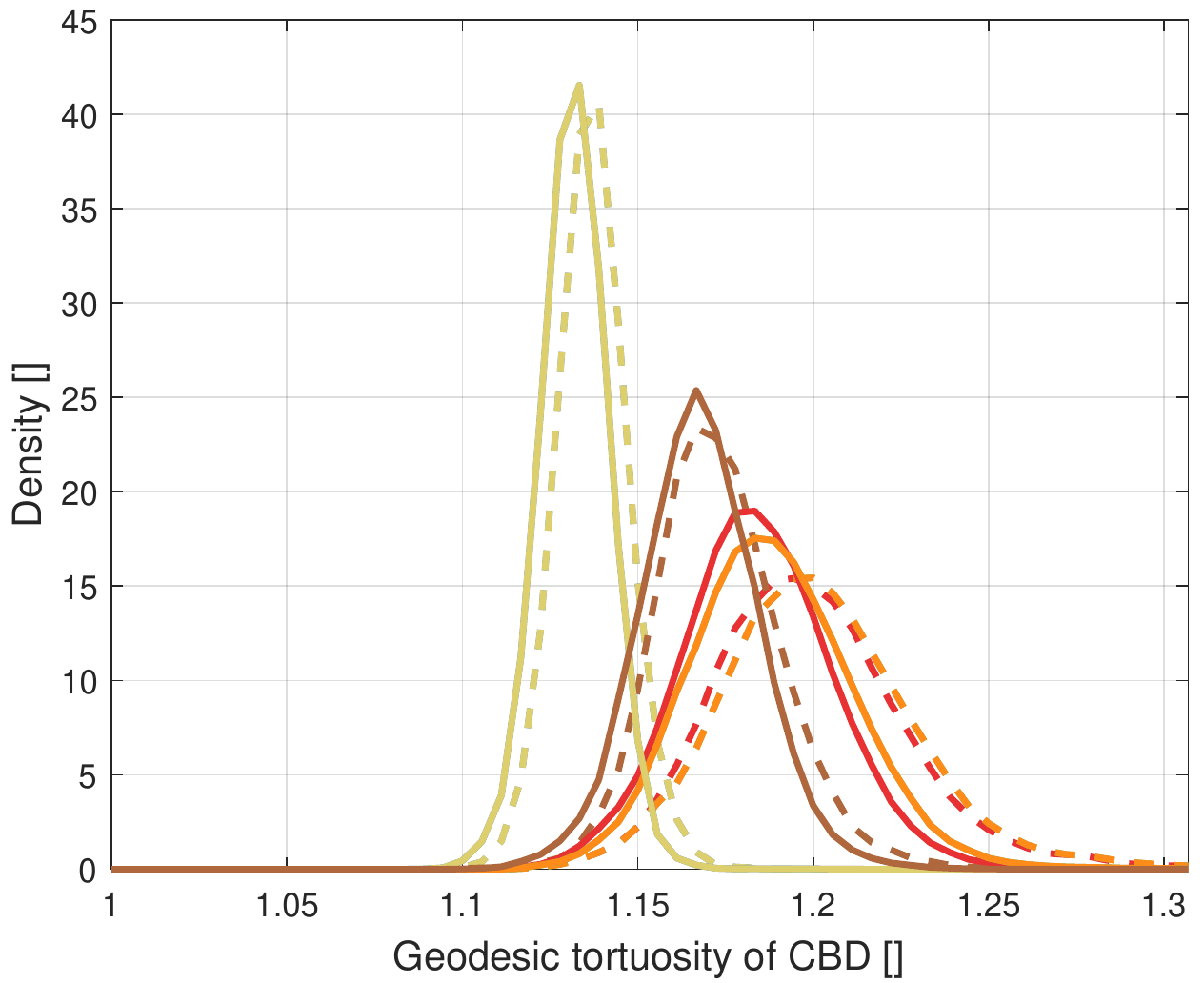}	
	\end{subfigure}
	\begin{subfigure}[c]{0.32\textwidth}
		\centering
		\includegraphics[width=0.9\textwidth]{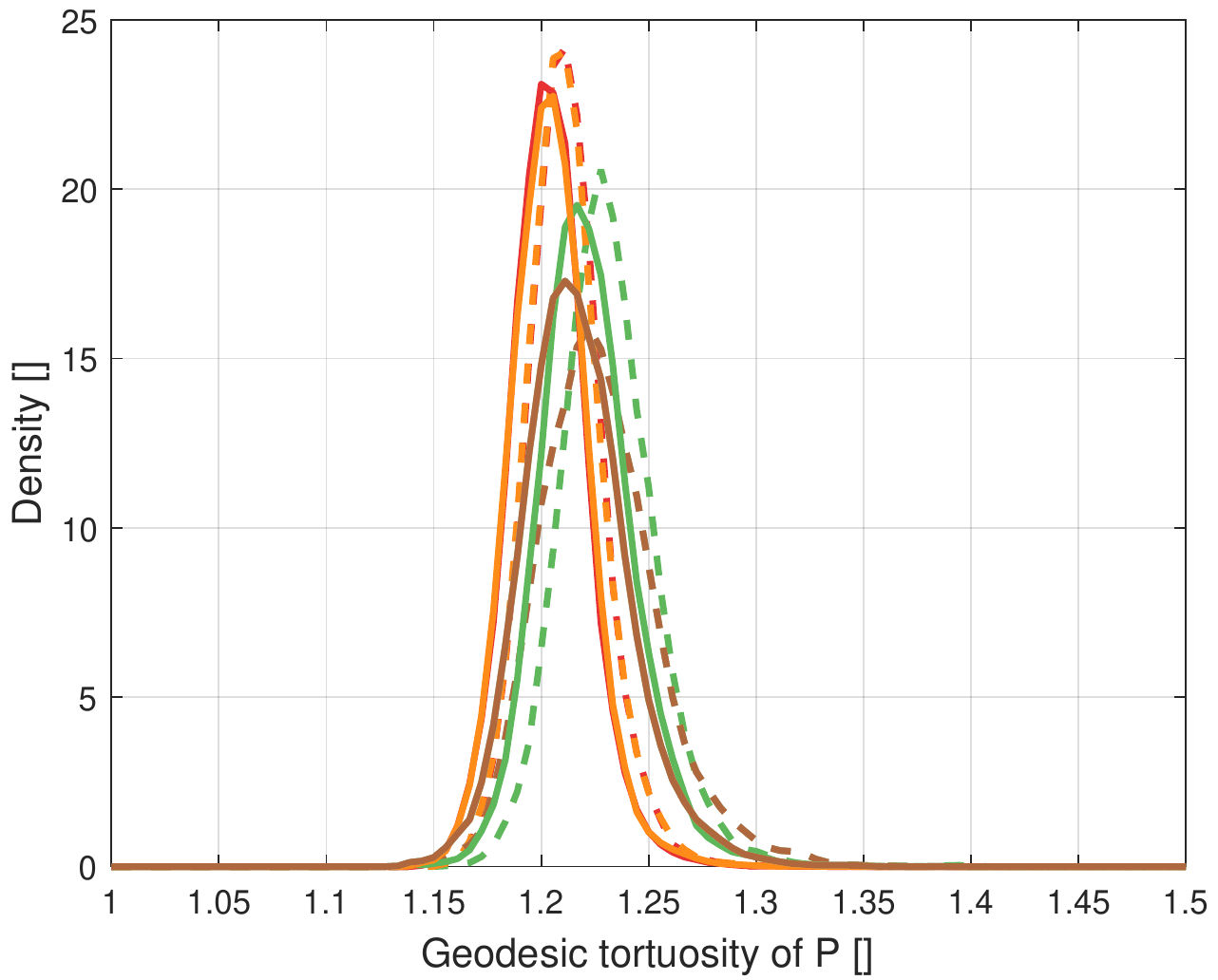}
	\end{subfigure}
	\caption{Geodesic tortuosity of active material (left), CBD (center) and pore space (right) for the two-layer cathode (dashed curves) and the single-layer cathode (solid curves).}
	\label{fig:geodesic_tortuosity_si}
\end{figure}

\begin{figure}[!htbp]
    \centering
	\begin{subfigure}[c]{0.32\textwidth}
		\centering
		\includegraphics[width=0.9\textwidth]{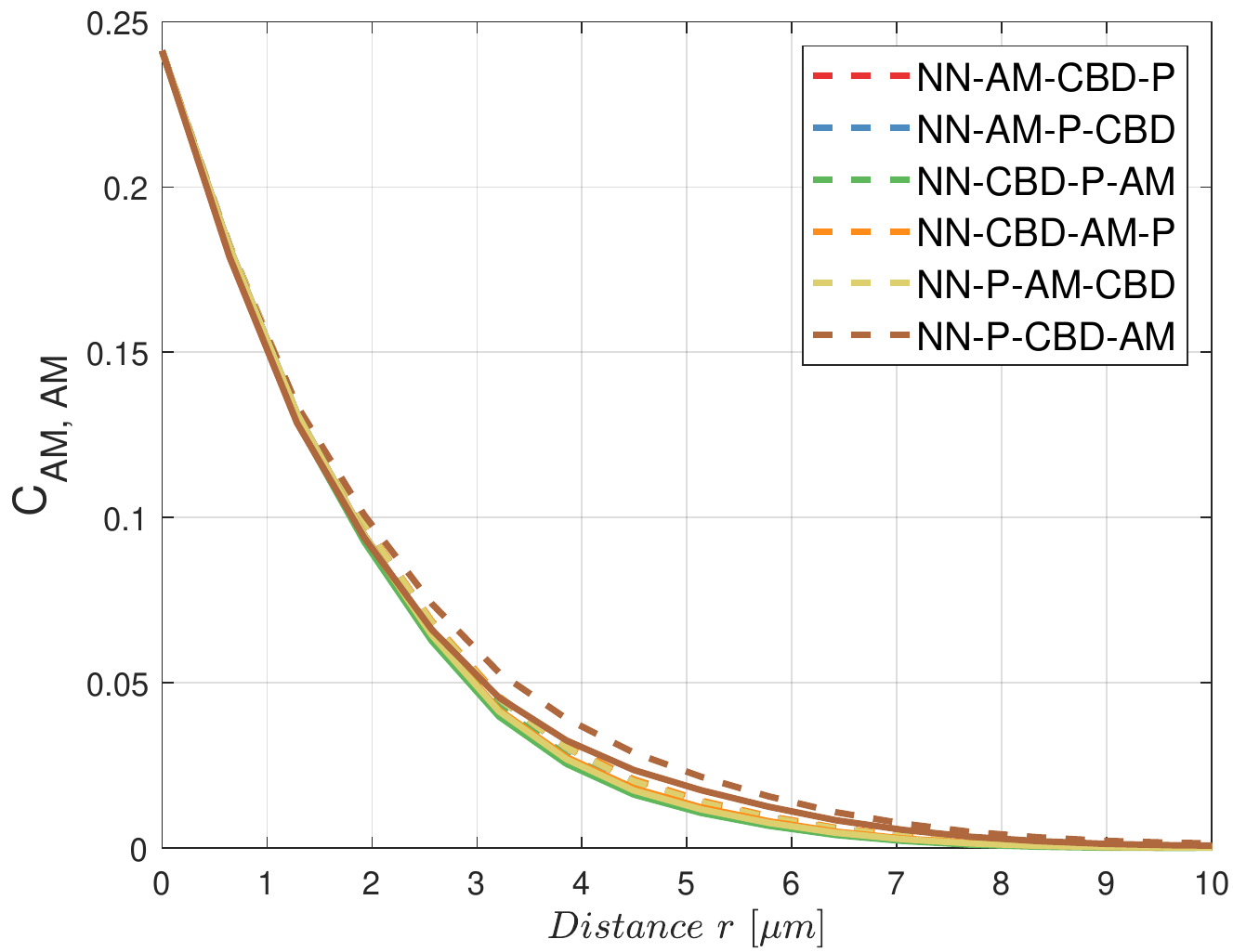}
	\end{subfigure}
	\hfill
	\begin{subfigure}[c]{0.32\textwidth}
		\centering
		\includegraphics[width=0.9\textwidth]{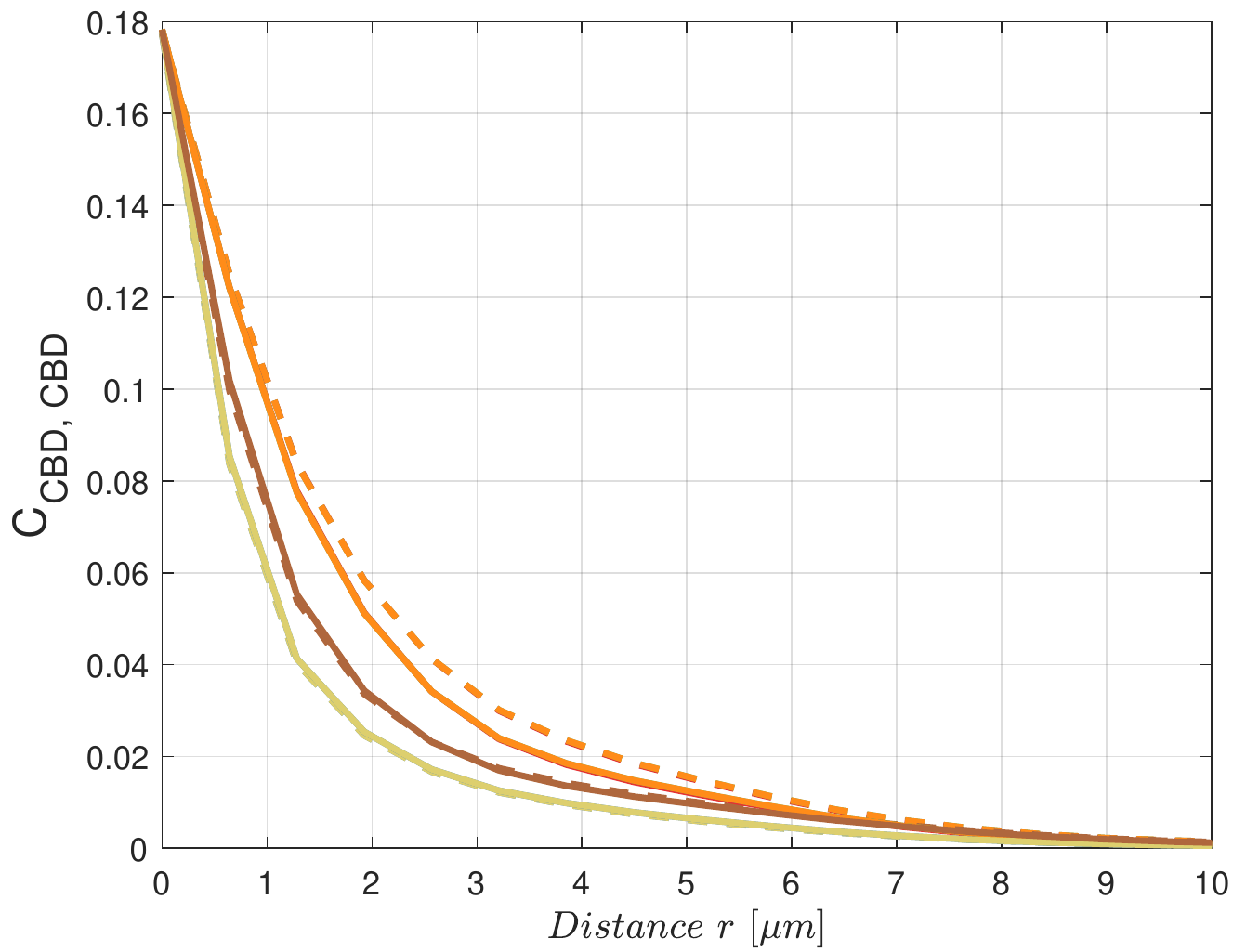}	
	\end{subfigure}
	\hfill
	\begin{subfigure}[c]{0.32\textwidth}
		\centering
		\includegraphics[width=0.9\textwidth]{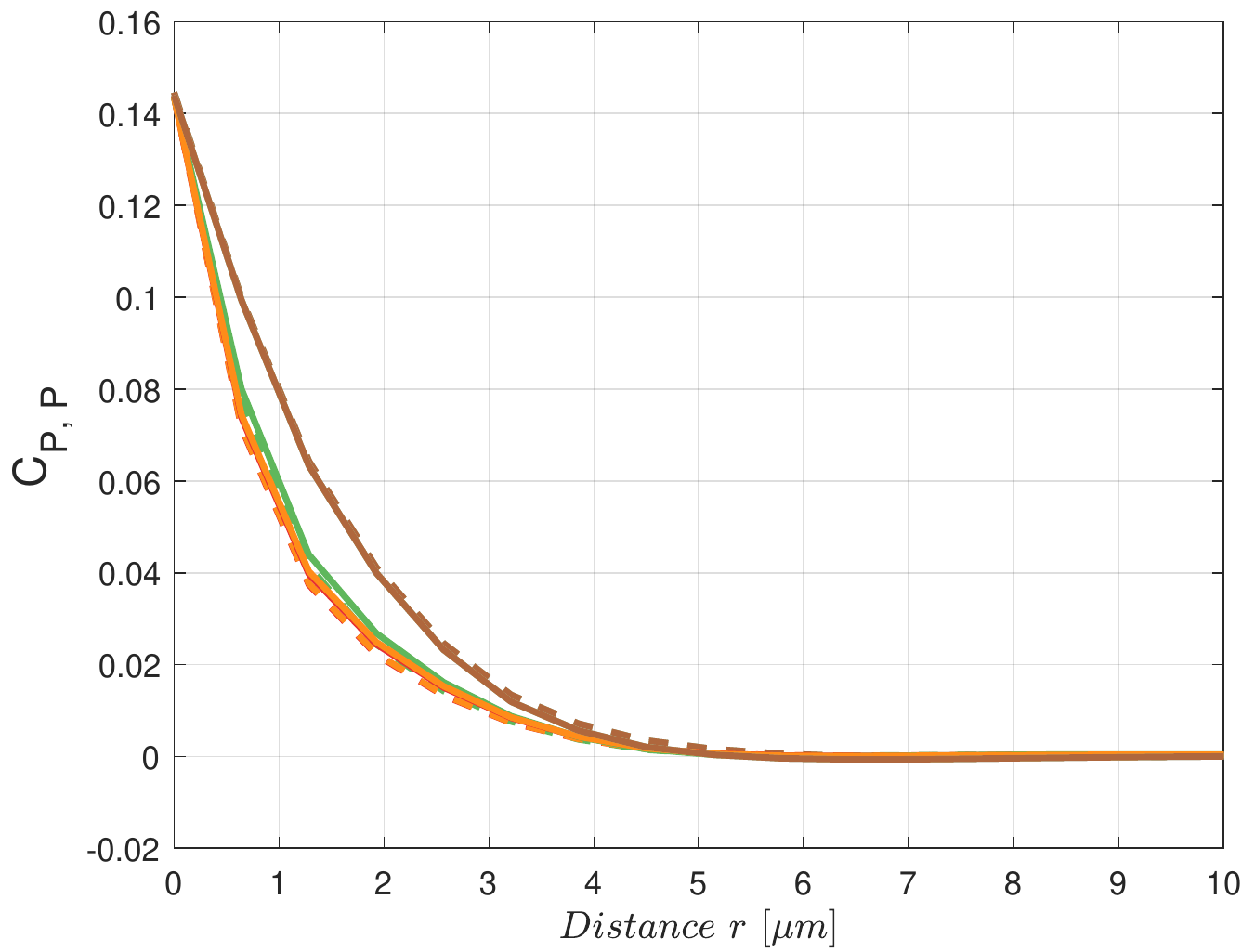}
	\end{subfigure}

	\begin{subfigure}[c]{0.32\textwidth}
		\centering
		\includegraphics[width=0.9\textwidth]{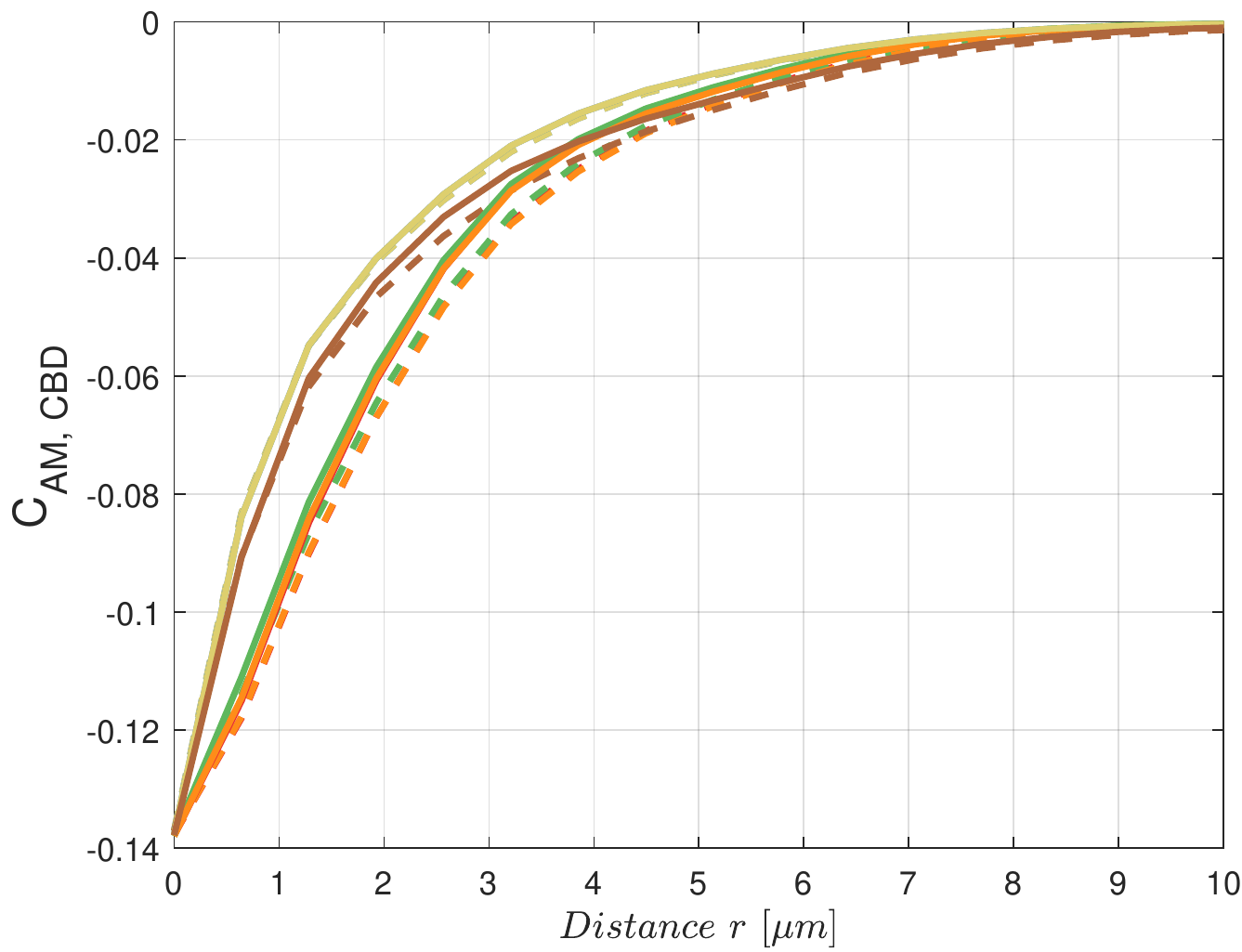}	
	\end{subfigure}
	\hfill
	\begin{subfigure}[c]{0.32\textwidth}
		\centering
		\includegraphics[width=0.9\textwidth]{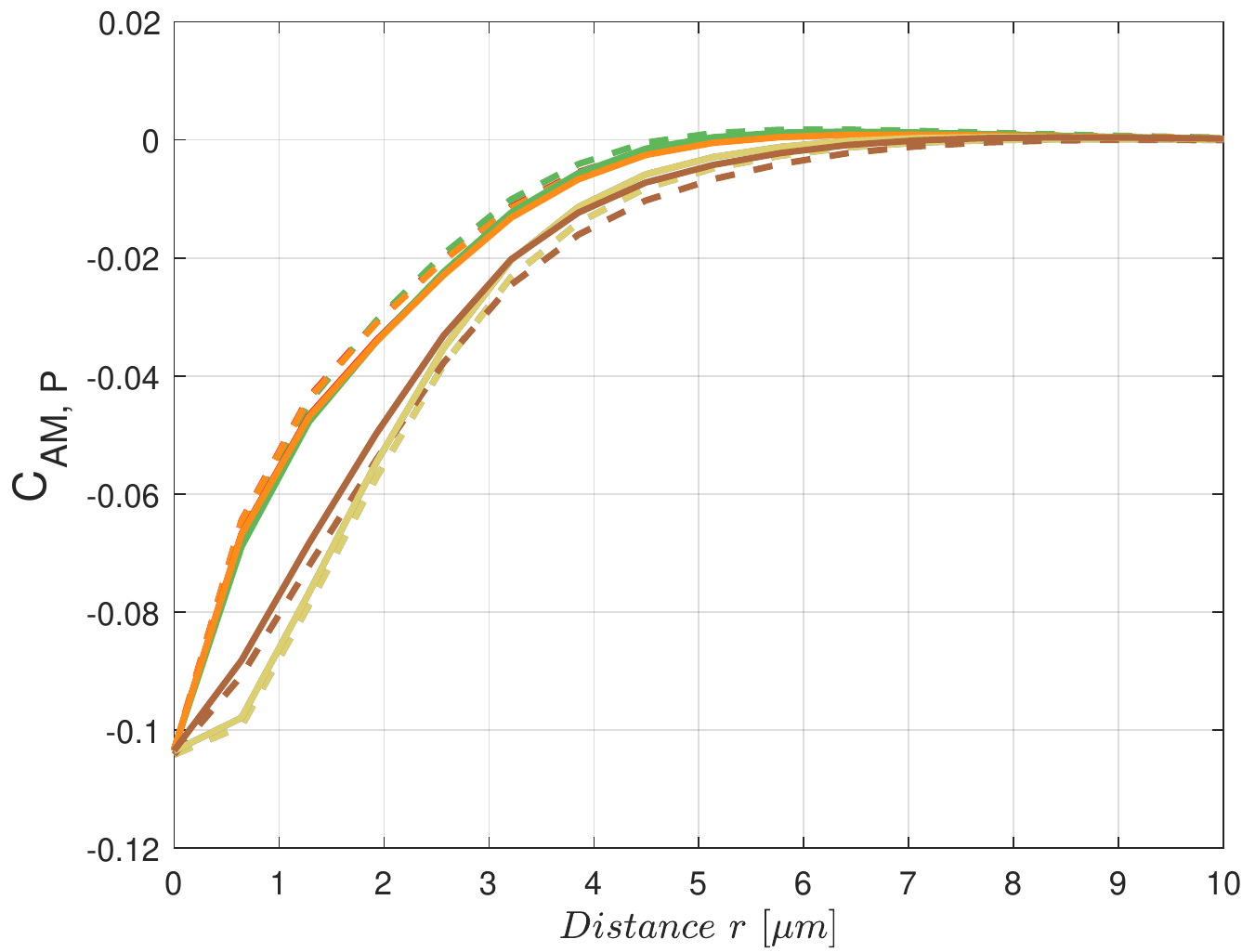}
	\end{subfigure}
	\hfill
	\begin{subfigure}[c]{0.32\textwidth}
		\centering
		\includegraphics[width=0.9\textwidth]{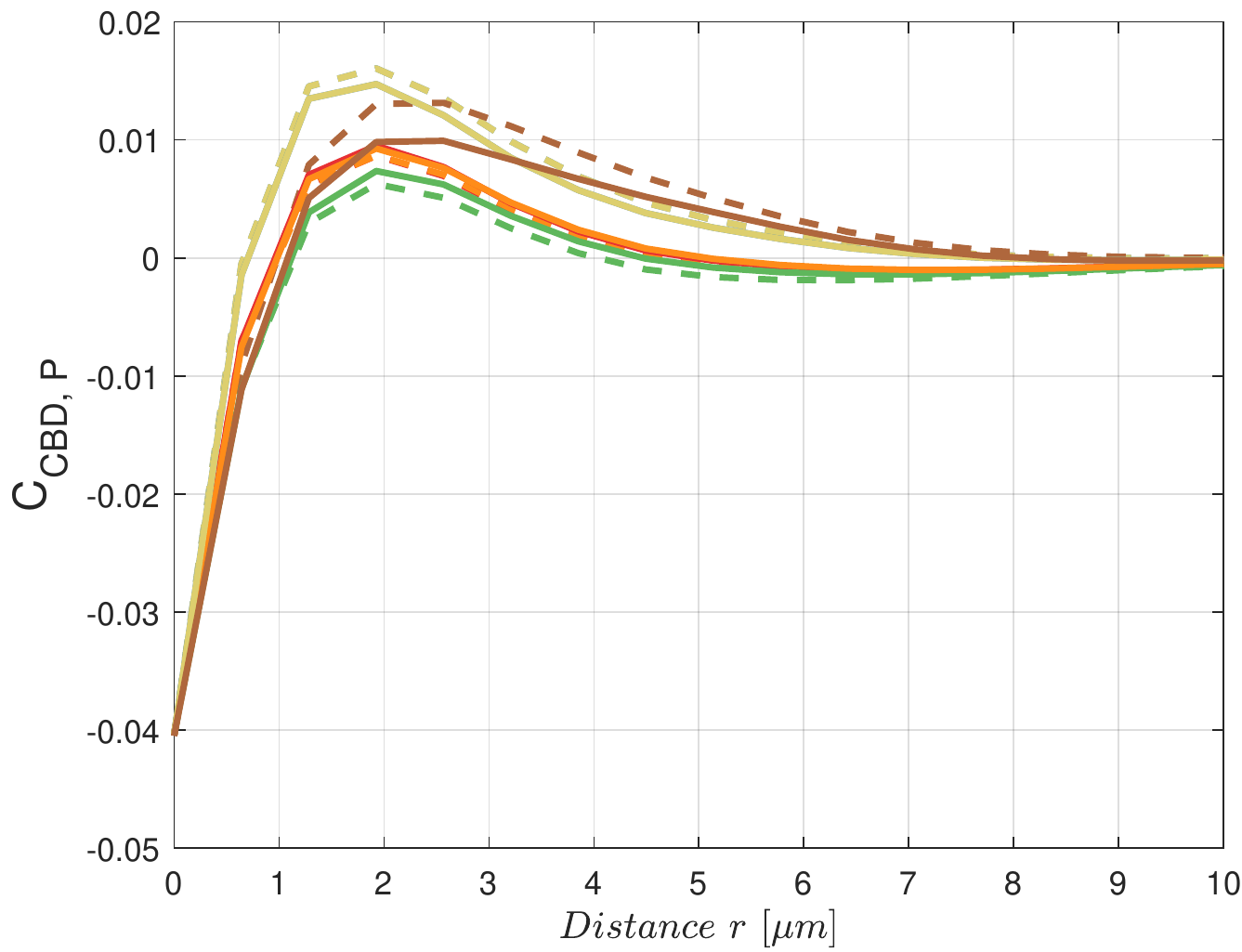}
	\end{subfigure}
	
	\caption{Top row: Centered two-point coverage probability function of active material, CBD and pore space (from left to right). Bottom row: Centered two-point coverage probability functions $C_{\text{AM,CBD}}$, $C_{\text{AM,P}}$ and $C_{\text{CBD,P}}$ (from left to right). Note that dashed curves are used for the two-layer cathode, whereas the solid curves correspond to the single-layer cathode.}
	\label{fig:tpp_si}
\end{figure}

\begin{figure}[!htbp]
	\begin{subfigure}[c]{0.32\textwidth}
		\centering
		\includegraphics[width=0.9\textwidth]{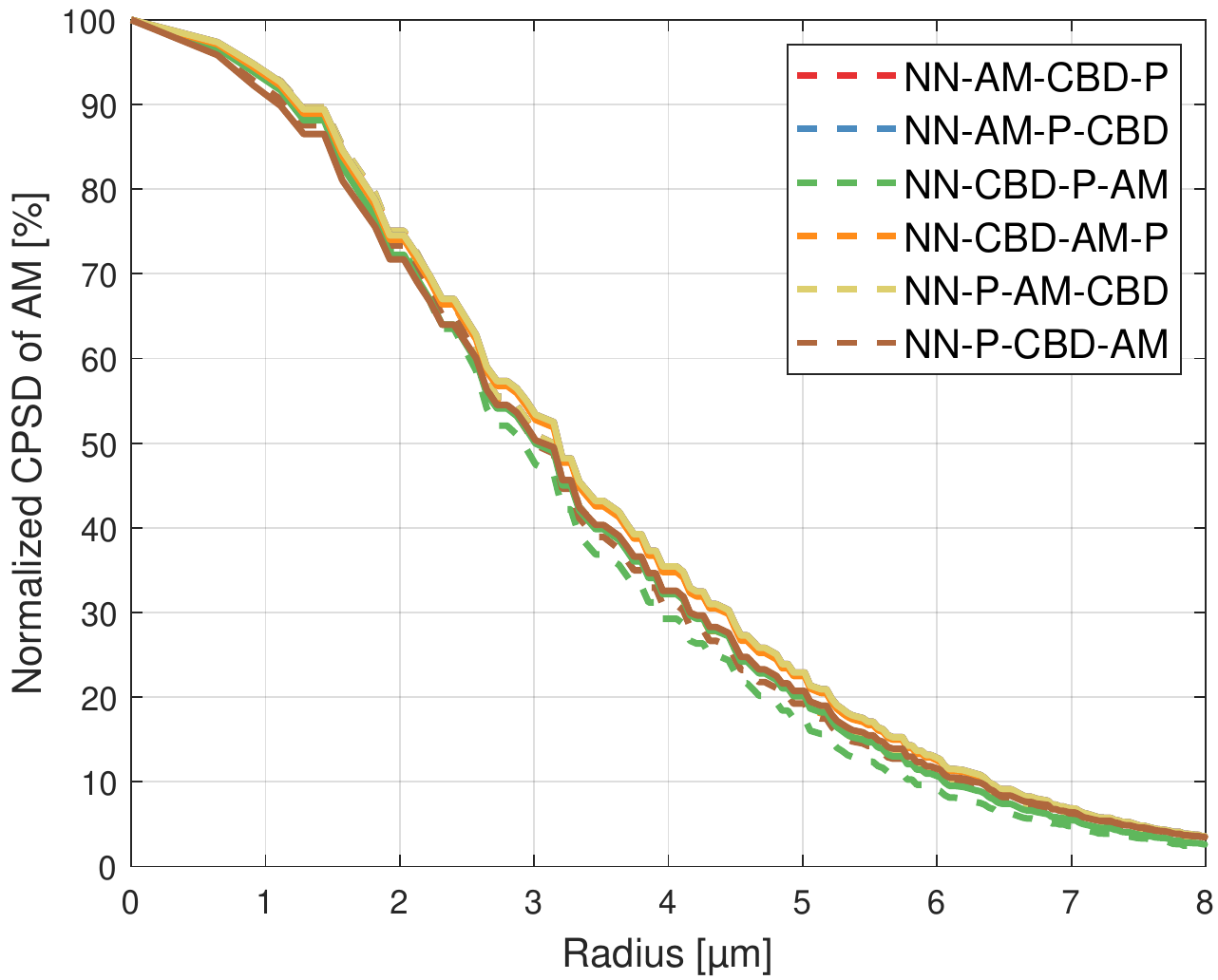}
	\end{subfigure}
	\hfill
	\begin{subfigure}[c]{0.32\textwidth}
		\centering
		\includegraphics[width=0.9\textwidth]{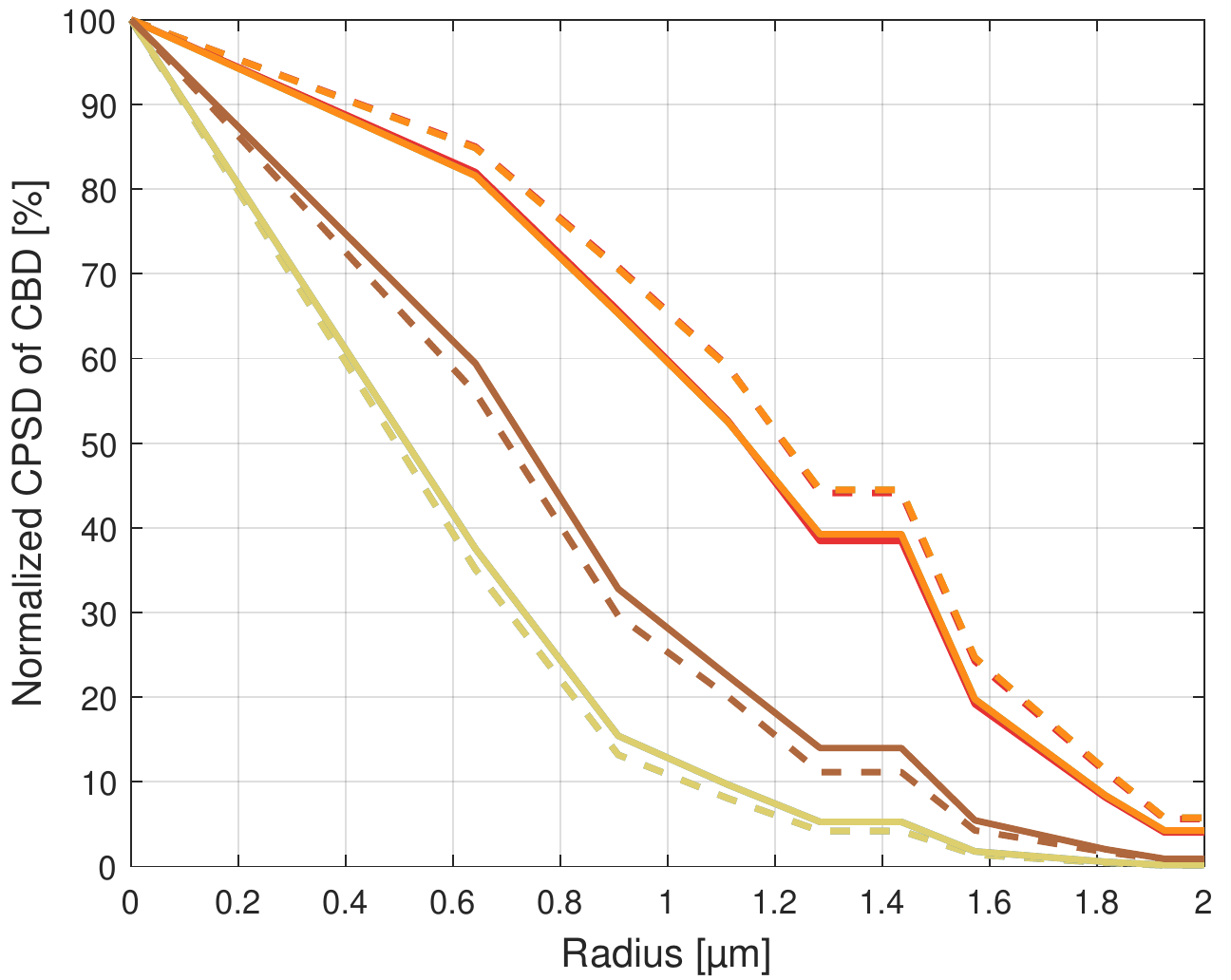}	
	\end{subfigure}
	\begin{subfigure}[c]{0.32\textwidth}
		\centering
		\includegraphics[width=0.9\textwidth]{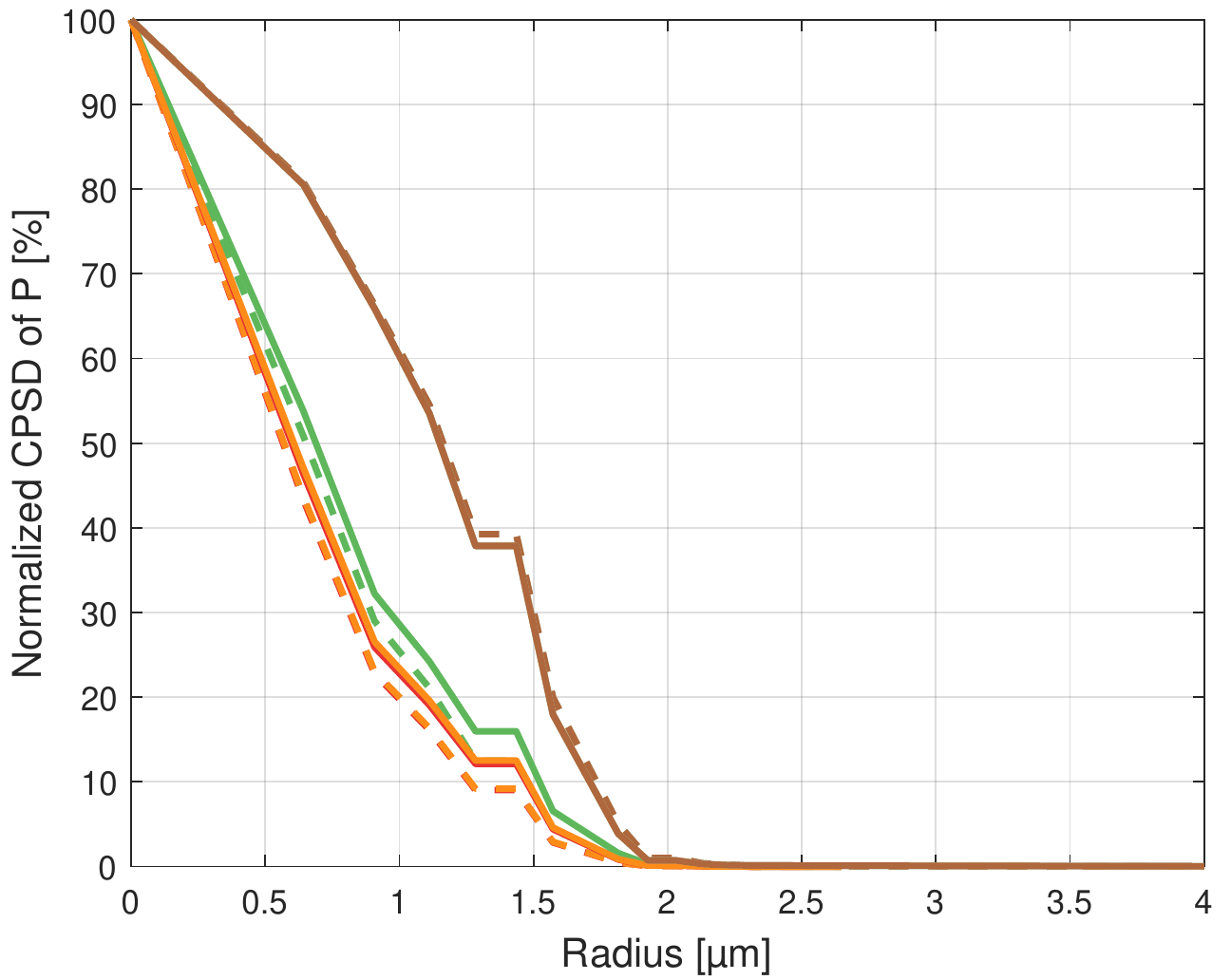}
	\end{subfigure}
	\caption{Continuous phase size distribution of active material (left), CBD (center), and pore space (right) for the two-layer cathode (dashed curves) and the single-layer cathode (solid curves).}
	\label{fig:cpsd_si}
\end{figure}

\begin{figure}[!htbp]
	\begin{subfigure}[c]{0.32\textwidth}
		\centering
		\includegraphics[width=0.9\textwidth]{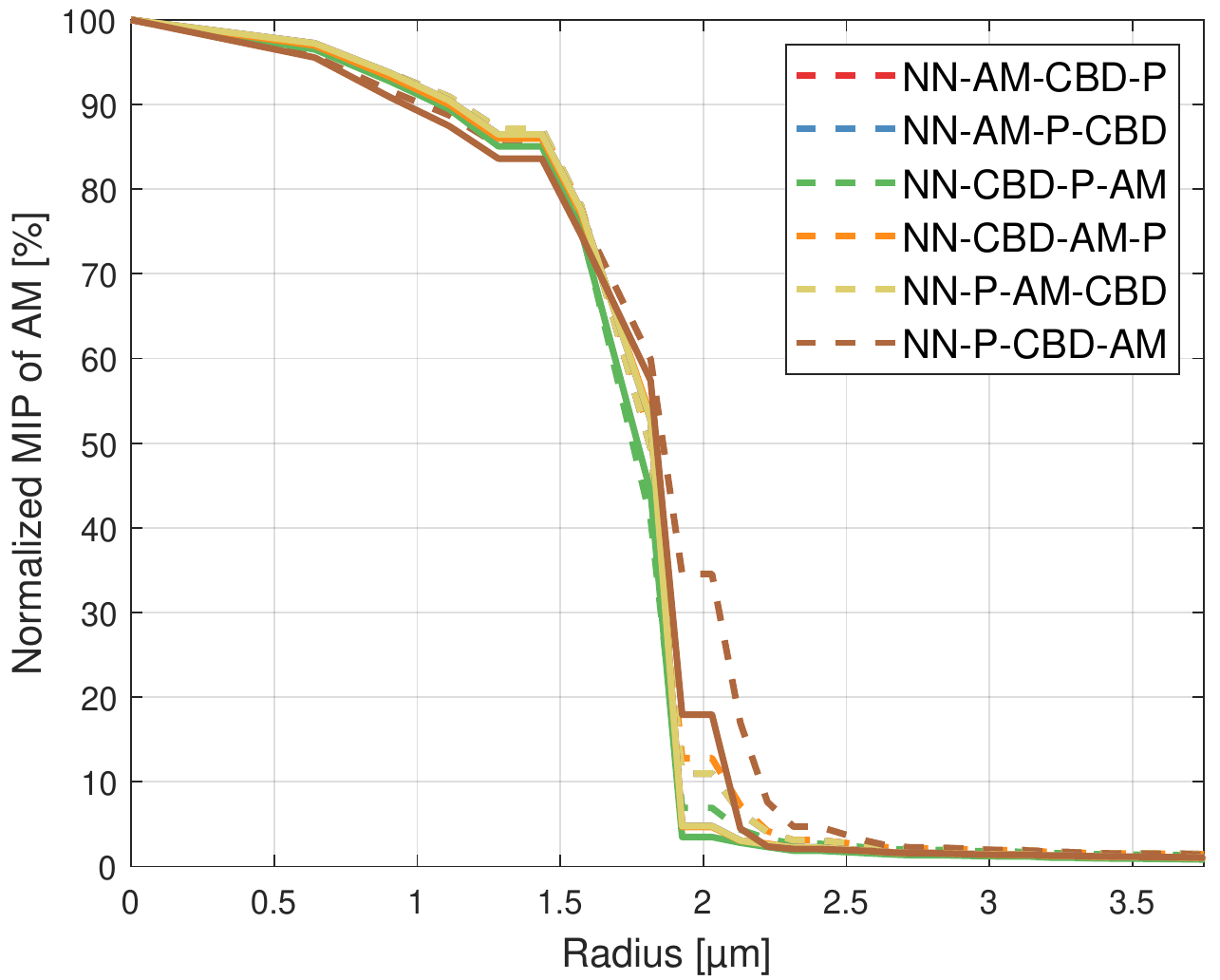}
	\end{subfigure}
	\hfill
	\begin{subfigure}[c]{0.32\textwidth}
		\centering
		\includegraphics[width=0.9\textwidth]{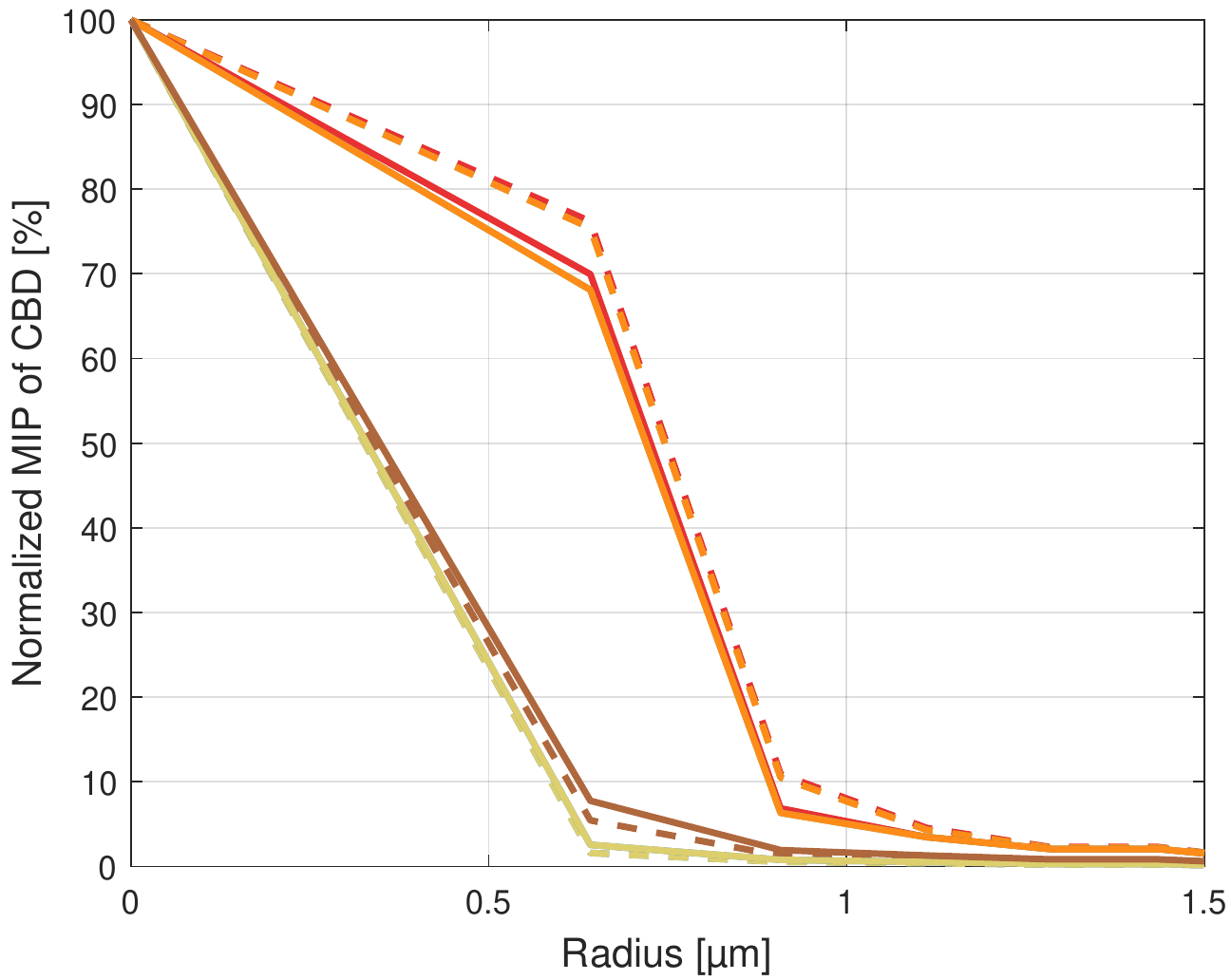}	
	\end{subfigure}
	\begin{subfigure}[c]{0.32\textwidth}
		\centering
		\includegraphics[width=0.9\textwidth]{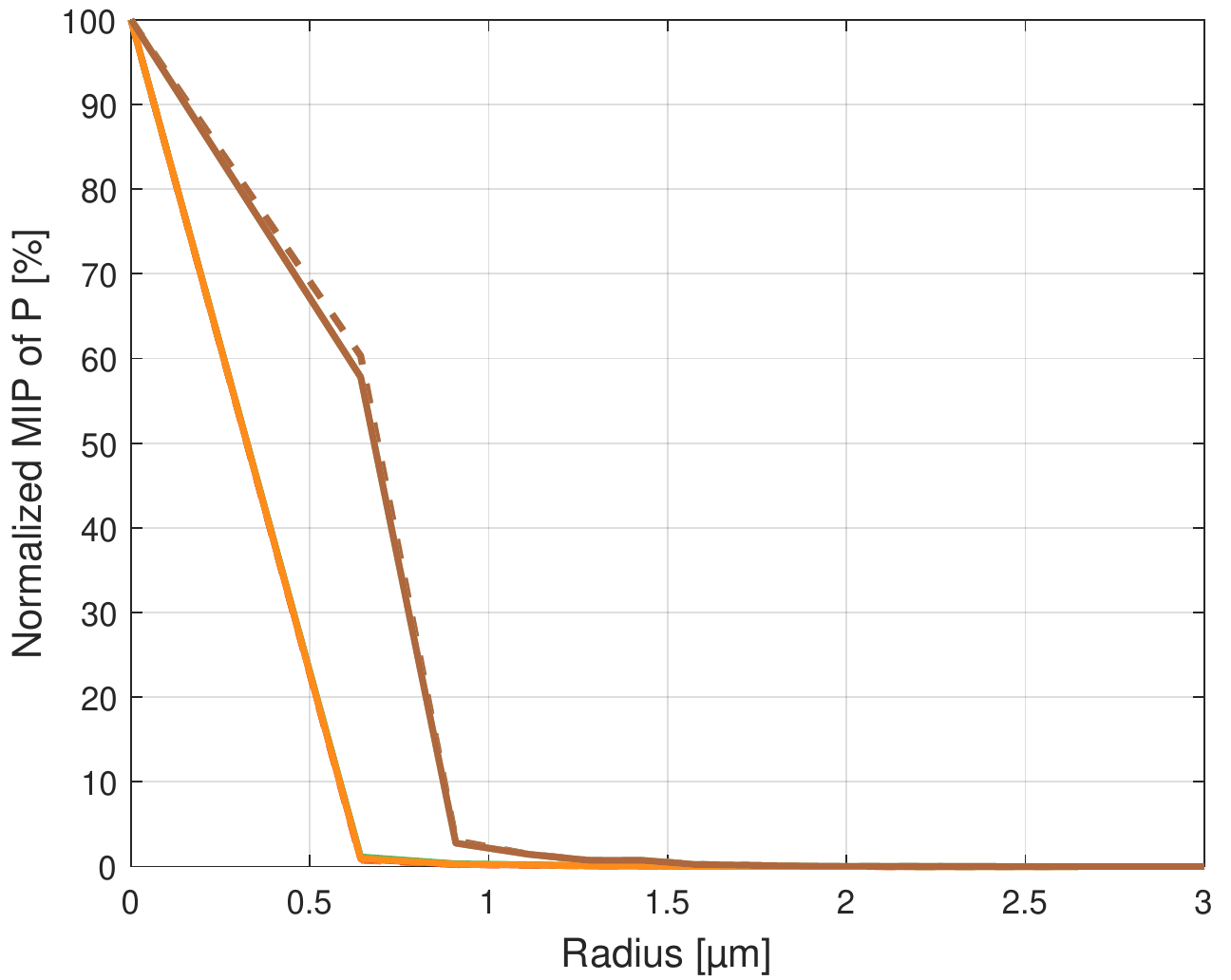}
	\end{subfigure}
	\caption{Simulated mercury intrusion porosimetry of active material (left), CBD (center) and pore space (right) for the two-layer cathode (dashed curves) and the single-layer cathode (solid curves).}
	\label{fig:mip_si}
\end{figure}

\begin{figure}[!htbp]
    \centering
	\begin{subfigure}[c]{0.32\textwidth}
		\centering
		\includegraphics[width=0.9\textwidth]{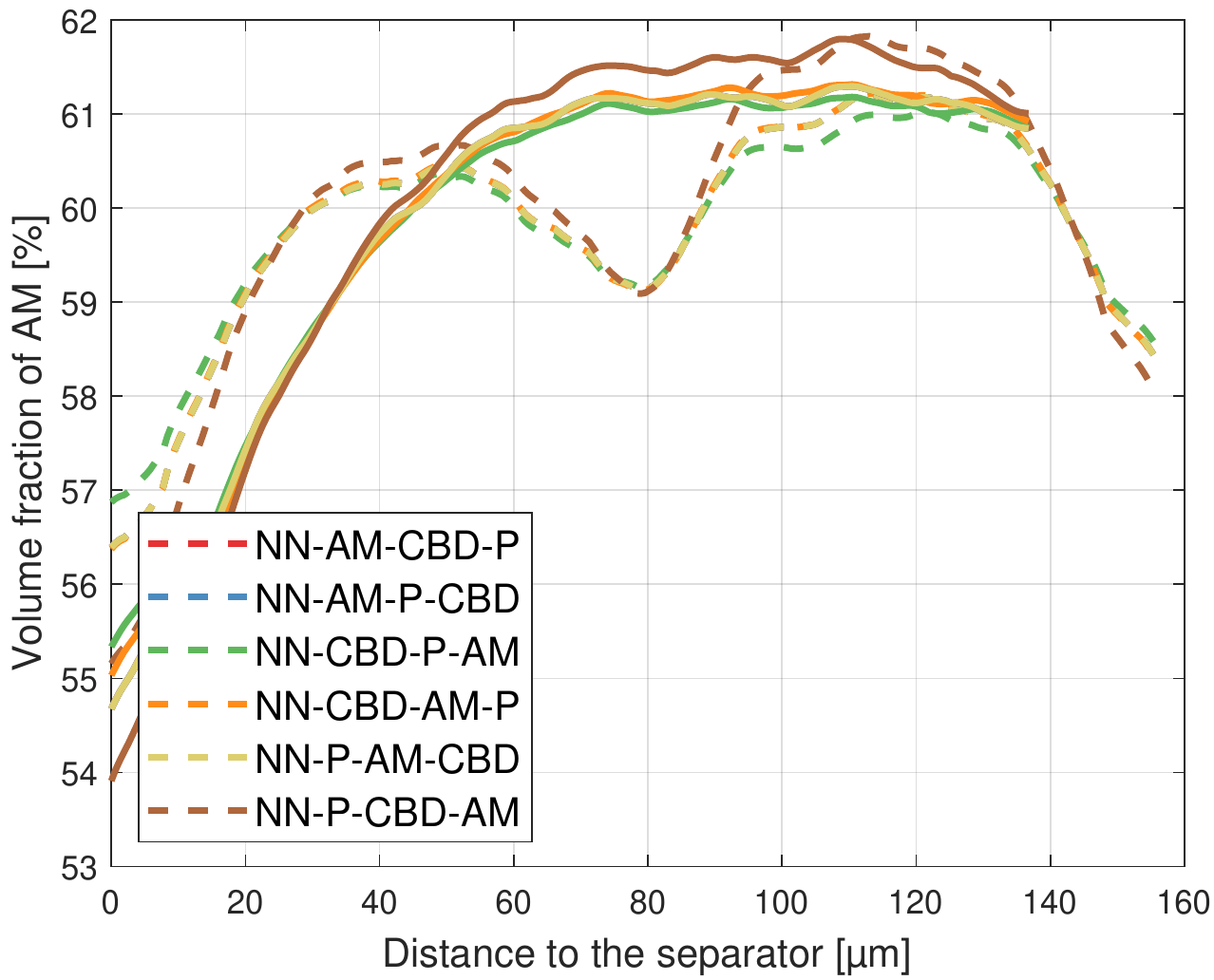}
	\end{subfigure}
	\hfill
	\begin{subfigure}[c]{0.32\textwidth}
		\centering
		\includegraphics[width=0.9\textwidth]{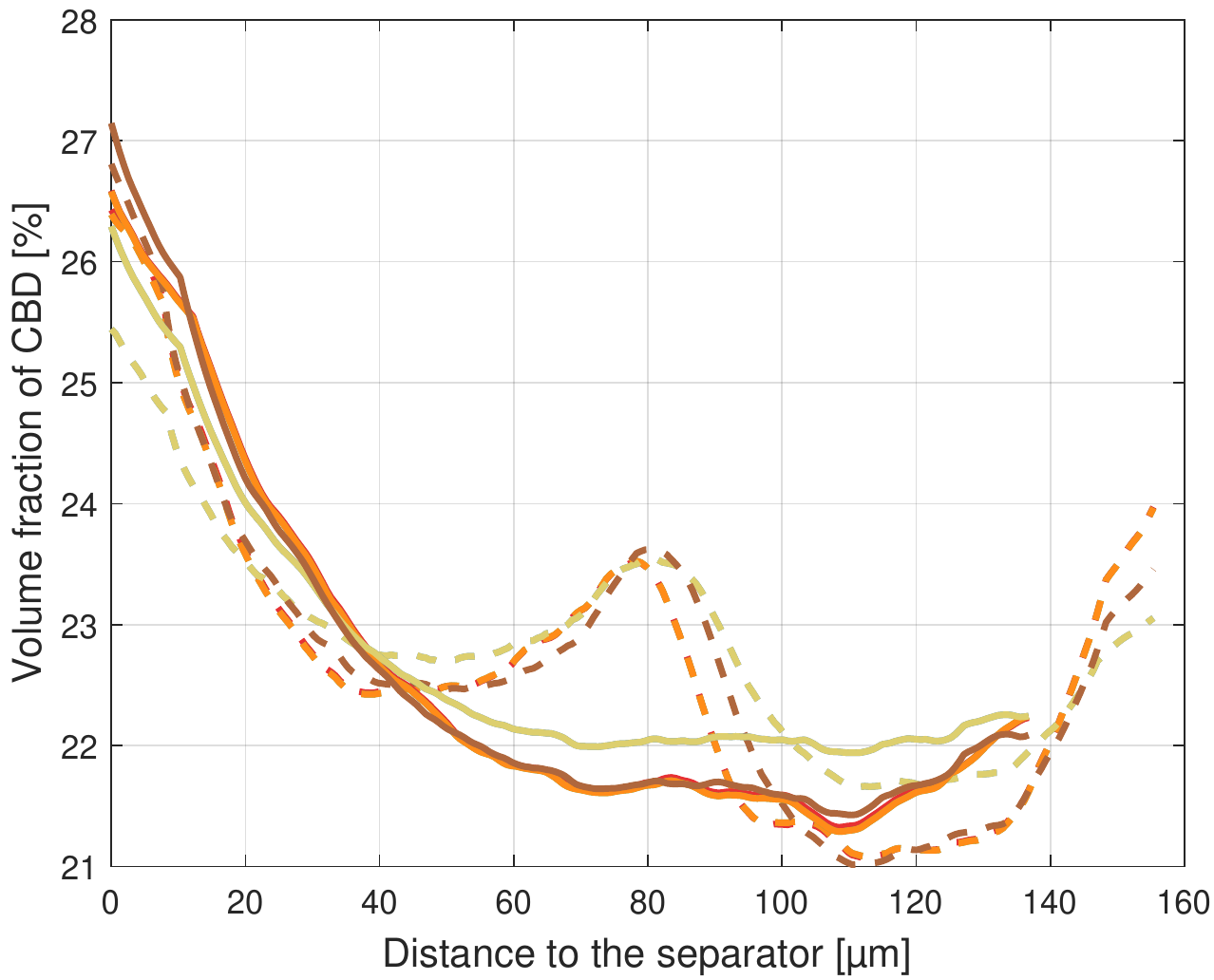}	
	\end{subfigure}
	\hfill
	\begin{subfigure}[c]{0.32\textwidth}
		\centering
		\includegraphics[width=0.9\textwidth]{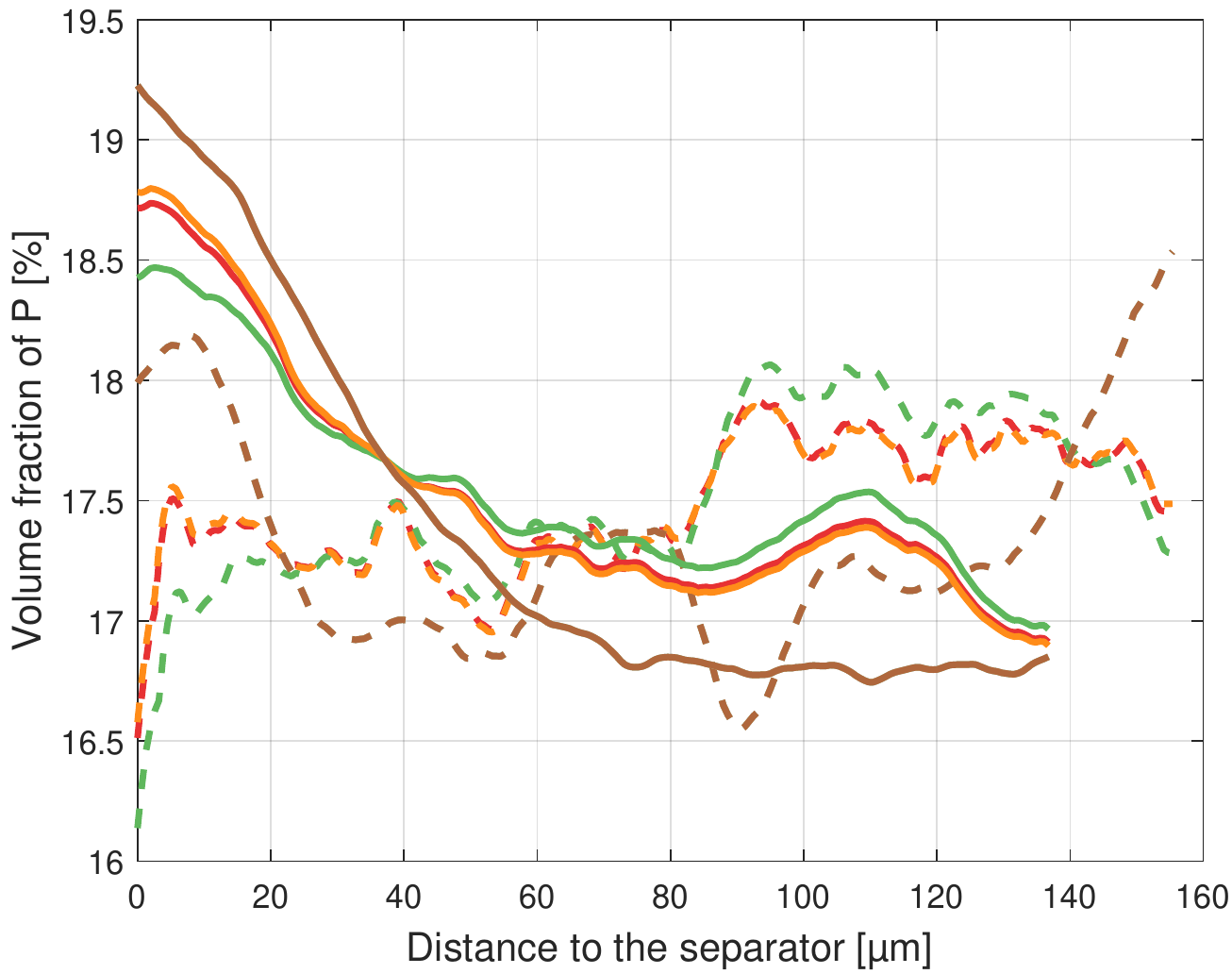}
	\end{subfigure}
	\caption{Volume fraction of active material (left), CBD (center) and pores (right) in dependence of the distance to the separator for the two-layer cathode (dashed curves) and the single-layer cathode (solid curves).}
	\label{fig:volfrac_dist_si}
\end{figure}

\subsection{Electrochemical transport model and parametrization}

The governing equations used for electrochemical simulations are listed in Table~\ref{tab:governing}.

\begin{table}[!htbp]
    \centering
    \begin{tabular}{lccc}
    \toprule
        domain & phase & equation & flux\\ \midrule
        \multirow{2}{*}{Electrolyte (Elyte)} & e & $\frac{\partial c_e}{\partial t}= - \vec{\nabla} \vec{N}_{e}^{elyte}\label{eq:el:matbal}$ & $\vec{N}_{e}^{elyte} = -D_{e}^{elyte}\cdot \vec{\nabla} c_{e} + \frac{t_+}{F} \vec{J}_{e}^{elyte}$ \\
        & e & $0 = -\vec{\nabla} \vec{J}_{e}^{elyte}$\label{eq:el:chacon}& $\vec{J}_{e}^{elyte} = -\kappa_{e}^{elyte}\cdot \vec{\nabla}\varphi_{e} + \kappa_{e}^{elyte} \frac{1-t_+}{F} \left( \frac{\partial \mu_{e}}{\partial c_{e}} \right)\vec{\nabla}c_{e}$ \\\midrule
        \multirow{2}{*}{Active material (AM)} & s & $\frac{\partial c_s}{\partial t}= - \vec{\nabla} \vec{N}_{s}\label{eq:so:matbal}$ & $\vec{N}_{s}= -D_{s}\cdot \vec{\nabla} c_{s}$\\
        & s &  $0 = -\vec{\nabla} \vec{J}_{s}^{AM}$\label{eq:so:chacon} & $\vec{J}_{s}^{AM}= -\sigma_{s}^{AM}\cdot \vec{\nabla} \Phi_{s}$\\\midrule
        \multirow{2}{*}{Separator} & e & $\frac{\partial c_e}{\partial t}= - \vec{\nabla} \vec{N}_{e}^{sep}\label{eq:sep:matbal}$ & $\vec{N}_{e}^{sep} = -D_{e}^{sep,eff}\cdot \vec{\nabla} c_{e} + \frac{t_+}{F} \vec{J}_{e}^{sep}$ \\
        & e & $0 = -\vec{\nabla} \vec{J}_{e}^{sep}$\label{eq:sep:chacon}& $\vec{J}_{e}^{sep} = -\kappa_{e}^{sep,eff}\cdot \vec{\nabla}\varphi_{e} + \kappa_{e}^{sep,eff} \frac{1-t_+}{F} \left( \frac{\partial \mu_{e}}{\partial c_{e}} \right)\vec{\nabla}c_{e}$ \\\midrule
        \multirow{3}{*}{CBD} & e & $\frac{\partial c_e}{\partial t}= - \vec{\nabla} \vec{N}_{e}^{CBD}\label{eq:cbd:matbal}$& $\vec{N}_{e}^{CBD} = -D_{e}^{sep,eff}\cdot \vec{\nabla} c_{e} + \frac{t_+}{F} \vec{J}_{e}^{sep}$ \\
        & e & $0 = -\vec{\nabla} \vec{J}_{e}^{CBD}$ & $\vec{J}_{e}^{CBD} = -\kappa_{e}^{CBD,eff}\cdot \vec{\nabla}\varphi_{e} + \kappa_{e}^{CBD,eff} \frac{1-t_+}{F} \left( \frac{\partial \mu_{e}}{\partial c_{e}} \right)\vec{\nabla}c_{e}$\\
        & s & $0 = -\vec{\nabla} \vec{J}_{s}^{CBD}$ & $\vec{J}_{s}^{CBD}= -\sigma_{s}^{CBD,eff}\cdot \vec{\nabla} \Phi_{s}$\\\midrule
        \multirow{1}{*}{Current collector (CC)} & s & $0 = -\vec{\nabla} \vec{J}_{s}^{CC}$\label{eq:cc:chacon} & $\vec{J}_{s}^{CC}= -\sigma_{s}^{CC}\cdot \vec{\nabla} \Phi_{s}$\\
        \bottomrule
    \end{tabular}
    \caption{List of governing equations used for the electrochemical simulations. The effective transport parameters in the CBD and separator are calculated according to Equation~(\ref{eq:effective}).}
    \label{tab:governing}
\end{table}

The battery system is solved in the solid phases for the electrical potential $\Phi_s$ and the lithium concentration $c_s$ and in the electrolyte phases for the electrochemical potential $\varphi_e$ and the lithium concentration $c_e$.
The governing equations in the different domains are connected using different interface and boundary conditions, see Table~\ref{tab:interfaceandboundary}).\\

\begin{table}[!htbp]
    \centering
    \begin{tabular}{ccc}
    \toprule
        \multicolumn{3}{c}{Interface conditions}\\\midrule
         & Domain 1: Electrolyte & Domain 2: Active material\\
        Lithium flux & $\vec{N}_{e}^{bulk} \cdot \vec{n}= \left(i_{react} + i_{DL} \right)/F$ & $\vec{N}_{s} \cdot \vec{n} = i_{react}/F$ \\
        Charge flux & $\vec{J}_{e}^{bulk} \cdot \vec{n}= i_{react} + i_{DL}$ & $\vec{J}_{s} \cdot \vec{n} = i_{react} + i_{DL}$\\\midrule
        & Domain 1: Separator & Domain 2: Active material\\
        Lithium flux & $\vec{N}_{e}^{sep} \cdot \vec{n} = \left(i_{react} + i_{DL} \right)/F \cdot \varepsilon_{e}^{sep}$  & $\vec{N}_{s} \cdot \vec{n} = i_{react}/F \cdot \varepsilon_{e}^{sep}$ \\
        Charge flux & $\vec{J}_{e}^{sep} \cdot \vec{n} = \left(i_{react} + i_{DL} \right) \cdot \varepsilon_{e}^{sep}$& $\vec{J}_{s} \cdot \vec{n} = \left(i_{react} + i_{DL} \right) \cdot \varepsilon_{e}^{sep}$\\\midrule
        & Domain 1: CBD & Domain 2: Active material\\
        Lithium flux & $\vec{N}_{e}^{CBD} \cdot \vec{n} = \left(i_{react} + i_{DL} \right)/F \cdot \varepsilon_{e}^{CBD}$& $\vec{N}_{s}^{AM} \cdot \vec{n} = i_{react}/F \cdot \varepsilon_{e}^{CBD}$ \\
        \multirow{2}{*}{Charge flux} & $\vec{J}_{e}^{CBD} \cdot \vec{n} = \left(i_{react} + i_{DL} \right) \cdot \varepsilon_{e}^{CBD}$& \multirow{2}{*}{$\vec{J}_{s}^{AM} \cdot \vec{n} = \left(i_{react} + i_{DL} \right) \cdot \varepsilon_{e}^{CBD} + j_{s}^{CBD}$}\\
         & $\vec{J}_{s}^{CBD} \cdot \vec{n} = j_{s}^{CBD} = - \sigma_{CBD,AM}^{eff} \vec{\nabla} \Phi_s$& \\\midrule
         & Domain 1: Electrolyte & Domain 2: CC\\
        Lithium flux & $\vec{N}_{e}^{bulk} \cdot \vec{n}= 0$ &- \\
        Charge flux & $\vec{J}_{e}^{bulk} \cdot \vec{n}= 0$ & $\vec{J}_{s}^{CC} \cdot \vec{n} = 0$\\\midrule
         & Domain 1: Separator & Domain 2: CC\\
        Lithium flux & $\vec{N}_{e}^{sep} \cdot \vec{n}= 0$ &- \\
        Charge flux & $\vec{J}_{e}^{sep} \cdot \vec{n}= 0$ & $\vec{J}_{s}^{CC} \cdot \vec{n} = 0$\\\midrule
         & Domain 1: CBD & Domain 2: CC\\
        Lithium flux & $\vec{N}_{e}^{CBD} \cdot \vec{n}= 0$ &- \\
        \multirow{2}{*}{Charge flux}& $\vec{J}_{e}^{CBD} \cdot \vec{n}= 0$ & \multirow{2}{*}{$\vec{J}_{s}^{CC} \cdot \vec{n} = - \sigma_{CBD,CC}^{eff} \vec{\nabla} \Phi_s$}\\
         & $\vec{J}_{s}^{CBD} \cdot \vec{n}= - \sigma_{CBD,CC}^{eff} \vec{\nabla} \Phi_s$ & \\\midrule
        & Domain 1: Active material & Domain 2: CC\\
        Lithium flux & $\vec{N}_{s}^{AM} \cdot \vec{n} = 0$ & - \\
        Charge flux &  $\vec{J}_{s}^{AM} \cdot \vec{n} = - \sigma_{AM,CC}^{eff} \vec{\nabla} \Phi_s$ & $\vec{J}_{s}^{CC} \cdot \vec{n} = - \sigma_{AM,CC}^{eff} \vec{\nabla} \Phi_s$\\\midrule
        \multicolumn{3}{c}{Boundary conditions}\\\midrule
        Side & operation mode & condition \\
        Anode side & all & $\Phi_{CC}^{Anode} = \Phi_{AM}\left(t=0\right)  = U_{0}^{Anode}\left(c_s^0\right)= fixed$ \\
        \multirow{2}{*}{Cathode side} & potentiostatic & $\Phi_{CC}^{Cathode} = \Phi_{CC}^{Anode} + U_{Applied}$ \\
          & galvanostatic & $\vec{J}_{CC}^{Cathode} \cdot \vec{n} = j_{applied}$ \\
        \bottomrule
    \end{tabular}
    \caption{Interface and boundary conditions of the governing equations for the different domains. The models used for the reaction and double current are listed in Table~\ref{tab:reaction}.}
    \label{tab:interfaceandboundary}
\end{table}

The used reaction models are listed in Table~\ref{tab:reaction}, whereas the electrochemical parameters of the materials are listed in Table~\ref{tab:parameters}.

\begin{table}[!htbp]
    \centering
    \begin{tabular}{cccc}
    \toprule
         Phase 1 & Phase 2 & type& equation\\\midrule
         Electrolyte & NMC & reaction & $i_{intercalation} = 2\cdot i_{00}^{intercalation} \cdot \sqrt{c_e \cdot c_s} \cdot \sinh\left(\frac{F}{2 R T}\eta_{intercalation}\right)$\\
         & & & $\eta_{intercalation} = \Phi_s - \varphi_e - U_0\left(c_s\right)$\\
         & & doublelayer & $i_{DL} = - C_{DL} \cdot \frac{d \Delta \Phi}{d t}\text{ with }\Delta\Phi\approx \Phi_s - \varphi_e$\\
         \midrule
         Electrolyte & Counter-Electrode & reaction &$i_{CE} = 2\cdot i_{00}^{CE}\cdot \sqrt{c_e}\cdot \sinh\left(\frac{F}{2 R T}\eta_{CE} \right)$\\
         & & & $\eta_{CE} = \Phi_s - \varphi_e - U_0^{CE}\text{ with }U_0^{CE}=0$\\
        \bottomrule
    \end{tabular}
    \caption{Reaction models used in this work.}
    \label{tab:reaction}
\end{table}

\begin{table}[!htbp]
    \centering
    \begin{tabular}{ccll}
    \toprule
        Parameter & Value & Description & Source \\\midrule
        \multicolumn{4}{c}{NMC}\\\midrule
        $c_{NMC}^0 \text{ / mol/cm}^3$ & $1.65\cdot 10^{-2}$ & initial value for lithiation simulation & calculated \\
        $c_{NMC}^0 \text{ / mol/cm}^3$ & $5.0218747\cdot 10^{-2} $ & initial value for impedance simulation & calculated \\
        $c_{NMC}^{max} \text{ / mol/cm}^3$ & $5.0451\cdot 10^{-2}$ & maximal lithium concentration in NMC622&\cite{kremer.2020}\\
        $D_{NMC} \text{ / cm}^2\text{/s}$ & See Equation (SI-8) \cite{kremer.2020} & Li-ion diffusion coefficient &\cite{kremer.2020}\\
        $\sigma_{NMC} \text{ / S/cm}$ & See Figure S 4 in \cite{kremer.2020} &  electronic conductivity  &\cite{kremer.2020}\\
        $U_0^{NMC} \text{ / V}$ & See Figure S 4 in \cite{kremer.2020}  & Open circuit potential of NMC &\cite{kremer.2020}\\
        $C_{DL}^{NMC} \text{ / F/cm}^2$ & $2.4\cdot 10^{-4}$& Doublelayer capacity of NMC & assumed\\\midrule
        \multicolumn{4}{c}{Electrolyte}\\\midrule
        $c_{e}^0 \text{ / mol/cm}^3$ & $1$ & Concentration of salt& \cite{landesfeind.2019, kremer.2020}\\
        $D_{e} \text{ / cm}^2\text{/s}$ & See Figure S 3 in \cite{kremer.2020}  & Li-ion diffusion coefficient& \cite{landesfeind.2019, kremer.2020}\\
        $\kappa_e \text{ / S/cm}$ & See Figure S 3 in \cite{kremer.2020}& ionic conductivity & \cite{landesfeind.2019, kremer.2020}\\
        $ f_{Li}^{Elyte}\text{ / -}$ & See Figure S 3 in \cite{kremer.2020} & activity factor used for $\left(\frac{\partial \mu_e}{\partial c_e}\right)$ & \cite{landesfeind.2019, kremer.2020}\\
        $t_+\text{ / -}$ & See Figure S 3 in \cite{kremer.2020} & transference number & \cite{landesfeind.2019, kremer.2020}\\\midrule
        \multicolumn{4}{c}{Counter-Electrode}\\\midrule
        $\sigma_{CE} \text{ / S/cm}$ & 100 &  electronic conductivity & assumed\\
        $U_0^{CE} \text{ / V}$ & 0 & Open circuit potential of Lithium & assumed\\\midrule
        \multicolumn{4}{c}{Current collector}\\\midrule
        $\sigma_{CC} \text{ / S/cm}$ & 100 & electronic conductivity & assumed\\\midrule
        \multicolumn{4}{c}{CBD}\\\midrule
        $\sigma_{CBD} \text{ / S/cm}$ & 10 &  electronic conductivity  & assumed\\
        \midrule
        \multicolumn{4}{c}{Kinetic parameters}\\\midrule
        $i_{00}^{inter}\text{A cm/mol}$ & $0.23047$& reaction rate of intercalation in NMC & \cite{kremer.2020}\\
        $i_{00}^{CE} \text{ / A/cm}^2$ & $0.06407$ & reaction rate at counter electrode& \cite{kremer.2020}\\
        \bottomrule
    \end{tabular}
    \caption{List of electrochemical parameters used for the simulations.}
    \label{tab:parameters}
\end{table}

The conductivity of the current collector, counter-electrode and conductive additive binder domain (CBD) are assumed values.
These values, which are smaller then the real conductivities of these materials, are selected to reduce the negative impact of theses regions on the numerical solution tolerance, and at the same time retain the high conductivity compared to the active material and electrolyte phase.

\subsection{Additional images regarding the influence of the effective parameter}

Symmetrical impedance spectra for the single-layer and two-layer electrode and the different trinarization techniques are shown in Figure~\ref{fig:EC:impedance}.
\begin{figure}[!htbp]
	\begin{subfigure}[c]{0.48\textwidth}
		\centering
		\includegraphics[width=0.9\textwidth]{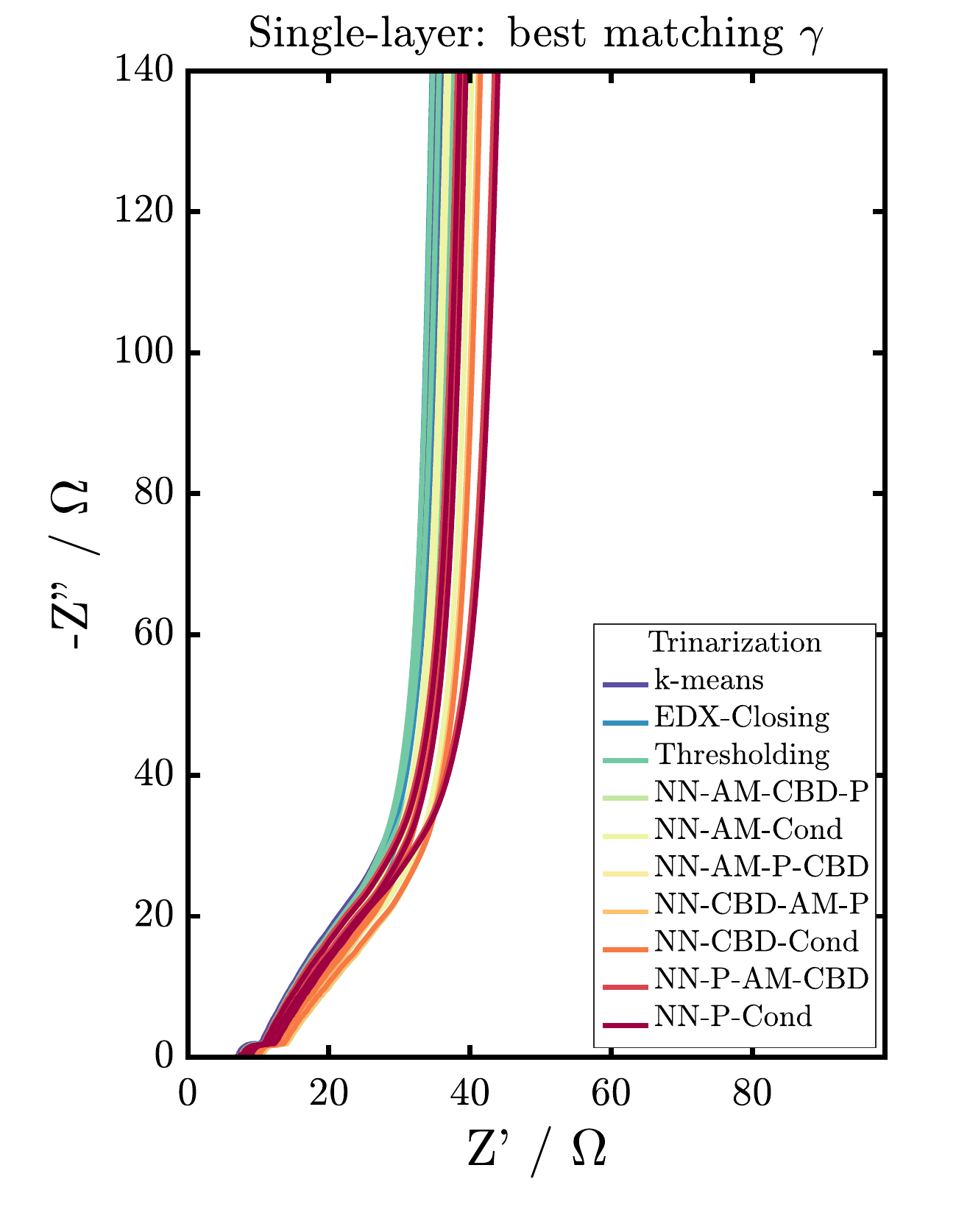}
		\caption{Impedance spectra for the single-layer cathode.}
	\label{fig:EC:imp:sl}
	\end{subfigure}
	\hfill
	\begin{subfigure}[c]{0.48\textwidth}
		\centering
		\includegraphics[width=0.9\textwidth]{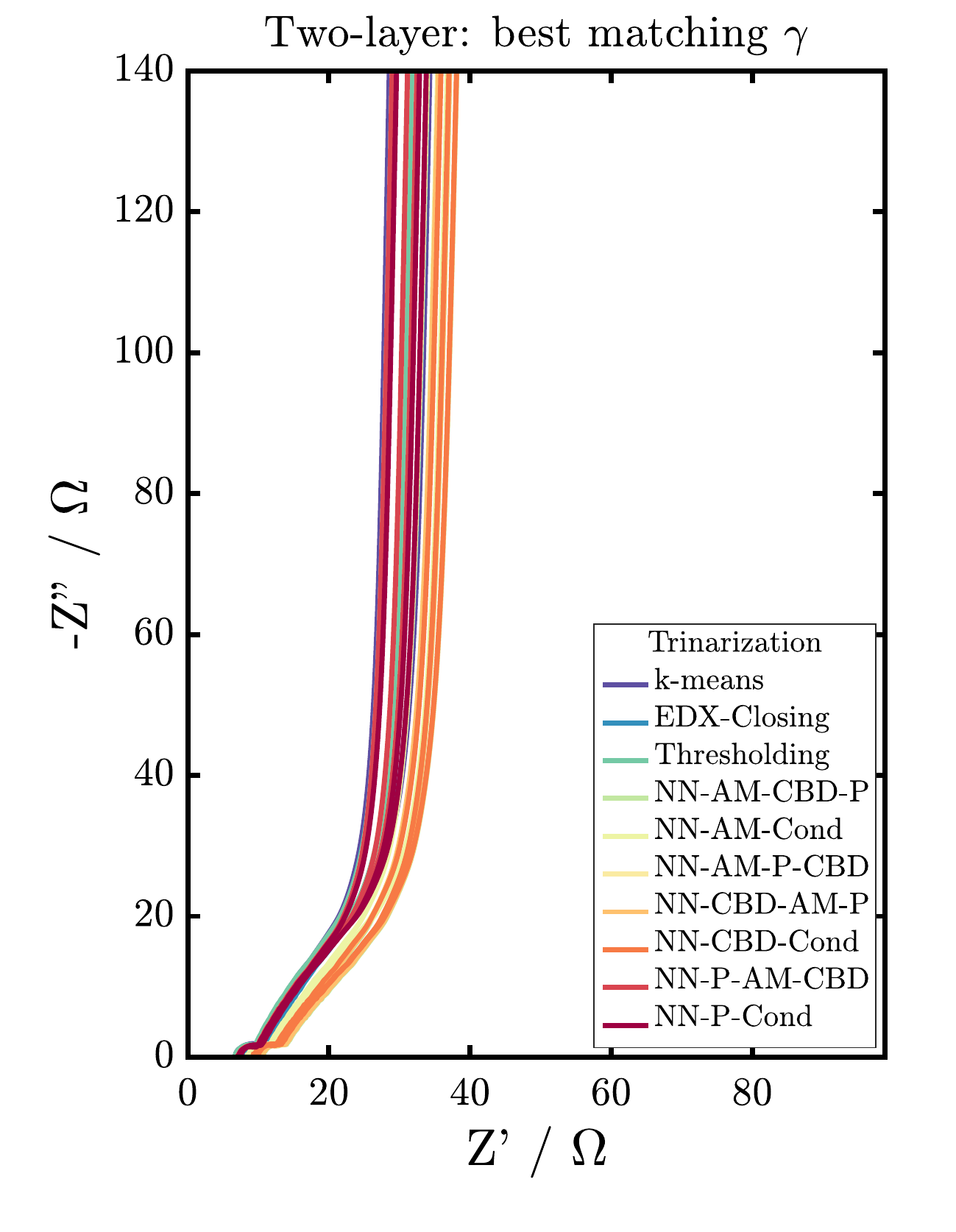}	
		\caption{Impedance spectra for the two-layer cathode.}
	\label{fig:EC:imp:tl}
	\end{subfigure}
	\caption{Symmetrical impedance spectra for all trinarization techniques for the two different electrodes with the best matching $\gamma$ as identified at $\SI{6}{\milli\ampere\per\cm^{-2}}$.}
	\label{fig:EC:impedance}
\end{figure}

The impact of the effective transport parameter $\gamma$ is on the symmetrical impedance is shown in Figure~\ref{SI:fig:impedance:gamma}.
\begin{figure}[htbp]
	\centering
	\includegraphics[width=0.48\textwidth]{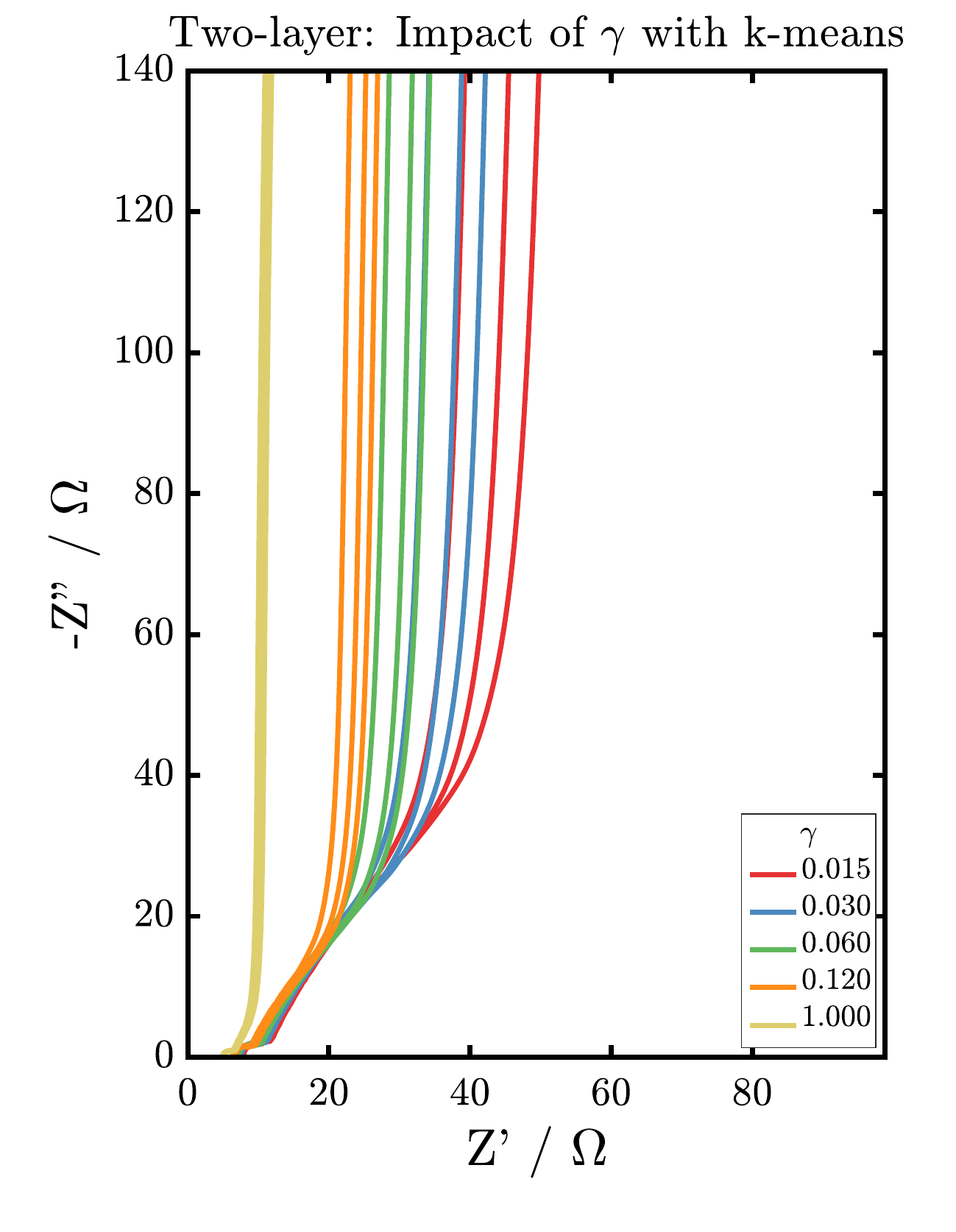}
	\caption{The symmetrical impedance spectra for the $k$-means approach for the two-layer electrode show the influence of the effective transport parameter $\gamma$.}
	\label{SI:fig:impedance:gamma}
\end{figure}

The effective tortuosity of the CBD is identified by comparing the lithiation simulations at $\SI{6}{\milli\ampere\per\cm^{2}}$ with the experimental data.
Figure~\ref{SI:fig:EC:variTau} contains the images used for the selection of the best-matching effective transport parameter $\tau_{\mathsf{CBD}}$ for the single-layer (left two columns) and the two-layer (right two columns).
\begin{figure}[ht]
	\begin{subfigure}[c]{0.24\textwidth}
		\centering
		\includegraphics[width=0.9\textwidth]{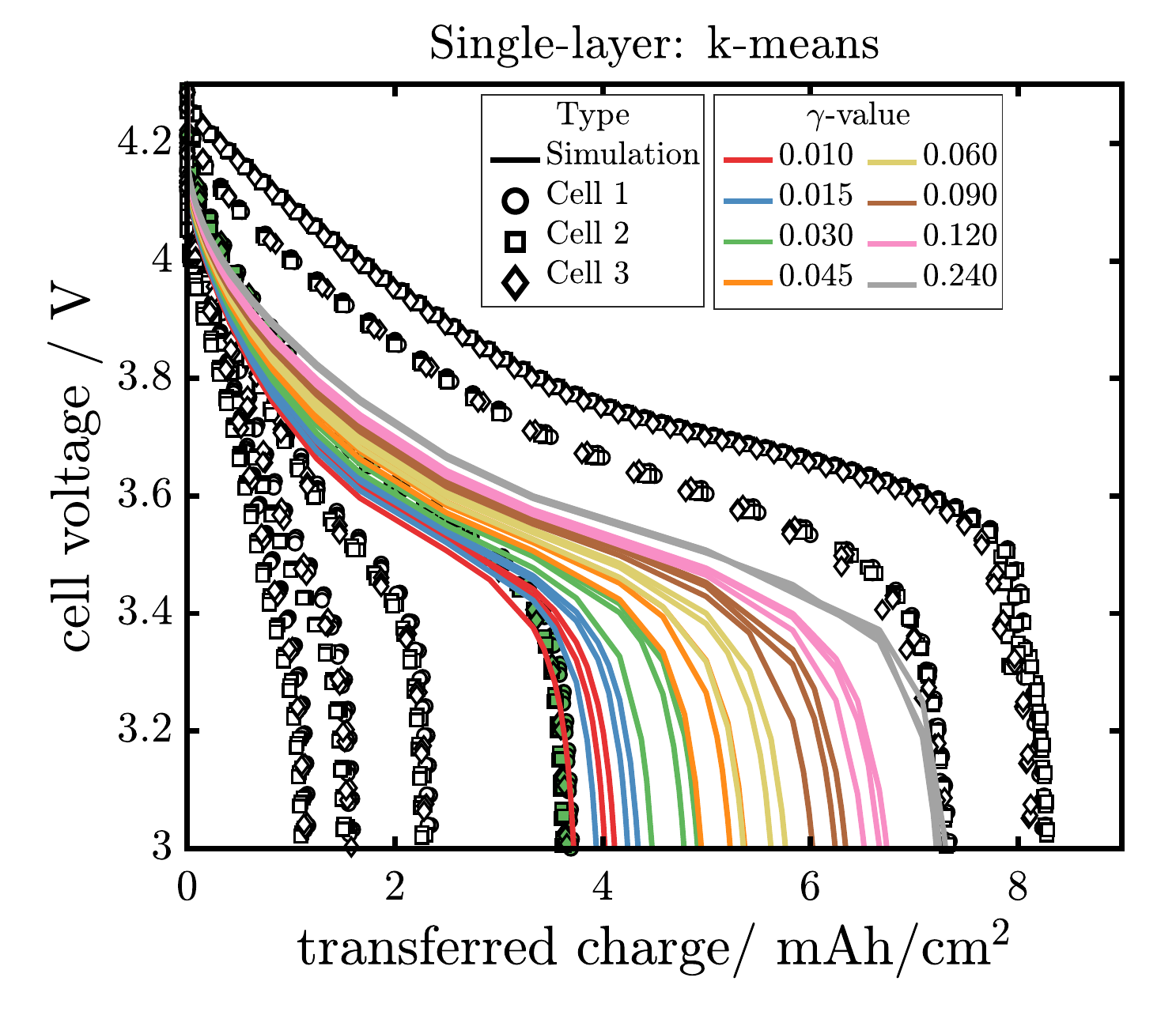}	
	\end{subfigure}
	\hfill
	\begin{subfigure}[c]{0.24\textwidth}
		\centering
		\includegraphics[width=0.9\textwidth]{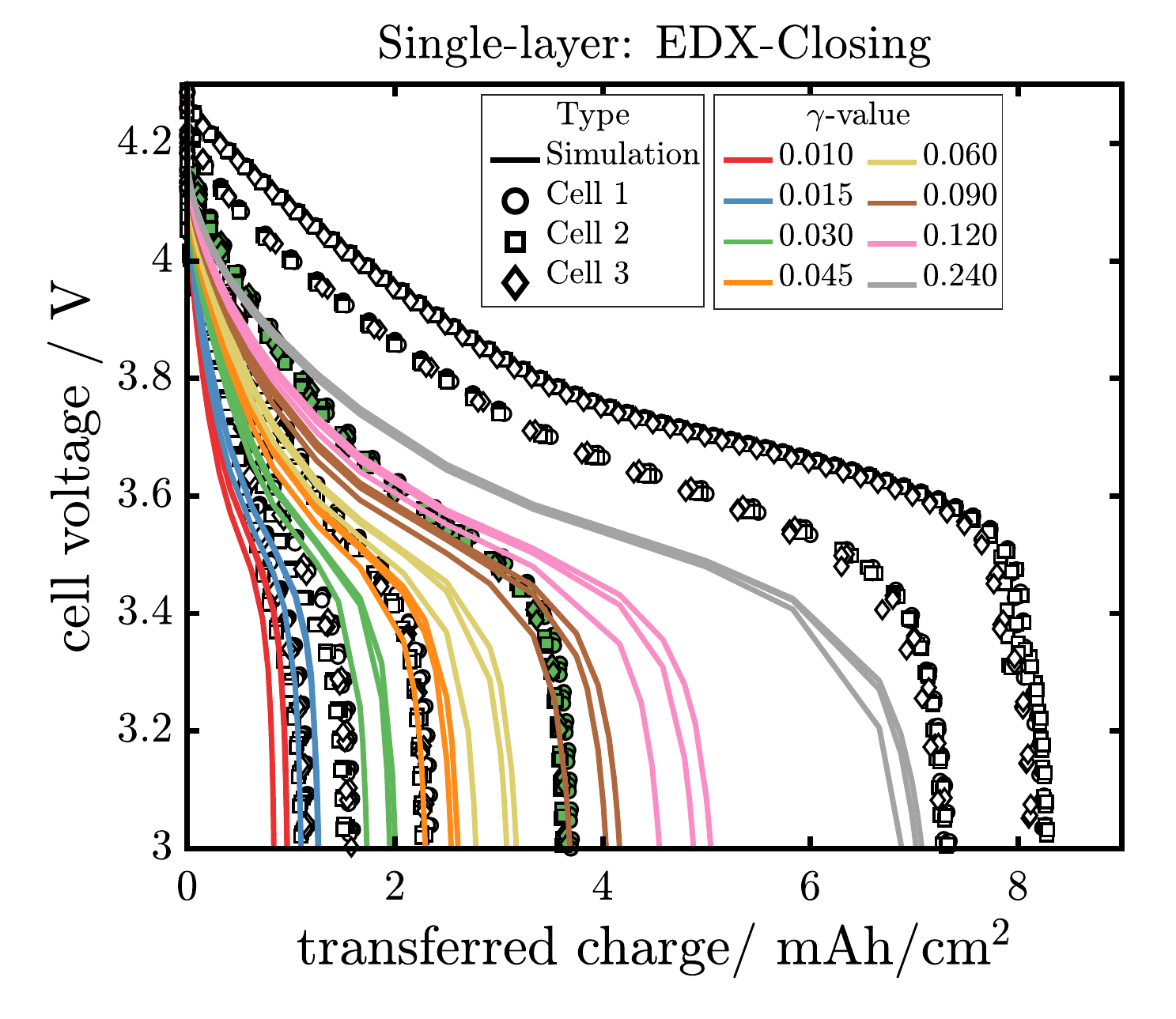}	
	\end{subfigure}
	\hfill
	\begin{subfigure}[c]{0.24\textwidth}
		\centering
		\includegraphics[width=0.9\textwidth]{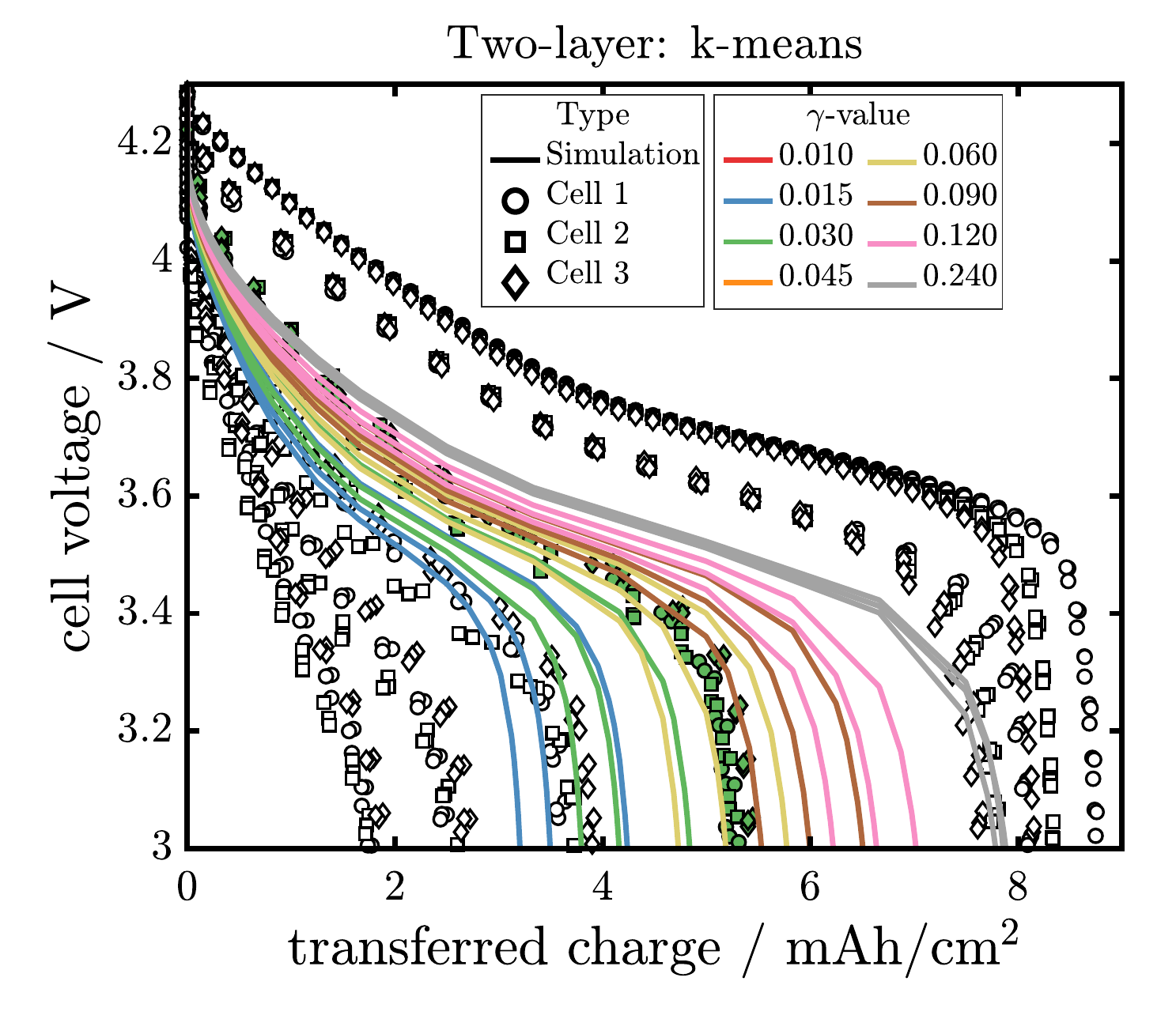}	
	\end{subfigure}
	\hfill
	\begin{subfigure}[c]{0.24\textwidth}
		\centering
		\includegraphics[width=0.9\textwidth]{imagesvorlaeufig/Lithiierung/TL_variTau_2_EDXClosing.pdf}
	\end{subfigure}
	\\
	\begin{subfigure}[c]{0.24\textwidth}
		\centering
		\includegraphics[width=0.9\textwidth]{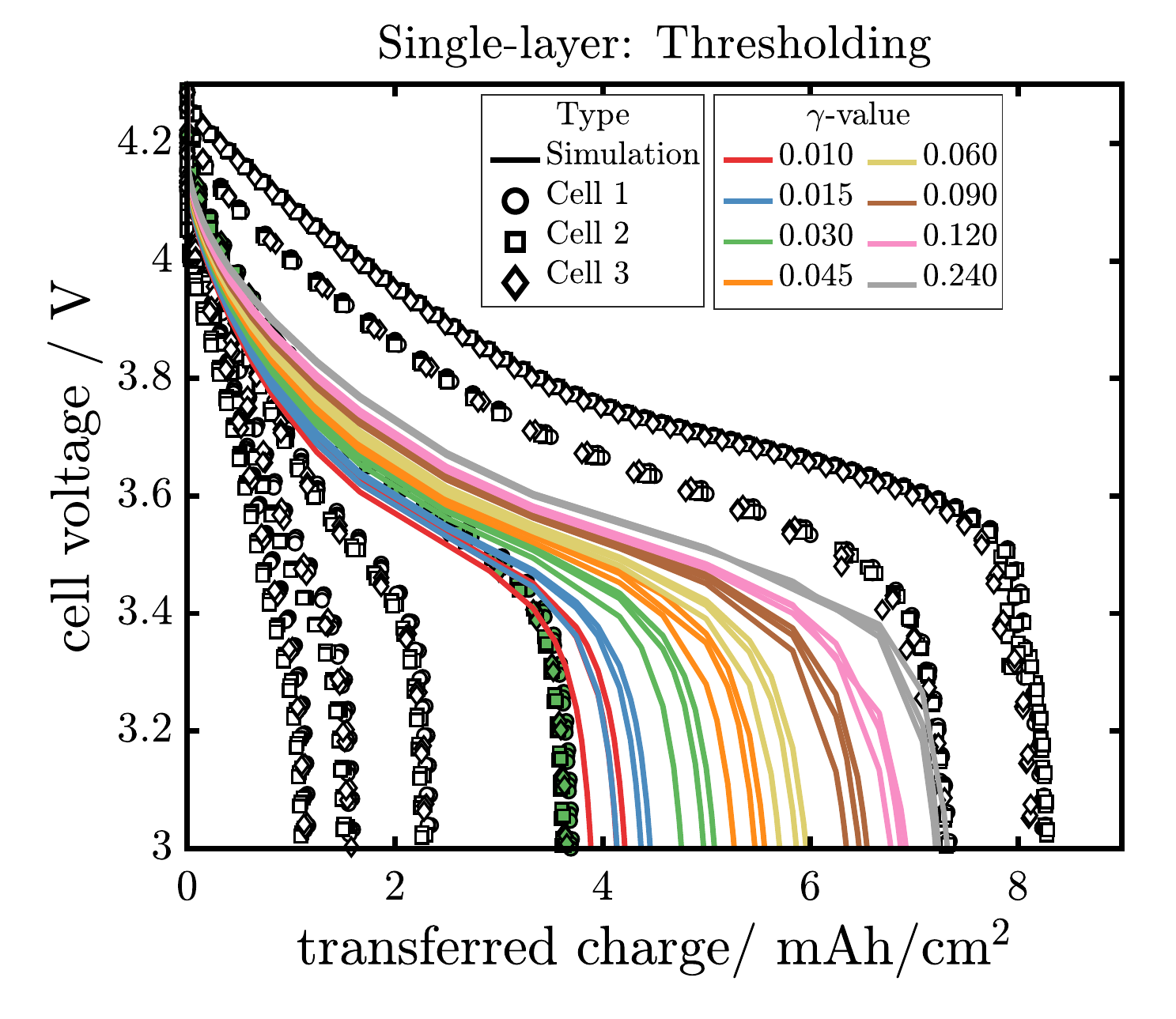}	
	\end{subfigure}
	\hfill
	\begin{subfigure}[c]{0.24\textwidth}
		\centering
		\includegraphics[width=0.9\textwidth]{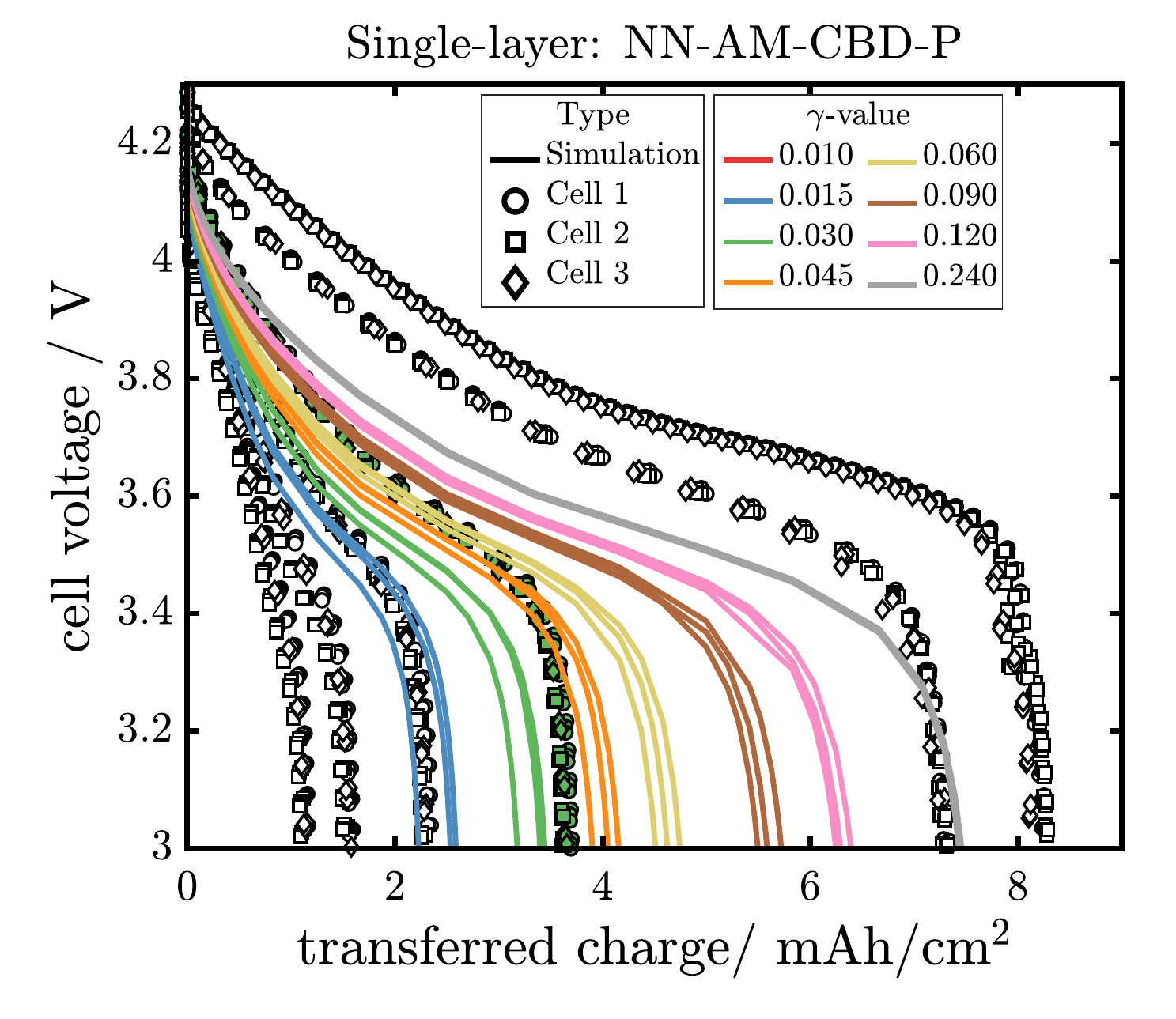}	
	\end{subfigure}
	\hfill
	\begin{subfigure}[c]{0.24\textwidth}
		\centering
		\includegraphics[width=0.9\textwidth]{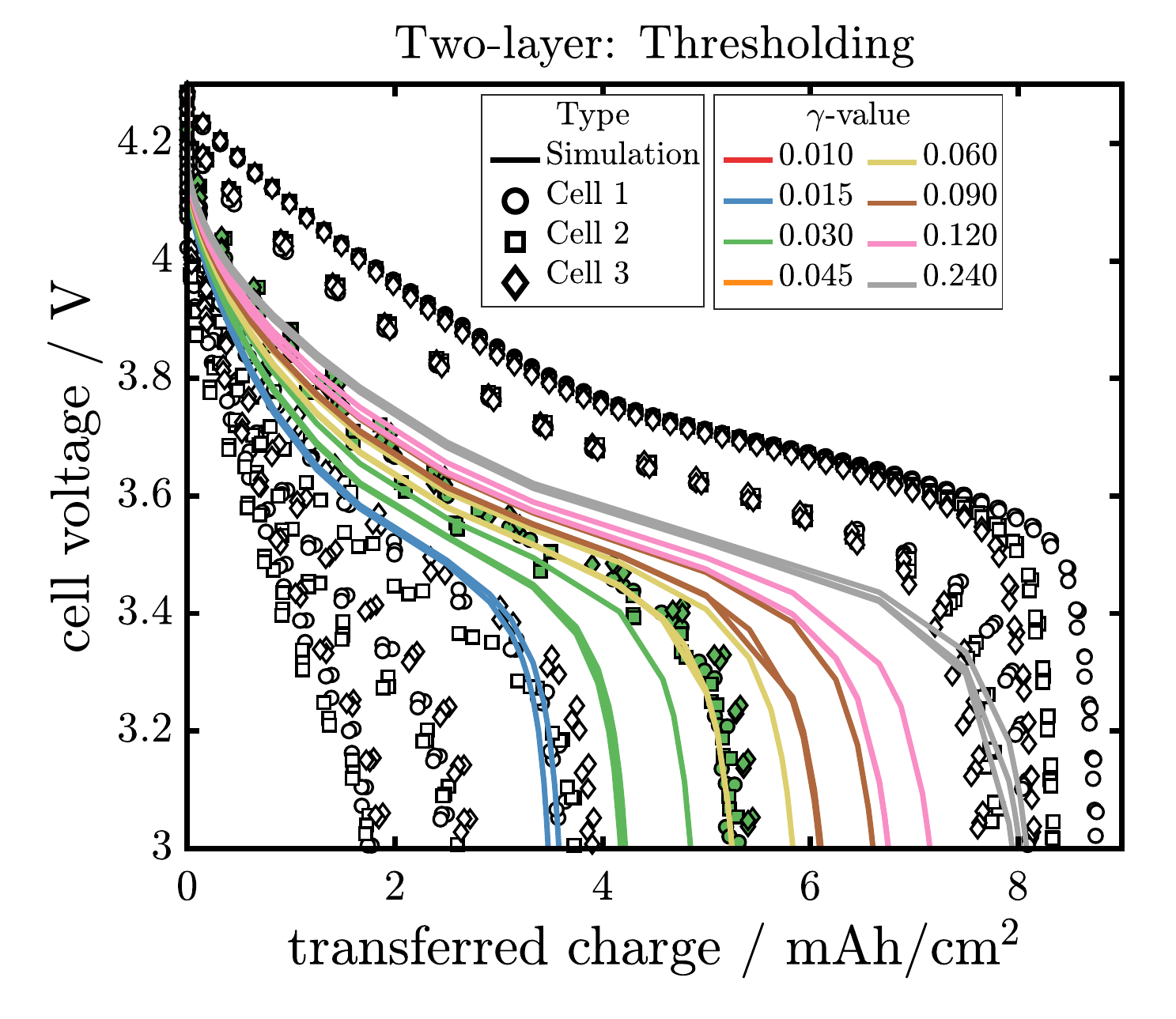}	
	\end{subfigure}
	\hfill
	\begin{subfigure}[c]{0.24\textwidth}
		\centering
		\includegraphics[width=0.9\textwidth]{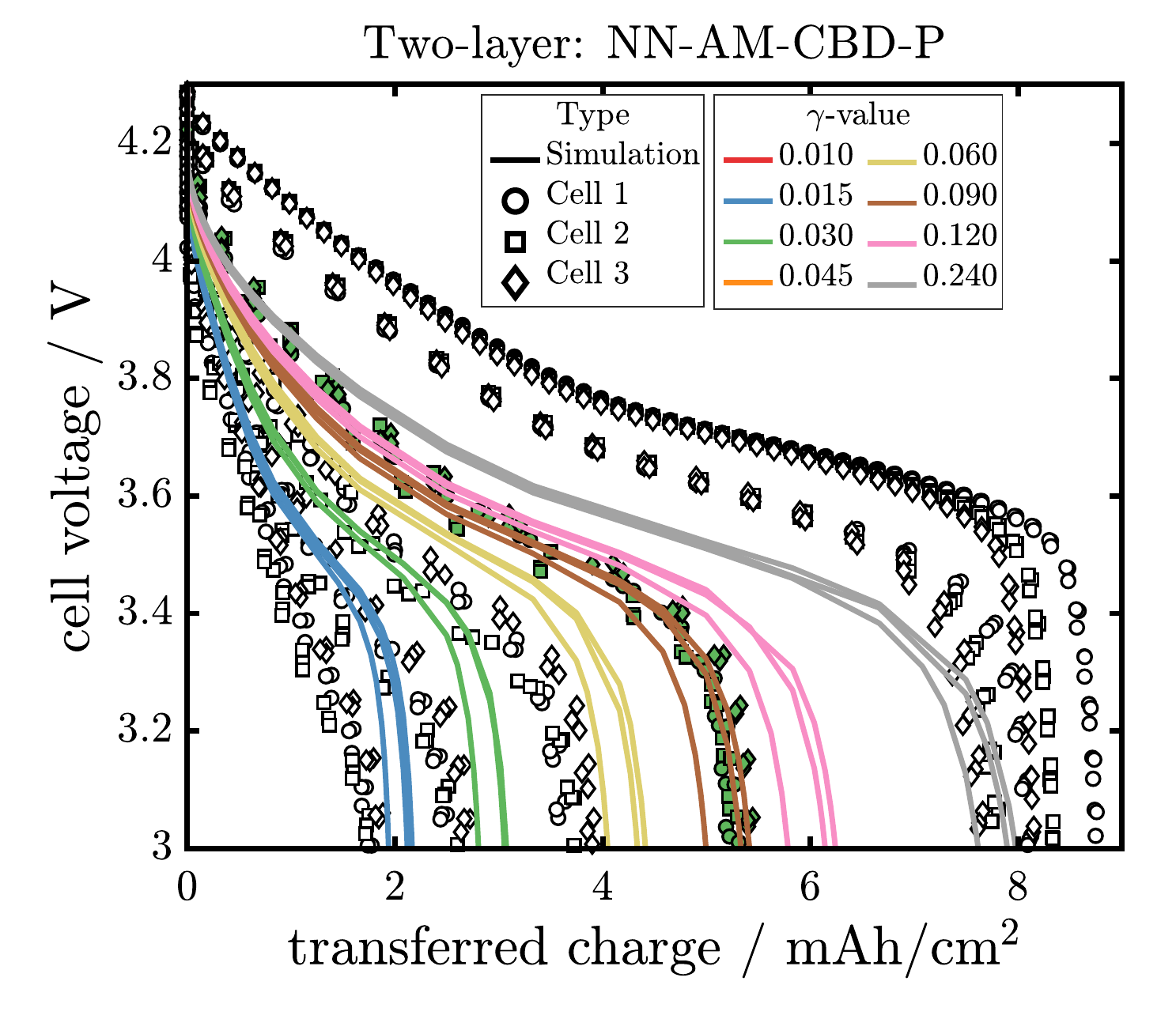}
	\end{subfigure}
	\\
	\begin{subfigure}[c]{0.24\textwidth}
		\centering
		\includegraphics[width=0.9\textwidth]{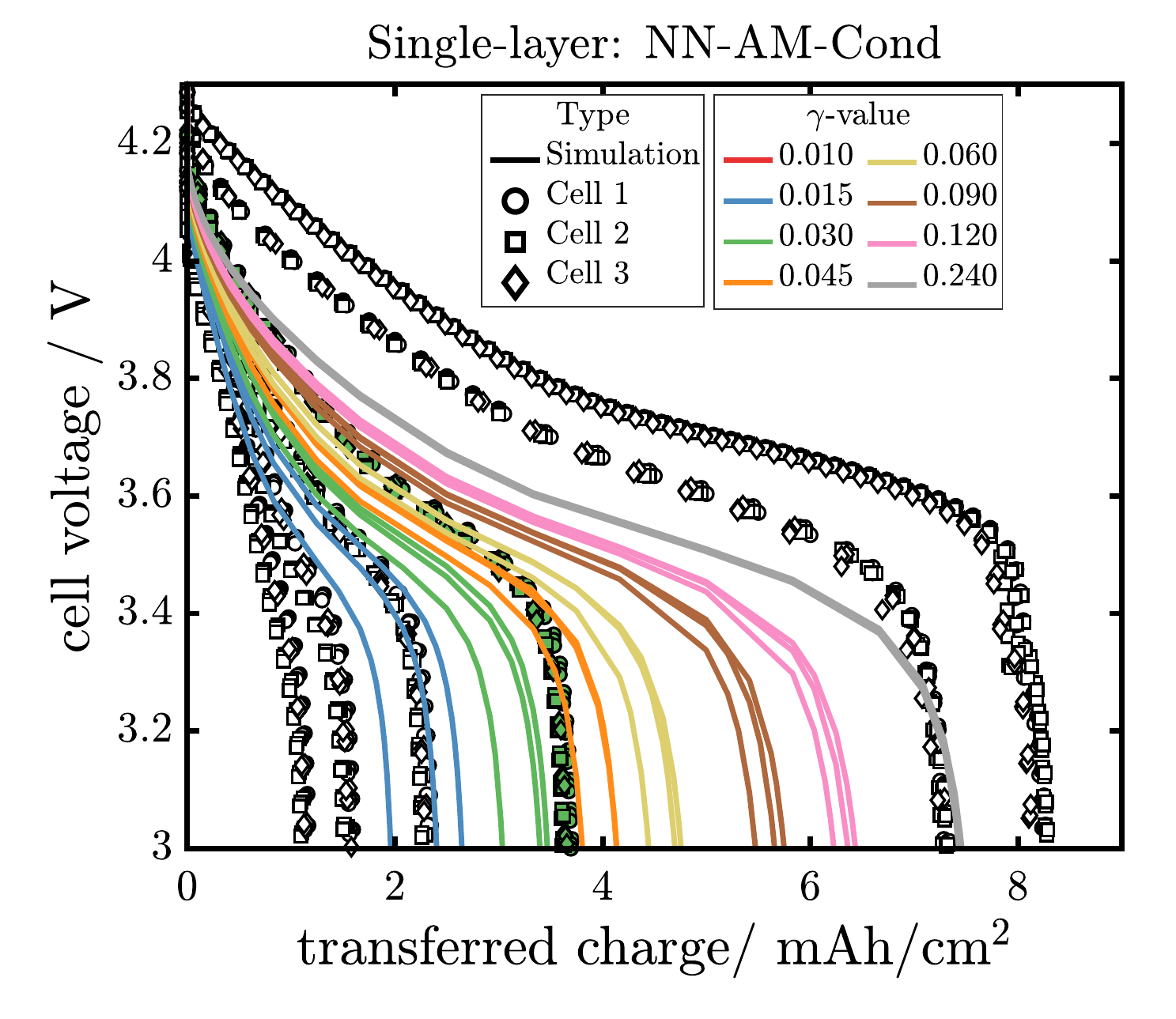}	
	\end{subfigure}
	\hfill
	\begin{subfigure}[c]{0.24\textwidth}
		\centering
		\includegraphics[width=0.9\textwidth]{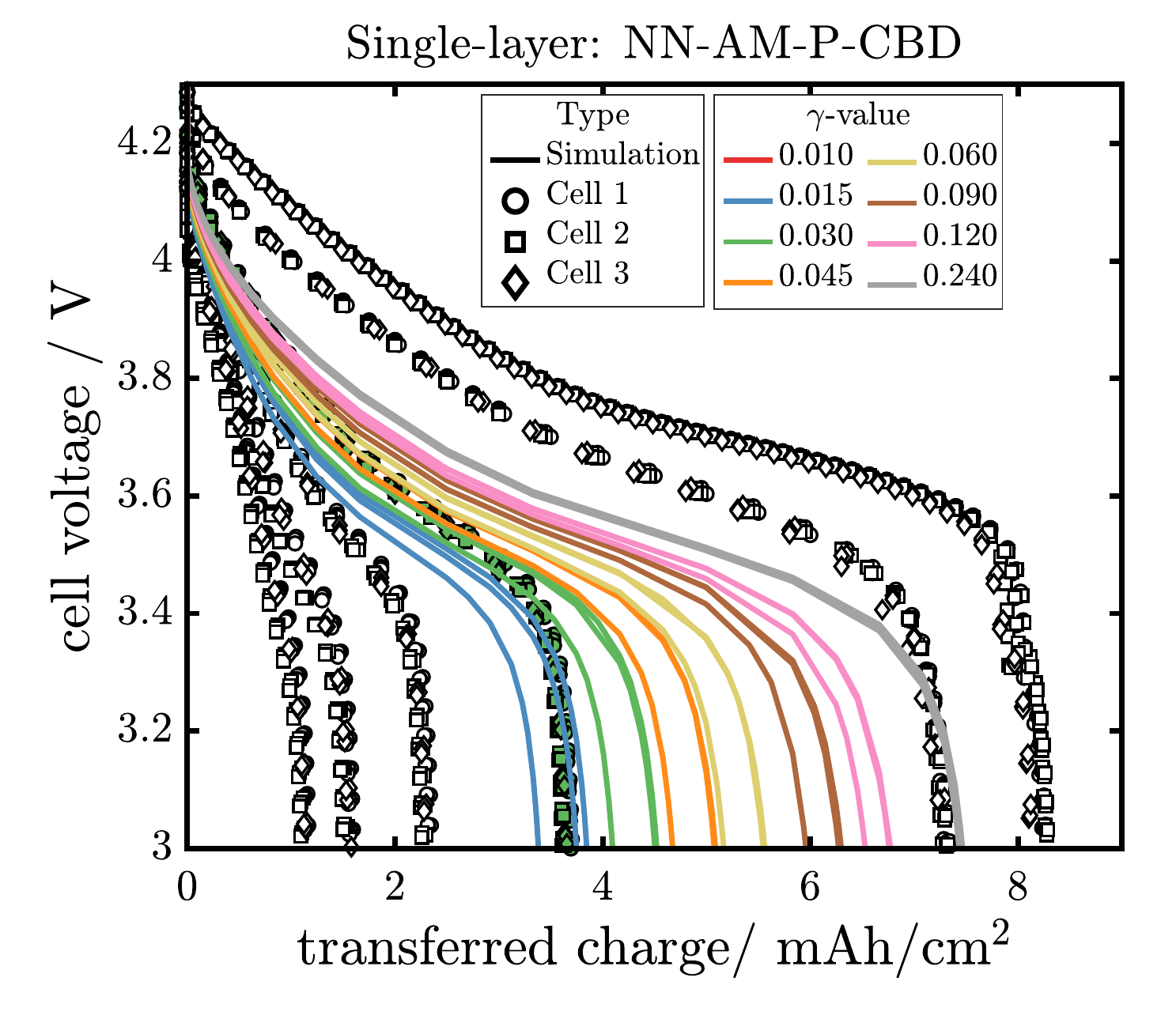}	
	\end{subfigure}
	\hfill
	\begin{subfigure}[c]{0.24\textwidth}
		\centering
		\includegraphics[width=0.9\textwidth]{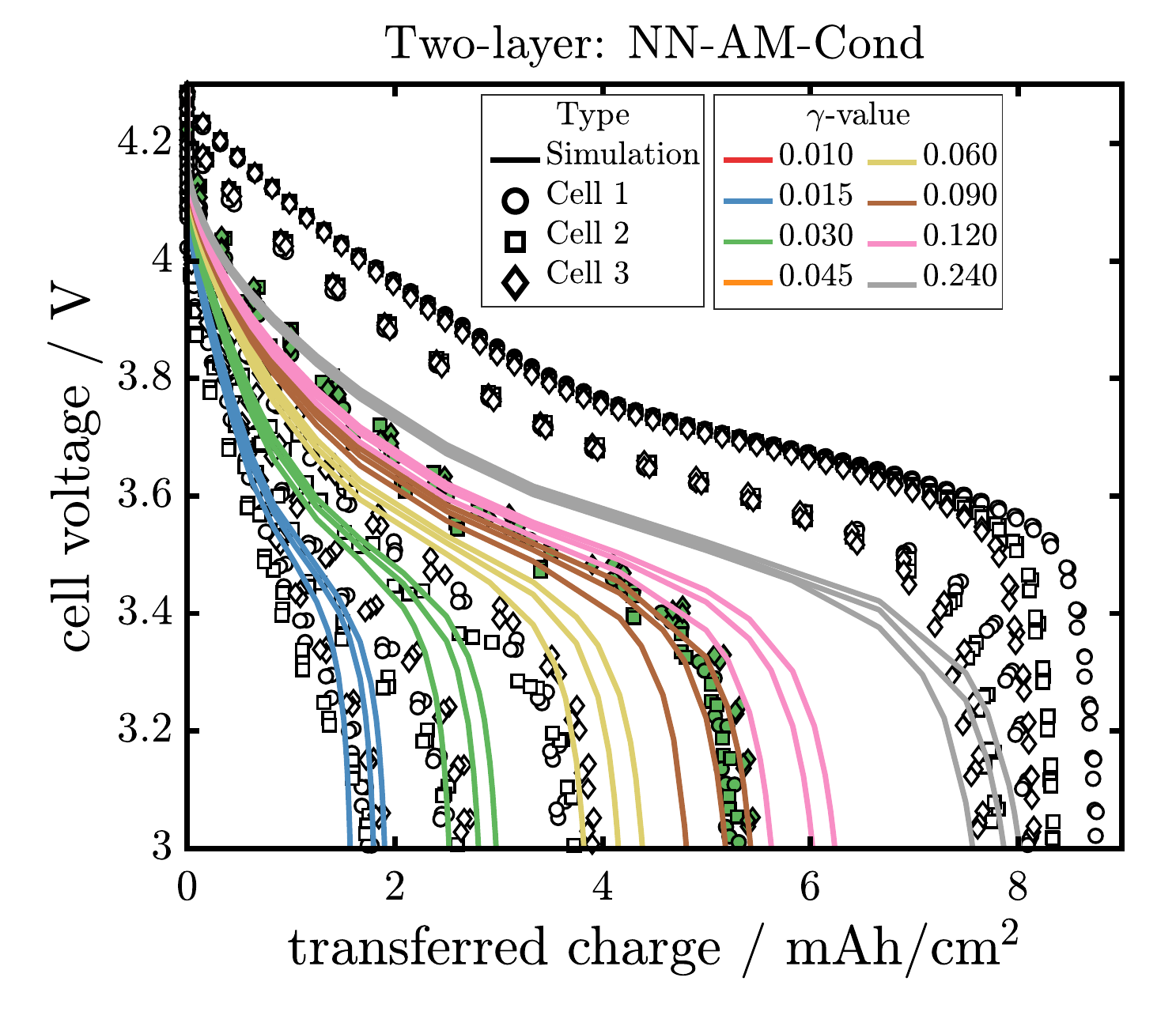}	
	\end{subfigure}
	\hfill
	\begin{subfigure}[c]{0.24\textwidth}
		\centering
		\includegraphics[width=0.9\textwidth]{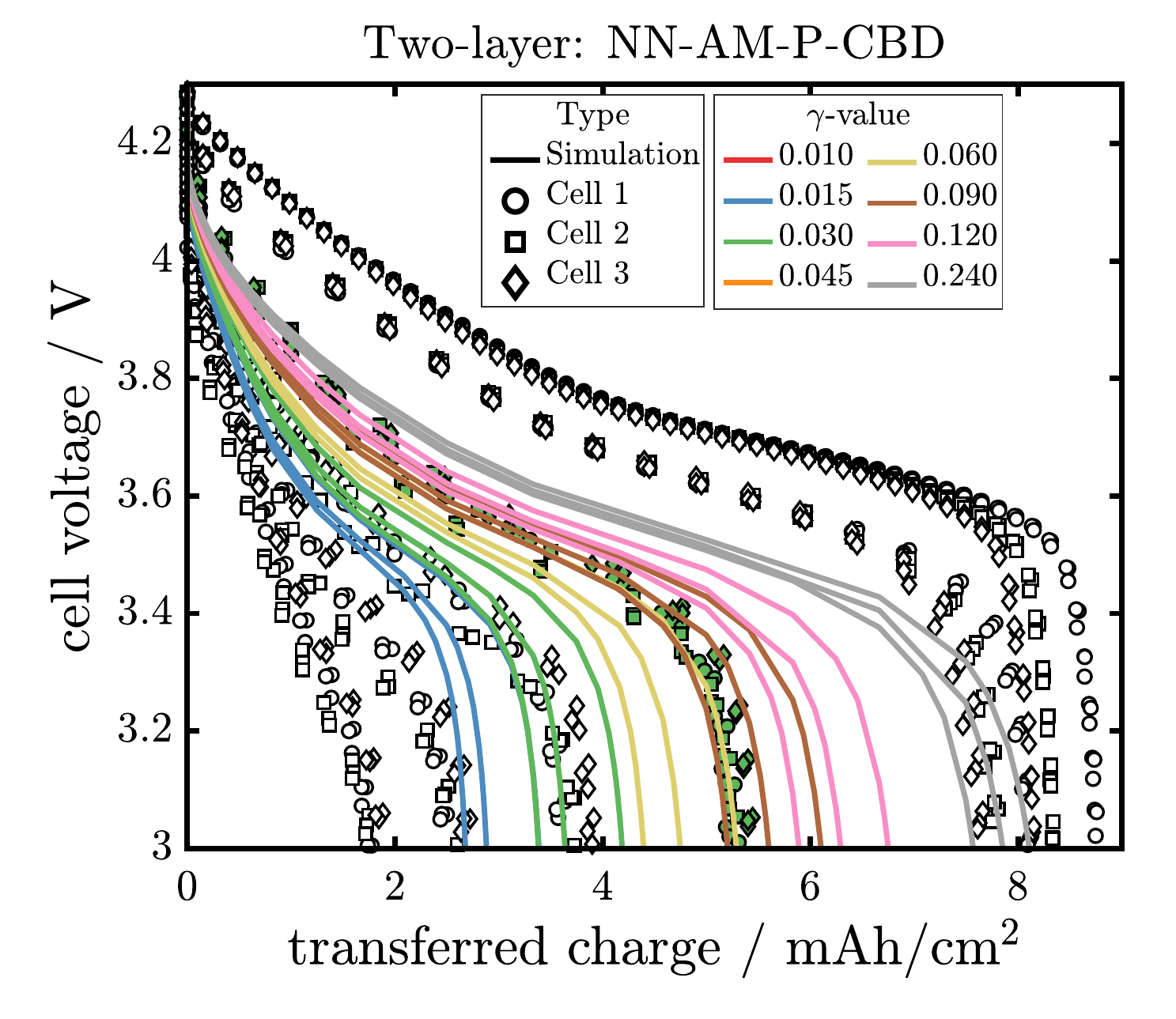}
	\end{subfigure}
	\\
	\begin{subfigure}[c]{0.24\textwidth}
		\centering
		\includegraphics[width=0.9\textwidth]{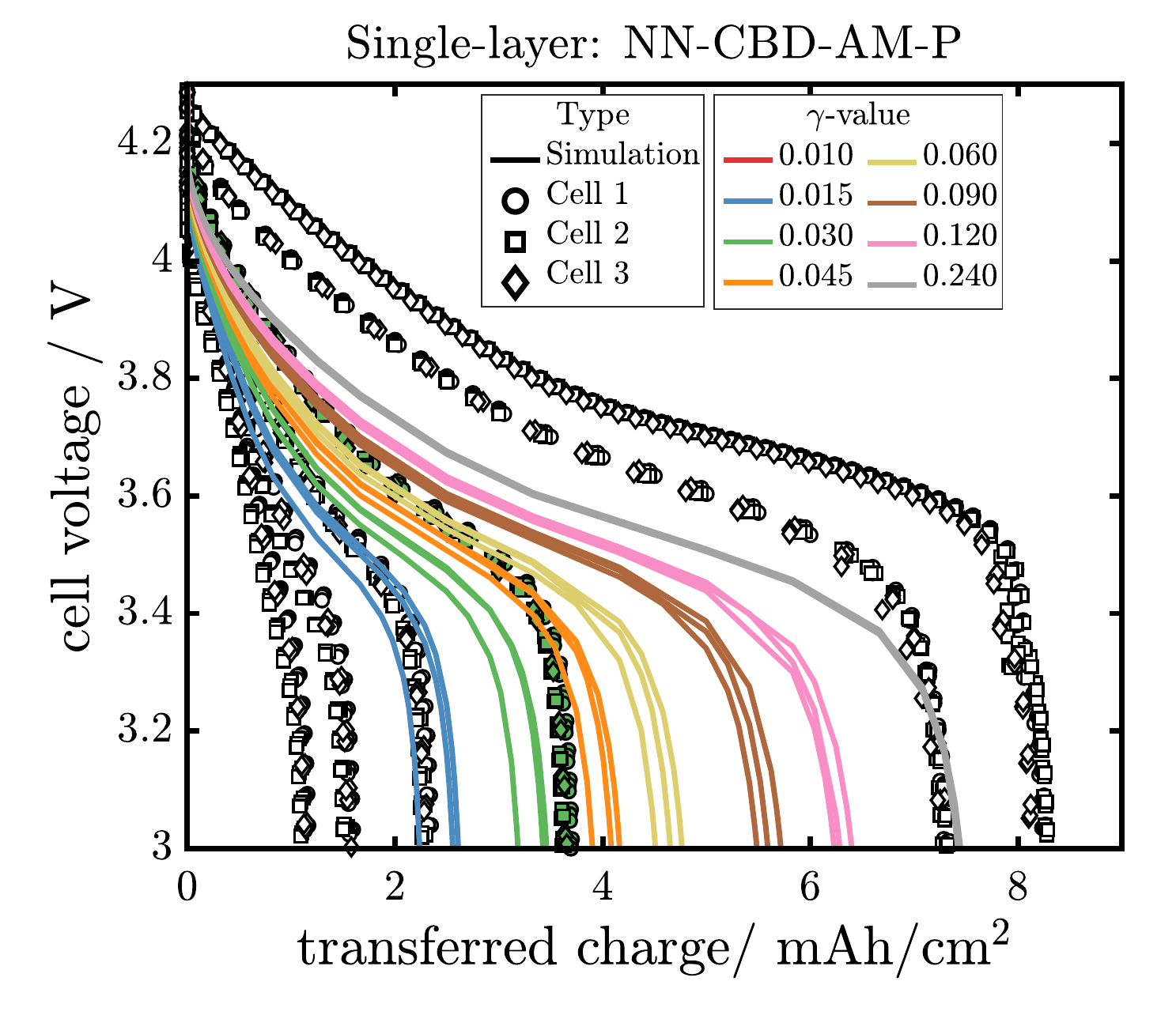}	
	\end{subfigure}
	\hfill
	\begin{subfigure}[c]{0.24\textwidth}
		\centering
		\includegraphics[width=0.9\textwidth]{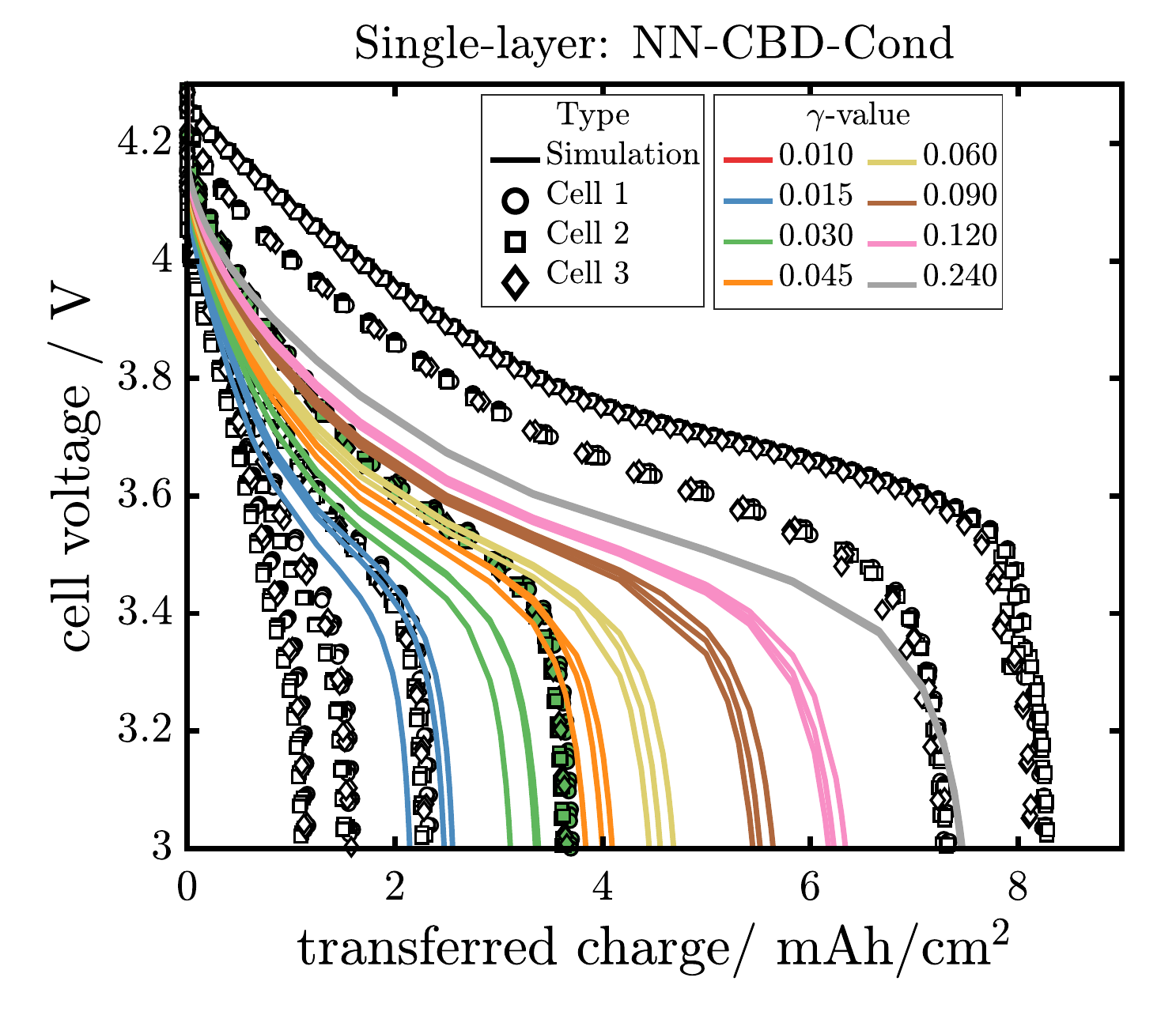}	
	\end{subfigure}
	\hfill
	\begin{subfigure}[c]{0.24\textwidth}
		\centering
		\includegraphics[width=0.9\textwidth]{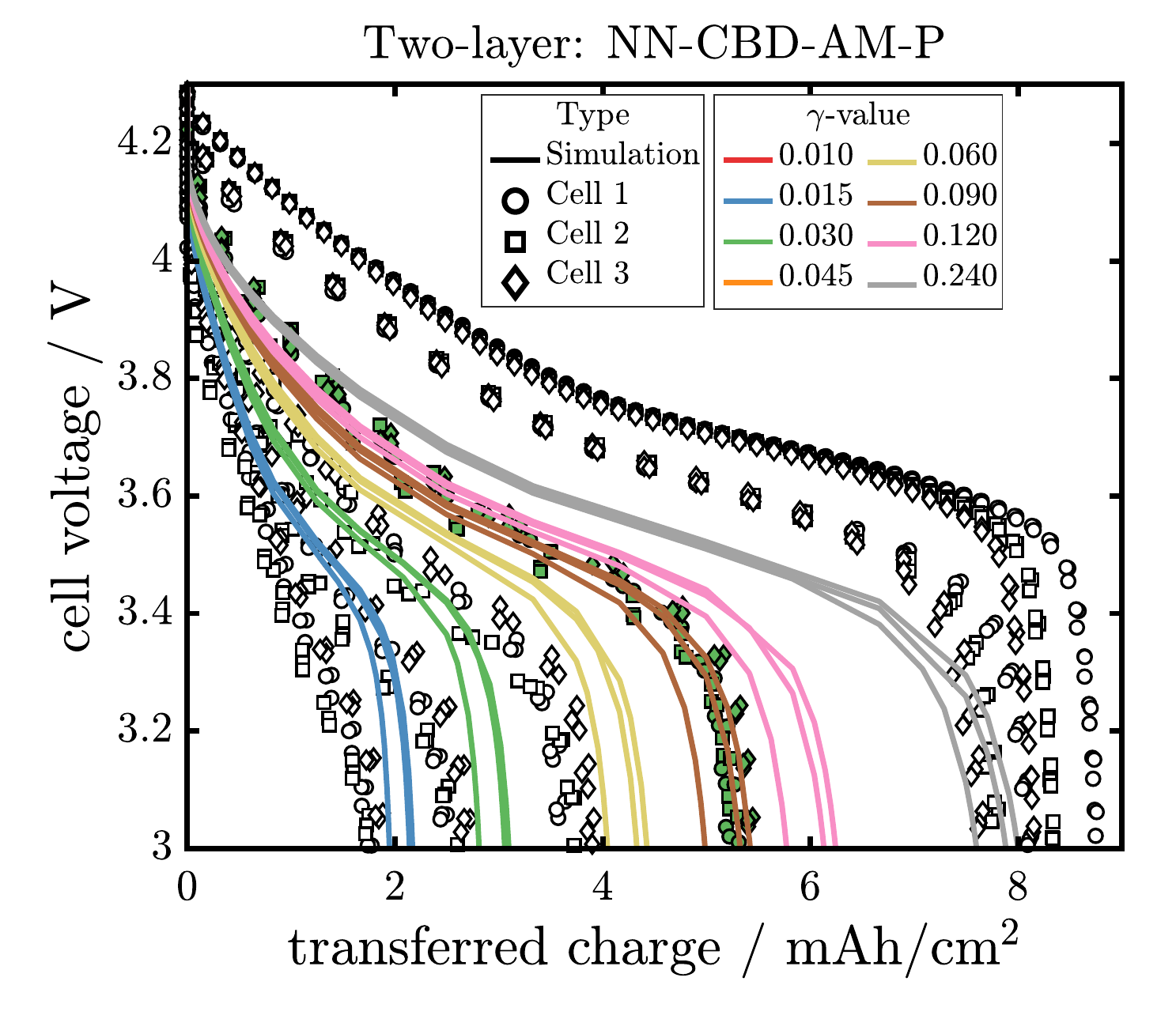}	
	\end{subfigure}
	\hfill
	\begin{subfigure}[c]{0.24\textwidth}
		\centering
		\includegraphics[width=0.9\textwidth]{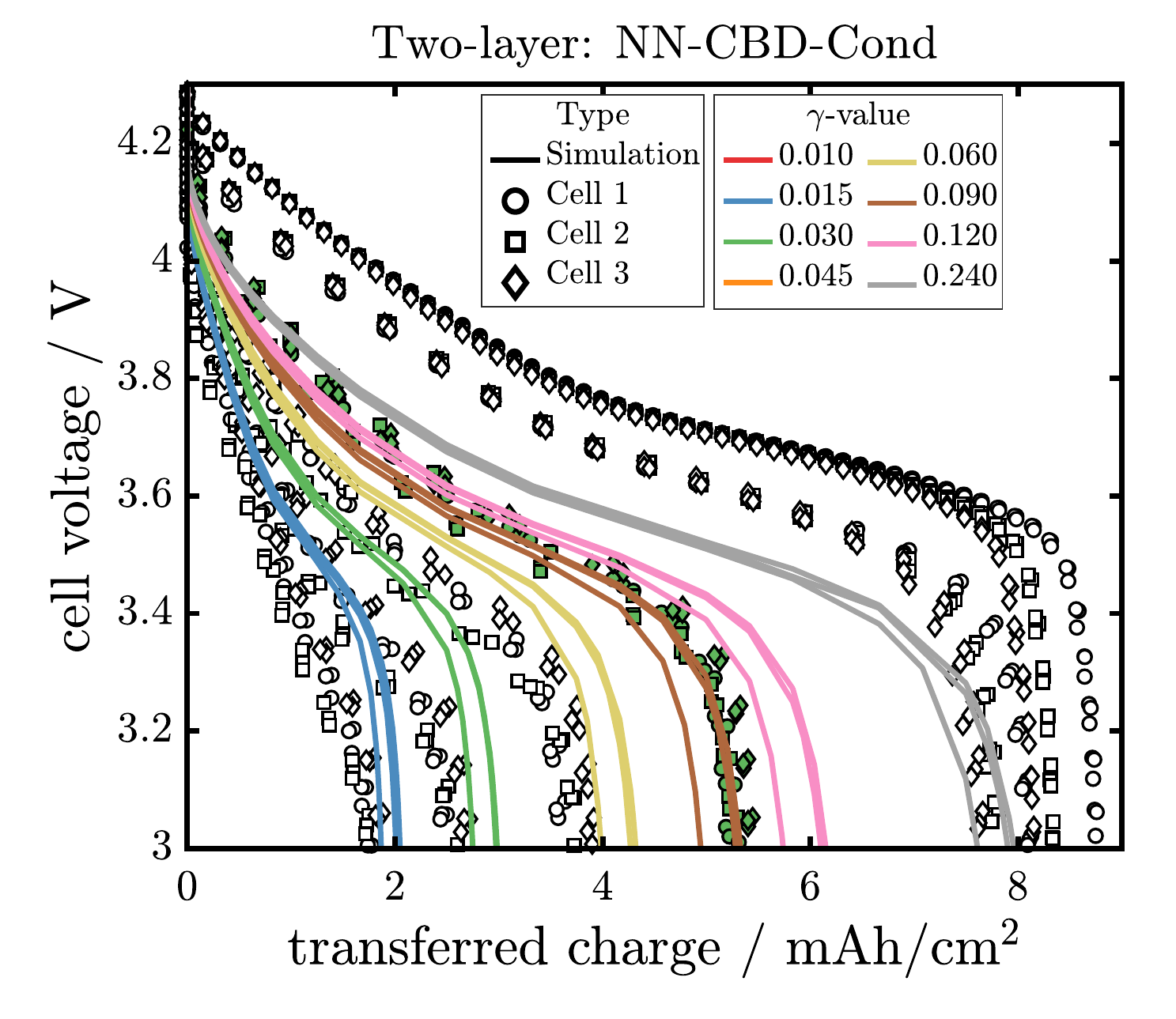}
	\end{subfigure}
	\\
	\begin{subfigure}[c]{0.24\textwidth}
		\centering
		\includegraphics[width=0.9\textwidth]{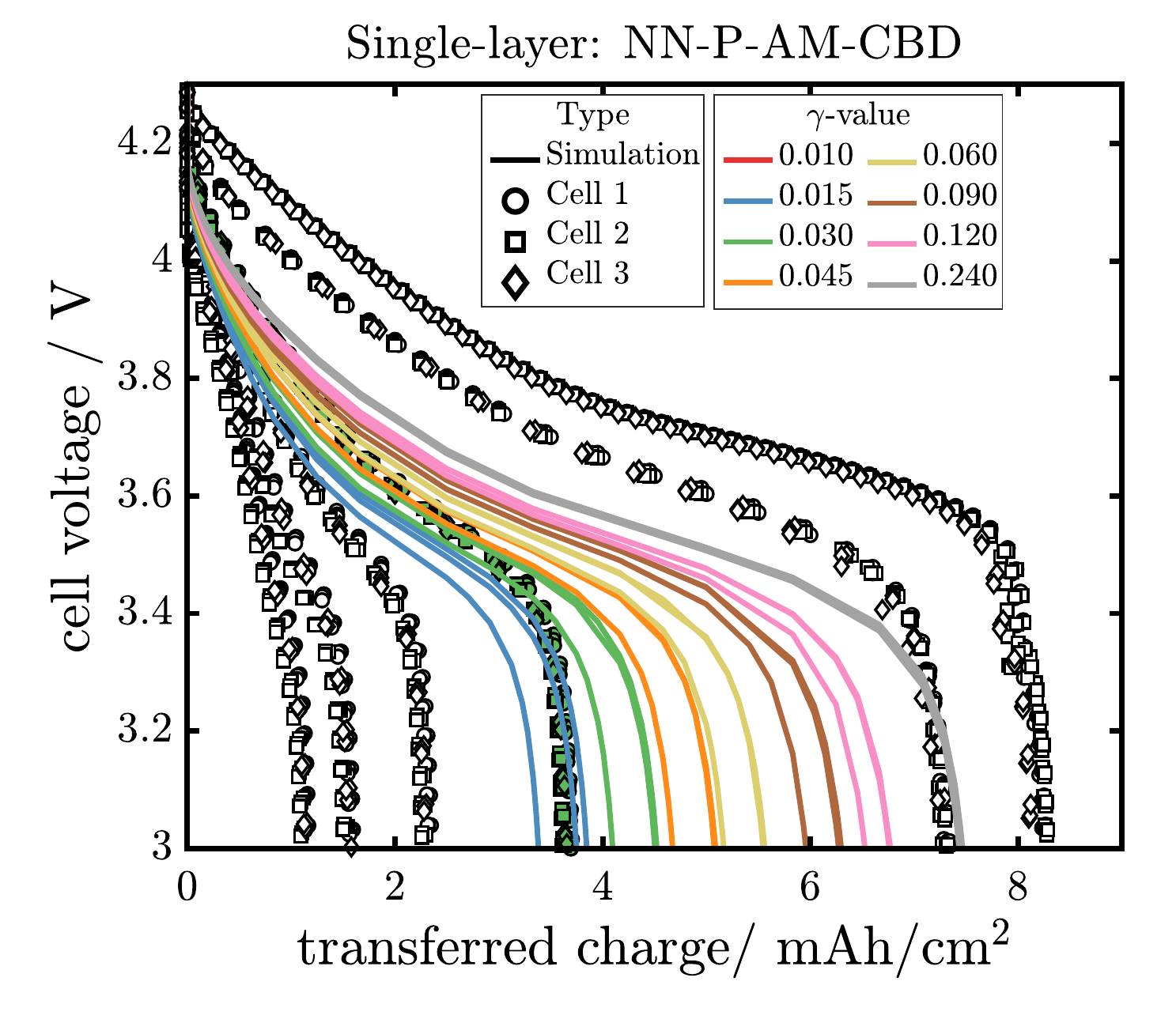}	
	\end{subfigure}
	\hfill
	\begin{subfigure}[c]{0.24\textwidth}
		\centering
		\includegraphics[width=0.9\textwidth]{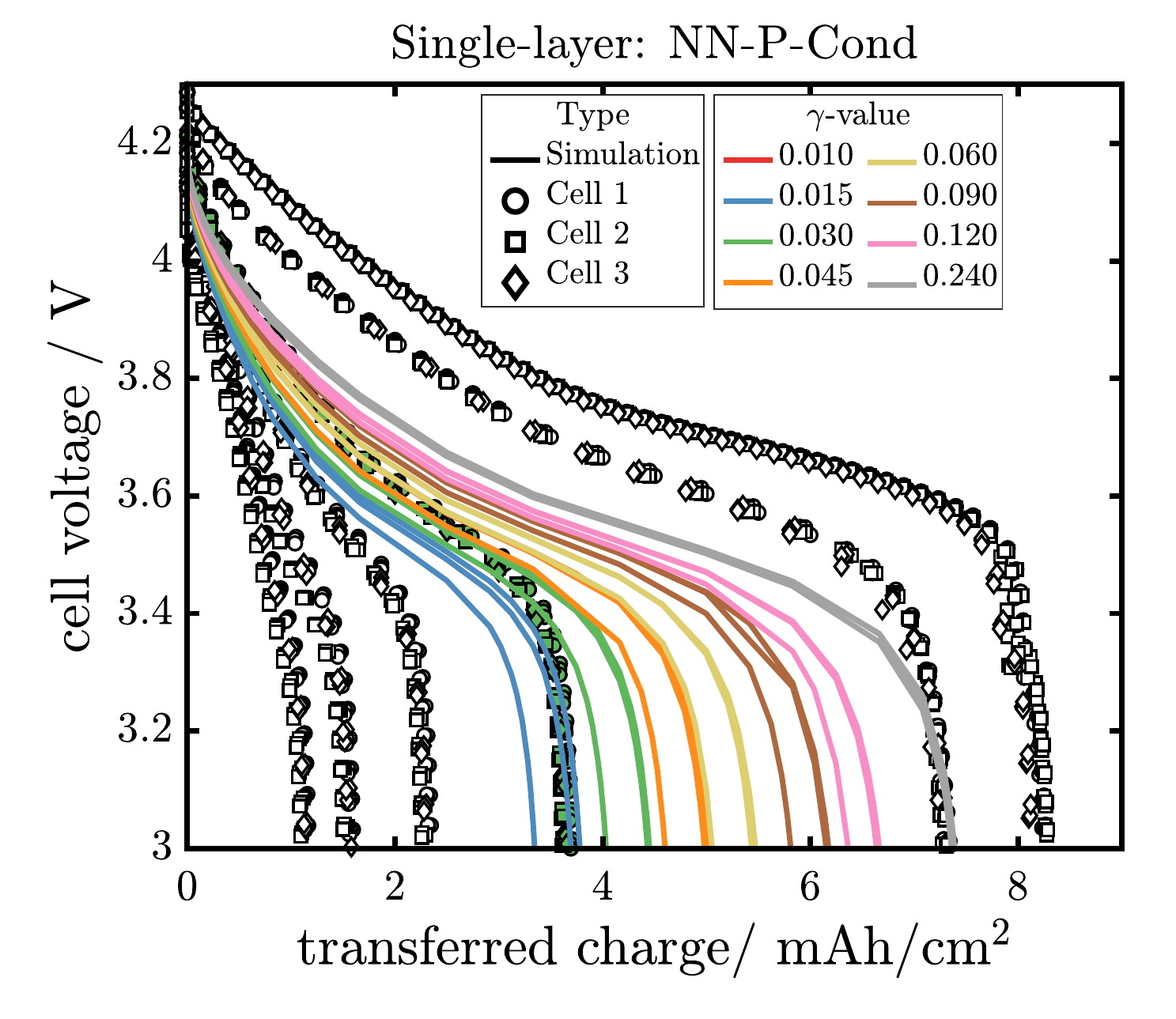}	
	\end{subfigure}
	\hfill
	\begin{subfigure}[c]{0.24\textwidth}
		\centering
		\includegraphics[width=0.9\textwidth]{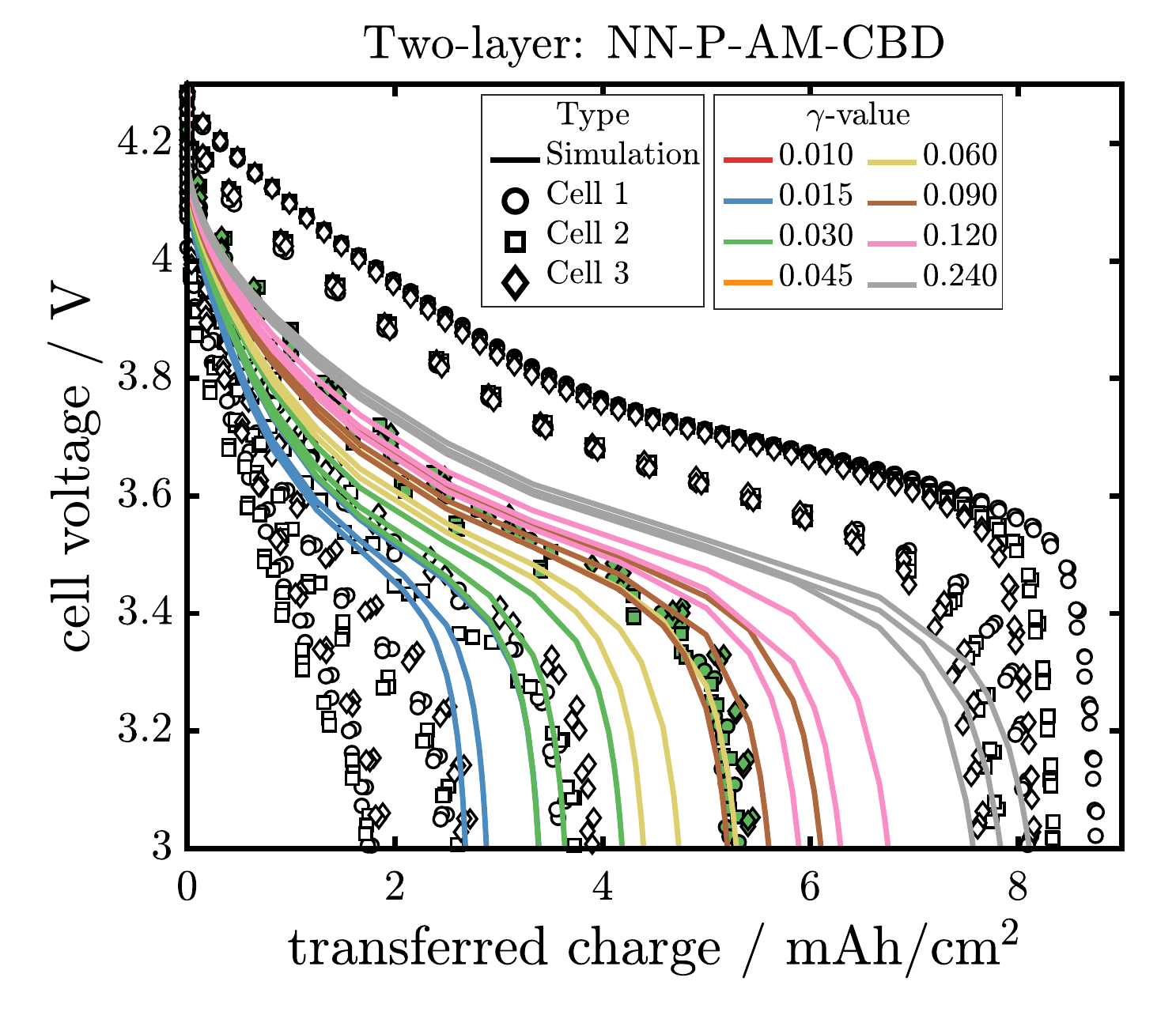}	
	\end{subfigure}
	\hfill
	\begin{subfigure}[c]{0.24\textwidth}
		\centering
		\includegraphics[width=0.9\textwidth]{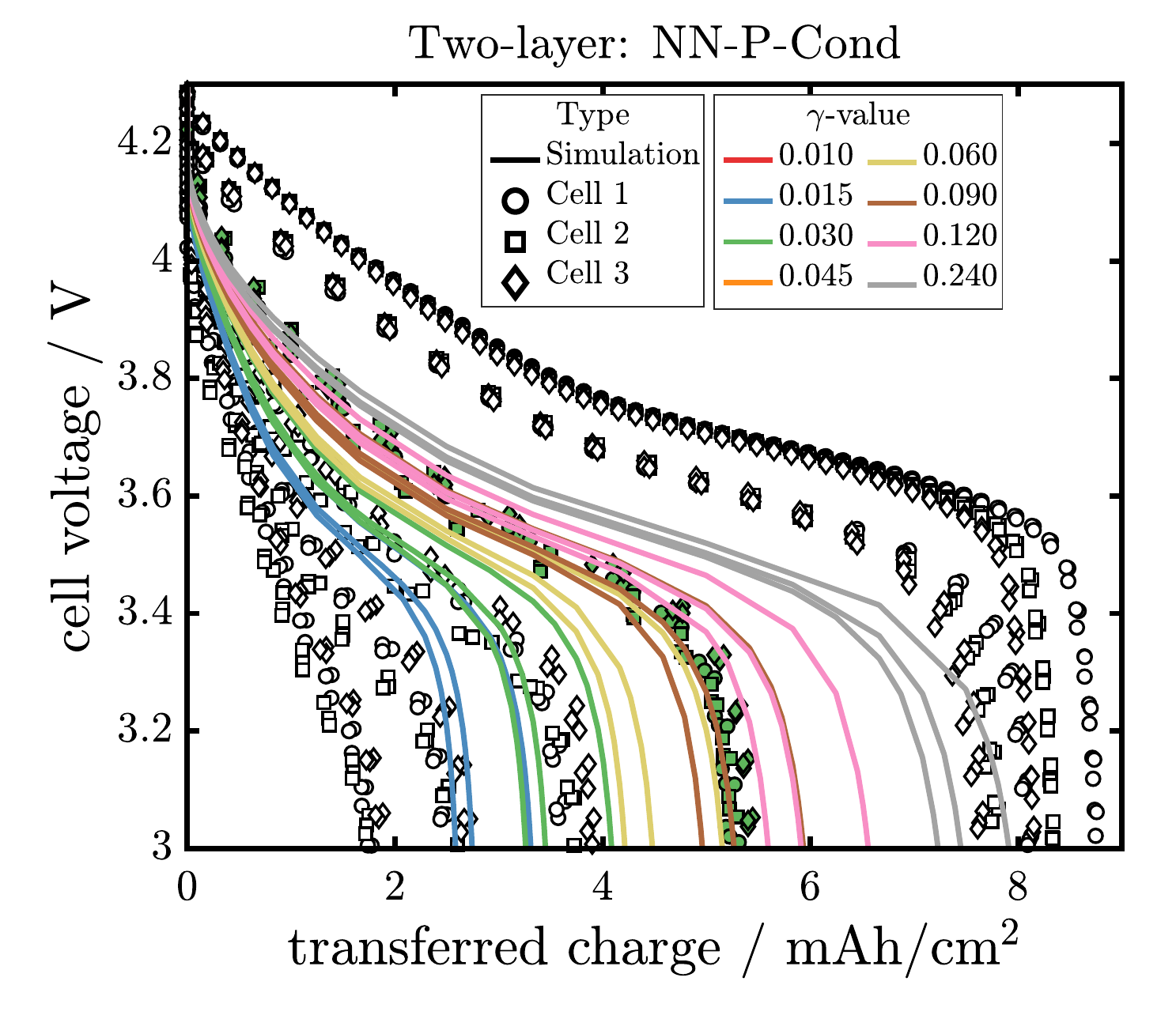}
	\end{subfigure}
	\caption{Variation of the effective transport parameter $\tau_{\mathsf{CBD}}$ for the single-layer cathode (left columns) and the two-layer cathode (right columns).}
	\label{SI:fig:EC:variTau}
\end{figure}

The lithiation simulations for the single-layer cathode (left two columns) and the two-layer cathode (right two columns) electrode for the selected effective CBD parameter (see Table~\ref{tab:tau}) are shown in Figure~\ref{SI:fig:EC:BestTau}.
\begin{figure}[ht]
	\begin{subfigure}[c]{0.24\textwidth}
		\centering
		\includegraphics[width=0.9\textwidth]{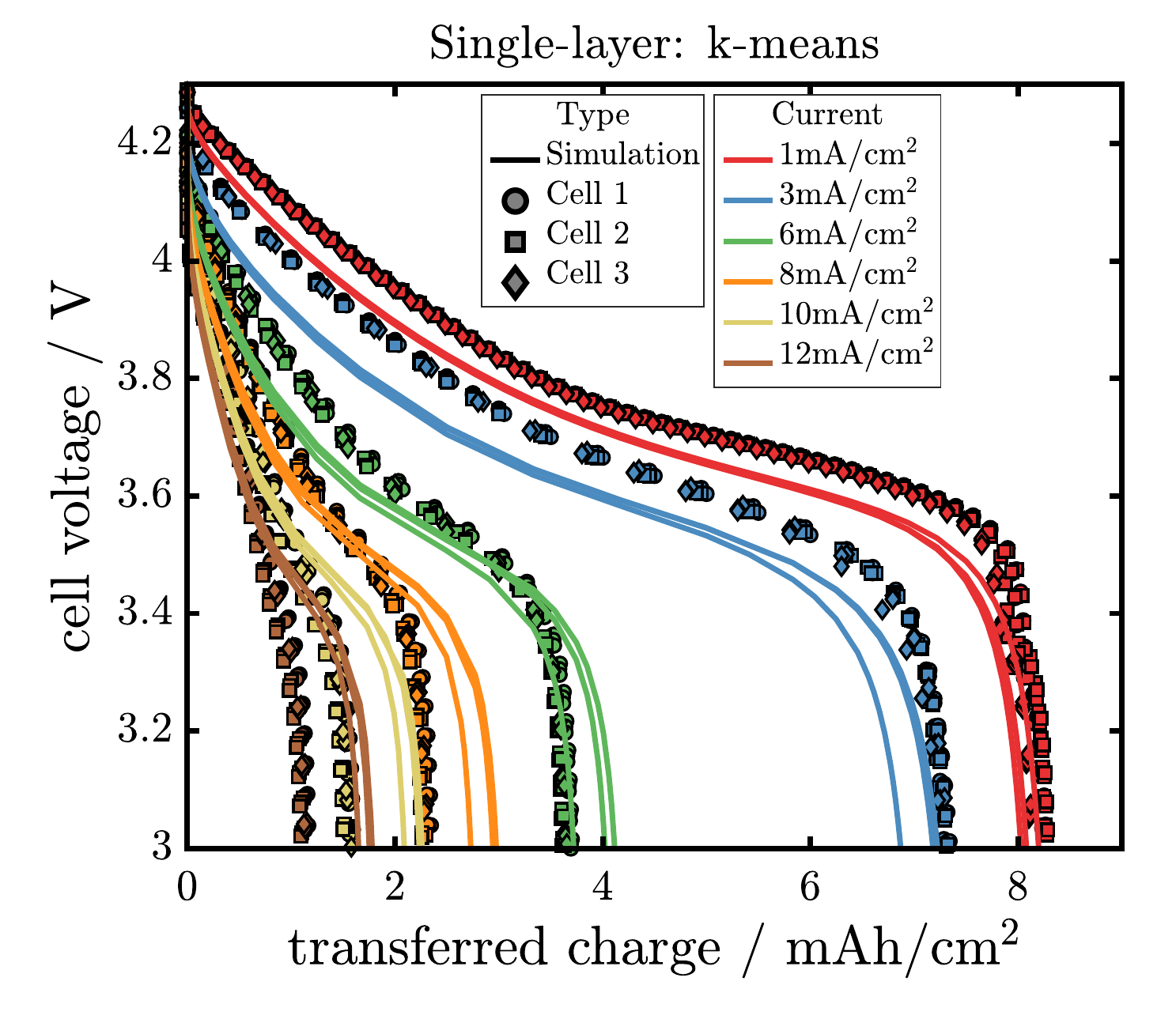}	
	\end{subfigure}
	\hfill
	\begin{subfigure}[c]{0.24\textwidth}
		\centering
		\includegraphics[width=0.9\textwidth]{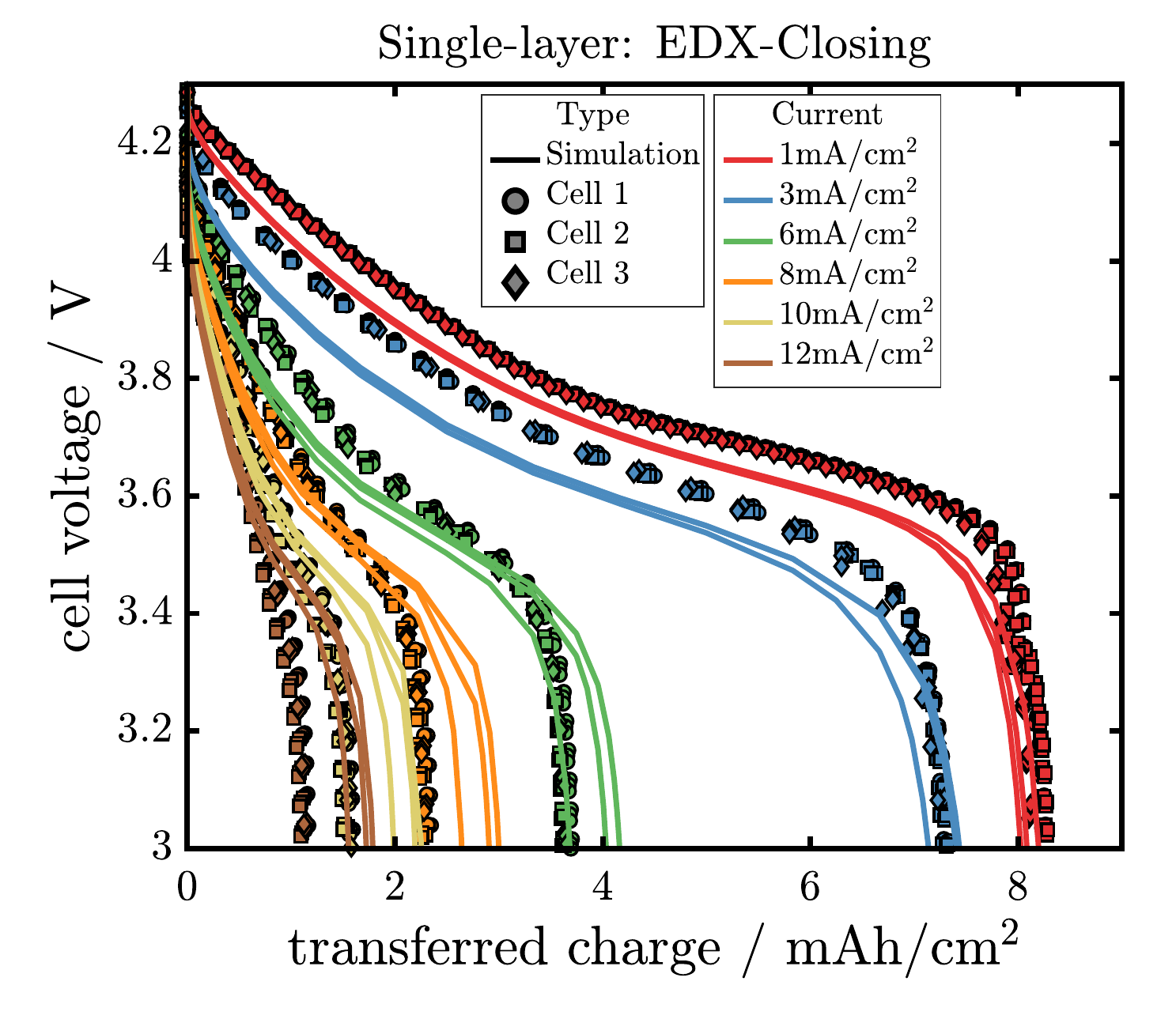}	
	\end{subfigure}
	\hfill
	\begin{subfigure}[c]{0.24\textwidth}
		\centering
		\includegraphics[width=0.9\textwidth]{imagesvorlaeufig/Lithiierung/TL_BestTau_1_KMEANS.pdf}	
	\end{subfigure}
	\hfill
	\begin{subfigure}[c]{0.24\textwidth}
		\centering
		\includegraphics[width=0.9\textwidth]{imagesvorlaeufig/Lithiierung/TL_BestTau_2_EDXClosing.pdf}
	\end{subfigure}
	\\
	\begin{subfigure}[c]{0.24\textwidth}
		\centering
		\includegraphics[width=0.9\textwidth]{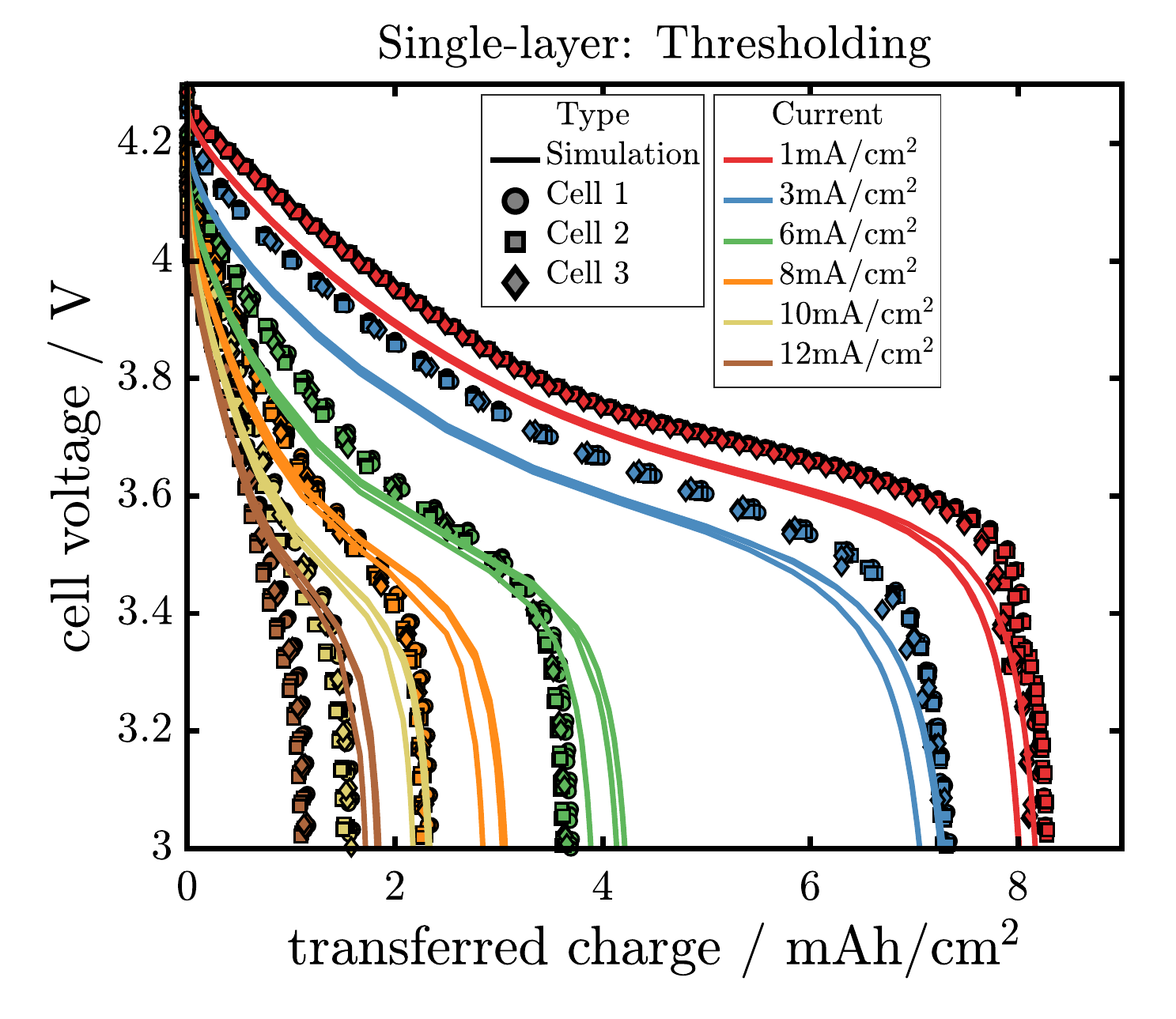}	
	\end{subfigure}
	\hfill
	\begin{subfigure}[c]{0.24\textwidth}
		\centering
		\includegraphics[width=0.9\textwidth]{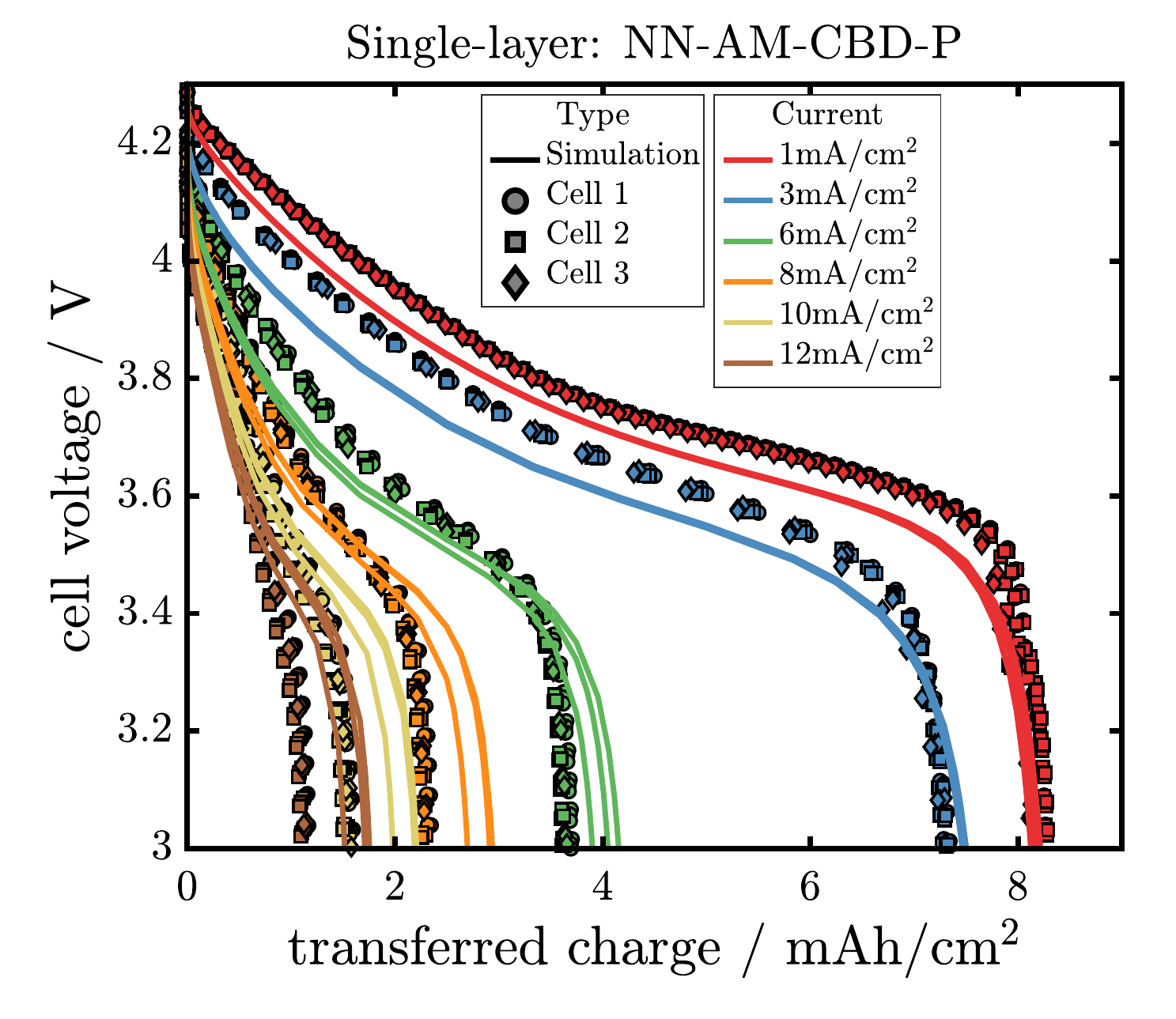}	
	\end{subfigure}
	\hfill
	\begin{subfigure}[c]{0.24\textwidth}
		\centering
		\includegraphics[width=0.9\textwidth]{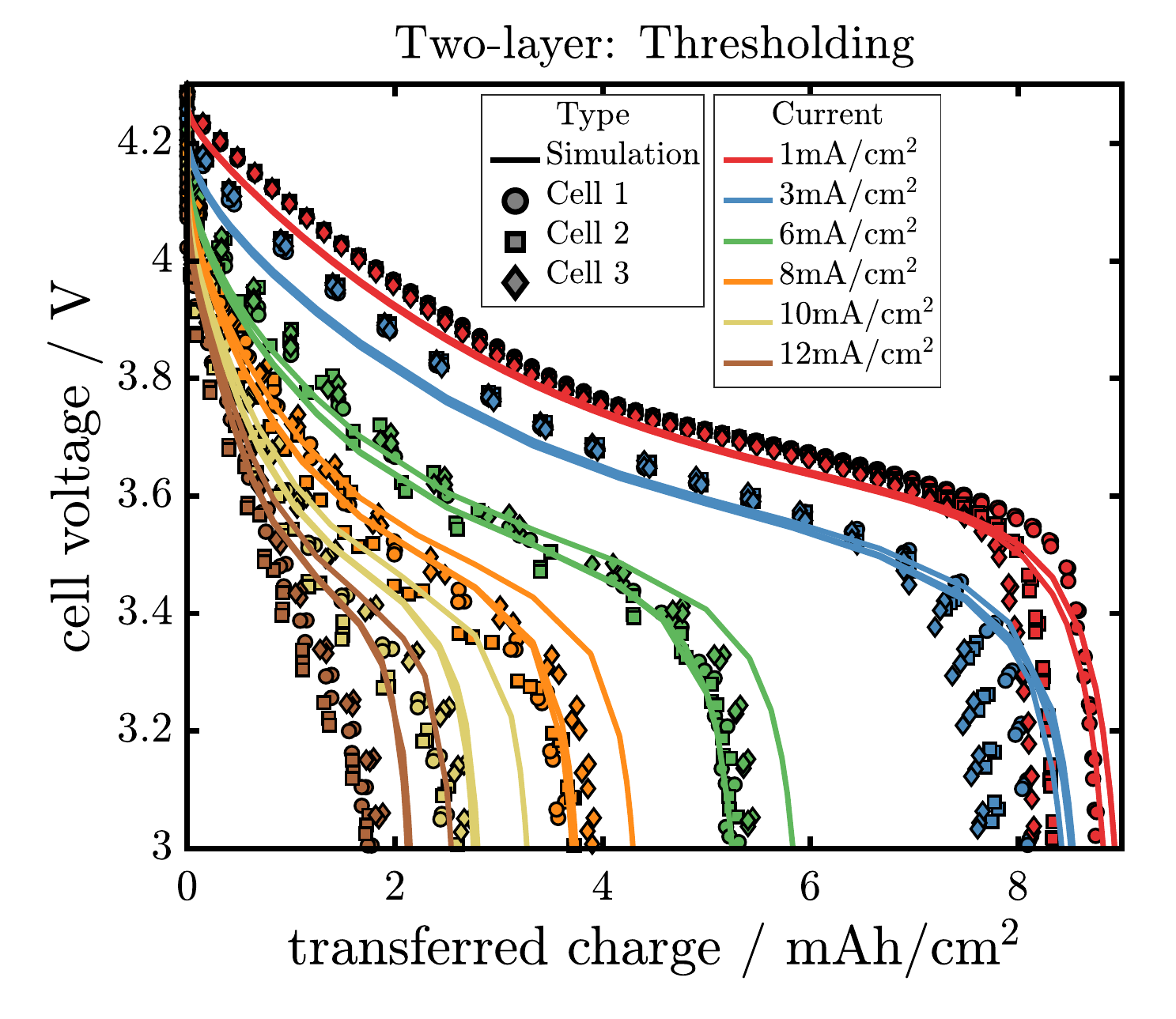}	
	\end{subfigure}
	\hfill
	\begin{subfigure}[c]{0.24\textwidth}
		\centering
		\includegraphics[width=0.9\textwidth]{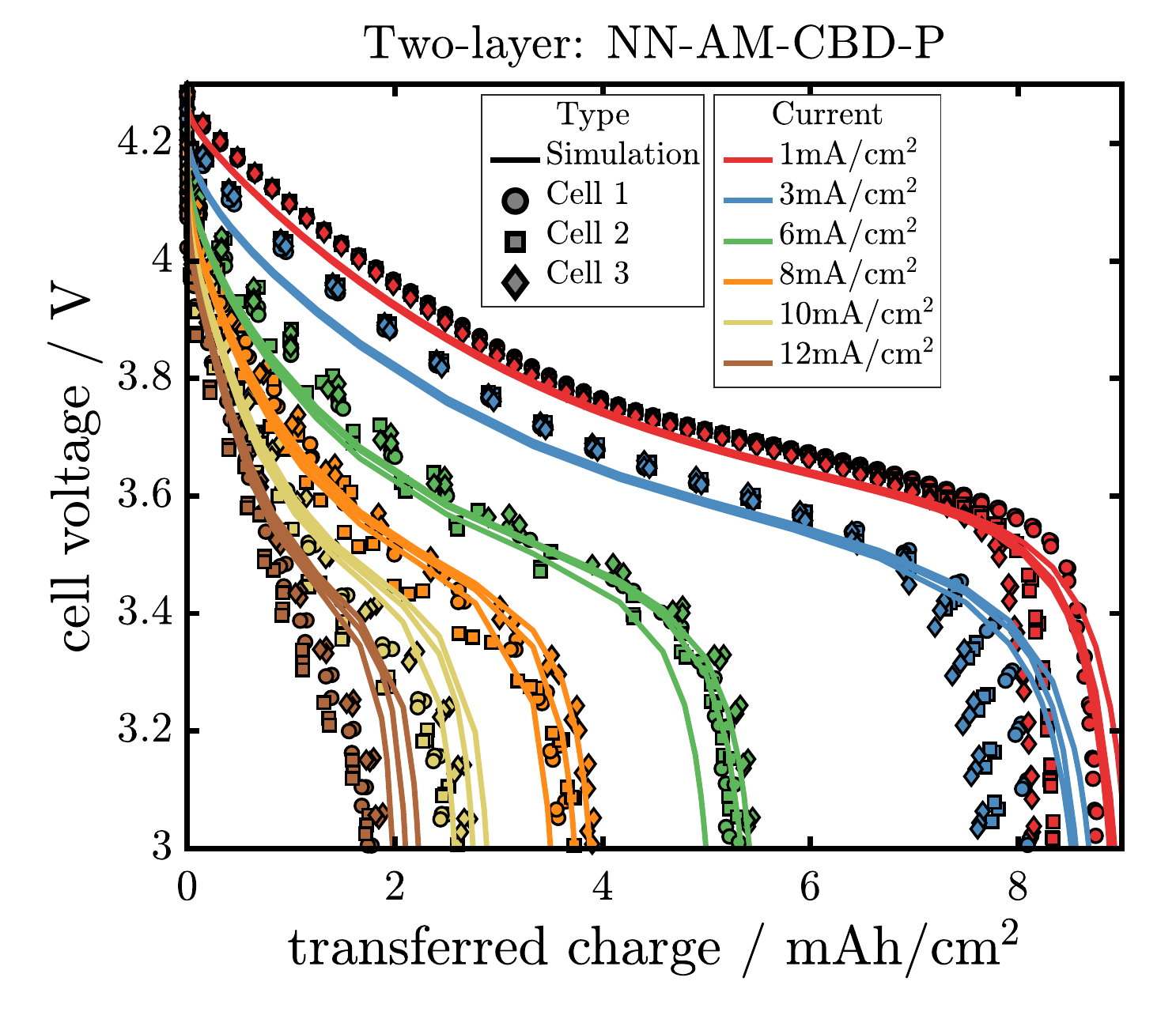}
	\end{subfigure}
	\\
	\begin{subfigure}[c]{0.24\textwidth}
		\centering
		\includegraphics[width=0.9\textwidth]{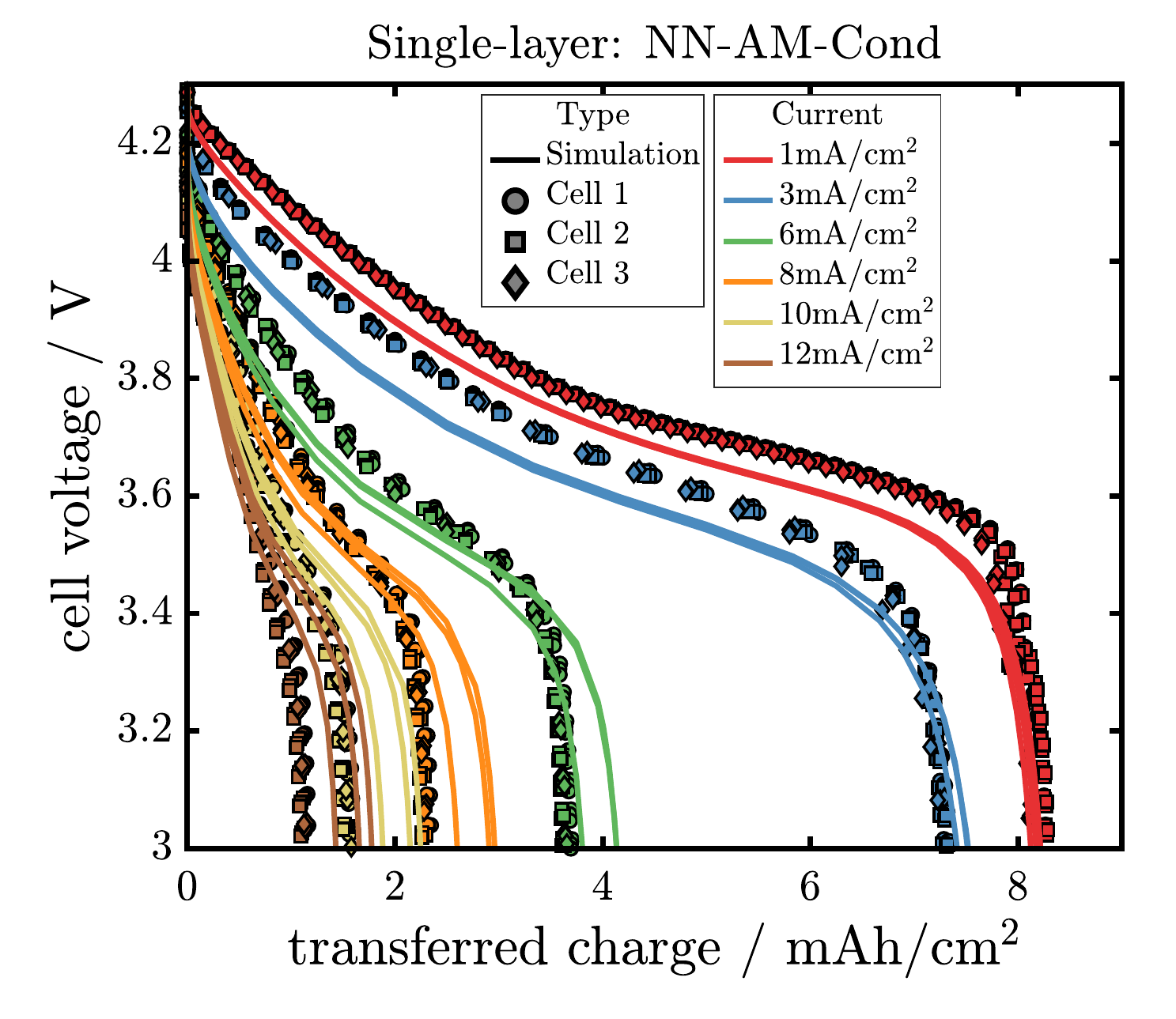}	
	\end{subfigure}
	\hfill
	\begin{subfigure}[c]{0.24\textwidth}
		\centering
		\includegraphics[width=0.9\textwidth]{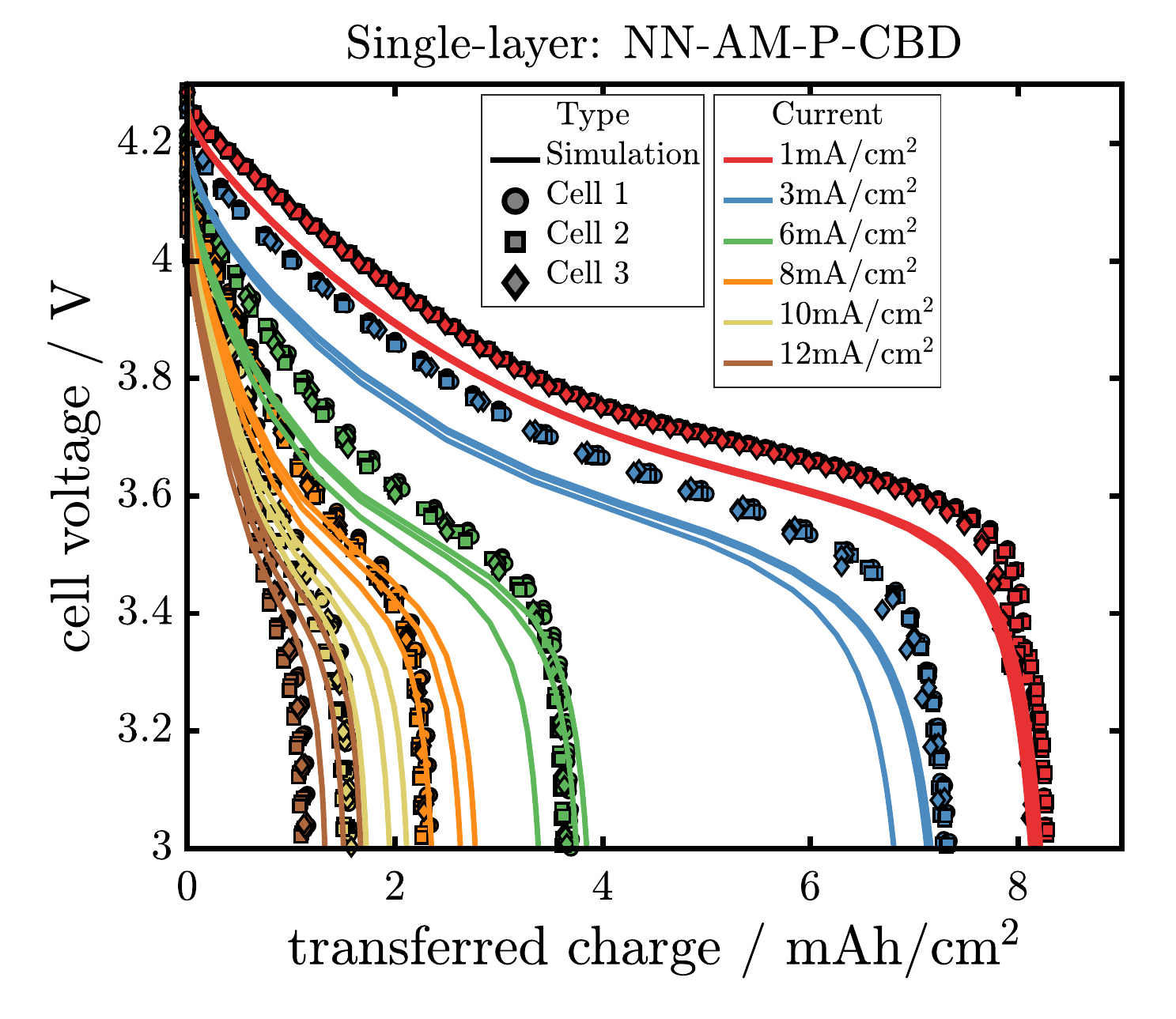}	
	\end{subfigure}
	\hfill
	\begin{subfigure}[c]{0.24\textwidth}
		\centering
		\includegraphics[width=0.9\textwidth]{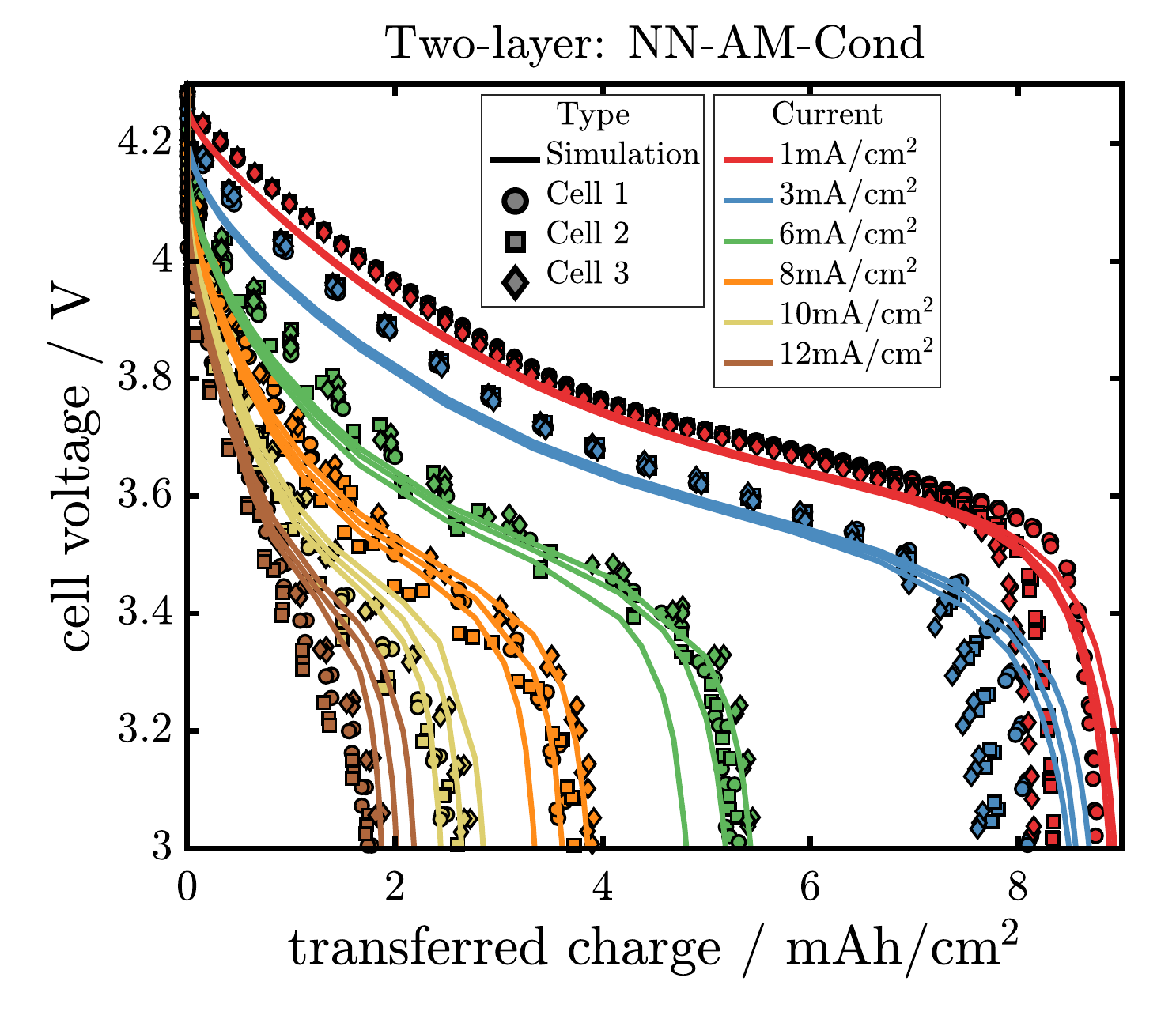}	
	\end{subfigure}
	\hfill
	\begin{subfigure}[c]{0.24\textwidth}
		\centering
		\includegraphics[width=0.9\textwidth]{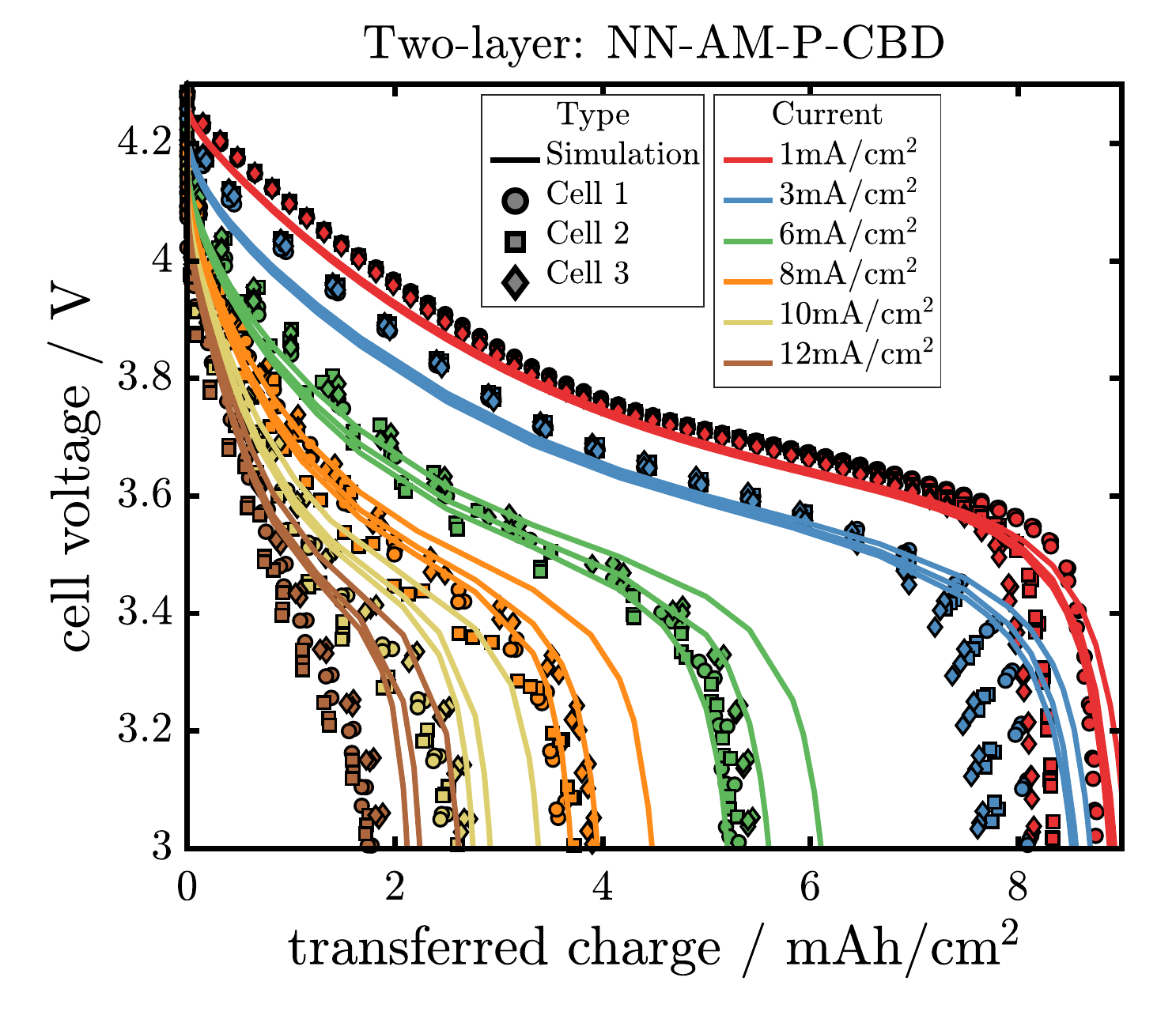}
	\end{subfigure}
	\\
	\begin{subfigure}[c]{0.24\textwidth}
		\centering
		\includegraphics[width=0.9\textwidth]{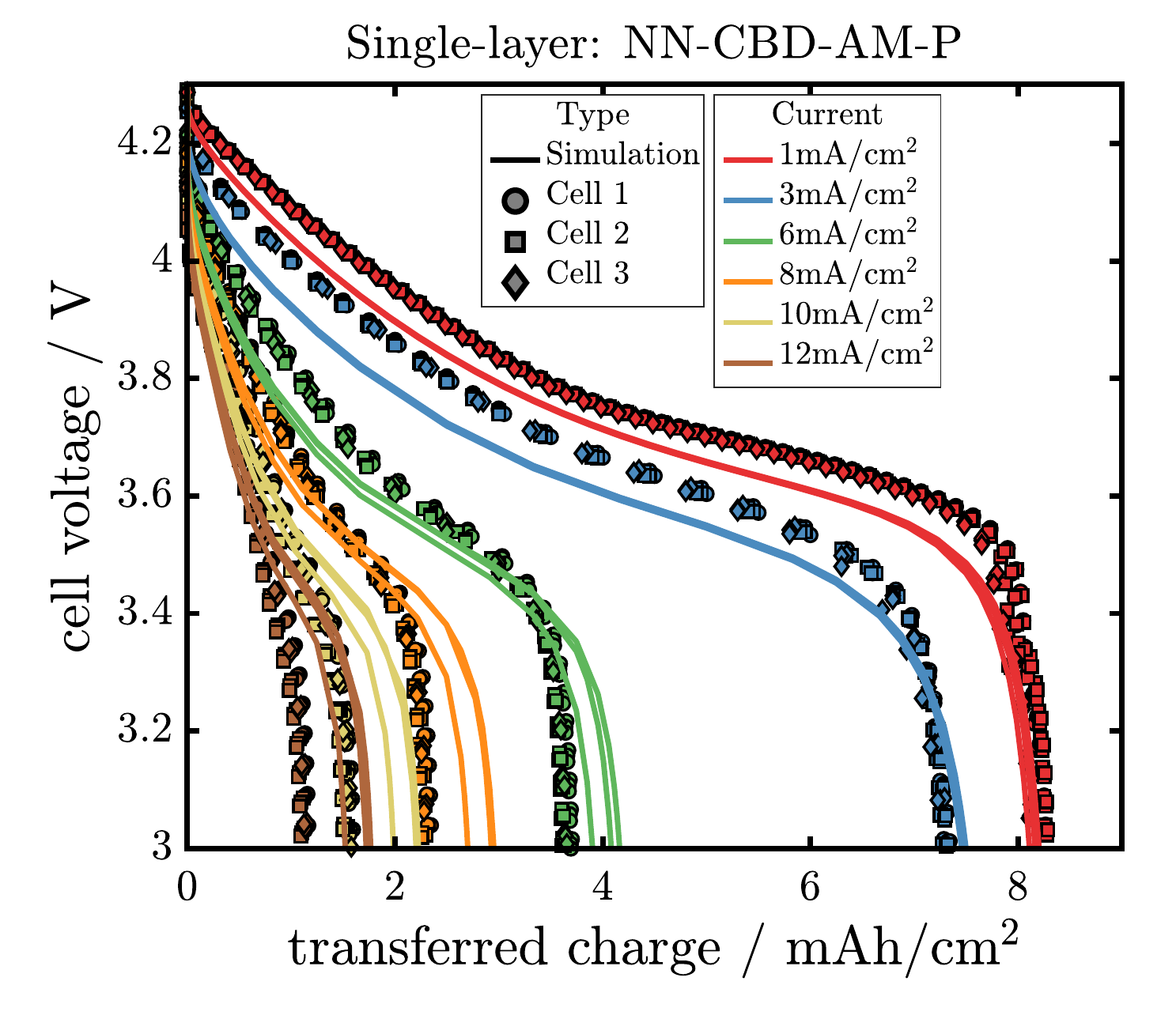}	
	\end{subfigure}
	\hfill
	\begin{subfigure}[c]{0.24\textwidth}
		\centering
		\includegraphics[width=0.9\textwidth]{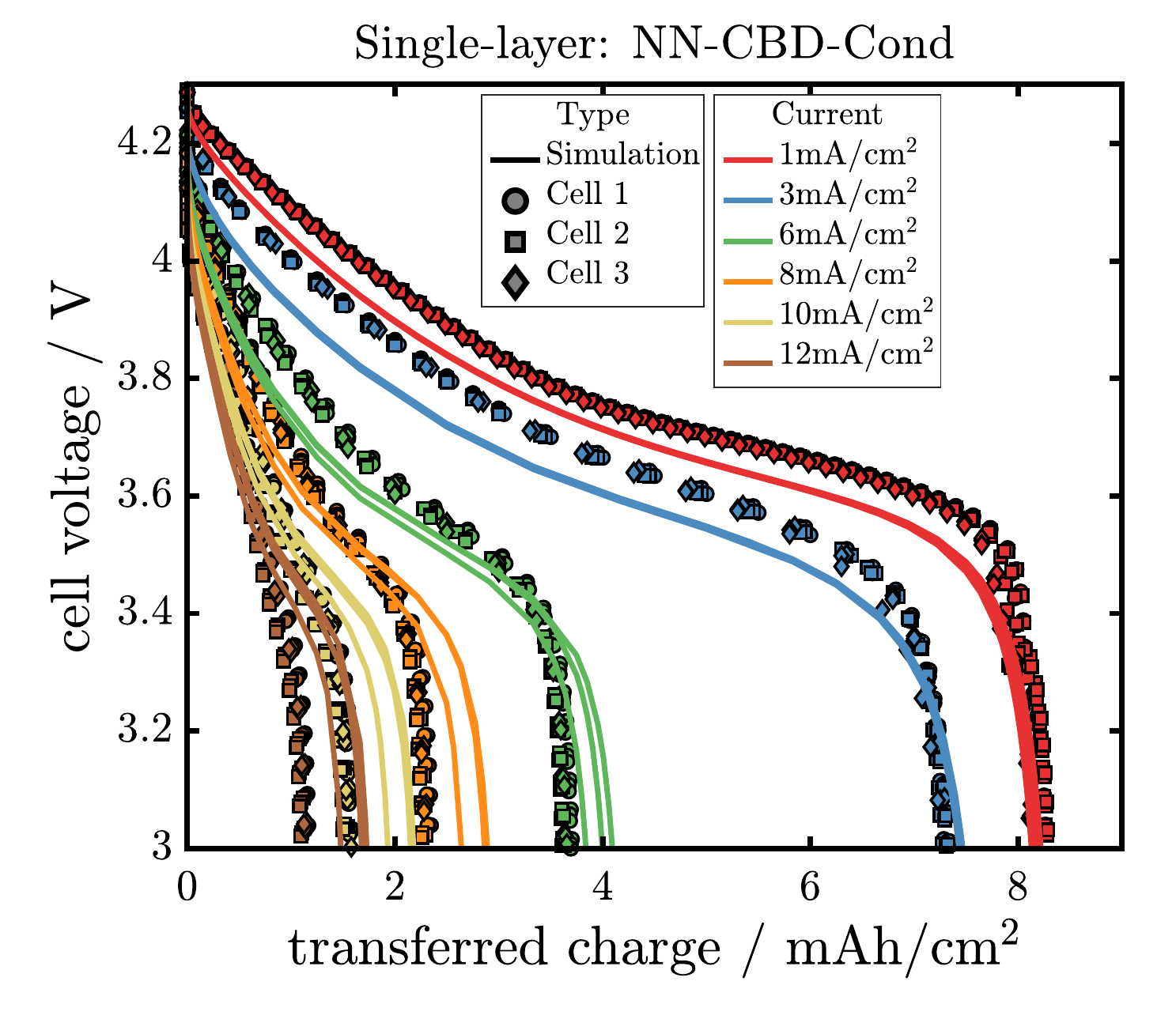}	
	\end{subfigure}
	\hfill
	\begin{subfigure}[c]{0.24\textwidth}
		\centering
		\includegraphics[width=0.9\textwidth]{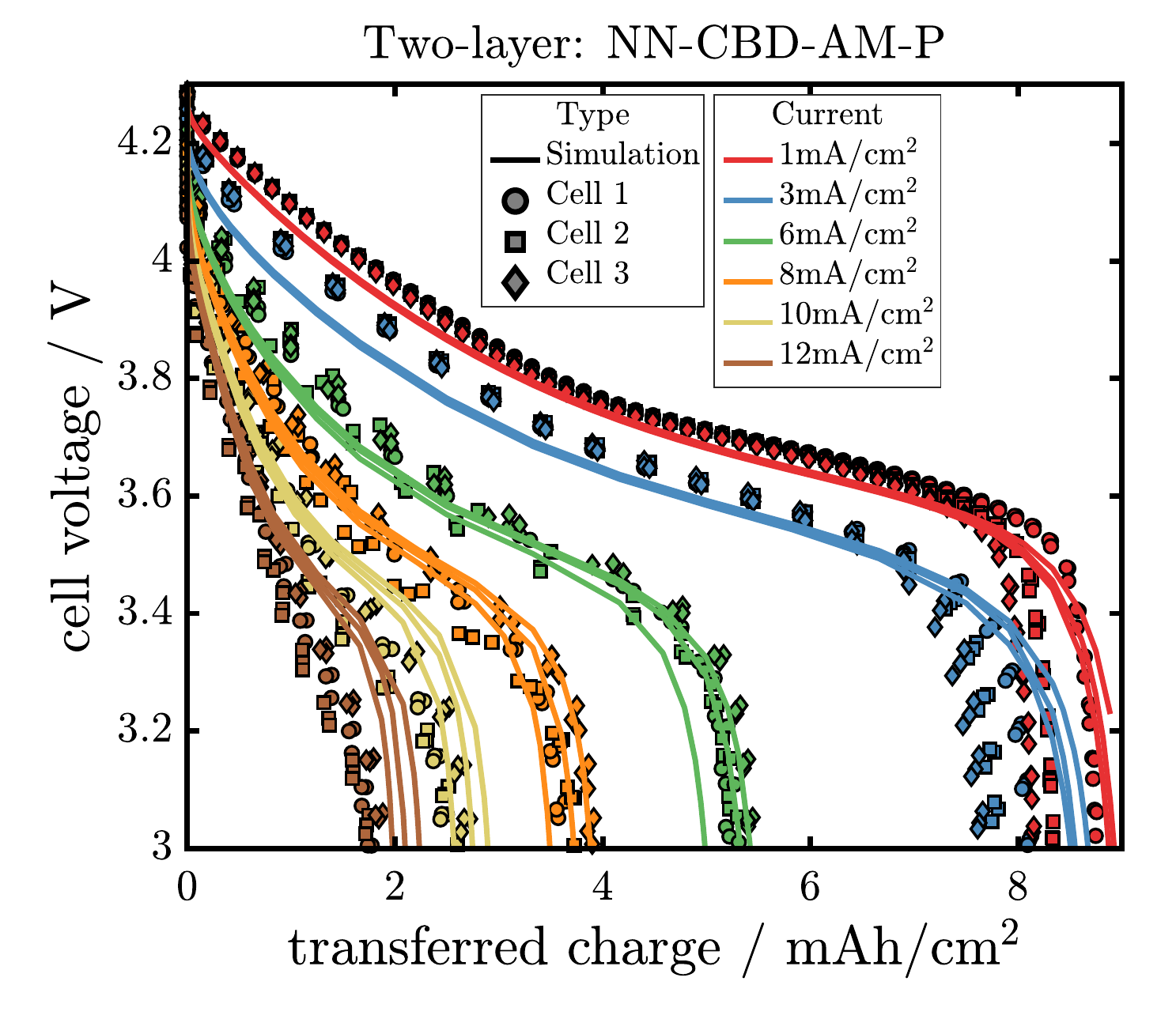}	
	\end{subfigure}
	\hfill
	\begin{subfigure}[c]{0.24\textwidth}
		\centering
		\includegraphics[width=0.9\textwidth]{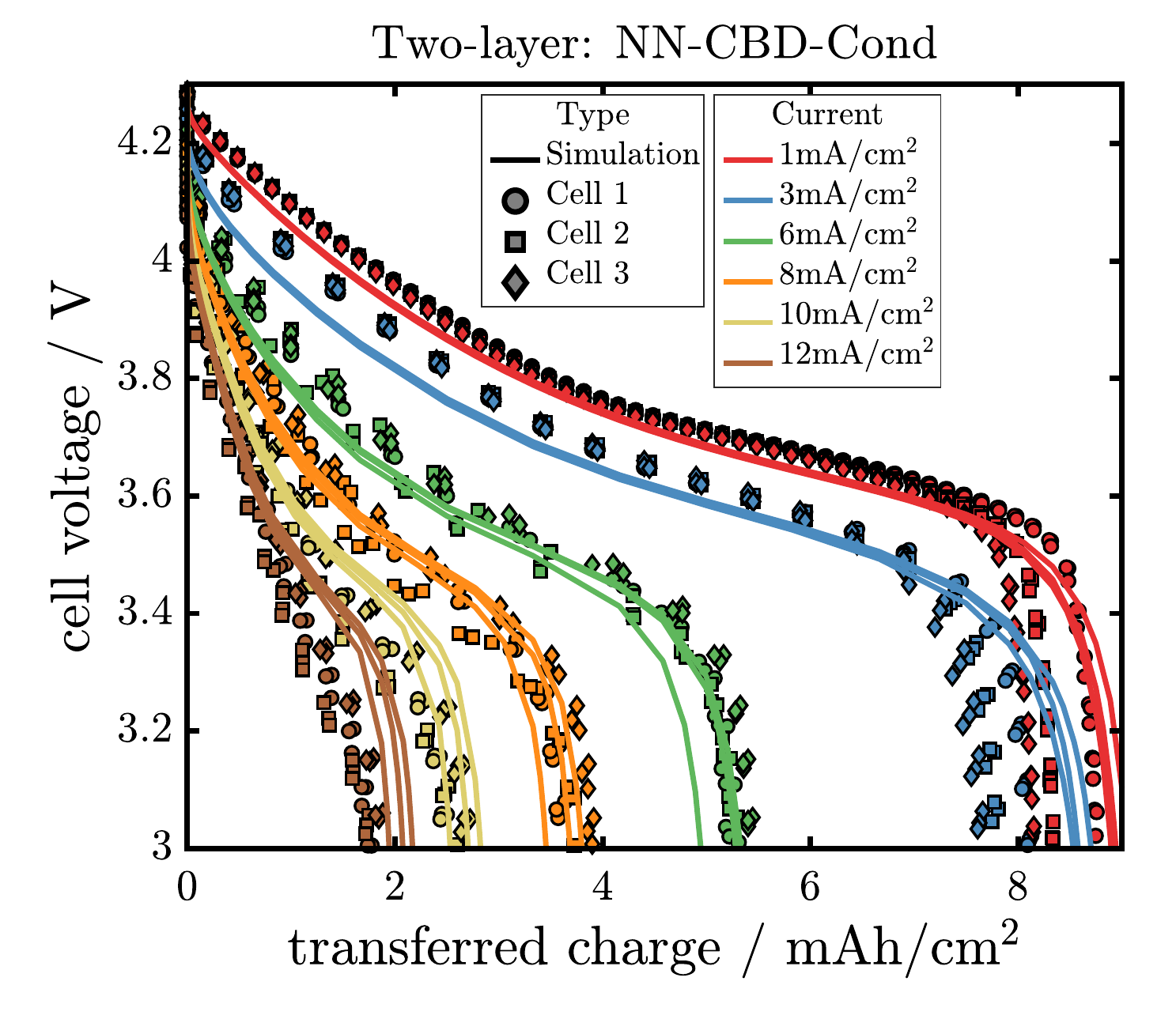}
	\end{subfigure}
	\\
	\begin{subfigure}[c]{0.24\textwidth}
		\centering
		\includegraphics[width=0.9\textwidth]{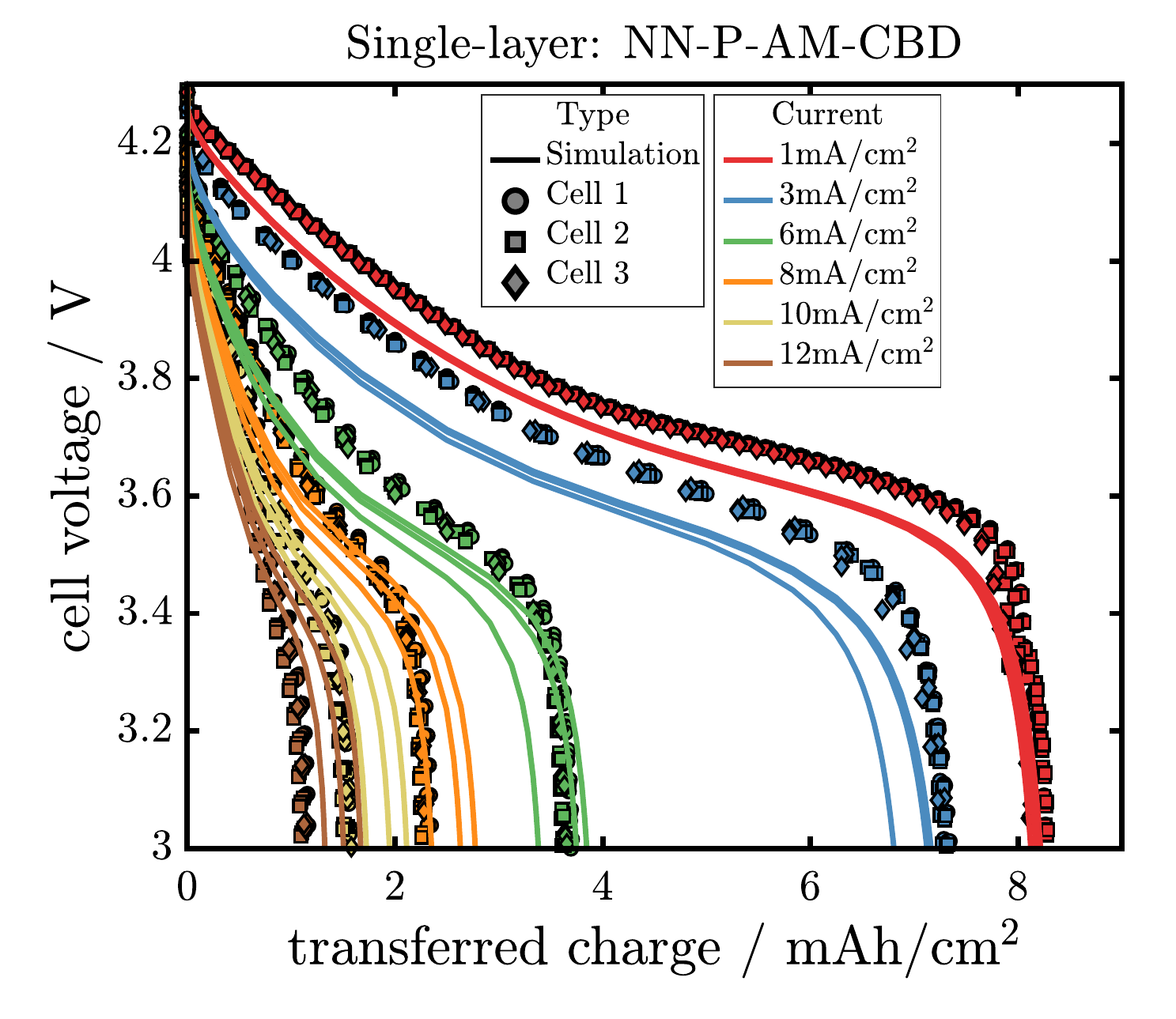}	
	\end{subfigure}
	\hfill
	\begin{subfigure}[c]{0.24\textwidth}
		\centering
		\includegraphics[width=0.9\textwidth]{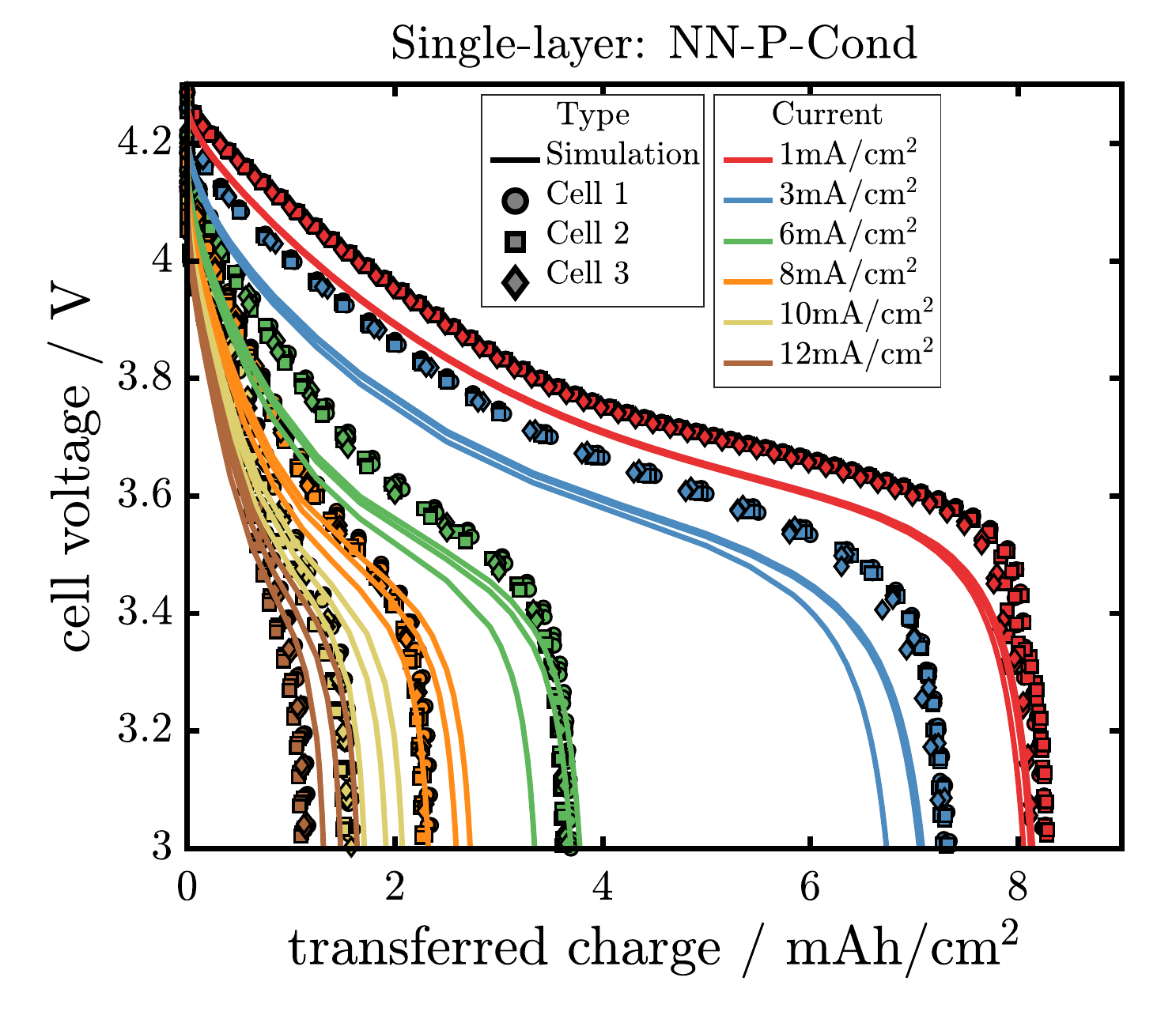}	
	\end{subfigure}
	\hfill
	\begin{subfigure}[c]{0.24\textwidth}
		\centering
		\includegraphics[width=0.9\textwidth]{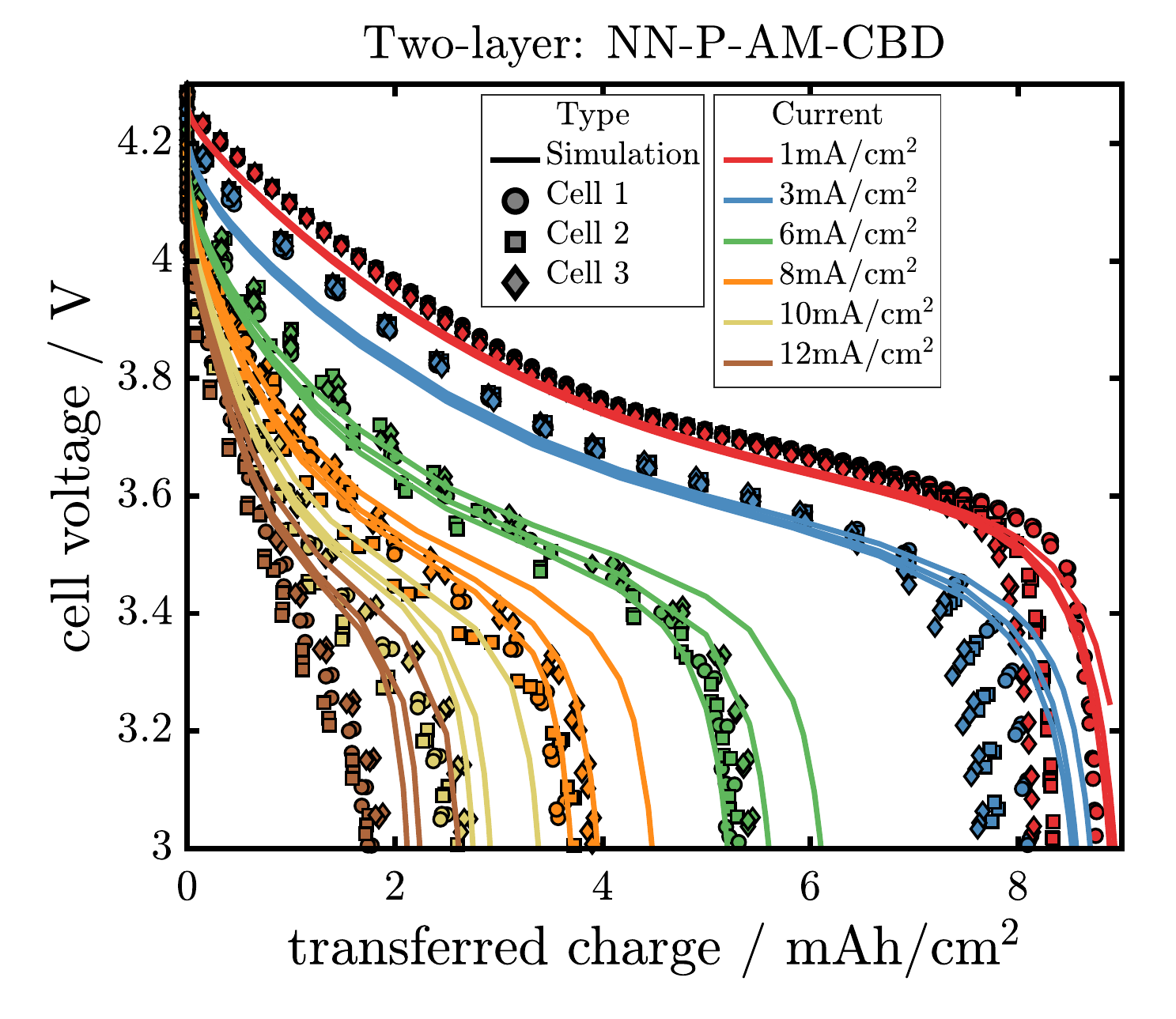}	
	\end{subfigure}
	\hfill
	\begin{subfigure}[c]{0.24\textwidth}
		\centering
		\includegraphics[width=0.9\textwidth]{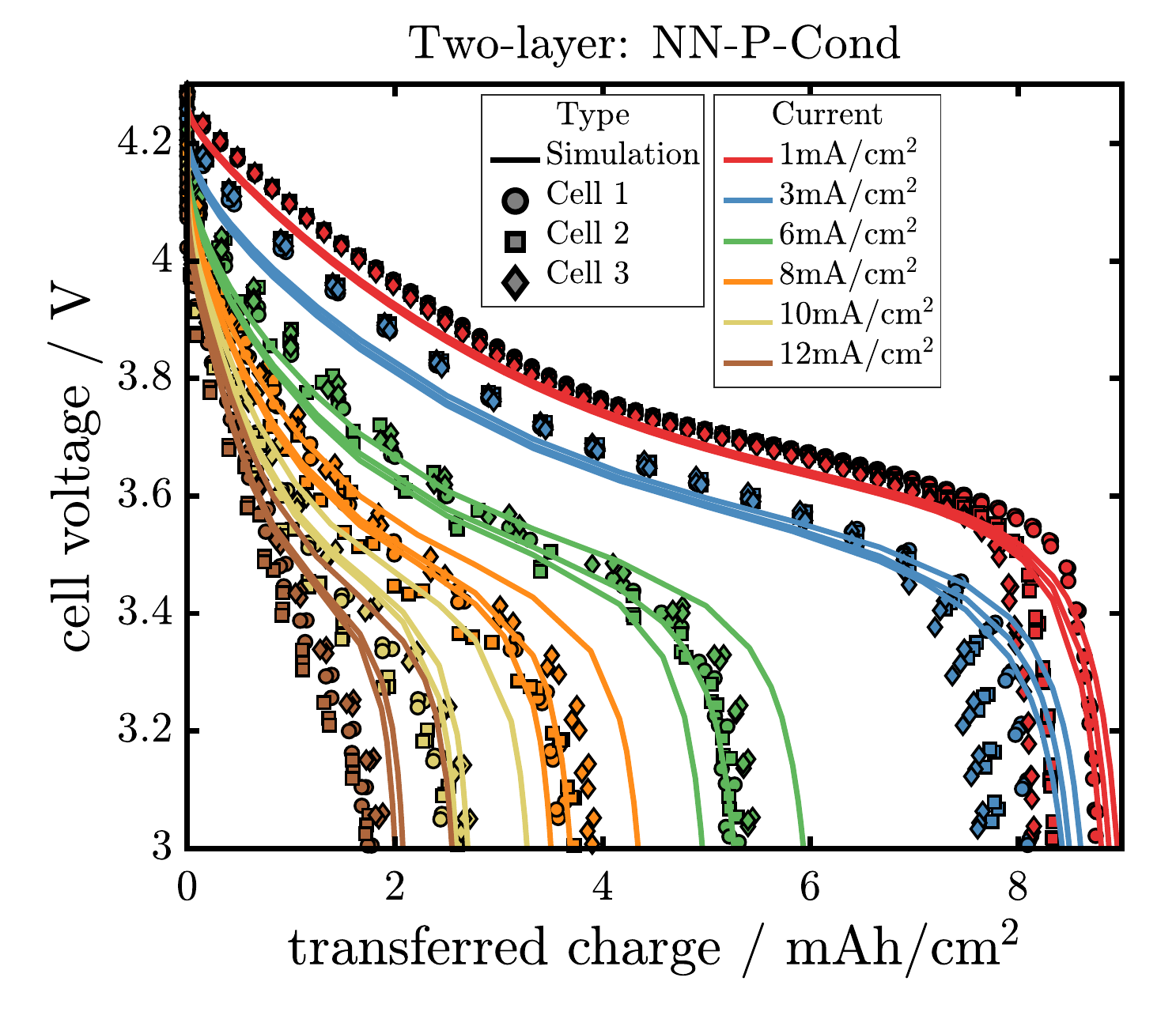}
	\end{subfigure}
	\caption{Best matching effective transport parameter $\tau_{\mathsf{CBD}}$ for the single-layer cathode (left columns) and two-layer cathode (right columns).}
	\label{SI:fig:EC:BestTau}
\end{figure}

\end{document}